\documentclass[11pt]{article}
\usepackage{amsmath,amsthm,latexsym,amssymb,amsfonts,epsfig}

\usepackage{amsfonts,amsmath,amssymb,graphics,psfrag}

\newcommand{\p}{\phi}
\newcommand{\te}{\theta}
\newcommand{\s}{\psi}

\makeatletter \@addtoreset{equation}{section} \makeatother

\def\be{\begin{equation}}
\def\ee{\end{equation}}
\def\ba{\begin{eqnarray}}
\def\ea{\end{eqnarray}}
\def\Nl{{\mathchoice
{\setbox0=\hbox{$\displaystyle\rm N$}\hbox{\hbox to0pt
{\kern0.4\wd0\vrule height0.9\ht0\hss}\box0}}
{\setbox0=\hbox{$\textstyle\rm N$}\hbox{\hbox to0pt
{\kern0.4\wd0\vrule height0.9\ht0\hss}\box0}}
{\setbox0=\hbox{$\scriptstyle\rm N$}\hbox{\hbox to0pt
{\kern0.4\wd0\vrule height0.9\ht0\hss}\box0}}
{\setbox0=\hbox{$\scriptscriptstyle\rm N$}\hbox{\hbox to0pt
{\kern0.4\wd0\vrule height0.9\ht0\hss}\box0}}}}
\def\Zl{{\mathchoice
{\setbox0=\hbox{$\displaystyle\rm Z$}\hbox{\hbox to0pt
{\kern0.4\wd0\vrule height0.9\ht0\hss}\box0}}
{\setbox0=\hbox{$\textstyle\rm Z$}\hbox{\hbox to0pt
{\kern0.4\wd0\vrule height0.9\ht0\hss}\box0}}
{\setbox0=\hbox{$\scriptstyle\rm Z$}\hbox{\hbox to0pt
{\kern0.4\wd0\vrule height0.9\ht0\hss}\box0}}
{\setbox0=\hbox{$\scriptscriptstyle\rm Z$}\hbox{\hbox to0pt
{\kern0.4\wd0\vrule height0.9\ht0\hss}\box0}}}}
\def\Ql{{\mathchoice
{\setbox0=\hbox{$\displaystyle\rm Q$}\hbox{\hbox to0pt
{\kern0.4\wd0\vrule height0.9\ht0\hss}\box0}}
{\setbox0=\hbox{$\textstyle\rm Q$}\hbox{\hbox to0pt
{\kern0.4\wd0\vrule height0.9\ht0\hss}\box0}}
{\setbox0=\hbox{$\scriptstyle\rm Q$}\hbox{\hbox to0pt
{\kern0.4\wd0\vrule height0.9\ht0\hss}\box0}}
{\setbox0=\hbox{$\scriptscriptstyle\rm Q$}\hbox{\hbox to0pt
{\kern0.4\wd0\vrule height0.9\ht0\hss}\box0}}}}
\def\Rl{{\mathchoice
{\setbox0=\hbox{$\displaystyle\rm R$}\hbox{\hbox to0pt
{\kern0.4\wd0\vrule height0.9\ht0\hss}\box0}}
{\setbox0=\hbox{$\textstyle\rm R$}\hbox{\hbox to0pt
{\kern0.4\wd0\vrule height0.9\ht0\hss}\box0}}
{\setbox0=\hbox{$\scriptstyle\rm R$}\hbox{\hbox to0pt
{\kern0.4\wd0\vrule height0.9\ht0\hss}\box0}}
{\setbox0=\hbox{$\scriptscriptstyle\rm R$}\hbox{\hbox to0pt
{\kern0.4\wd0\vrule height0.9\ht0\hss}\box0}}}}
\def\Cl{{\mathchoice
{\setbox0=\hbox{$\displaystyle\rm C$}\hbox{\hbox to0pt
{\kern0.4\wd0\vrule height0.9\ht0\hss}\box0}}
{\setbox0=\hbox{$\textstyle\rm C$}\hbox{\hbox to0pt
{\kern0.4\wd0\vrule height0.9\ht0\hss}\box0}}
{\setbox0=\hbox{$\scriptstyle\rm C$}\hbox{\hbox to0pt
{\kern0.4\wd0\vrule height0.9\ht0\hss}\box0}}
{\setbox0=\hbox{$\scriptscriptstyle\rm C$}\hbox{\hbox to0pt
{\kern0.4\wd0\vrule height0.9\ht0\hss}\box0}}}}
\def\Hl{{\mathchoice
{\setbox0=\hbox{$\displaystyle\rm H$}\hbox{\hbox to0pt
{\kern0.4\wd0\vrule height0.9\ht0\hss}\box0}}
{\setbox0=\hbox{$\textstyle\rm H$}\hbox{\hbox to0pt
{\kern0.4\wd0\vrule height0.9\ht0\hss}\box0}}
{\setbox0=\hbox{$\scriptstyle\rm H$}\hbox{\hbox to0pt
{\kern0.4\wd0\vrule height0.9\ht0\hss}\box0}}
{\setbox0=\hbox{$\scriptscriptstyle\rm H$}\hbox{\hbox to0pt
{\kern0.4\wd0\vrule height0.9\ht0\hss}\box0}}}}
\def\Ol{{\mathchoice
{\setbox0=\hbox{$\displaystyle\rm O$}\hbox{\hbox to0pt
{\kern0.4\wd0\vrule height0.9\ht0\hss}\box0}}
{\setbox0=\hbox{$\textstyle\rm O$}\hbox{\hbox to0pt
{\kern0.4\wd0\vrule height0.9\ht0\hss}\box0}}
{\setbox0=\hbox{$\scriptstyle\rm O$}\hbox{\hbox to0pt
{\kern0.4\wd0\vrule height0.9\ht0\hss}\box0}}
{\setbox0=\hbox{$\scriptscriptstyle\rm O$}\hbox{\hbox to0pt
{\kern0.4\wd0\vrule height0.9\ht0\hss}\box0}}}}

\title{{\sf Semiclassical analysis}\\
{\sf of the}\\
{\sf Loop Quantum Gravity volume operator:} \\
{\sf Area Coherent States}
}

\author{{\sf C. Flori$^1$ }\thanks{{\sf cecilia.flori@aei.mpg.de}, {\sf ceciliaflori18@googlemail.com}},
\\
\\
{\sf $^1$ MPI f. Gravitationsphysik, Albert-Einstein-Institut,} \\
{\sf Am M\"uhlenberg 1, 14476 Potsdam, Germany}\\
}
\date{{\small\sf Preprint AEI-2009-026}}

\begin{document}

\maketitle

\begin{abstract}
We continue the semiclassical analysis of the Loop Quantum Gravity
(LQG) volume operator that was started in the companion paper
\cite{30b}. In the first paper we prepared the technical tools, in
particular the use of complexifier coherent states that use
squares of flux operators as the complexifier. In this paper, the
complexifier is chosen for the first time to involve squares of
area operators.

Both cases use coherent states that depend on a graph. However,
the basic difference between the two choices of complexifier is
that in the first case the set of surfaces involved is discrete,
while, in the second it is continuous. This raises the important
question of  whether the second set of states has improved
invariance properties with respect to relative orientation of the
chosen graph in the set of surfaces on which the complexifier
depends. In this paper,  we examine this question in detail,
including a semiclassical analysis.

The main result is that we obtain the correct semiclassical
properties of the volume operator for i) artificial rescaling of
the coherent state label; and ii) particular orientations of the
4- and 6-valent graphs that have measure zero in the group SO(3).
Since such requirements are not present when analysing dual cell
complex states, we conclude that coherent states whose
complexifiers are squares of area operators are not an appropriate
tool with which to analyse the semiclassical properties of the
volume operator. Moreover, if one intends to go further and sample
over graphs in order to obtain embedding independence, then the
area complexifier coherent states should be ruled out altogether
as semiclassical states.
%
%
\end{abstract}
 \vspace{.5in}
 \begin{center}\emph{I would like to dedicate this paper to my parents Elena Romani and Luciano Flori}\end{center}
 \newpage

\newpage

\tableofcontents

\newpage

\section{Introduction}
\label{s1}

The volume operator in Loop Quantum Gravity\footnote{See
\cite{1a,1b} for recent books and \cite{2} for recent reviews.}
(LQG) plays a central role in the quantum dynamics of the theory.
Without it, it is not possible to define the Hamiltonian
constraint operators \cite{3}, the Master constraint operators
\cite{4}, or the physical Hamiltonian operators \cite{5}. The same
applies to truncated models of LQG such as Loop Quantum Cosmology
(LQC) \cite{6} which is believed to describe well the homogeneous
sector of the theory. In LQC  the quantum evolution of operators
that correspond to classically-singular Dirac observables
typically remains finite. This  can be traced back to the
techniques used to quantize inverse powers of the local volume
that enter into the expression for the triad, co-triad and other
types of coupling between geometry and geometry, or geometry and
matter \cite{3}. Specifically, such derived operators arise as
commutators between holonomy operators and powers of the volume
operator.

In view of its importance, it is of considerable interest to
verify whether the classical limit of the LQG volume operator
coincides with the classical volume. By this we mean that (i) the
expectation value of the volume operator with respect to suitable
semiclassical states which are peaked at a given point in phase
space, coincides with the value of the classical volume at that
phase space point, up to small corrections; and (ii) its
fluctuations are small.

It should be remarked that there are actually two versions of the
volume operator that have been discussed in the literature
\cite{7,8}. These come from inequivalent regularisations of the
products of operator-valued distributions that appear at
intermediate stages. However, only the operator in \cite{8}
survives the consistency test of \cite{9}, namely, writing the
volume in terms of triads which then are quantised using the
commutator mentioned above, gives the same operator up to
$\hbar$-corrections as that obtained from direct quantisation.
This consistency check is important as otherwise we could not
trust the triad and co-triad quantisations that enter the quantum
dynamics.

A semiclassical analysis of the volume operator has not yet been
carried out, although, in principle, suitable semiclassical (even
coherent) states for LQG are available \cite{10}. This is because
the spectral decomposition (projection-valued measure) of the
volume operator cannot be computed exactly in closed form.
However, this is needed for exact, practical calculations. More
precisely, the volume operator is the fourth root of a positive
operator, $Q$, whose matrix elements can be computed in closed
form \cite{11} but which cannot be diagonalised analytically.

In more detail, the volume operator has a discrete (that is,
pure-point) spectrum and it attains a block-diagonal form where
the blocks are labelled by the graphs and spin quantum numbers
(labelling the edges of the graph) of spin-network functions
(SNWF) \cite{12}. SNWF form a convenient basis of the LQG Hilbert
space \cite{13} which carries a unitary representation of the
spatial diffeomorphism group and the Poisson$^\ast-$algebra of the
elementary flux and holonomy variables \cite{14}. The blocks turn
out to be finite-dimensional matrices whose matrix elements can be
expressed in terms of polynomials of 6j symbols. Fortunately,
these complicated quantities can be avoided\footnote{Notice that
Racah's formula provides a closed expression for the 6j symbol but
it also involves implicit sums and factorials that involve large
integers. These quickly becomes unmanageable, even numerically.}
by using a telescopic summation technique \cite{15} related to the
Elliot-Biedenharn identity \cite{16}, so that a manageable closed
expression results. However, the size of these matrices grows
exponentially with increasing spin quantum numbers and, since the
expression for coherent states is a coherent superposition of
SNWF's with arbitrarily large spin quantum numbers, a numerical
computation of the expectation value using the numerical
diagonalisation techniques, that are currently being developed
\cite{17}, is still a long way off.\footnote{The coherent
superposition contains a damping factor that suppresses large
spins, and thereby large matrices, so that the complicated
infinite series over spin quantum numbers can be truncated at
finite values, making only negligible errors. However, the
computational effort required is currently too high, even for
supercomputers; for example, see the estimates of computation time
reported in \cite{17}.}

One way forward is to use the semiclassical perturbation theory
developed in \cite{18}, and applied already in \cite{19,20}. The
basic idea is quite simple. In practical calculations one needs
the expectation value of $Q^q$ where $q$ is a rational number in
the range $0<q\le \frac{1}{4}$. In order to attain that, one has
to introduce the `perturbation operator', $X:=\frac{Q}{<Q>}-1$,
where the expectation value $<Q>$ of the positive operator $Q$ is
exactly computable. Notice that $X$ is bounded from below by $-1$.
Then trivially $<Q^q>=<Q>^q\;<[1+X]^q>$. Now we exploit the
existence of positive numbers $p$ such that the classical
estimates $1+qx-px^2\le (1+x)^q\le 1+q x$ are valid for all $x\ge
-1$. Finer estimates of this form involving arbitrary powers of
$x$ are also available \cite{19}. By virtue of the spectral
theorem, this classical estimate survives at the quantum level and
we have $Y_-\le Y\le Y_+$ where $Y_+=1+qX,\;Y_-=Y_+ -p X^2,\;
Y=(1+X)^q$. It follows that $<Y>\in [<Y_+>-p<X^2>,<Y_+>]$.
However, $<Y_+>=1$ and $<X^2>=[<Q^2>-<Q>^2]/<Q>^2$ is proportional
to the relative fluctuations of $Q$, which are of order of $\hbar$
\cite{10}. It follows that to zeroth-order in $\hbar$ we may
replace $<Q^q>$ by $<Q>^q$ which is computable, as are the  error
estimates in the way shown above.

The exact computation of $<Q>,\;<Q^2>,\;..$ is feasible, but it is
still quite involved. At this point it is convenient to use the
fact that (see \cite{10}), to zeroth-order in $\hbar$,  the
computation of these expectation values (more generally, the
expectation values of low-order polynomials in the flux operators)
for $SU(2)$ spin-network states coincides with the corresponding
calculations for $U(1)^3$ spin networks. On the latter, the volume
operator is even diagonal. Hence we conclude that, as long as we
are only interested in the zeroth-order in $\hbar$ contribution,
we may evaluate the expectation value of the volume operator for a
fictitious theory in which the non-Abelian group $SU(2)$ is
replaced by the Abelian group $U(1)^3$. This dramatically
simplifies all the calculations.

The coherent states developed so far for LQG have all been
constructed using the complexifier method reviewed in \cite{21}.
This  generalises the coherent-state construction for phase spaces
that are cotangent bundles over a compact group; see
\cite{22}\footnote{See also \cite{23} for related ideas valid for
Abelian gauge theories such as Maxwell theory or linearised
gravity.}. This construction involves computing the heat-kernel
evolution of the delta distribution (which is the matrix element
of the unit operator in the Schr\"odinger (position)
representation) with respect to a generalized Laplace operator,
called the `complexifier', followed by a certain analytic
continuation.

Now the unit operator in LQG can be written as a resolution of
unity in terms of SNWF's and, although the heat kernel is a
damping operator, since the SNWF are not countable (the LQG
Hilbert space is not separable), the resulting expression is not
normalisable. However, it does give a well-defined distribution
(in the algebraic dual of the finite linear span of SNWF's) which
can be conveniently written as a sum over cut--off states labelled
by finite graphs. These states, called ``shadows'' in \cite{22},
{\it are} normalisable, and can be used to perform semiclassical
calculations. Of course, one expects that the  good semiclassical
states will only be those cut--off states that are labelled by
graphs which are sufficiently fine with respect to the classical
three-metric to be approximated.

Thus the input in the semiclassical calculations consists in the
choice of a complexifier and the choice of a graph. One may wonder
why the complexifier has to be chosen \emph{a priori} rather than
being dictated by the dynamics of the theory: that is, why do the
coherent states remain coherent under quantum time evolution? For
example, for the harmonic oscillator, the complexifier is just the
Laplacian on the real line.

The reason is two-fold. On the one hand, one could perform a
constraint quantisation so that we are working at the level of the
kinematical Hilbert space on which the quantum constraints have
not yet been imposed. In this situation, all that is needed is to
ensure that the volume operator, which is not gauge invariant (and
thus does not preserve the physical Hilbert space), has a good
classical limit on the kinematical Hilbert space. We would
certainly not trust a constraint quantization for which even that
was not true.

On the other hand, one could also work at the level of the
physical Hilbert space, as outlined in \cite{5}. Then, one would
expect that the time evolution of any reasonable choice of
complexifier coherent states with respect to the physical
Hamiltonian should remain coherent (and peaked on the classical
trajectory) for sufficiently short time intervals. Note that for
interacting theories, even the simple example of the anharmonic
oscillator, globally stable coherent states have yet to be found.

The choice of the complexifier will be guided by practical
considerations: namely (i) it is diagonal on SNWF's; and (ii) it
is a damping operator that renders normalisable the heat-kernel
evolution of the delta distribution restricted to a graph.
Moreover, it should be gauge invariant under the $SU(2)$ gauge
group. As for the choice of graph, for practical reasons one will
use graphs that are topologically regular, that is, have a
constant valence for each vertex. Indeed, the semiclassical
calculations performed in \cite{19,20} were done using graphs of
cubic topology, with good results.

The existing literature on such coherent states for LQG can be
divided into two classes, depending on a certain structure that
defines them. The first  are {\it gauge-covariant flux} coherent
states, which depend on collections of surfaces and path systems
inside them \cite{10}. The second are {\it area} coherent states,
which also depend on collections of surfaces but involve area
operators rather than flux operators \cite{21}. In our companion
paper \cite{30b} we have reviewed and generalised both
constructions. In its most studied incarnations, the collection of
surfaces involved in \cite{10} is defined by a polyhedronal
partition of the spatial manifold, while the collection used in
\cite{21} uses a plaquette of foliations of the spatial manifold.

The main objective of this series of two papers is to analyse the
semiclassical properties of the volume operator with respect to
both classes of coherent state. In \cite{30b} we employed the
states of \cite{10}.  In this paper we employ the states of
\cite{21}. This exhausts all known coherent states for LQG. The
result of our analysis is that, if we use the states \cite{10},
{\it the correct semiclassical limit is attained with these states
for $n=6$ only}. If instead we use the states \cite{21}, the
correct semiclassical limit is attained \emph{only} for: 1)
artificial rescaling of the coherent state label; and 2)
particular embeddings of the 4-valent and 6-valent graphs with
respect to the set of surfaces on which the complexifier depends.
However, the combinations of Euler angles for which such
embeddings are attained have measure zero in SO(3), and are
therefore negligible. Thus the {\it area complexifier coherent
states are not the correct tools with which to analyse the
semiclassical properties of the volume operator}.

If one wants to obtain embedding independence, a possible strategy
is to sample over graphs (Dirichlet-Voronoi sampling \cite{40}) as
outlined in \cite{10}. What this strategy amounts to is that,
instead of singling out one particular coherent state
$\psi_{\gamma, m}$---as defined in terms of a single graph
$\gamma$---one considers an ensemble of coherent states
constructed by averaging the one-dimensional projections
$\hat{P}_{\gamma, m}$ onto the states $\psi_{\gamma, m}$ over a
subset $\Gamma_m$ of the set of all allowed graphs. In other
words, one considers a mixed state (with an associated density
matrix) rather than a single coherent state. In such a way, if the
subset $\Gamma_m$ is big enough, it can be shown (\cite{10}) that
it is possible to eliminate the embedding dependence (the
`staircase problem'\footnote{Roughly the staircase problem can be
stated as follows. Consider an area operator $\hat{A}_S$ for a
surface $S$. If we compute the expectation value for $\hat{A}_S$
with respect to a coherent state $\psi_{\gamma, m}$, such that the
surface $S$ intersects transversely one and only one edge $e$ of
$\gamma$, then the expectation value of the area operator
coincides with the classical value $A(m)$. However, if the surface
$S$ lies transversally to the edges, then we do not obtain the
correct classical limit.}).

It is straightforward to deduce that the area complexifier
coherent states cannot be used to construct embedding-independent,
mixed coherent states because of condition 2) above. We thus claim
that \emph{area complexifier coherent states should be ruled out
as semiclassical states altogether if one wants to attain
embedding independence}. Instead one should use the flux coherent
states, as was done in \cite{10}. For such states we will show
that {\it the correct semiclassical limit is attained only for
$n=6$ }. In other words, {\it up to now, there are no
semiclassical states known other than those with cubic-graph
topology!}

Thus the implication of our result for LQG is that the
semiclassical sector of the theory is spanned by SNWF that are
based on cubic graphs. This has some bearing for spin-foam models
\cite{25} which are supposed to be---but, so far, have not been
proved to be---the path-integral formulation of LQG. Spin foams
are certain state-sum models that are based on simplicial
triangulations of four manifolds whose dual graphs are therefore
5-valent. The intersection of this graph with a boundary
three-manifold is 4-valent, and therefore we see that spin-foam
models, based on simplicial triangulations, correspond to boundary
Hilbert spaces spanned by spin-network states based on 4-valent
graphs only\footnote{ As an aside, whether this boundary Hilbert
space of spin foams really can be interpreted as the 4-valent
sector of LQG is a subject of current debate, even with the recent
improvements \cite{26} in the Barrett--Crane model \cite{27}.
There are two problems: first, the boundary connection predicted
by spin foams does not coincide with the LQG connection \cite{28}.
Secondly, the 4-valent sector of the LQG Hilbert space is not a
superselection sector for the holonomy flux algebra of LQG. In
fact, the LQG representation is known not only to be cyclic but
even irreducible \cite{28a}. Therefore the 4-valent sector is not
invariant under the LQG algebra.}. However, we have proved that
the correct semiclassical states for analysing the semiclassical
properties of the volume operator are the {\it gauge covariant
flux} states. For such states, only those of cubic topology give
the correct semiclassical value of the volume operator.

Even if the mismatches between the 4-valent sector of LQG and the
boundary Hilbert space of spin foams could be surmounted, the
result of our analysis seems to be that {\it the boundary Hilbert
space of current spin-foam models does not contain any
semiclassical states!} This apparently contradicts recent findings
that the graviton propagator derived from spin-foam models is
correct \cite{29}. However, it is notable that these latter
results only show that the propagator has the correct fall-off
behaviour: the correct tensorial structure has not yet been
verified.

One straightforward way of possibly repairing this situation is to
generalise spin-foam models to allow for arbitrary---in
particular, cubic---triangulations, as suggested in \cite{30,30a}.

In the present paper we will perform detailed calculations of the
expectation value of the volume operator for the 4-, 6-, and
8-valent graphs using the area coherent states reviewed in
\cite{30b}. The main difference between the flux coherent states
and the area coherent states is that the former depend on a
discrete set of flux operators while the latter depend on a
continuous family of area operators. In particular, the surfaces
on which the area operators depend fill all of space. Hence the
question arises whether the area coherent states depend less
severely on the relative orientation of the chosen cut-off graph
with respect to the system of surfaces. Specifically, it would be
interesting to know if the calculations of the expectation values
are insensitive to Euclidean motions of the graph (translations
and rotations) when the spatial metric to be approximated is the
Euclidean one.

The structure of the paper is as follows. In Section \ref{s4} we
calculate the expectation value for a general graph. The
expectation value depends on the edge metric, already encountered
in \cite{30b}, which is close to diagonal when the edge lengths,
$\delta$, are much larger than the lengths, $l$, of the boundaries
of the surfaces. In order to compute the off-diagonal metric
entries we stick to the graphs defined in our companion paper as
otherwise we encounter severe `bookkeeping' problems. These graphs
are dual to tetrahedronal, cubical and octahedronal
triangulations, all of which derive from a cubulation of the
spatial manifold, as discussed in \cite{30b}. One can expand the
expectation value into a convergent power series in $l/\delta$. We
establish that, contrary to the results in
\cite{30b}, to zeroth-order in both $\hbar$ and $l/\delta$ the
correct expectation value is obtained for $n=6$ and $n=4$, but
\emph{only} for specific embeddings of the graph with respect to
the surfaces on which the area operators depend. However, we show
that such embeddings have measure \emph{zero} in $SO(3)$.
Moreover, by using the coherent states in \cite{10} we also find
that the graphs (4-, 6, and 8-valent) are all Euclidean invariant,
up to sets of measure zero in $SO(3)$ and up to linear order in
$l/\delta$. Nonetheless, the embeddings with non-zero measure all
fail to give the correct semiclassical value for the volume
operator.

This shows that coherent states  \cite{10} are not the correct
tools for analysing the semiclassical properties of the volume
operators.

In Section 3 we  perform an explicit calculation of the
expectation value of the volume operator for 4-, 6-, and 8-valent
graphs, including the higher-order effects in $l/\delta$. This
contribution is only Euclidean-invariant up to linear order. For
each of these valences we first consider a graph constructed in
terms of regular simplicial, cubical and octahedron cell
complexes, as done in \cite{30b}.
After that, we study the effects of rotations and translations.

In Section 4 we summarise our results and draw appropriate
conclusions.

The more technical calculations have been transferred to an
appendix in order not to distract from the line of argument in the
main text.

\section{Volume Operator Expectation Values for Area Coherent States}
\label{s4}

In this Section we compute the expectation value of the operator
$\hat{V}_{\gamma,v}$ for an arbitrary $n$-valent vertex, $v$, for
the stack family coherent states using the replacement of $SU(2)$
by $U(1)^3$. This uses the calculational tools developed in
\cite{30b}, to which we refer for our notation. We may, therefore,
replace the $SU(2)$ right-invariant vector fields by $U(1)^3$
right-invariant vector fields $X^j_{e_I(v)}=i
h^j_I\partial/\partial h^j_I$ acting on $h^j_I:=A^j(e_I(v))$. The
crucial simplification is that these vector fields mutually
commute. Their common eigenfunctions are the spin-network
functions which, for $U(1)^3$, take the explicit form \be
\label{4.0.5} T_{\gamma,n}(A)=\prod_{e\in E(\gamma)}\;
\prod_{j=1}^3\; [A^j(e)]^{n_j^e} \ee We will refer to them as
`charge network states' because the $n_j^e\in \mathbb{Z}$ are
integer valued. Using the spectral theorem we may immediately
write down the eigenvalues of $\hat{V}_{\gamma,v}$ on
$T_{\gamma,n}$ as \be \label{4.0.6} \lambda_{\gamma,n,v}=\ell_P^3
\sqrt{\Big|\frac{1}{8} \sum_{{v\in e_I\cap e_J\cap e_K } \atop
{1\le I\leq J\leq K\leq N}} \epsilon^{ijk}\epsilon(e_I,e_J,e_K)
\big[n^{e_I}_i\; n^{e_J}_j \; n^{e_K}_k\big]\Big|} \ee

What follows is subdivided into four parts. We begin by performing
the calculation for a general graph. This leads to the inverse of
the edge metric which, for large graphs, is beyond analytical
control. In the second part we restrict the class of graphs, which
lets us perform perturbative computations of the inverse of the
edge metric. This gives a good approximation of the actual
expectation value. In the third and fourth parts we consider the
dependence of our results on the relative orientation of the graph
with respect to the family of stacks.

\subsection{General expectation value of the volume operator}
\label{s4.1}

In this Section we drop the graph label and set
$t^{jk}_{ee'}:=\delta^{jk}\; t l^\gamma_{e e'},\;t:=\ell_P^2/a^2$.
This gives a positive, symmetric bilinear form on vectors
$n:=(n^e_j)_{e\in E(\gamma),\;j=1,2,3}$ that is defined by
$t(n,n'):=\sum_{e,e',j,k} n^e_j n^{\prime e'}_k t^{jk}_{e e'}=n^T
\cdot t\cdot n'$. We also set $Z^T \cdot n:=\sum_{e, j} Z^j(e)
n^e_j$.

The coherent state associated with an $n$-valent graph in which
more than one edge intersects a given plaquette $S$, is as
follows: \be \label{4.1} \psi_{Z,\gamma} =\sum_n\;e^{-\frac{1}{2}
n^T\cdot t\cdot n}\; \; e^{Z^T \cdot n}\; T_{\gamma, n} \ee The
norm of the coherent states is given by \be \label{4.2}
||\psi_{Z,\gamma}||^2 =\sum_n \; e^{-n^T \cdot t \cdot n}\; e^{2
P\cdot n} \ee where \be \label{4.2a} P^j(e)=i\int_e
(Z-A)=:\frac{1}{b^2} E^e_j \ee The length parameter, $b$, that
appears here is generally different from the parameter, $a$, that
enters the classicality parameter $t=\ell_P^2/a^2$, as explained
in \cite{30b}.

The expectation value for the volume operator is \be \label{4.3}
<\hat{V_v}>_{Z,\gamma}=
\frac{\langle\psi_{Z,\gamma}\hat{V_{\gamma,v}}
\psi_{Z,\gamma}\rangle}{||\psi_{Z,\gamma}||^2} =\frac{\sum_n \;
e^{-n^T \cdot t \cdot n}\; e^{2 P\cdot n}
\lambda_{\gamma,v}(n)}{||\psi_{Z,\gamma}||^2} \ee where \be
\label{4.4} \lambda_{\gamma,v}(n)=\ell_P^3
\sqrt{\Big|\frac{1}{48}\sum_{e\cap e^{\prime}\cap
e^{\prime\prime}=v}\;\epsilon(e,e^{\prime},e^{\prime\prime})\;
\det_{e e' e^{\prime\prime}}(n)\Big| } \ee are the eigenvalues of
the volume operator $\hat{V}_{\gamma,v}$. We have introduced the
notation \be \label{4.5} \det_{e e' e^{\prime\prime}}(n):=
\epsilon^{jkl}\;n^e_j n^{e'}_k n^{e^{\prime\prime}}_l \ee

The semiclassical limit of the volume operator is obtained from
(\ref{4.3}) in the limit of vanishing $t$. That is, it is the
zeroth-order in $t$ of the expansion of (\ref{4.3}) in powers of
$t$. Since (\ref{4.3}) converges slowly for small values of $t$,
we will perform a Poisson transform which replaces $t$ by
$\frac{1}{t}$ which converges quickly. To this end, analogous to
\cite{19}, we introduce the following variables \be \label{4.6}
t_e:=t_{ee},\; T_e:=\sqrt{t_e},\;x^e_j:= T_e n^e_j,\;
y^e_j(x):=x^e_j/T_e,\; C^e_j:=P^e_j/T_e,\;w^e_j(n):=n^e_j/T_e,\;
A_{e e'}^{jk}:= A_{e e'} \delta^{jk}:=\frac{t_{ee'}}{\sqrt{t_e
t_{e'}}} \ee Notice that the diagonal entries of $A$ equal unity.
The off-diagonal ones, however, are bounded from above by unity by
the Schwarz inequality applied to the scalar product defined by
$A$ and are restricted to the six-dimensional subspace restricted
to vectors with non-zero entries for the $e,e'$ components only.

Then (\ref{4.3}) turns into \be \label{4.7} <\hat{V_v}>_{Z,\gamma}
=\frac{\sum_n \; e^{-x^T \cdot A \cdot x}\; e^{2 C\cdot x}
\lambda_{\gamma,v}(y(x))} {\sum_n \; e^{-x^T \cdot A \cdot x}\;
e^{2 C\cdot x}} \ee Applying the Poisson transform to (\ref{4.7})
we obtain (recall $N:=|E(\gamma)|$) \be \label{4.8}
<\hat{V_v}>_{Z,\gamma} = \frac{\sum_n \; \int_{\mathbb{R}^{3 N}}\;
d^{3N}x\; e^{-2\pi i w(n)^T \cdot x} e^{-x^T \cdot A \cdot x}\;
e^{2 C\cdot x} \lambda_{\gamma,v}(y(x))} {\sum_n \;
\int_{\mathbb{R}^{3 N}}\; d^{3N}x\; e^{-2\pi i w(n)^T \cdot x}
e^{-x^T \cdot A \cdot x}\; e^{2 C\cdot x}} \ee

In order to perform the Gaussian integrals in (\ref{4.8}) we
notice that, by construction, $A$ is a positive-definite,
finite-dimensional matrix, so that its square root, $\sqrt{A}$,
and its inverse are well defined via the spectral theorem. Hence
we introduce as new integration variables \be \label{4.9}
z^e_j:=\sum_{e',k} (\sqrt{A})^{ee'}_{jk} x^{e'}_k,\;
y^e_j(z)=\frac{1}{T_e} x^e_j = \frac{1}{T_e}\sum_{e' k}
\big([\sqrt{A}]^{-1}]\big)^{ee'}_{jk} z^{e'}_k \ee The Jacobian
from the change of variables drops out in the fraction, and
(\ref{4.8}) becomes \be \label{4.10} <\hat{V_v}>_{Z,\gamma} =
\frac{\sum_n \; \int_{\mathbb{R}^{3 N}}\; d^{3N}z\; e^{2(C-2\pi i
w(n))^T \cdot \sqrt{A}^{-1} \cdot z} e^{-z^T \cdot z}\;
\lambda_{\gamma,v}(y(z))} {\sum_n \; \int_{\mathbb{R}^{3 N}}\;
d^{3N}z\; e^{2(C-2\pi i w(n))^T \cdot \sqrt{A}^{-1} \cdot z}
e^{-z^T \cdot z} } \ee

Now one would like to shift $z$ into the complex domain by
$\sqrt{A}^{-1}(C-i \pi w)$ and then perform the ensuing Gaussian
integral. This is unproblematic for the denominator of (\ref{4.8})
which is analytic in $z$; however, the numerator is not. The
careful analysis  in \cite{19} shows the existence of branch cuts
in $\mathbb{C}^{3N}$ of the fourth-root function involved. In
turn, this shows  that, in the semiclassical limit, both numerator
and denominator are dominated by the $n=0$ term while the
remaining terms in the series are of order $\hbar^\infty$ (i.e
they decay as $\exp(-k_n/t),\;t=\ell_P^2/a^2$ for some $k_n>0, \;
\lim_{n\to \infty} k_n=\infty$). See \cite{19} for the details.

The upshot is that to any polynomial order in $\hbar$ we may
replace (\ref{4.10}) by \be \label{4.11} <\hat{V_v}>_{Z,\gamma} =
\frac{\int_{\mathbb{R}^{3 N}}\; d^{3N}z\; e^{-z^T \cdot z}\;
\lambda_{\gamma,v}(y(z+\sqrt{A}^{-1} C))} {\int_{\mathbb{R}^{3
N}}\; d^{3N}z\; e^{-z^T \cdot z} } \ee which is now defined
unambiguously because the argument of the fourth root is the
square of a real number.

The denominator of (\ref{4.11}) simply equals $\sqrt{\pi}^{3N}$.
Therefore, the only $\hbar$-dependence of (\ref{4.11}) lies in the
numerator in the function $\lambda_{\gamma,v}$. We now note that
the eigenvalues of the volume operator come with a factor of
$\ell_P^3$, as displayed in (\ref{4.4}). Pulling it under the
square root into the modulus, and noticing that the modulus is a
third-order polynomial in $y$, we see that \be \label{4.12}
\lambda_{\gamma,v}\big(y(z+\sqrt{A}^{-1} C)\big) = \sqrt{
\Big|\frac{1}{48}\sum_{e\cap e'\cap e^{\prime\prime}=v}
\epsilon(e, e',e^{\prime\prime}) \det_{e,
e',e^{\prime\prime}}\big(\ell_P^2 y(z+\sqrt{A}^{-1} C)\big)\Big| }
\ee where we have \ba \label{4.13} \big[\ell_P^2
y(z+\sqrt{A}^{-1})C\big]^e_j &=& \frac{\ell_P^2}{T_e}\sum_{e',k}
\big[\sqrt{A}^{-1}\big]^{e e'}_{jk} \big[z+\sqrt{A}^{-1}
C\big]^{e'}_k
\nonumber\\
&=& \frac{\sqrt{t}}{T_e}\sum_{e',k}\;\Big( a^2 \sqrt{t}
\big[\sqrt{A}^{-1}\big]^{e e'}_{jk} z^{e'}_k
+\frac{\sqrt{t}}{T_{e'}} \big[A^{-1}\big]^{e e'}_{jk}
\Big(\frac{a}{b}\Big)^2 E^{e'}_k\Big) \ea where (\ref{4.2a}) has
been used in the last line.

To extract the leading order in $t$ of (\ref{4.11}) is now easy.
First note that the matrix elements of  both $A$ and
$T_e/\sqrt{t}$ are of order unity. Since $a$ is some macroscopic
length scale, the first term that is proportional to $z$ in
(\ref{4.13}) is therefore of order $\sqrt{t}$, while the second is
of zeroth-order in $t$. Therefore
$F:=\lambda_{\gamma,v}\big(y(z+\sqrt{A}^{-1} C)\big)$ is of the
form \be \label{4.14} F(z\sqrt{t})=\root 4 \of{Q+P(z\sqrt{t})} \ee
where $P$ is a certain sixth-order polynomial in $z \sqrt{t}$ with
no zeroth-order term, while $Q$ is independent of $z$. Moreover,
$Q+P(z\sqrt{t})$ is non-negative for all $z$ because it is the
square of a third-order polynomial in $z$. In particular, this
holds at $z=0$, and therefore $Q$ is also a non-negative number.
Then,  provided $Q>0$, we can define \be \label{4.15}
f(z\sqrt{t}):=\frac{F(z\sqrt{t})}{\root 4\of{Q}}=:\root 4
\of{1+R(z\sqrt{t})} \ee where $R$ is a sixth-order polynomial with
no zeroth-order term which is bounded from below by $-1$. Now, as
in \cite{18}, we exploit the existence of  $r>0$ such that \be
\label{4.16} 1+\frac{1}{4} R-r R^2\le f\le 1+\frac{1}{4} R \ee for
all $R\ge -1$. Inserting this estimate into (\ref{4.11}) we can
bound the integral from above and below because the Gaussian is
positive. The integrals over $R$ and $R^2$ are finite and are at
least of order $t$ because odd powers of $z$ do not contribute to
the Gaussian integral.

It follows that to zeroth-order in $t$ we have \be \label{4.17}
<\hat{V_v}>_{Z,\gamma} =\sqrt{\Big|\frac{1}{48} \sum_{e\cap e'\cap
e^{\prime\prime}=v} \; \epsilon(e,e',e^{\prime\prime}) \;
\det_{e,e',e^{\prime\prime}}(Y)\Big|} \ee where \be \label{4.18}
Y^e_j:=\Big(\frac{a}{b}\Big)^2\sum_{e',k}\; \frac{t}{T_e T_{e'}}
\; [A^{-1}]^{e e'}_{jk} E^{e'}_k \ee

This is as far as we can go with the calculation for a general
graph. Notice that the inverse of the edge metric appears in this
expression and, for a general graph, this is beyond analytical
control. Therefore we will now make restrictions on the graph so
as to analyse (\ref{4.18}) further.

\subsection{Simplifying assumptions}
\label{s4.2}

\begin{figure}[hbt]
\begin{center}
 \includegraphics[scale=0.3]{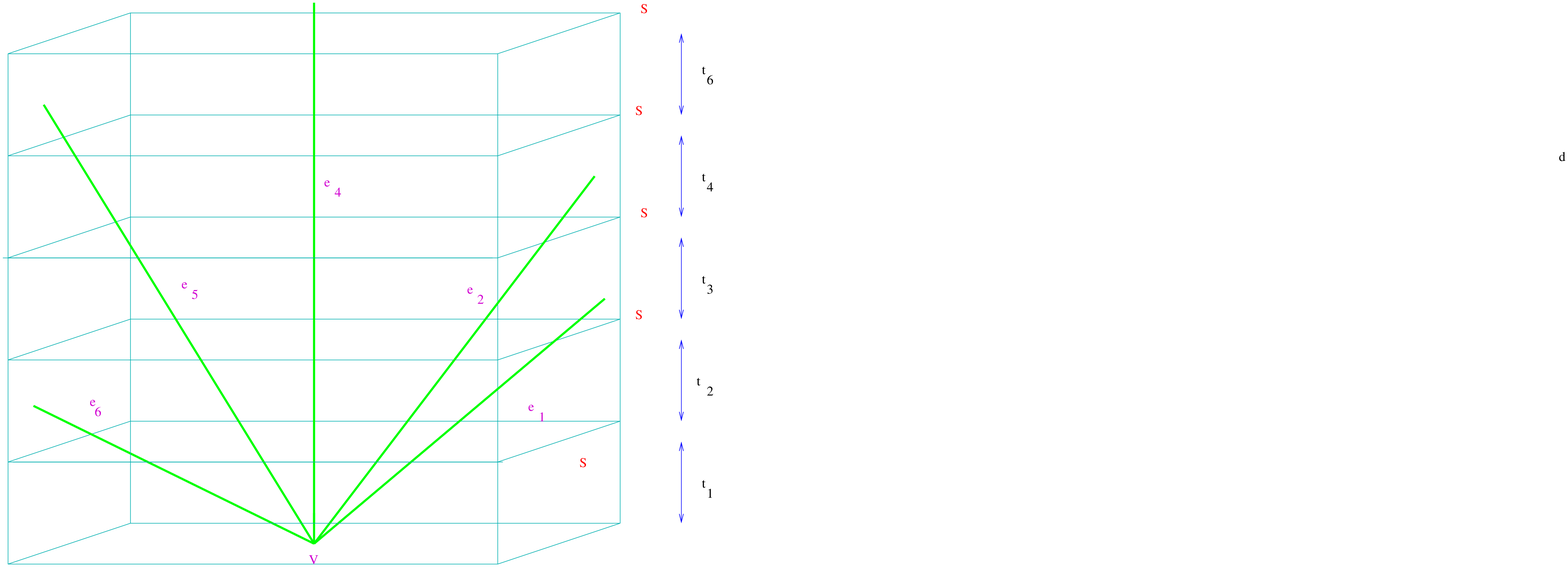}
\caption{Example of a tame, course and non-aligned
graph\label{fig0}}
\end{center} \end{figure}

The assumptions about the class of graphs to be considered are as
follows:
\begin{itemize}
\item[1.] {\it Coordinate Chart}\\
The graph, the region $R$ and the families of stacks lie in a
common coordinate chart $X:\;\mathbb{R}^3\to \sigma$. This is not
a serious restriction because the general situation may be reduced
to this one by appropriately restricting attention to the various
charts of an atlas that covers $\sigma$.
\item[2.] {\it Tame Graphs}\\
We assume that the graph is {\it tame} with respect to the stacks.
By this we mean that for each direction $I$, and each stack
$\alpha$, a given edge, $e$, of the graph enters and leaves that
stack at most once. This means that the graph does not `wiggle'
too much on the scale of the plaquettes. Analytically, it means
that $l^{I \alpha \gamma}_N$ vanishes whenever $|N_e|\ge 2$ for
any $e$, and that, for given $e$, the number $l^{I \alpha
\gamma}_N$ is non-vanishing at most for either $N_e=+1$ or
$N_e=-1$ but, not both, and independently of $\alpha$. Finally, it
means that the sets $S^{I \alpha \gamma}_N$ are connected.
\item[3.] {\it Coarse Graphs}\\
We assume that the graph is much coarser than the plaquettation in
the sense that any edge intersects many different stacks in at
least one direction $I$.
\item[4.] {\it Non-Aligned Graphs}\\
We exclude the possibility that distinct edges are `too aligned'
with each other in the sense that the number of stacks that they
commonly traverse is much smaller than the number of stacks that
they individually traverse .
\end{itemize}
Pictorially, the situation therefore typically looks as in figure
\ref{fig0}.
A consequence of the tameness, coarseness and `alignedness'
assumption is that $|t^\gamma_{e,e'}|\ll t^\gamma_e=t^\gamma_{e
e}, \; t^\gamma_{e'}=t^\gamma_{e' e'}$ for all $e,e'$, as is
immediately obvious from the formulae displayed in our companion
paper
because the number of stacks with $|N_e|=|N_{e'}|=1$ will be very
much smaller than the number of stacks with $|N_e|=1, N_{e'}=0$ or
$|N_{e'}|=1, N_e=0$. Hence the edge metric will be almost
diagonal. This is important because we need its inverse, which can
only be calculated with good approximation (that is, for large,
 semiclassically relevant graphs) if it is almost
diagonal. The graphs that we will eventually consider are
embeddings of subgraphs dual to tetrahedronal, cubical or
octahedronal triangulations of $\mathbb{R}^3$. These correspond to
embeddings of regular $4$-,$6$-,$8$-valent lattices, which ensures
the non-alignedness property.

Thus, without loss of generality, we may  choose the stacks and
plaquettes as follows. Using the availability of the chart
$X:\;\mathbb{R}^3\to \sigma;\;s\to X(s)$ we consider the
foliations $F^I$ defined by the leaves
$L_{It}:=X^I_t(\mathbb{R}^2)$ where for $\epsilon_{IJK}=1$ we set
$X^I_t(u^1,u^2):=X(s^I:=t,\;s^J:=u^1,s^K:=u^2)$. The stacks are
labelled by $\alpha=(\alpha^1,\alpha^2)\in \mathbb{Z}^2$. The
corresponding plaquettes are given by $p^I_{\alpha
t}=\{X^I_t([\alpha+u]l);\;u\in [0,1)^2\}$ where $l>0$ is a
positive number.

Likewise, using the availability of the chart, we take the edges
of the graph to be embeddings of straight lines in $\mathbb{R}^3$
(with respect to the Euclidean background metric available there),
that is, $e(t)=X(s_e+v_e \delta t)$ where $v_e$ is a vector in
$\mathbb{R}^3$ and $e(0)=X(s_e)$ defines the beginning point of
the edge.

After these preparations, we can now analyse (\ref{4.17}) and
(\ref{4.18}) further. Recall that \be \label{4.19}
E^e_j=\sum_{\alpha,I}\; \int\; dt\; E_j(p^{\alpha I}_t)
\sigma(p^{\alpha I}_t,e) \ee and \be \label{4.20}
t^\gamma_{ee'}=\sum_{\alpha I} \int\; dt\;
 \sigma(p^{\alpha I}_t,e)
\; \sigma(p^{\alpha I}_t,e') \ee By the assumption about the
graphs made above, the signed intersection number takes at most
the numbers $\pm 1$ and independently of $\alpha$, so that
$\sigma(p^{\alpha I}_t,e)^2=\sigma^I_e \sigma(p^{\alpha I}_t,e)$
for certain $\sigma^I_e=\pm 1$ which takes the value $+1$ if the
orientation of $e$ agrees with that of the leaves of the
foliation, $-1$ if it disagrees, and $0$ if it lies inside a leaf.
If we assume that the electric field $E^a_j$ is slowly varying at
the scale of the graph (and hence at the scale of the plaquettes
as well) then we may write \be \label{4.21} E^e_j \approx \sum_I
t^\gamma_{e e} \sigma^I_e E_j(p^I_v) \ee where
$p^I_v=p^{\alpha_I(v) I}_{t_I(v)}$ and $v$ is the vertex at which
$e$ is adjacent and which is under consideration in
$V_{\gamma,v}$. It follows that (\ref{4.18}) can be written as \be
\label{4.22} Y^e_j=\Big(\frac{a}{b}\Big)^2\sum_{e',k}\;
\frac{t}{T_e T_{e'}} \; \big[A^{-1}\big]^{e e'}_{jk} \sum_I
t^\gamma_{e' e'} \sigma^I_e E_k(p^I_v)
=\Big(\frac{a}{b}\Big)^2\sum_{e'}\; \sqrt{\frac{t^\gamma_{e'
e'}}{t^\gamma_{e e}}} \; \big[A^{-1}\big]^{e e'}\sum_I \sigma^I_e
E_j(p^I_v) \ee where we have used $T_e^2=t t^\gamma_{e e}$.

Now, by construction, $A=1+B$ with $B$ off-diagonal and with small
entries \be \label{4.23} B_{ee'}=\frac{t^\gamma_{e
e'}}{\sqrt{t^\gamma_{ee} t^\gamma_{e' e'}}} \ee which are of the
order of $l/\delta$ since two distinct edges will typically only
remain in the same stack for a parameter length $l$, while the
parameter length of an edge is $\delta$. Now notice that under the
assumptions we have made, we have $l^\gamma_{e e'}=0$ if $e,e'$
are not adjacent. Define $S_e$ to be the subset of edges which are
adjacent to $e$. Then \ba \label{4.24} ||B x||^2 &=&
\sum_e\sum_{e', e^{\prime\prime}\in S_e} x_{e'} B_{e' e} B_{e
e^{\prime\prime}} x_{e^{\prime \prime}}
\nonumber\\
& \le & [\sup_{e,e'} B_{e e'}^2]\; \sum_e \; \big[\sum_{e'\in S_e}
\;x_{e'}\big]^2
\nonumber\\
&\le& \Big(\frac{l}{\delta}\Big)^2 \sum_e\; \Big[ \big(\sum_{e'\in
S_e} 1^2\big)^{1/2} \; \big(\sum_{e'\in S_e}
x_{e'}^2\big)^{1/2}\Big]^2
\nonumber\\
&\le & \Big(\frac{l}{\delta}\Big)^2\; M\; \sum_e\; \sum_{e'\in
S_e} x_{e'}^2
\nonumber\\
&=& \Big(\frac{l}{\delta}\Big)^2\; M\; \sum_{e'} x_{e'}^2 \;
\sum_{e} \chi_{S_e}(e')
\nonumber\\
&\le & \Big(\frac{l}{\delta}\Big)^2\; M^2\; ||x||^2 \ea Here, in
the second step we estimated the matrix elements of $B$ from
above; in the third step we applied the Schwarz inequality; in the
fourth step we estimated $|S_e|\le M$, where $M$ is the maximal
valence of a vertex in $\gamma$; and in the sixth step we
exploited the symmetry \be \label{4.25} \chi_{S_e}(e') = \left\{
\begin{array}{cc}
1 & :\;\;e'\cap e\not=\emptyset \\
0 & :\;\;e'\cap e=\emptyset
\end{array}
\right. = \chi_{S_e'}(e) \ee as well as the definition of the norm
of $x$.

It follows that  for $l/\delta< M$, $B$ is bounded from above by
unity.  Therefore, the geometric series
$A^{-1}=1+\sum_{n=1}^\infty (-B)^n$ converges in norm. Hence we
are able to consider the effects of a non-diagonal edge metric up
to arbitrary order, $n$, in $l/\delta$. Here we will consider
$n=1$ only and write $A^{-1}=1-(A-1)=2\cdot 1-A$. But before
considering corrections from the off-diagonal nature of $A$ notice
that, to zeroth-order in $l/\delta$, equation (\ref{4.22}) becomes
simply \be \label{4.25a} Y^e_j =\Big(\frac{a}{b}\Big)^2 \sum_I
\sigma^I_e E_j(p^I_v) \ee Inserting (\ref{4.25a}) into
(\ref{4.17}) we find \be \label{4.26} <\hat{V_v}>_{Z,\gamma}\,
\approx\, \Big(\frac{a}{b}\Big)^3 \sqrt{\big|\det(E)(v)\big|}\;
\sqrt{\Big|\frac{1}{48} \sum_{e\cap e'\cap e^{\prime\prime}=v} \;
\epsilon(e,e',e^{\prime\prime}) \;
\det_{e,e',e^{\prime\prime}}(\sigma)\; \det(p_v)\Big|} \ee where
\be \label{4.27}
\det(p_v)=\frac{1}{3!}\epsilon_{IJK}\epsilon^{abc} \;
\int_{[0,1)^2}\; d^2u\; n_{a t_I(v)}^{\alpha_I(v) I}(u) \;
\int_{[0,1)^2}\; d^2u\; n_{b t_J(v)}^{\alpha_J(v) I}(u) \;
\int_{[0,1)^2}\; d^2u\; n_{c t_K(v)}^{\alpha_K(v) I}(u) \ee On
recalling that $n^{\alpha I}_{a t}=\epsilon_{abc} X^{I b}_{\alpha
t, u^1} X^{I c}_{\alpha t, u^2}$ with $X^{Ia}_{\alpha
t}(u)=X^a(s^I=t,s^J=\alpha^1+l u^1, s^K=\alpha^2+l u^2)$ for
$\epsilon_{IJK}=1$, we find \be \label{4.28} \det(p_v)\approx l^6
\big[\det(\partial X(s)/\partial s)\big]_{X(s)=v} \ee Hence
(\ref{4.26}) becomes \be \label{4.29} <\hat{V_v}>_{Z,\gamma}
\approx \Big(\frac{a l}{b}\Big)^3 \sqrt{\big|\det(E)(v)\big|}\;
\Big|\big[\det(\partial X(s)/\partial s)\big]_{X(s)=v}\Big|\;
\sqrt{\Big|\frac{1}{48} \sum_{e\cap e'\cap e^{\prime\prime}=v} \;
\epsilon(e,e',e^{\prime\prime}) \;
\det_{e,e',e^{\prime\prime}}(s)\Big|} \ee

We can draw an important conclusion from expression (\ref{4.29}).
Namely, the first \emph{three} factors approximate the classical
volume $V_v(E)$ as determined by $E$ of an embedded cube with
parameter volume $(al/b)^3$. When we sum (\ref{4.29}) over the
vertices of $\gamma$, which have a parameter distance, $\delta$,
from each other where $l\ll \delta$ by assumption, then the volume
expectation value only has a chance to approximate the classical
volume when the graph is such that $\delta=a l/b$, or
$\delta/l=a/b$. This could never have been achieved for $b=a$ and
explains why we had to rescale the labels of the coherent states
by $(a/b)^2$ while keeping the classicality parameter at
$t=\ell_P^2/a^2$. See our companion paper for a detailed
discussion. There we also explained why one must have $\delta/l$
actually equal to $a/b$ and not just of the same order: While one
could use this in order to favour other valences of the volume
operator,  the expectation value of other geometrical operators
such as area and flux would be incorrect.

Assuming $\delta/l=a/b$ we write (\ref{4.29}) as \be \label{4.30}
<\hat{V_v}>_{Z,\gamma}=:V_v(E)\; G_{\gamma,v} \ee thereby
introducing the graph geometry factor $G_{\gamma,v}$. It does not
carry any information about the phase space, only about the
embedding of the graph relative to the leaves of the
three-foliations. From the fact that (\ref{4.30}) reproduces the
volume of a cube up to a factor, we may already anticipate that
the geometry factor will be close to unity for, at most, a cubic
graph. Whether this holds for an arbitrary orientation of the
graph with respect to the stack family will occupy a large part of
the analysis which follows.

\subsection{Preliminary analysis of the graph geometry factor}
\label{s4.3} We start by investigating the behaviour of the graph
geometry factor $G_{\gamma,v}$ under diffeomorphisms, $\varphi$,
of $\sigma$, that is, under $G_{\gamma,v}\mapsto
G_{\varphi(\gamma),\varphi(v)}$ while the linearly-independent
families of stacks are left untouched. This will answer the
question of how much the geometry factor depends on the relative
orientation of the graph with respect to the stacks.

In fact, the orientation factor $\epsilon(e,e',e^{\prime\prime})$
is invariant under diffeomorphisms of the spatial manifold
$\sigma$. The signature factor \be \label{4.31}
\det_{e,e'e^{\prime\prime}}(\sigma)=\epsilon_{IJK} \sigma^I_e
\sigma^J_{e'} \sigma^K_{e^{\prime\prime}} \ee is obviously
invariant under any diffeomorphism that preserves the foliations
$F^I$, i.e. which map leaves onto leaves, because \be \label{4.32}
\sigma^I_e=\frac{1}{2}\int_e dx^a\, \epsilon_{abc} \int_{L_{It}}
dy^b\wedge dy^c \delta(x,y) \ee where $L_{It}$ is any leaf in $t$
which intersects $e$ transversely. Since we consider graphs whose
edges are embedded lines in $\mathbb{R}^3$ with the same embedding
that defines the stacks, it follows that the geometry factor is
invariant under any embedded global translations in
$\mathbb{R}^3$.

Next, since global rescaling in $\mathbb{R}^3$ preserves the
foliations and the topological invariant (\ref{4.32}), the
geometry factor is also invariant under embedded global rescalings
of $\mathbb{R}^3$. Finally, any embedded global rotations of
$\mathbb{R}^3$ that preserves all the orientation factors
$\sigma^I_e$ will leave the geometry factors invariant. Since the
orientation factors only take the values $+1,-1, 0$ (depending on
whether an edge agrees, disagrees with the orientation of the
leaves, or lies within a leaf), there will be a vast range of
Euler angles for which this condition is satisfied, if the graph
is an embedded, regular lattice of constant valence\footnote{In
fact, for a random graph we may also have rotational invariance on
large scales.}. Hence, in order to check whether the geometry
factor is rotationally invariant under any rotation we need only
worry about those rotations which lead to changes in the
$\sigma^I_e$. Likewise, if we rotate a graph which is dual to a
polyhedronal complex, we expect that the expectation value remains
invariant as long as the graph remains dual to the complex.

Fortunately, using the explicit formulae derived for the edges and
vertices for $n=4$-, $6$-, $8$-valent graphs displayed in
\cite{30b} we can  calculate the $\sigma^I_e$ for each edge $e$.
Intuitively, it is clear, that whenever many of the $\sigma^I_e$
change from $+1$ to $-1$,  we can expect a drastic change of the
expectation value. However, one has to take into account the
combined effect of these changes,  and this is what makes
rotational invariance possible. As a first step we determine the
action of a rotation on the sign factors.

\subsection{Explicit calculation of the $\sigma^I_e$ terms}
\label{s4.4}

 In what follows we  discuss the cases that show a drastic change in
 the value of
 $\det_{e,e^{'},e^{''}}(\sigma)=\epsilon_{IJK}\sigma_e^I\sigma_{e^{'}}^J
 \sigma_{e^{''}}^k$ caused by a change in the values of $\sigma^I_e$.
To carry out this calculation we will perform a rotation of each
of the three different types of lattice analysed so far: namely,
the $4$-, $6$- and $8$-valent lattices. These rotations will be
parameterised by Euler angles and will be centred at a particular
vertex of the lattice, for example $V_0$. The effects of a
rotation will depend on the distance of the vertices from the
centre of the rotation. In fact, the position of each vertex in
the lattice after rotation will depend on both the distance from
the centre of the rotation and the Euler angles used in
 the rotation. Fortunately, the values of the terms $\sigma^I_e$ will
 not depend on the former
 but only on the latter.

This is easy to see since the value of $\sigma^I_e$ can be either
1, -1 or 0 depending on whether the edge is outgoing, ingoing, or
lies on the plaquette in the direction $I$. Thus it will only
depend on the angle the edge makes with the perpendicular to the
plaquette in any given direction, i.e. it will depend on the
angles the edge makes with respect to a coordinate system centred
at the vertex at which the edge is incident. Clearly, only the
values of the Euler angles of the rotation  will affect the angles
each edge has with respect to the vertex at which it is incident.
In particular, since the graph we are using is regular, following
the rotation, all edges which were parallel to each other will
remain such, and thus will have the same angles with respect to
the vertex at which they are incident. This implies that in order
to compute the values of the terms $\sigma^I$, we can consider
each vertex separately and apply the same rotation to each vertex
individually.

\begin{figure}[htb]
\begin{center}
\psfrag{b}{$e$}\psfrag{c}{$e^{'}$}\psfrag{a}{$t_{e,e^{'}}$} 
\includegraphics[scale=0.7]{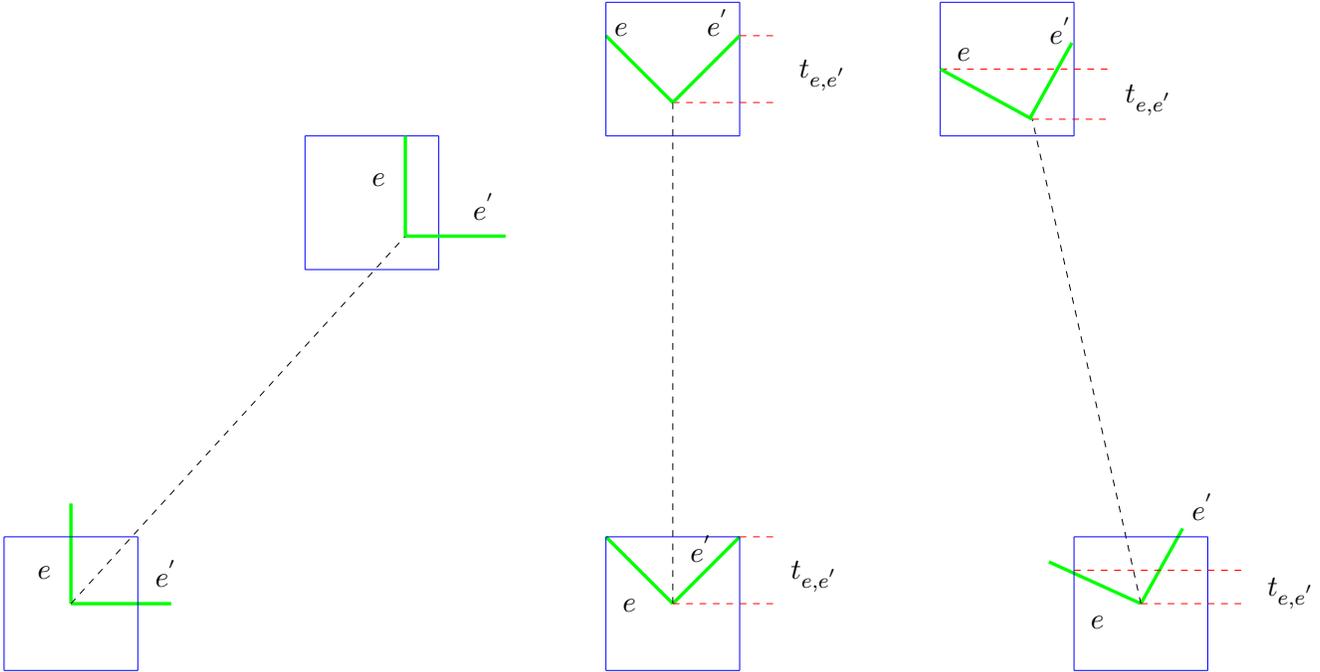}
\caption{Example of translation and rotation \label{fig1}}
\end{center}
 \end{figure}

On the other hand, the distance from the centre of the rotation
affects the position of each vertex with respect to the plaquette
structure, and thereby affects both the values of the terms
$t_{ee^{'}}$ and the number of them that are different from zero.
These effects  can be easily understood with the aid of the
two-dimensional diagram (Figure \ref{fig1}).

It is clear that, for any two parallel edges, the angle each of
them has with respect to the vertex at which they are incident, is
independent of the distance of the edge from the centre of
rotation. On the other hand, the values of the $t_{ee^{'}}$ will
depend on both the rotation and the distance of the centre of
rotation, since the position of the rotated vertex with respect to
the plaquette  depends on both these parameters. Therefore, we can
tentatively assume that two different geometric factors will be
involved in the computation of the volume operator:
\begin{enumerate}
\item[i)] $G_{\gamma,V}$, which indicates how the terms $\sigma^I$ are
affected by rotation. This geometric factor affects all orders of
approximation of the expectation value of the volume operator.
\item[ii)] $C_{\gamma, V}$, which indicates the  effect of rotation on the
terms $t_{ee^{'}}$. This term affects only the first- and
higher-order approximations of the expectation value of the volume
operator, not the zeroth-order.
\end{enumerate}

In what follows we will analyse the geometric term $G_{\gamma,
V}$: i.e., we will analyse the changes in the values of the
$\sigma_e^{I}$ due to a rotation applied at each vertex
independently. We will do this for the 4-, 6- and 8-valent graphs
separately. The geometric factor $C_{\gamma, V}$ will be analysed
in subsequent Sections.

As we will not see, our calculations show that for all 4-, 6- and
8-valent graphs, the rotations that produce drastic change in the
values of $\det_{e,e^{'},e^{''}}(\sigma)
 =\epsilon_{IJK}\sigma_e^I\sigma_{e^{'}}^J\sigma_{e^{''}}^k$ have
 measure zero in $SO(3)$ since they occur for specific Euler angles rather
 than for a range of them.

\begin{figure}[htbp]
\begin{center} \includegraphics[scale=0.7]{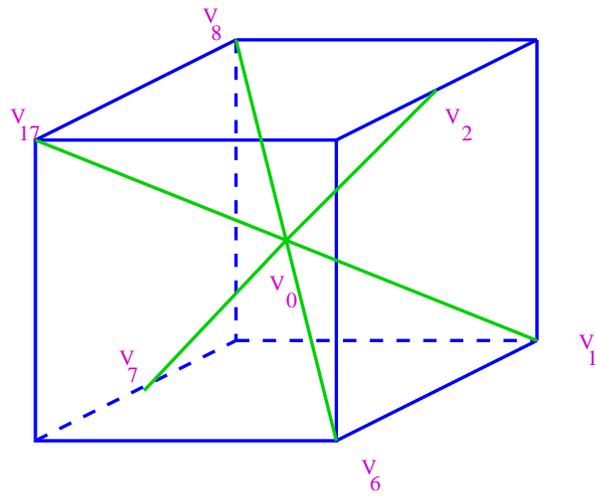}
\caption{6-valent vertex \label{fig6}}
\end{center}
 \end{figure}

Let us start with the 6-valent graph (Figure \ref{fig6}).
From the discussion above, we need only consider the effects of
the rotation on one vertex, $V_0$.

In order to compute the change in the values of the individual
$\sigma_e^{I}$, we will divide into eight small sub-cubes the cube
formed by the intersection of the plaquettes in the three
directions and containing the vertex we are analysing ($V_0$).
\begin{figure}[htbp]
\begin{center}
 \includegraphics[scale=0.7]{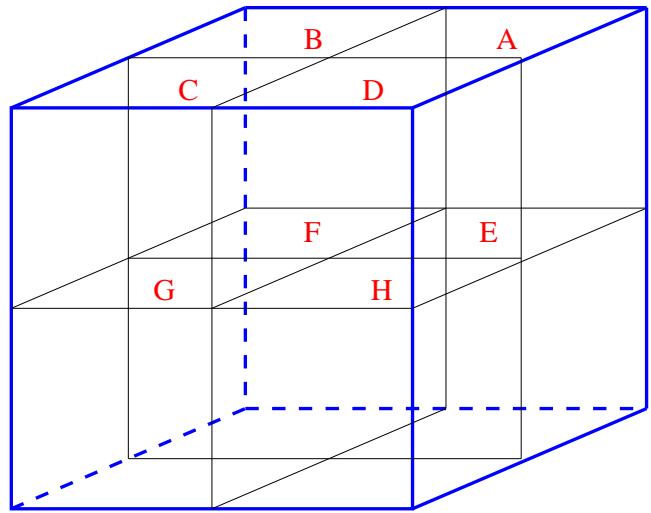}
\caption{Division of the cube in 8 sub-cubes \label{figdivision}}
\end{center}
 \end{figure}
It is then easy to see that for each edge $e$, the corresponding
value of $\sigma_e^I$  depends on  the
 sub-cube in which it lies. In particular, we have the
following table for the values of $\sigma^I_e$.
\begin{center}
\begin{tabular}{l|l|l|l|l|l|l|l|l}
& A & B & C & D & E & F & G & H\\ \hline $\sigma_e^x$& + & - &- &
+& + & - & - & +\\ \hline $\sigma_e^y$& + & + &- & - & + & + & - &
-      \\ \hline $\sigma_e^z$& + & + &+& +&- & -& -& - \\ \hline
\end{tabular}
\end{center}

From the above table it is clear that when an edge moves from one
of the eight cubes to another, the values of each of the
$\sigma_e$ changes accordingly. Given any 6-valent vertex, each of
the six edges incident at a vertex will be in one distinct cube.
Moreover, since any two edges incident at a vertex can be either
co-planar or perpendicular (in the abstract pull-back space with
Euclidean metric), there are only certain combinations of allowed
positions. For instance, for the edges $e_1,e_2,e_3,e_4,e_5,e_6$
only the combinations \be C_1B_2D_3E_4F_5H_6\mbox{ , }
C_1D_2A_3F_4G_5E_6 \mbox{ , }A_1B_2C_3H_4E_5G_6 \mbox{ ,
}B_1A_2D_3H_4G_5F_6 \ee are allowed (here the notation $A_i$ means
that the edge i lies in the cube A);  the combination
$A_1B_2C_3D_4G_5F_6$ is not allowed.

Because of the highly symmetric structure of the 6-valent graph we
do not have to analyse all possible combinations of all the six
edges incident at a vertex, since different combinations are
related by symmetry arguments. For example, the combination in
which edges $e_1,e_2,e_3,e_4,e_5,e_6$ lie in the cubes
$C_1D_2A_3F_4G_5E_6$, and the combination in which they lie in the
cubes $F_1G_2H_3C_4D_5A_6$, lead to the same value of
$|\det_{e,e^{'},e^{''}}(\sigma)|=
|\epsilon_{IJK}\sigma_e^I\sigma_{e^{'}}^J\sigma_{e^{''}}^k|$, and
an equal number of $\det_{e,e^{'},e^{''}}(\sigma)>0$ and
$\det_{e,e^{'},e^{''}}(\sigma)<0$, but obtained from different
triplets $e,e^{'},e^{''}$. In particular, any consistent
relabelling of the edges will produce the same overall result for
the determinants of the triplets. These symmetries reduce
considerably the number of cases that need to be analysed.

\begin{figure}[htbp]
\begin{center}
 \includegraphics[scale=0.7]{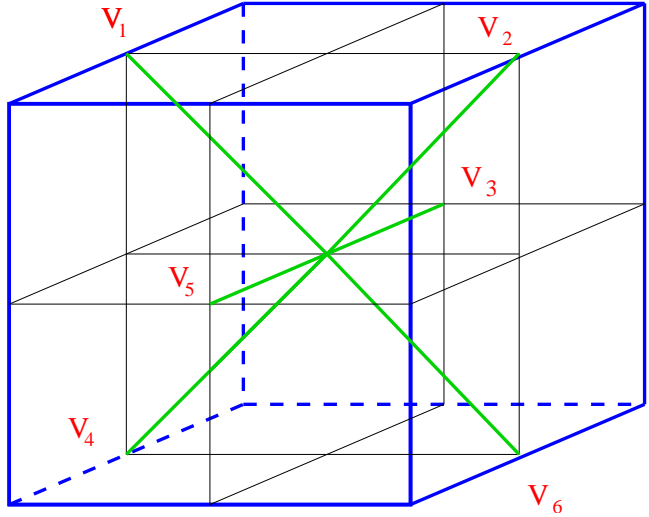}
\caption{Example of configuration with zero measure in
$SO(3)$\label{fig:9} }
\end{center}
\end{figure}

\begin{figure}[htbp]
\begin{center}
 \includegraphics[scale=0.7]{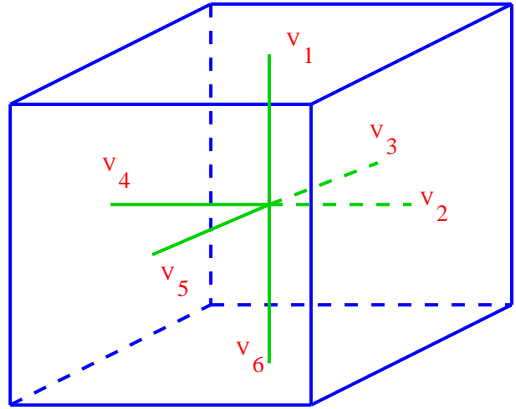}
\caption{Regular 6-valent graph \label{fig:orig}}
\end{center}
\end{figure}

In what follows, we consider the cases for which the edges
$e_1,e_2,e_3,e_4,e_5,e_6$ lie in the following combinations of
cubes: \be C_1B_2D_3E_4F_5H_6\mbox{ , }C_1D_2A_3F_4G_5E_6\mbox{ ,
}A_1B_2C_3H_4E_5G_6\mbox{ , }B_1A_2D_3H_4G_5F_6\ee For each of
these cases there will be sub-cases according to whether one edge
or more lie in a particular plaquette, or are parallel to a given
direction $I,J,K$. These sub-cases are the following:
\begin{enumerate}
\item No edge lies in any plaquette, or is parallel to any of the
directions.

In this case we obtain $|\det_{e,e^{'},e^{''}}(\sigma)|=4$ for all
triplets, but four of these triplets will have
$\det_{e,e^{'},e^{''}}(\sigma)=-4$ while the remaining four will
have $\det_{e,e^{'},e^{''}}(\sigma)=4$.

\item Only one edge lies in a particular plaquette (say the $J$ direction)
(see Figure \ref{fig6}). This edge and its co-linear edge will
have $\sigma^J_e$ equal to zero (J being the direction of the
plaquette in which the edge lies.)

In this case we obtain $\det_{e,e^{'},e^{''}}(\sigma)=-4$ for four
triplets, and $\det_{e,e^{'},e^{''}}(\sigma)=4$ for the remaining
four triplets.\footnote{Note that the geometric factor associated
to this edge orientation will coincide with the geometric factor
as derived from case 1). In this sense, case 2) can be seen as a
limiting case of 1)}

\item Two edges lie in two different plaquettes such that each of
these two edges and their respective co-linear edges will have
$\sigma^I_e$ equal to zero in the direction of the plaquette in
which they lie. In this case, because of the geometry of the
6-valent lattice, the remaining edges will each be parallel to a
given direction $J$ such that all but the $\sigma^J_e$ are zero.

In this case we obtain $\det_{e,e^{'},e^{''}}(\sigma)=2$ for four
triplets while the remaining four will have
$\det_{e,e^{'},e^{''}}(\sigma)=-2$. (See Figure \ref{fig:9}).

\item Each edge is parallel to a given direction such that all the $\sigma^I_e$ (for any $I,J,K$) are equal to zero
except for the one in the direction to which the edge is parallel.
In this case we obtain $\det_{e,e^{'},e^{''}}(\sigma)=1$ for four
triplets and $\det_{e,e^{'},e^{''}}(\sigma)=-1$ for the remaining
four. (See Figure \ref{fig:orig})
\end{enumerate}
Only sub-cases 3 and 4 might lead to a change of value for the
geometric factor $G_{\gamma,V}$. However cases 2, 3 and 4 have
measure zero in $SO(3)$.

As a demonstrative calculation on how this is derived we will
choose case 3. In particular, we select the configuration depicted
in Figure \ref{fig:9}
 which can be obtained by a rotation of the original configuration
in Figure \ref{fig:orig}.
Let us consider the linearly-independent triples comprised of the
edges that connect the barycenter of the cube to the vertices
$V_1$, $V_2$ and $V_3$. In the original configuration, the
coordinates of these vertices  are (in what follows we will denote
the length of an edge, $e$, by $\delta_e=\delta$) \be
V_1=(0,0,\delta)\mbox{ , } V_2=(\delta,0,0)\mbox{ , }
V_3=(0,\delta,0)\ee By applying a general Euler rotation, whose
matrix representation is given in (\ref{a1}), the coordinates of
the rotated vertices become: \be V_1=(R_{13}\delta, R_{23}\delta,
R_{33}\delta)\mbox{ , } V_2=(R_{11}\delta, R_{21}\delta,
R_{31}\delta)\mbox{ , } V_3=(R_{12}\delta, R_{22}\delta,
R_{32}\delta)\ee

 Our task now is to determine which Euler angles
would give rise to the configuration in Figure \ref{fig:9}. Since
in such a configuration the edges $e_{01}$ (the edge joining the
barycenter of the cube to vertex $V_1$) and $e_{03}$ lie in the
plane $x$--$y$, while the edge $e_{01}$ is parallel to the
$z$-direction,
 the coordinates of the rotated vertices are
constrained by the following set of equations:
\begin{align}
&x_{V_1}=R_{13}= \sin\s \sin\te<0\hspace{.2in} y_{V_1}=R_{23}=
\cos\s \sin\te=0\hspace{.2in} z_{V_1}=R_{33}=\cos\te >0\\
& x_{V_2}=R_{11}=\cos \s \cos \p -\cos \te \sin\p \sin\s<0
\hspace{.2in} y_{V_2}=R_{21}-\sin\s \cos\p- \cos\te \sin\p \cos\te=0\\
&z_{V_2}=R_{31}=\sin\te \sin\p>0\\
&x_{V_3}=R_{12}\cos\s \sin\p +\cos\te \cos\p
\sin\s=0\hspace{.2in}y_{V_3}=R_{22}=-\sin\s \sin\te+\cos\te \cos\p
\cos\s>0\\ &z_{V_3}=R_{32}=-\sin\te \cos\p=0
\end{align}
By solving this set of equations we find that the Euler angles
 $\s$, $\p$ and $\te$ that give rise to the configuration in Figure
 \ref{fig:9} are
\begin{enumerate}
\item[i)] $\te=(n+1)\frac{\pi}{2}$ and $\s=(p+1)\frac{\pi}{2}$ for $n$=odd, $p$=even and $0<\te<\frac{\pi}{2}$
\item[ii)] $\te=(n+1)\frac{\pi}{2}$ and $\s=(p+1)\frac{\pi}{2}$ for $n$=even, $p$=odd and $\frac{3\pi}{2}<\te<2\pi$
\end{enumerate}
It follows that the arrangement of edges under scrutiny has
measure zero in $SO(3)$.

By a similar method it can be shown that whenever an edge lies in
a plaquette, or is parallel to a plaquette, one of the Euler
angles will have to be equal to $\frac{n}{\pi}$ for $n$ odd or
even. Therefore, that arrangement will have measure zero. This is
not so for the general arrangement (number 1) delineated above.
However, for any such arrangement, the values for the orientation
factor and subsequently the geometric factor $G_{\gamma, V}$ will
always be the same and, in zeroth-order, will not lead to any
changes  of the expectation value of the volume operator.

Hence, the only cases of interest---i.e. the cases with measure
different from zero---will not lead to a rotational dependence of
the expectation value of the volume operator in zeroth-order. This
should not come as a surprise since the geometry of a regular
6-valent graph is such that to each edge there corresponds a
co-linear one. Thus whenever the term $\sigma^I_e$ for edge $e$
changes from -1 to 1, the term $\sigma^I_{e^{'}}$ of the co-linear
edge undergoes the inverse transformation. As a consequence there
will always be the same number of
$\det_{e,e^{'},e^{''}}(\sigma)=-4$ and
$\det_{e,e^{'},e^{''}}(\sigma)=4$, although the triplets involved
will be different in each case. It follows that  the overall value
of the geometric factor $G_{\gamma,V}$ remains constant.

A similar reasoning holds for the 8-valent graph since here too
each edge has a corresponding co-linear edge. Therefore, there
will always be an equal number of $\sigma^I_e=1$ and
$\sigma^I_e=-1$. This implies that, as in the case for 6-valent
graph, when no edge lies on a plaquette, the value of the
expectation value of the volume operator for each 8-valent vertex
will be rotationally invariant. On the other hand, the orientation
of edges in an 8-valent graph in which one or more edges lie in a
plaquette, or an edge is parallel to a given direction, have
measure zero in $SO(3)$, as was the case for the 6-valent graph.
However, as previously stated, it is precisely such cases that
lead to a change in the value of the geometric factor $G_{\gamma,
V}$.

For the 4-valent case the situation is somewhat different since
there are no co-planar edges. Those arrangements of edges with
respect to the stacks of plaquettes that cause drastic changes in
the values of the orientation factor are the following:
\begin{enumerate}
\item No edge lies in any plaquette. In this case we obtain $|\det_{e,e^{'},e^{''}}(\sigma)|=4$ for all linearly-independent triplets.
\item Each edge lies in a given plaquette. This gives
$|\det_{e,e^{'},e^{''}}(\sigma)|=2$ for all linearly-independent
triplets.
\item One edge is aligned with a given plaquette, one edge lies in a given plaquette, and  the remaining edges do not lie in---and are not aligned to---any plaquette. In this case we obtain $|\det_{e,e^{'},e^{''}}(\sigma)|=1$ for two triplets , $|\det_{e,e^{'},e^{''}}(\sigma)|=2$ for one triplet, and
$|\det_{e,e^{'},e^{''}}(\sigma)|=4$ for the remaining triplet.
\end{enumerate}
Similar calculations to those for the 6-valent graph then show
that the cases 1 and 2  above have measure zero in $SO(3)$.

In summary,  the discussion above shows that for all 4-, 6- and
8-valent graphs, those orientations of the edges with respect to
the stacks that cause a drastic change in the orientation factor
have measure zero in $SO(3)$. Therefore, up to measure zero in
$SO(3)$, the geometric factor $G_{\gamma, V}$ for these graphs is
rotationally invariant.

\subsubsection{Computation of the geometric factor for 4-, 6- and 8-valent graphs }
\label{s 2.4.1} In this Section we will compute the geometric
factor $G_{\gamma,v}$ for the 4-, 6- and 8-valent graphs. We
recall from equation (\ref{4.29}) that the expression for the
geometric factor is \ba\label{ali:g}
G_{\gamma,v}:&=&\sqrt{\Big|\frac{1}{48}\sum_{e\cap e^{'}\cap
e^{''}=v}\epsilon(e, e^{'}, e^{''})\det_{e, e^{'},
e^{''}}(\sigma)\Big|}\\\nonumber
&=&\sqrt{\Big|\frac{1}{8}\sum_{1\leq i\le j\le k\leq
N}\epsilon(e_i, e_{j}, e_{k})\det_{e_i, e_{j},
e_{k}}(\sigma)\Big|} \ea where $N$ is the valence of the vertex
and $\det_{e_i, e_{j}, e_{k}}(\sigma)=\epsilon_{IJK}
\sigma^I_{e_i} \sigma^J_{e_j} \sigma^K_{e_k}$. In what follows we
will calculate $G_{\gamma, v}$ for the 4-, 6- and 8-valent graphs
respectively. In particular (for each valence) we will analyse
each of the cases discussed in the previous Section which lead to
different values of orientation factor. Any sub-case of these
cases will lead to the same geometric factor.

\paragraph*{4-valent graph:}

We now compute the geometric factor for the 4-valent vertex for
different embeddings of the graph in the stack of surfaces.
\begin{enumerate}
\item
The most general situation is one in which none of the edges is
aligned to, or lies in, a given plaquette. Thus, for example,
consider the situation in which the edges $e_1$, $e_2$ $e_3$ and
$e_4$ are in the octants A, C, H and F respectively (see Figure
\ref{fig:4general}). Such a combination has a non-zero measure in
$SO(3)$.

The values for $\det_{e_i, e_{j}, e_{k}}(\sigma)$ relative to this
case are given by
 \begin{align}\label{ali:g4}
 &\det_{e_1,e_{2},e_{3}}(\sigma)=\det_{e_1,e_{3},e_{4}}(\sigma)=-4\\\nonumber
 &\det_{e_1,e_{2},e_{4}}(\sigma)=\det_{e_2,e_{3},e_{4}}(\sigma)=4
 \end{align}
Inserting these values in (\ref{ali:g}) gives \ba
G_{\gamma,v}&=&\sqrt{\Big|\frac{1}{8}\sum_{1\leq i\le j\le k\leq
4}\epsilon(e_i, e_{j}, e_{k})\det_{e_i, e_{j},
e_{k}}(\sigma)\Big|}\\\nonumber
&=&\sqrt{\Big|\frac{1}{8}\big(-4\epsilon(e_1, e_{2},
e_{3})-4\epsilon(e_1, e_{3}, e_{4})+4\epsilon(e_1, e_{2},
e_{4})+4\epsilon(e_2, e_{3}, e_{4})\big)\Big|}\\\nonumber
&=&\sqrt{\frac{1}{8}16} \ea
\begin{figure}[hbtp]
 \begin{center}
  \includegraphics[scale=0.7]{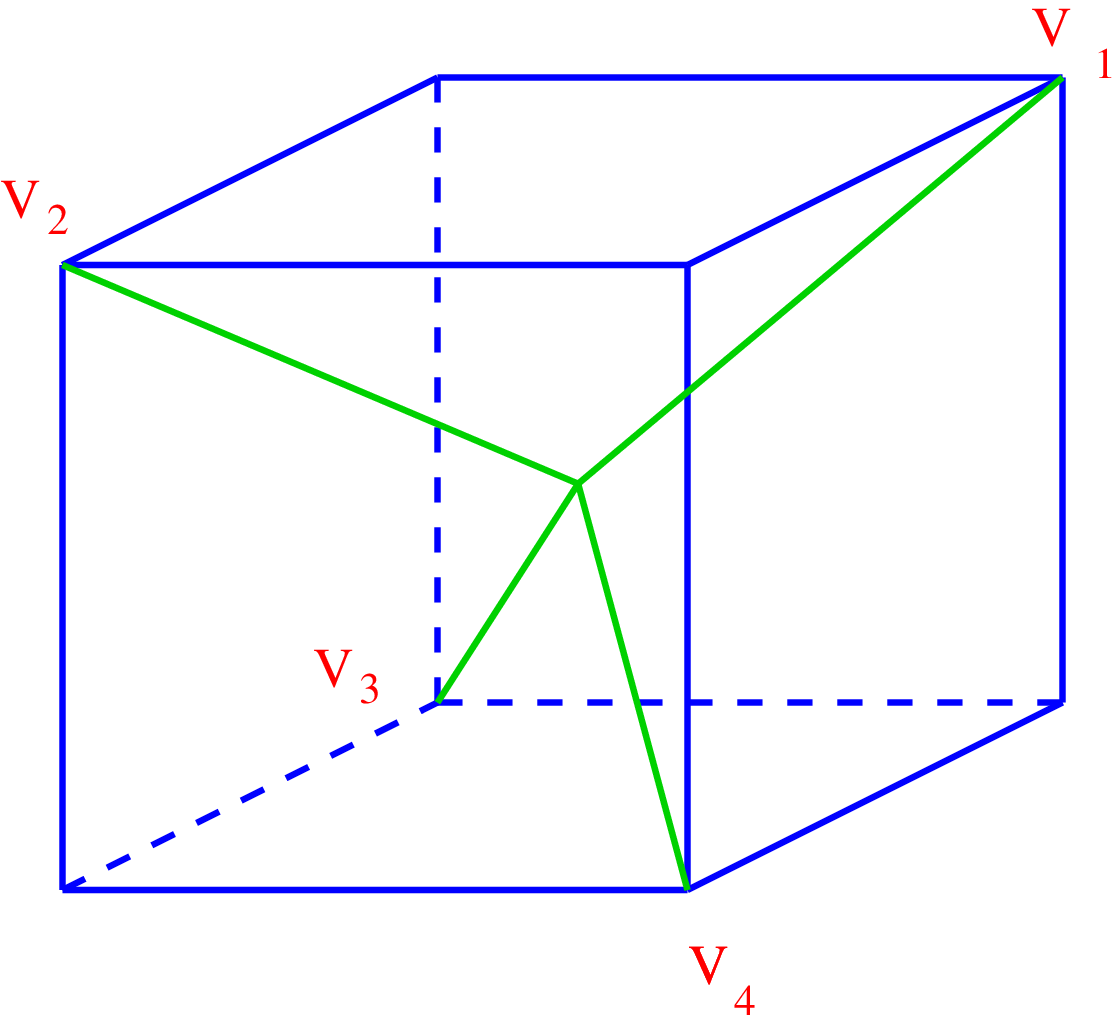}
 \caption{General  4-valent vertex }\label{fig:4general}
 \end{center}
 \end{figure}
\item If instead we consider the case in which each of the edges lies in a plaquette as, for example, is depicted in Figure \ref{fig:4aligned}, then the value for the geometric factor is
\ba G_{\gamma,v}&=&\sqrt{\Big|\frac{1}{8}\sum_{1\leq i\le j\le
k\leq 4}\epsilon(e_i, e_{j}, e_{k})\det_{e_i, e_{j},
e_{k}}(\sigma)\Big|}\\\nonumber
&=&\sqrt{\Big|\frac{1}{8}\big(-2\epsilon(e_1, e_{2},
e_{3})-2\epsilon(e_1, e_{3}, e_{4})+2\epsilon(e_1, e_{2},
e_{4})+2\epsilon(e_2, e_{3}, e_{4})\big)\Big|}\\\nonumber
&=&\sqrt{\frac{1}{8}8} \ea
\begin{figure}[hbtp]
 \begin{center}
  \includegraphics[scale=0.7]{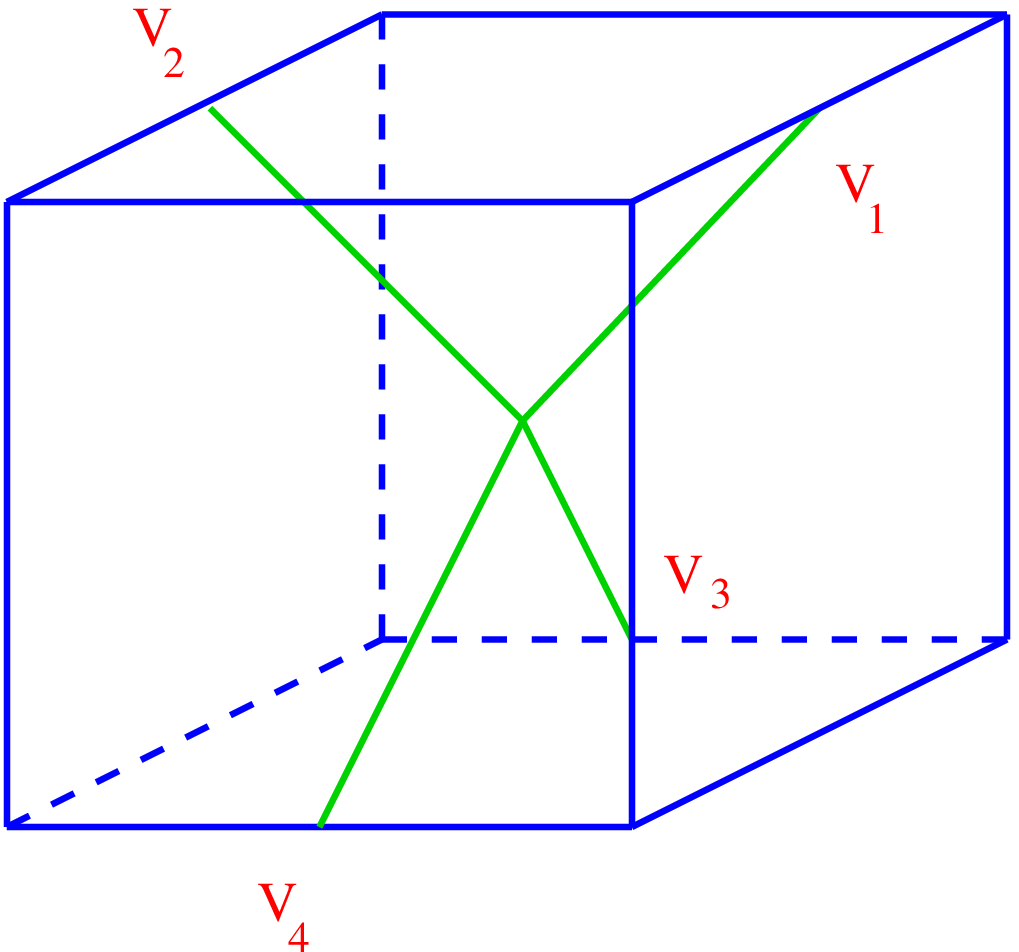}
 \caption{General 4-valent vertex }\label{fig:4aligned}
 \end{center}
 \end{figure}
\item For the situation in which one edge lies in a plaquette and another edge is aligned with a plaquette in another direction (Figure \ref{fig:43}), we obtain\ba
G_{\gamma,v}&=&\sqrt{\Big|\frac{1}{8}\sum_{1\leq i\le j\le k\leq
4}\epsilon(e_i, e_{j}, e_{k})\det_{e_i, e_{j},
e_{k}}(\sigma)\Big|}\\\nonumber
&=&\sqrt{\Big|\frac{1}{8}\big(-1\epsilon(e_1, e_{2},
e_{3})-4\epsilon(e_1, e_{3}, e_{4})+1\epsilon(e_1, e_{2},
e_{4})+2\epsilon(e_2, e_{3}, e_{4})\big)\Big|}\\\nonumber
&=&\sqrt{\frac{1}{8}8} \ea
 \begin{figure}[hbtp]
 \begin{center}
  \includegraphics[scale=0.7]{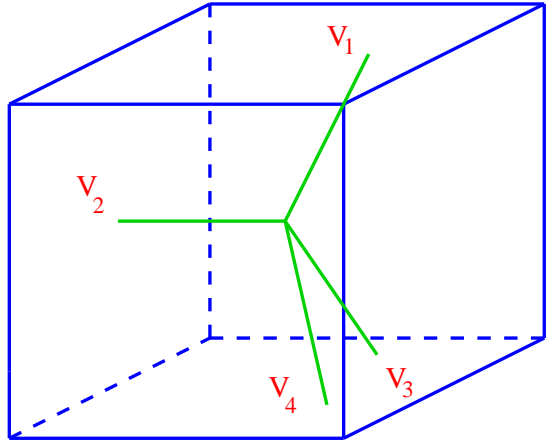}
 \caption{General  4-valent vertex }\label{fig:43}
 \end{center}
 \end{figure}
\end{enumerate}
However, we proved above that the embeddings of the vertex with
respect to the stack depicted in cases $2)$ and $ 3) $ have
measure zero in $SO(3)$.


\paragraph*{6-valent graph:}
We now compute the geometric factor for the 6-valent vertex in the
cases 1 to 4 described in the previous Section and which lead to
different values of the signature factor.
\begin{enumerate}
\item We start with the most general embedding of a 6-valent vertex with respect to the stacks. For example, consider the case in which the edges $e_1$, $e_2$ $e_3$, $e_4$ $e_5$ and $e_6$ are in the octants A H E B C and G respectively. We then obtain the following value for the geometric factor:
 \ba
G_{\gamma,v}&=&\sqrt{\Big|\frac{1}{8}\sum_{1\leq i\le j\le k\leq
4}\epsilon(e_i, e_{j}, e_{k})\det_{e_i, e_{j},
e_{k}}(\sigma)\Big|}\\\nonumber
&=&\sqrt{\Big|\frac{1}{8}\big(4\epsilon(e_1, e_{2},
e_{3})+4\epsilon(e_1, e_{3}, e_{4})+4\epsilon(e_1, e_{4},
e_{5})-4\epsilon(e_1, e_{2}, e_{5})-4\epsilon(e_2, e_{3},
e_{4})+4\epsilon(e_2, e_{5}, e_{6}})\\\nonumber &
&\overline{-4\epsilon(e_3, e_{4}, e_{6})-4\epsilon(e_4, e_{5},
e_{6})\big)\Big|}\\\nonumber &=&\sqrt{\frac{1}{8}4\times 8} \ea

\item For the geometric factor when only \emph{one} edge and its co-planar edge lie in a plaquette (Figure \ref{fig6}) we obtain: $G_{\gamma,v}=2$.

\item For the case in which \emph{two} edges and their co-planar edge lie in two different plaquettes in two different directions, while the remaining edge and its co--planar edge are aligned with the plaquette in the third direction (Figure \ref{fig:9}), we obtain $G_{\gamma,v}=\sqrt{2}$.

\item For the case in which all the edges are aligned with the stacks
(Figure \ref{fig:orig}) we obtain  $G_{\gamma,v}=1$, since in that
case $|\det_{e_i,e_j,e_k}|=1$ for all linearly independent
triplets $e_i$, $e_j$, $e_k$. \end{enumerate} However, we have
proved above that cases 2), 3) and 4) have measure zero in
$SO(3)$.

\paragraph*{8-valent graph:}We now compute the geometric factor for the 8-valent vertex for
different embeddings of the graph with respect to the stack of
surfaces.
\begin{enumerate}
\item
In the most general case, none of the edges lie in, or are aligned
to, a given plaquette: for example, when the edges $e_1$, $e_2$
$e_3$, $e_4$ $e_5$, $e_6$ $e_7$ and $e_8$ are in the octants $B$,
$C$, $A$, $D$, $H$, $E$, $G$ and $F$ respectively. This leads to
the following result \ba
G_{\gamma,v}&=&\sqrt{\Big|\frac{1}{8}\sum_{1\leq i\le j\le k\leq
4}\epsilon(e_i, e_{j}, e_{k})\det_{e_i, e_{j},
e_{k}}(\sigma)\Big|}\\\nonumber
&=&\sqrt{\Big|\frac{1}{8}\big(4\epsilon(e_1, e_{2},
e_{3})+4\epsilon(e_1, e_{2}, e_{4})-4\epsilon(e_1, e_{2},
e_{8})-4\epsilon(e_1, e_{2}, e_{7})-4\epsilon(e_1, e_{3},
e_{4})+4\epsilon(e_1, e_{3}, e_{6})}\\\nonumber &
&\overline{+4\epsilon(e_1, e_{3}, e_{8})+4\epsilon(e_1, e_{4},
e_{6})-4\epsilon(e_1, e_{4}, e_{7})+4\epsilon(e_1, e_{6},
e_{7})+4\epsilon(e_1, e_{6}, e_{8})-4\epsilon(e_1, e_{7},
e_{8})}\\\nonumber
& &\overline{-4\epsilon(e_2, e_{3}, e_{4})-4\epsilon(e_2, e_{3}, e_{5})\cdots\big)\Big|}\\
&=&\sqrt{\frac{1}{8}4\times 32} \ea

\item A more restricted case is when one edge and its co-planar edge are aligned with a plaquette in a given, different direction, while the remaining three edges and their co-planar edge lie in a given plaquette. Here  we obtain $G_{\gamma, V}=\sqrt{5}$.

\item A special case is when  each edge lies in a given plaquette: this gives $G_{\gamma, V}=2\sqrt{2}$.
\end{enumerate}
Similarly to the 4- and 6-valent vertex above, arrangement $2)$
and $3)$ have measure zero in $SO(3)$.

From the discussion above of the geometric factor we can already
deduce that, ignoring off-diagonal entries of the edge metric $A$,
the expectation value of the volume operator gives the correct
semiclassical value only for combinations of edges that have
measure zero in $SO(3)$.

In fact, in zeroth-order in $\frac{l}{\delta}$ the expectation
value of the volume operator is given by \be
<\hat{V_v}>_{Z,\gamma} \approx \Big(\frac{a l}{b}\Big)^3
\sqrt{\big|\det(E)(v)\big|}\; \Big|[\det(\partial X(s)/\partial
s)]_{X(s)=v}\Big|\; \sqrt{\Big|\frac{1}{48} \sum_{e\cap e'\cap
e^{\prime\prime}=v} \; \epsilon(e,e',e^{\prime\prime}) \;
\det_{e,e',e^{\prime\prime}}(\sigma)\Big|} \ee where $(\frac{a
l}{b})^3 \sqrt{\big|\det(E)(v)\big|}\big|[\det(\partial
X(s)/\partial s)]_{X(s)=v}\big|$ approximates the classical volume
$V_v(E)$ as determined by $E$ of an embedded cube with parameter
volume $(al/b)^3$. It is  straightforward to see that the correct
semiclassical behaviour is attained for $G_{\gamma, V}=1$.

The fact that the correct semiclassical behaviour of the volume
operator is attained \emph{only} for cases in which the graph is
aligned to the plaquettation (6-valent case), or each edge lies in
a given plaquette (the 4-valent case), seems rather puzzling
since, both cases, have measure zero in $SO(3)$. This makes one
question the\textit{ prima facie} validity of utilizing the area
coherent states to compute the expectation value of the volume
operator. However, it is interesting to note that case $4)$ of the
6-valent graph is precisely what one gets when constructing such a
graph as the dual of a cubical cell complex.


We will now proceed to compute the  higher,
$\frac{l}{\delta}$-order dependence of the expectation value of
the volume operator for 4-, 6- and 8-valent graphs respectively.

\section {The Higher, $\frac{l}{\delta}$-order Dependence of the Expectation
Value of the Volume Operator} \label{s7}

In this Section we compute the expectation value of the volume
operator for the 4-, 6-, and 8-valent graphs. The construction of
these graphs was discussed in
 \cite{30b} in terms of regular simplicial, cubical and octahedronal
cell complexes respectively.

The following Section is subdivided into four parts. In the first
we explain the general method to be applied in the subsequent
Sections. In the second, third and fourth parts we apply this
method to our 4-, 6- and 8-valent graphs respectively. Each of
these subsections is itself subdivided into three parts: in the
first,  the stack family and the cubulation that defines the
platonic-body cell complex dual to the graph are aligned (see
\cite{30b}); in the second we study the effect of a rotation; and
in the third we study the effect of a translation.

\subsection{Initial preparations}
\label{s7.1}

As a first step towards computing the expectation value of the
volume operator, we must calculate the values of the quantities
$t^{\gamma}_e$ and $t^{\gamma}_{ee^{'}}$ defined in \cite{30b},
which indicate the number of surfaces, $s^I_{\alpha t}$, that the
edge $e$ intersects, and the number of surfaces, $s^I_{\alpha t}$,
which are intersected by both edges $e$ and $e^{'}$. Both these
quantities depend explicitly on how the graph is embedded in the
stack family $S^I$ (see \cite{30b}). In fact, the conditions for
two or more edges to intersect a common surface are the following:
\begin{enumerate}
\item[1)] Two edges $e_i$ and $e_j$ intersect the same plaquette,
$s^z_{\alpha
t}$, iff $0< \phi_i,\phi_j<\frac{\pi}{2}$ or $\frac{\pi}{2}< \phi_i,\phi_j<\pi$\\

\item[2)] Two edges $e_i$ and $e_j$ intersect the same plaquette,
$s^x_{\alpha t}$, iff
$-\frac{\pi}{2}<\theta_i,\theta_j<\frac{\pi}{2}$ or
$\frac{\pi}{2}<\theta_i,\theta_j<\frac{3\pi}{2}$\\

\item[3)] Two edges $e_i$ and $e_j$ intersect the same plaquette,
$s^y_{\alpha
t}$, iff $0< \theta_i,\theta_j<\pi$ or $\pi< \theta_i,\theta_j<2\pi$\\

\item[4)] If we have equalities in any of the above conditions, such that the angles of each of the two edges correspond
to a different limiting case, we obtain
$t^I_{e_ie_j}=\{t_i\in\Rl|S^{I}_{t}\cap
e_k\neq\emptyset, k=i,j\}=\emptyset$.\\
\end{enumerate}
We can also have situations in which two or more edges intersect a
common plaquette in more than one stack. The
conditions for such occurrences are the following:\\[-5pt]
\begin{enumerate}
\item[a)] Given condition (1), two edges $e_i$ and $e_j$ will
intersect more than one $z$-stack iff $|\theta_i|+|\theta_j|<
\pi/2$ and such that
 $n\frac{\pi}{4}<\theta_i,\theta_j<(n+1)\frac{\pi}{4}$,
 where $n=\{1,2,3,4,5,6,7,8\}$.

\item[b)] Given condition (2), two edges  $e_i$ and $e_j$ will
intersect more than one $x$-stack iff condition (1) above is
satisfied and
$n\frac{\pi}{4}<\theta_i,\theta_j<(n+1)\frac{\pi}{4}$, where
$n=\{1,2,3,4,5,6,7,8\}$.

\item[c)] Given condition (3), two edges  $e_i$ and $e_j$ will
intersect more than one $y$-stack iff condition (1) above is
satisfied and
$n\frac{\pi}{4}<\theta_i,\theta_j<(n+1)\frac{\pi}{4}$, where
$n=\{1,2,3,4,5,6,7,8\}$.
\end{enumerate}

The conditions above imply that rotating the graph will  change
the values of the $t^I_{e_ie_j}$ and also the number of the
$t^I_{e_ie_j}$ that are non-zero.

We will now briefly explain, with the aid of an easy example, the
strategy we use to compute the terms
$\frac{t^I_{e_ie_j}}{\sqrt{t_{e_i}^It_{e_j}^I}}$ that are used in
the calculations of the expectation value of the volume operator
for the 4-,6- and 8-valent graphs. To this end, consider an edge,
$e\in\gamma$, of a generic graph whose length is given by
$\delta$. This edge will intersect the stacks of plaquettes in
each direction a certain number of times. In particular, given a
length $l$ of a plaquette, each edge will have $n$ intersections
with the stacks of any given direction, where $n$ is identified
with the Gauss bracket $[\frac{c_i}{l}]$ and $c_i$ is proportional
to $\delta$, where $c_i$ for $i\in\{x,y,z\}$ are the coordinates
of the edge.

\begin{figure}[htbp]
\begin{center}
\psfrag{V}{$V_1$}
\includegraphics[scale=0.9]{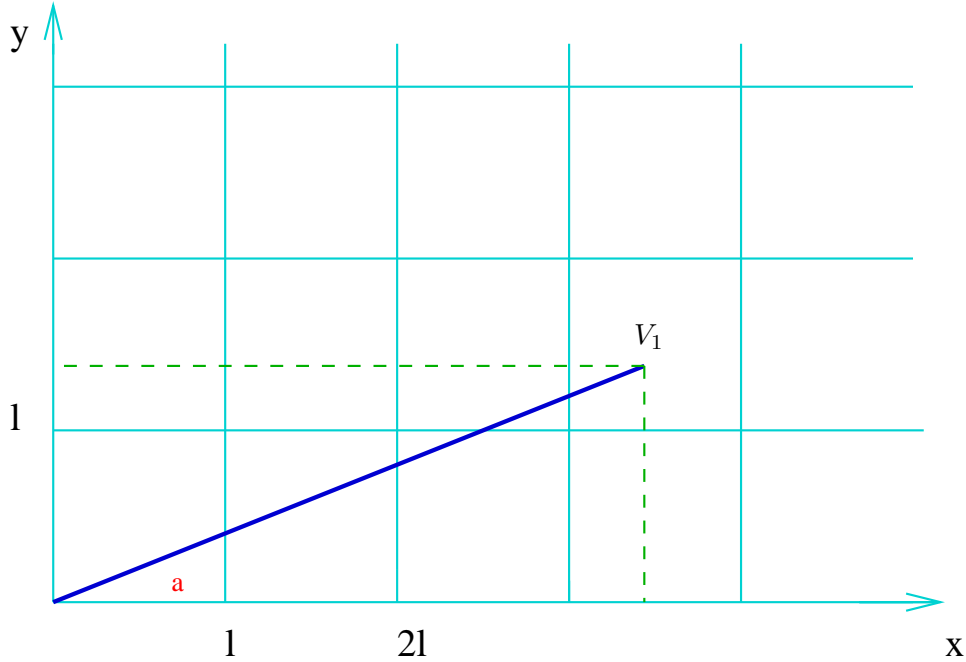}
\caption{Intersections of an edge with the plaquettes in 2-dimensions}\label{fig:7}
\end{center}
\end{figure}

For example, in the two-dimensional case of Figure \ref{fig:7},
the values of $n$ in any given direction for vertex $V_1$ (or
equivalently the edge $e_{0,1}$ of length $\delta$), whose
coordinates are
$V_1=(\delta cos(a),\delta sin(a))$, would be $n_x=[\frac{\delta
cos(a)}{l}]$ and $n_y =[\frac{\delta sin(a)}{l}]$.

The values $n_i$, $i\in\{x,y,z\}$, depend on both the angle $a$
and the ratio $\frac{l}{\delta}$. Concomitantly, the expectation
value of the volume operator will also depend on such parameters.

The rotational dependence will be dealt with later. In the present
Section we will focus on the $\frac{l}{\delta}$ dependence. We
need to consider three different sub-cases:
\begin{enumerate}
\item $\frac{\delta}{l}>1$
\item $\frac{\delta}{l}=1$
\item $\frac{\delta}{l}<1$
\end{enumerate}
and determine which of the them leads to consistent solutions.

However, to obtain an expansion of $\sqrt{A}^{-1}$ we need to
perform a Taylor series. The condition for applying such an
expansion is that $\Vert A-1\Vert<1$. From the expression for the (square) matrix $A$
(see (\ref{4.23})) it is clear that the condition above is
satisfied iff
$m(m-1)\frac{t_{ee^{\prime}}}{\sqrt{t_et{_e^{\prime}}}}<1$ where
$m$ is the dimension of the matrix. As we will show,
$\frac{t_{ee^{\prime}}}{\sqrt{t_et{_e^{\prime}}}}=C\times
\frac{l^{\prime}}{\delta_e}$ where C=constant and $l^{\prime}<l$;
thus the condition
$m(m-1)\frac{t_{ee^{\prime}}}{\sqrt{t_et{_e^{\prime}}}}<1$ becomes
$\l<<\delta_e$, i.e. we need to choose the parquet to be much
finer than the edge length (see Section \ref{s4.2}). If this
requirement is satisfied, then we can perform a Taylor expansion
of $\sqrt{A}$ obtaining
$\sqrt{A}=1+\frac{1}{2}(A-1)+\frac{1}{8}(A-1)^2+\mathcal{O}(A-1)^3$.
Actually, we are only interested in first-order terms, and so we
shall only consider the approximation $\sqrt{A}\simeq 1+\frac{1}{2}(A-1)$ whose inverse,
in first-order, is simply $(\sqrt{A})^{-1}\simeq 1-\frac{1}{2}(A-1)$.
Since the parquet length must be much finer than the edge length,
in the remainder of the paper we will consider only case (1) and
analyse whether it gives the correct semiclassical limit.

The first step in the calculation is to determine the range of
allowed positions for each vertex, $V_i$, of the graph with
respect to the plaquette. Since the graphs we consider are
regular, determining the position of one vertex suffices to derive
the positions of the remaining vertices in the graph.

\begin{figure}[htbp]
\begin{center}
\psfrag{a}{$e_i$}\psfrag{b}{$e_j$}
\includegraphics[scale=0.8]{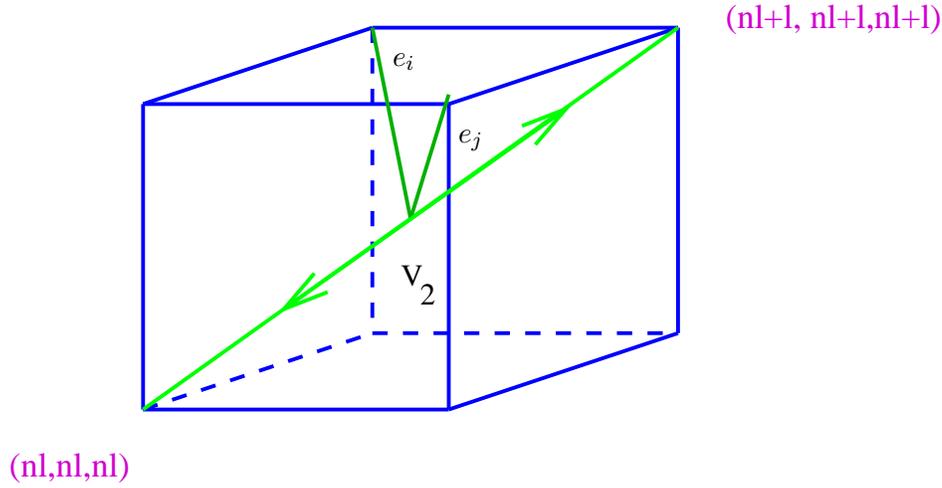}
\caption{Allowed positions of a vertex \label{fig:allowedpos}}
\end{center}
\end{figure}

As an explanatory example let us consider a regular 4-valent graph
$\gamma$ whose vertex $V_0$ coincides with the point $(0,0,0)$ of
the plaquettation and whose vertex $V_2$ (equivalently the edge
$e_{0,2}$) has coordinates
$(\frac{-\delta}{\sqrt{3}},\frac{\delta}{\sqrt{3}},\frac{\delta}{\sqrt{3}})$.
It follows that the range of allowed positions of $V_2$ is from
(nl, nl, nl) to (nl+l, nl+l, nl+l) where
$n_x=n_y=n_z=n=[\frac{\delta}{\sqrt{3}l}]$, as depicted in Figure
\ref{fig:allowedpos}.
\\It is straight-forward to understand that different positions of $V_2$ will determine different values of $t_{e_ie_j}$ for any two edges $e_i$ and $e_j$ incident at $V_2$. A detailed analysis shows that the terms $t_{e_ie_j}$ differ according to which of the following conditions is satisfied:
\begin{enumerate}
\item[I)]   $|x_{V_2}|>|nl+\frac{l}{2}|$
\item[II)]  $|x_{V_2}|<|nl+\frac{l}{2}|$
\item[III)] $|x_{V_2}|=|nl+\frac{l}{2}|$
\end{enumerate}
Similar conditions apply for all vertices in $\gamma$.

Since the position of $V_2$ will determine the positions of all
other
 vertices, it is possible to establish which positions of $V_2$ will lead to different values of the terms $\frac{t_{e_ie_j}}{\sqrt{t_{e_i}t_{e_j}}}$ for all edges of all vertices of the graph $\gamma$. Such positions of $V_2$ for a regular 4-valent graph are:
\begin{enumerate}
\item[a)]  $|nl|\leq|x_{V_2}|\leq|nl+\frac{l}{6}|$
\item[b)]  $|nl+\frac{l}{6}|\leq|x_{V_2}|\leq|nl+\frac{l}{4}|$
\item[c)]  $|nl+\frac{l}{4}|\leq|x_{V_2}|\leq|nl+\frac{l}{2}|$
\item[d)]  $|nl+\frac{l}{2}|\leq|x_{V_2}|\leq|nl+\frac{l}{3}|$
\item[e)]  $|nl+\frac{l}{3}|\leq|x_{V_2}|\leq|nl+\frac{3l}{4}|$
\item[f)]  $|nl+\frac{3l}{4}|\leq|x_{V_2}|\leq|nl+2l|$
\end{enumerate}
For each such condition it is possible to derive the respective
conditions for both the $y$- and the $z$-coordinates in the
three-dimensional case. It turns out that similar relations hold
for the 6- and 8-valent graphs as well.

To explicitly compute the terms $t_{e_ie_j}$ we must choose one of
the above conditions ($a\rightarrow f$), each of which will lead
to different values for each $t_{e_ie_j}$. However, the
computation procedures are the same. In the calculations of
Sections \ref{s7.2} we will choose case (a).

To describe the method for computing the values of $t_{e_ie_j}$,
we go  back to a very simple example in two dimensions. We will
then give the general outline of how this calculation can be
generalised to the 3-dimensional case.

\begin{figure}[htbp]
\begin{center}
\psfrag{t}{$t^y_{e_i}$}\psfrag{s}{$t^y_{e_j}$}
\includegraphics[scale=0.9]{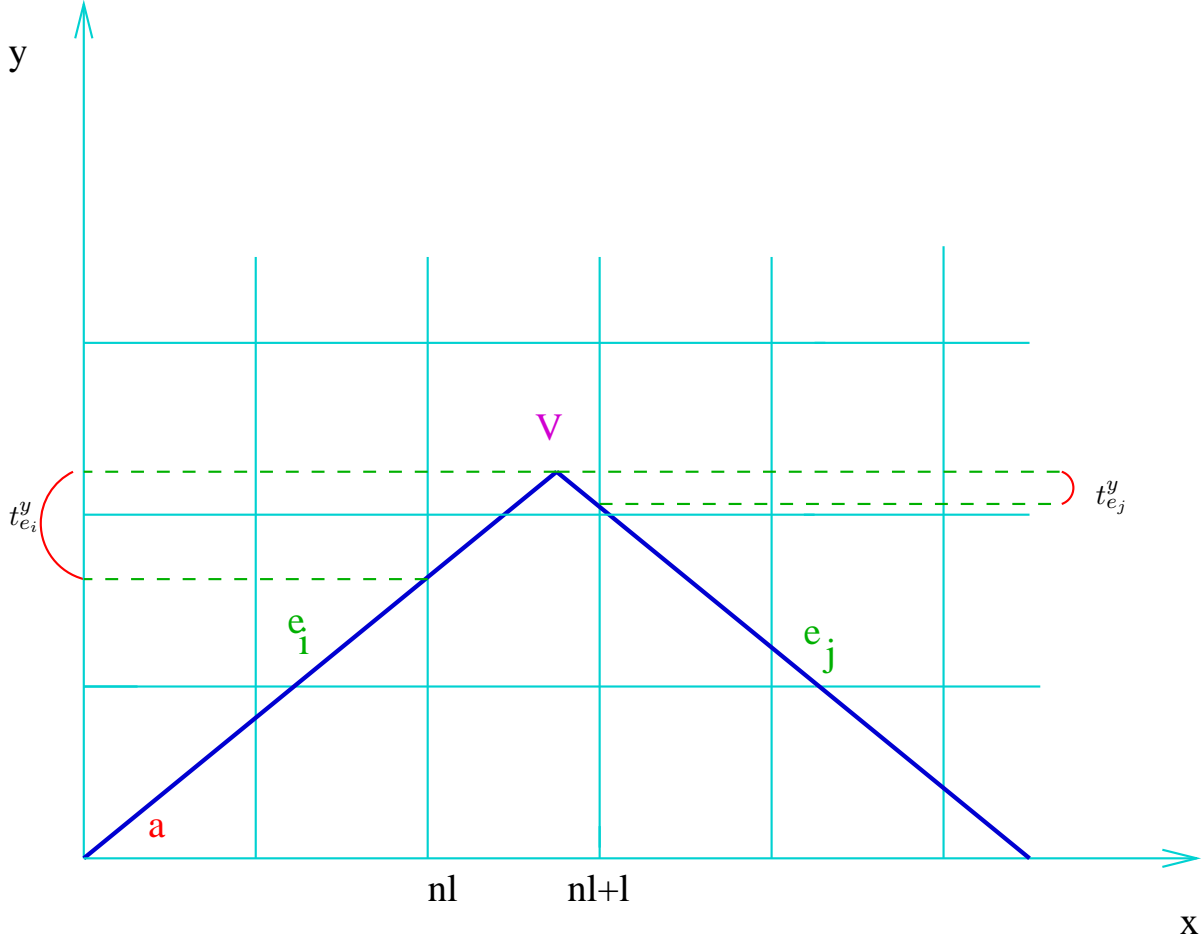}
\caption{Example in two dimensions \label{fig:2d}}
\end{center}
\end{figure}

Let us consider Figure \ref{fig:2d},
where we chose $x_{v_1}> nl+\frac{l}{2}$. For simplicity we are
assuming that the vertex is symmetric with respect to the axis,
i.e., the angles, $\phi$, made by the two edges with respect to the
$y$-axis are the same.

We now want to compute the values of
$t_{e_ie_j}=t^y_{e_ie_j}+t^x_{e_ie_j}+t^y_{e_ie_j}$ where for each
$t^k_{e_ie_j}$, $k=\{x,y,z\}$ we have $t^k_{e_ie_j}=t^k_{e_i}\cap
t^k_{e_j}$.

As a first step we compute for each edge, $e_i$, the value of
$t^y_{e_i}$ in the $y$-direction, obtaining \be
t_{e_j}^y=(nl+l-x_{v_1})\cot a\hspace{.5in}\text{ and
}\hspace{.5in} t_{e_i}^y=(x_{v_1}-nl)\cot a\ee Since the two edges
commonly intersect only one $y$ stack, in order to define the
value of $t^y_{e_ie_j}=t^y_{e_i}\cap t^y_{e_j}$ we need to
establish which of the two terms $t^y_{e_i}$ or $t^y_{e_j}$ is the
smallest. Thus, for example, \be
x_{v_1}-nl>nl+l-x_{v_1}\hspace{.5in}\text{ iff }\hspace{.5in}
x>nl+\frac{l}{2}\ee Since we have chosen $x_{v_1}> nl+\frac{l}{2}$
it follows that $t_{e_i}^y<t_{e_j}^y$ which implies that
$t^y_{e_ie_j}=t_{e_j}^y=(nl+l-x_{v_1})\cot a$. 
As can be seen from  Figure \ref{fig:2d}, there are no
intersections in the $x$ stacks, therefore we obtain \be
t_{e_ie_j}=t^y_{e_ie_j}+t^x_{e_ie_j}=t^y_{e_ie_j}\ee

 We now want to determine the values for $\frac{t_{e_{i}e_{j}}} {\sqrt{t_{e_{i}}t_{e_{j}}}}$ where, in this situation, $t_{e_{i}}=\tau_{e_{i}}^x+\tau_{e_{i}}^y=\delta\sin a+\delta\cos a=t_{e_{j}}$; therefore, $\frac{t_{e_{i}e_{j}}} {\sqrt{t_{e_{i}}t_{e_{j}}}}=\frac{(nl+l-x_{v_1})\cot a}{\sqrt{(\delta\sin a+\delta\cos a)^2}}$.

\begin{figure}[htbp]
\begin{center}
\psfrag{a}{$t^{y_1}_{e_i,e_j}$}\psfrag{b}{$t^{y_2}_{e_i,e_j}$}
\psfrag{f}{$e_j$}\psfrag{e}{$e_i$}
\includegraphics[scale=0.8]{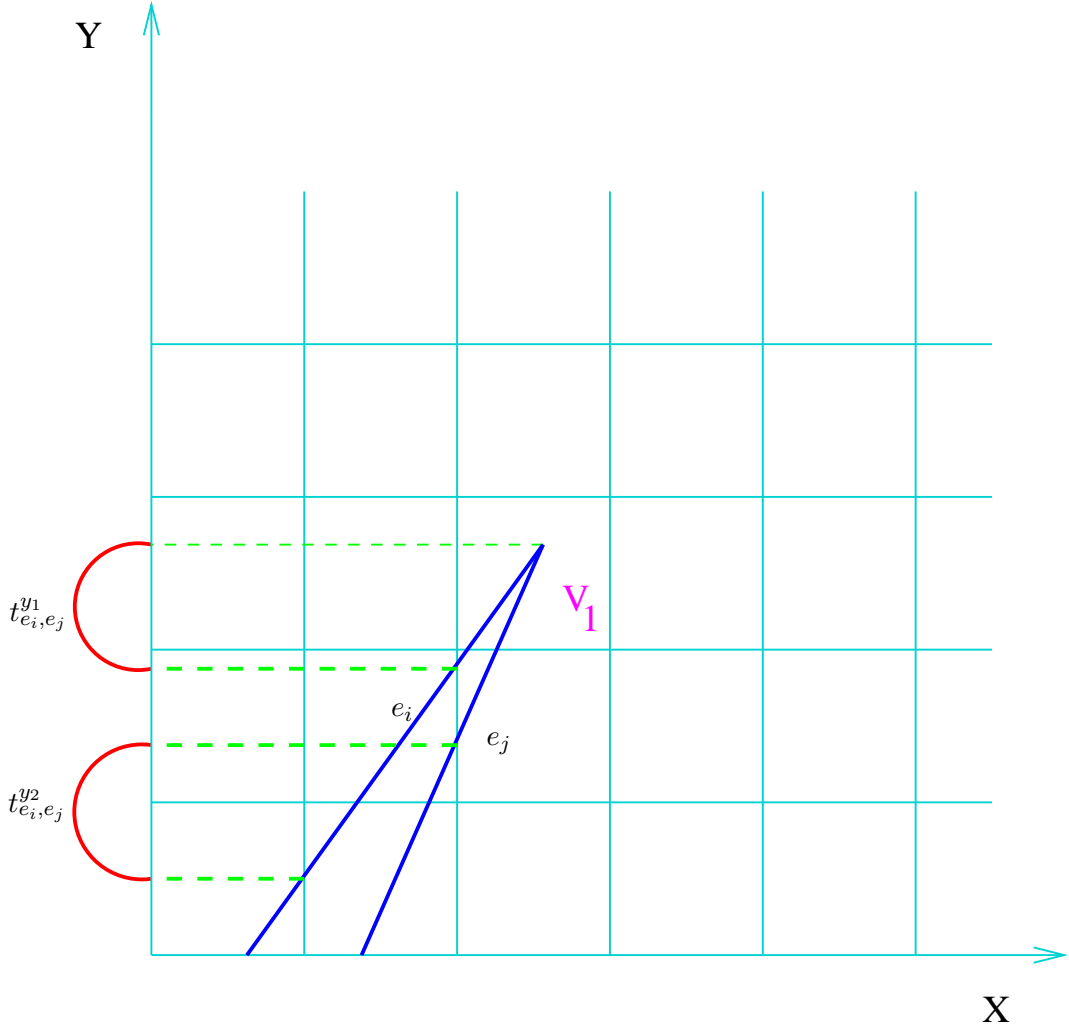}
\caption{The second example in two dimensions \label{pic:2examp}}
\end{center}
\end{figure}

This calculation is very simple since the intersection of the two
edges occurs only in one $y$ stack. But it could well be the case
that the angle between two edges is such that they intersect more
than one stack in a given direction. For example, consider Figure
\ref{pic:2examp}, always in two dimensions.
In this case we would have
$t^y_{e_ie_j}=t^{y_1}_{e_ie_j}+t^{y_2}_{e_ie_j}$.

Since in analysing the expectation value for the volume operator
we will be considering graphs formed by regular 4-, 6- and
8-valent lattice, it turns out that the angles $\theta_i$---the
angle formed by the projection on the edge on the $x$--$y$-plane
and the $x$-axis---and the angle, $\phi_i$, with respect to the
$z$-axis for any edge, are such that two or more edges can only
commonly intersect at most one plaquette in a given direction.

\begin{figure}[htbp]
\begin{center}
\psfrag{t}{$t^z_{e_i,e_j}$}\psfrag{a}{$e_i$}\psfrag{b}{$e_j$}
\includegraphics[scale=0.8]{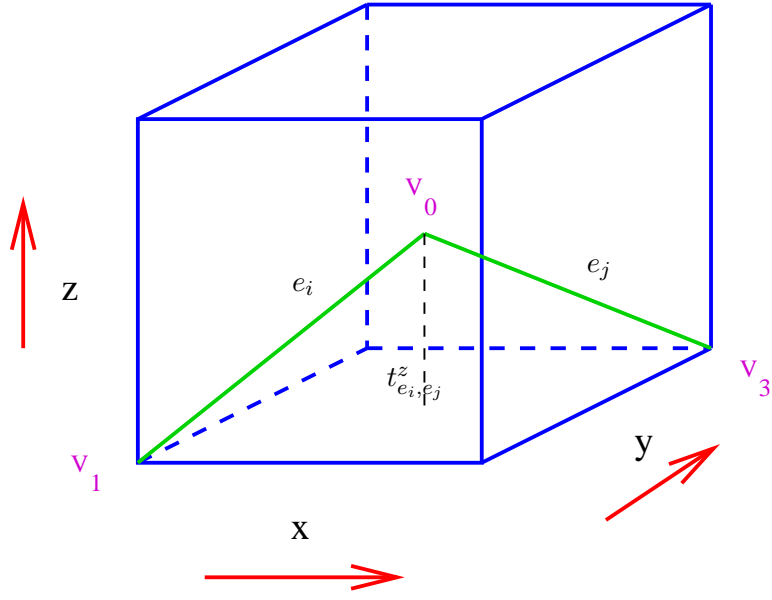}
\caption{Example in three dimensions \label{fig:3d}}
\end{center}
\end{figure}

When generalising the procedure described above for calculating
the values of $\frac{t_{e_{i}e_{j}}} {\sqrt{t_{e_{i}}t_{e_{j}}}}$
to the 3-dimensional case, some extra care is needed. In fact,
consider Figure \ref{fig:3d}.
It is clear that the values for $t^z_{e_i}$ can be computed with
respect to both the $x$- and the $y$-coordinates as follows
\begin{equation}\label{equ:te}
^xt^z_{e_i}=a\times\cot\phi\frac{1}{\cos\theta}=a\times
Z^x_{e_i}\hspace{.5in}
^yt^z_{e_i}=c\times\cot\phi\frac{1}{\sin\theta}=c\times Z^y_{e_i}
\end{equation}
where
\begin{equation*}
a=\begin{cases}
x_{V_j}-n_{x_{V_j}}l & \text{iff the edge points in the negative x direction} \\
n_{x_{V_j}}l+l-x_{V_j} & \text{iff the edge points in the positive
x direction}
\end{cases}
\end{equation*}
and
\begin{equation*}
c=\begin{cases}
y_{V_j}-n_{y_{V_j}}l & \text{iff the edge points in the negative y direction} \\
n_{y_{V_j}}l+l-y_{V_j} & \text{iff the edge points in the positive
y direction}
\end{cases}
\end{equation*}

The term $^xt^z_{e_i}$ in these equations represents the value of
$t^z_{e_i}$ as computed with respect to the $x$-coordinate, while
$^yt^z_{e_i}$ is the value of $t^z_{e_i}$ as computed with respect
to the $y$-coordinate. The non-uniqueness of the computation of
the values $t^z_{e_i}$ implies that there is an extra difficulty
in the three-dimensional case. We will illustrate this
with the aid of an example.
\begin{figure}[htbp]
\begin{center}
\psfrag{a}{$^yt^{z}_{e_i}$}\psfrag{b}{$^xt^{z}_{e_i}$}
\psfrag{e}{$e_i$}\psfrag{c}{$V_j$}
\includegraphics[scale=0.8]{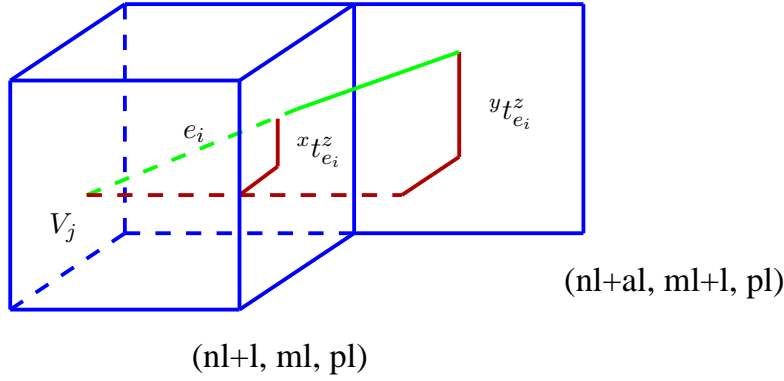}
\caption{Computation for the terms $t^z_{e_i}$ in 3-dimensions
\label{fig:difficulty}}
\end{center}
\end{figure}

Consider the edge $e_i$ in figure \ref{fig:difficulty}. The value
of $t^z_{e_i}$ can be computed with respect to both the $x$ and
the $y$ coordinate, thus obtaining $^xt^z_{e_i}$ or $^yt^z_{e_i}$
respectively. However, it is clear from the diagram that the
intersection of the edge $e_i$ with the stack of plaquettes in the
$z$ direction containing the vertex $V_j$ is given by
$^xt^z_{e_i}$. On the other hand $^yt^z_{e_i}$
defines the intersection of the edge $e_i$ with the stacks of plaquettes in the $z$ direction containing the vertex \emph{plus} the stack in the $z$ direction delimited, in the $x$ direction, by the values $nl+l$ and $nl+al$.

This example shows that, given the values
$^xt^z_{e_i}$ and $^yt^z_{e_i}$, the intersection of the edge
$e_i$ with the stacks of plaquettes in the $z$ direction which
contain the vertex $V_j$ is given by the smallest term, i.e.,
$t^z_{e_i}=^xt^z_{e_i}\cap ^yt^z_{e_i}$. It follows that, given
two edges $e_i$ and $e_j$, in order to find
$t^z_{e_i,e_j}:=t^z_{e_i}\cap t^z_{e_j}$ we first need to
establish whether $t^z_{e_i}=^xt^z_{e_i}$ or
$t^z_{e_i}=^yt^z_{e_i}$ and similarly for the edge $e_j$.
Once the value of the terms $t^z_{e_{i}}$ and $t^z_{e_{j}}$ is
determined, we can proceed as for the two-dimensional case and
identify $t^z_{e_{i}e_{j}}$ with the smallest $t^z$, i.e.
$t^z_{e_i,e_j}:=t^z_{e_i}\cap t^z_{e_j}$.

For intersections in the $x$ and $y$ stacks the procedure for
computing the values of $t_{e_{i}e_{j}}$ is essentially the same.
However, the formulae for the values of the individual terms,
$^jt^k_{e_i}$, are different. Specifically, for the $x$-direction
we have:
\begin{equation}
\label{equ:te2} ^zt^x_{e_i}=d\times\tan\phi\cos\theta=d\times
F^x_{e_i}\hspace{.5in}^yt^x_{e_i}=c\times\cot\theta=c\times
T^x_{e_i}
\end{equation}
 where
\begin{equation*}
d=\begin{cases}
z_{V_j}-n_{z_{V_j}}l & \text{iff the edge points upwards} \\
n_{z_{V_j}}l+l-z_{V_j} & \text{iff the edge points downwards}
\end{cases}
\end{equation*}
and $c$ is defined as above. For the $y$-direction we have
\begin{equation}
\label{equ:te3} ^zt^y_{e_i}=d\times\tan\phi\sin\theta=d\times
F^y_{e_i}\hspace{.5in}^xt^y_{e_i}=a\times\tan\theta=a\times
T^y_{e_i}
\end{equation}
 where $a$ and $d$ are defined as above.

When computing the values of $\frac{t_{e_{i}e_{j}}}
{\sqrt{t_{e_{i}}t_{e_{j}}}}$ in three dimensions, as for the
two-dimensional case we need to compute the values for
$t_{e_{i}}$, which in this case are simply
$t_{e_{i}}=t^x_{e_{i}}+t^y_{e_{i}}+t^z_{e_{i}}=
\delta_{e_i}\cos(90-\phi_{e_i})\cos\theta_{e_i}+\delta_{e_i}
\cos(90-\phi_{e_i})\sin\theta_{e_i}+\delta_{e_i}\cos\phi_i$. The
explicit values of the terms $\frac{t_{e_{i}e_{j}}}
{\sqrt{t_{e_{i}}t_{e_{j}}}}$ obtained for the 4-, 6-, and 8-valent
graph which satisfies condition (a) above, namely
$|nl|\leq|x_{V_2}|\leq|nl+\frac{l}{6}|$, for the 4-valent graph
and an equivalent condition for the 6- and 8-valent graphs are
given in the Appendix.

Since for all 4-, 6- and 8-valent graphs we are dealing with
symmetric lattices, after a certain number of vertices the values
for the terms $\frac{t_{e_{i}e_{j}}} {\sqrt{t_{e_{i}}t_{e_{j}}}}$
will  repeat, i.e., there will be a periodicity in the values of
the terms $\frac{t_{e_{i}e_{j}}} {\sqrt{t_{e_{i}}t_{e_{j}}}}$.
Therefore, in computing these values we need only consider those
vertices which comprise the periodicity cell, i.e., those vertices
for which the values of the term $\frac{t_{e_{i}e_{j}}}
{\sqrt{t_{e_{i}}t_{e_{j}}}}$ cannot be obtained through symmetry
arguments. As we will see later, this periodicity is different for
graphs of different valency.

We now proceed to compute the expectation value of the volume
operator for the 4-, 6- and 8-valent cases, utilising the values
of the terms $\frac{t_{e_{i}e_{j}}} {\sqrt{t_{e_{i}}t_{e_{j}}}}$
given in the Appendix.

\subsection{Analysis of the expectation value of the volume operator
for a 4-valent graph} \label{s7.2}

In this Section we will compute the expectation value of the
volume
 operator as applied to a 4-valent graph. We will first take into
 consideration the non-rotated graph. In establishing rotational
 and translational dependence of the expectation value we will perform
 both a rotation by arbitrary
 Euler angles and a translation and, then, recalculate the
 expectation value. We will see
 that the contributions that come from the terms $\frac{t_{e_{i}e_{j}}}
 {\sqrt{t_{e_{i}}t_{e_{j}}}}$, which comprise the off-diagonal elements
 of the matrix $\sqrt{A}^{-1}$, are not trivial, thereby producing a strong
 rotational and translational dependence in the expectation value of the
volume operator in higher order in $\frac{l}{\delta}$.

\subsubsection{Expectation value of the volume operator for a
4-valent graph} \label{s7.2.1} 
\begin{figure}[htbp]
\begin{center}
 \includegraphics[scale=0.6]{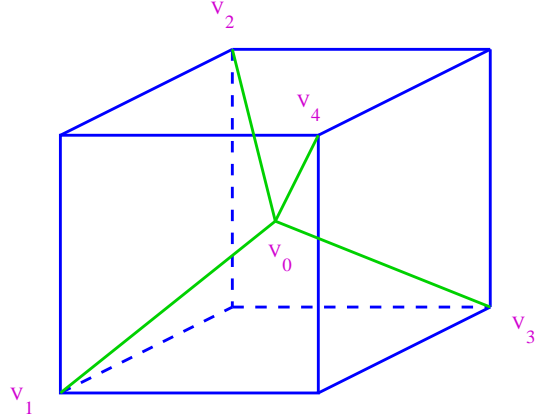}
\caption{General vertex of a 4-valent graph \label{fig:46a}}
\end{center}
\end{figure}

To calculate the expectation value
of the volume operator we will consider a 4-valent graph
constructed from the simplicial cell complex as discussed in
\cite{30b}. We choose the vertex $V_0$ to be $V_0=(0,0,0)$, and
the angles $\phi_e=\cos^{-1}(\frac{1}{\sqrt{3}})$ and
$\theta_e=45^{\circ}$ for all $e\in\gamma$ such that we obtain the
configuration depicted in picture \ref{fig:46a}.
The periodicity cell for a 4-valent graph contains four vertices,
including $V_0$. The coordinates of the remaining three vertices
are $V_2 = \big(\frac{\delta}{\sqrt3},
\frac{\delta}{\sqrt3},\frac{\delta}{\sqrt3}\big)$; $V_8 =\big
(\frac{\delta}{\sqrt3}, 3\frac{\delta}{\sqrt3}
,\frac{\delta}{\sqrt3}\big)$; $V_{13} = \big(0,2
\frac{\delta}{\sqrt3} ,2\frac{\delta}{\sqrt3}\big)$.

It follows that the edges $e_{0,1}$, $e_{0,2}$, $e_{0,3}$ and
$e_{0,4}$ lie in the octants $G$, $B$, $E$ and $D$ respectively.
This implies that the geometric factor for the vertex $V_0$ will
be $G_{\gamma, V}=\sqrt{2}$. Because of the geometry of a regular
4-valent graph, it turns out that all the vertices comprising the
periodicity cell, $V_2$, $V_8$ and $V_{13}$ will have $G_{\gamma,
V_i}=\sqrt{2}$.

The following table gives the values obtained for the terms
$\frac{t_{e_{i}e_{j}}} {\sqrt{t_{e_{i}}t_{e_{j}}}}$ for the
4-valent graph that satisfies condition (a) as defined in the
previous Section, namely $|nl|\leq|x_{V_2}|\leq|nl+\frac{l}{6}|$.

It should be noted that, because of the geometry of the 4-valent
graph, the terms
$T^{\{x,y\}}_{e_i}$,$F^{\{x,y\}}_{e_i}$,$Z^{\{x,y\}}_{e_i}$ in
equations \ref{equ:te}, \ref{equ:te2} and \ref{equ:te3} are all
equal to $1$ for each edge $e_i$:
\begin{center}
\begin{tabular}{l|l}
$V_0$& $\text{six terms } \frac{t_{e_{i}e_{j}}}
{\sqrt{t_{e_{i}}t_{e_{j}}}}=0$\\ \hline $V_2$& $\text{six terms
}\frac{t_{e_{i}e_{j}}}
{\sqrt{t_{e_{i}}t_{e_{j}}}}=\big(\frac{\delta}{\sqrt{3}}-nl\big)\frac{1}{\delta\sqrt{3}}$\\
\hline $V_8$& $\text{six terms }\frac{t_{e_{i}e_{j}}}
{\sqrt{t_{e_{i}}t_{e_{j}}}}=\big(\frac{\delta}{\sqrt{3}}-nl\big)\frac{1}{\delta\sqrt{3}}$\\
\hline $V_{13}$& $\text{two terms }\frac{t_{e_{i}e_{j}}}
{\sqrt{t_{e_{i}}t_{e_{j}}}}=\big(\frac{2\delta}{\sqrt{3}}-2nl\big)\frac{1}{\delta\sqrt{3}}$\\
\hline
\end{tabular}
\end{center}
Here, $\delta$ is the length of the edge $e$. In order to apply
equation (\ref{4.17}), we first need to determine the values of
the term $\det_{e,e^{'}e^{''}}(\sqrt{A}^{-1})$ for each triplet of
linearly-independent edges $e,e^{'},e^{''}$. Using the fact that,
in first-order approximation,
$(\sqrt{A})^{-1}=1-\frac{1}{2}(A-1)$, the explicit expression for
$(\sqrt{A})^{-1}$ for the 4-valent graph under consideration is
\[ \left( \begin{array}{c|ccccccccccccccc}
&e_{0,1}&e_{0,2}&e_{0,11}&e_{0,12}&e_{2,7}&e_{2,5}&e_{2,6}&e_{13,9}&e_{13,8}&e_{13,10}&e_{13,11}&e_{8,13}&e_{8,14}&e_{8,15}&e_{8,5}\\\hline
e_{0,1}&1 & 0 & 0 & 0 & 0 & 0 & 0 & 0 & 0 & 0 & 0 & 0 & 0 & 0 & 0  \\
e_{0,2}&0 & 1 & 0 & 0 & \frac{1}{2}\alpha & -\frac{1}{2}\alpha & -\frac{1}{2}\alpha & 0 & 0 & 0 & 0 & 0 & 0 & 0 & 0 \\
e_{0,11}&0 & 0 & 1 & 0 & 0 & 0 & 0 & 0 & 0 & 0 & 0 & 0 & 0 & 0 & 0  \\
e_{0,12}&0 & 0 & 0 & 1 & 0 & 0 & 0 & 0 & 0 & 0 & 0 & 0 & 0 & 0 & 0  \\
e_{2,7}&0 & -\frac{1}{2}\alpha & 0 & 0 & 1 & -\frac{1}{2}\alpha & -\frac{1}{2}\alpha & 0 & 0 & 0 & 0 & 0 & 0 & 0 & 0  \\
e_{2,5}&0 &-\frac{1}{2} \alpha & 0 & 0 & -\frac{1}{2}\alpha & 1 & -\frac{1}{2}\alpha & 0 & 0 & 0 & 0 & 0 & 0 & 0 & 0  \\
e_{2,6}&0 & -\frac{1}{2}\alpha & 0 & 0 & -\frac{1}{2}\alpha & -\frac{1}{2}\alpha & 1 & 0 & 0 & 0 & 0 & 0 & 0 & 0 & 0 \\
e_{13,9}&0 & 0 & 0 & 0 & 0 & 0 & 0 & 1 & -\frac{1}{2}\alpha & -\frac{1}{2}\alpha & -\frac{1}{2}\alpha & 0 & 0 & 0 & 0 \\
e_{13,8}&0 & 0 & 0 & 0 & 0 & 0 & 0 & -\frac{1}{2}\alpha & 1 & -\frac{1}{2}\alpha & -\frac{1}{2}\alpha & 0 & 0 & 0 & 0 \\
e_{13,10}&0 & 0 & 0 & 0 & 0 & 0 & 0 & -\frac{1}{2}\alpha & -\frac{1}{2}\alpha & 1 & -\frac{1}{2}\alpha & 0 & 0 & 0 & 0  \\
e_{13,11}&0 & 0 & 0 & 0 & 0 & 0 & 0 & -\frac{1}{2}\alpha & -\frac{1}{2}\alpha & -\frac{1}{2}\alpha & 1 & 0& 0 & 0 & 0 \\
e_{8,13}&0 & 0 & 0 & 0 & 0 & 0 & 0 & 0 & 0 & 0 & 0 & 1 & 0 &  -\alpha & 0  \\
e_{8,14}&0 & 0 & 0 & 0 & 0 & 0 & 0 & 0 & 0 & 0 & 0 & 0 & 1 & 0 &  -\alpha  \\
e_{8,15}&0 & 0 & 0 & 0 & 0 & 0 & 0 & 0 & 0 & 0 & 0 &  -\alpha & 0 & 1 & 0 \\
e_{8,5}&0 & 0 & 0 & 0 & 0 & 0 & 0 & 0 & 0 & 0 & 0 & 0 & -\alpha & 0 & 1  \\
 \end{array} \right)\]
where $\alpha=\frac{t_{e_{i}e_{j}}}
{\sqrt{t_{e_{i}}t_{e_{j}}}}=\big(\frac{\delta}{\sqrt{3}}-nl\big)\frac{1}{\delta\sqrt{3}}$
and the terms $t_{e_{i}e_{j}}$, $t_{e_{i}}$ and $t_{e_{j}}$ are
computed using the techniques defined in the previous Section.

Now that we have an expression for the inverse of the matrix
$\sqrt{A}$ we can compute the expectation value of the volume
operator for each of the four vertices in the periodicity cell
and, then, sum their contributions.

We start with the vertex $V_0$. First consider the sub-matrix of
the matrix $(\sqrt{A})^{-1}$ formed by all the edges incident at
$V_0$. This is
\[\left( \begin{array}{c|ccccc}
&e_{0,1}&e_{0,2}&e_{0,11}&e_{0,12}\\\hline
e_{0,1}&1 & 0 & 0 & 0  \\
e_{0,2}&0 & 1 & 0 & 0  \\
e_{0,11}&0 & 0 & 1 & 0  \\
e_{0,12}&0 & 0 & 0 & 1 \\
 \end{array} \right)\]
Because of the geometry of the 4-valent graph, at each vertex
there are four triplets of linearly-independent edges.  Keeping
this in mind and computing the determinant of the matrices formed
by each such set of triplets, we obtain the following expression
for the expectation value of the volume operator at $V_0$:
\begin{equation}
\delta^3\sqrt{\frac{1}{8}}\sqrt{\det\big(E^a_j(u)\big)}\;\Big|16\big|\det\Big(\frac{\partial
X_S^a}{\partial(s,u^1,u^2)}\Big)\big|^2\Big|^{\frac{1}{2}}
\end{equation}
By a similar procedure for vertices $V_2$ and $V_3$ we obtain
\begin{equation}
\delta^3\sqrt{\frac{1}{8}}\sqrt{\det\big(E^a_j(u)\big)}\;\Big|16\Big(1-\frac{3\alpha^2}{4}-
\frac{\alpha^3}{4}\Big)\,\big|\det\Big(\frac{\partial
X_S^a}{\partial(s,u^1,u^2)}\Big)\big|^2\Big|^{\frac{1}{2}}
\end{equation}

In both cases, the sub-matrix of $\sqrt{A}^{-1}$ we consider is
\[\left( \begin{array}{cccc}
1 &  -\frac{1}{2}\alpha &  -\frac{1}{2}\alpha &  -\frac{1}{2}\alpha  \\
 -\frac{1}{2}\alpha & 1 &  -\frac{1}{2}\alpha &  -\frac{1}{2}\alpha  \\
 -\frac{1}{2}\alpha &  -\frac{1}{2}\alpha & 1 &  -\frac{1}{2}\alpha  \\
 -\frac{1}{2}\alpha &  -\frac{1}{2}\alpha &  -\frac{1}{2}\alpha & 1 \\
 \end{array} \right)\]
and then we compute the determinant of all the sub-matrices formed
by linearly-independent triplets of edges.

For the vertex $V_{4}$ we obtain
\begin{equation}
\delta^3\sqrt{\frac{1}{8}}\sqrt{\det(E^a_j(u)\big)}\;
\Big|16(1-\alpha^2)\Big|\det\Big(\frac{\partial
X_S^a}{\partial(s,u^1,u^2)}\Big)\Big|^2\Big|^{\frac{1}{2}}
\end{equation}
where we have used the sub-matrix
\[\left( \begin{array}{cccc}
1 & 0 & -\alpha & 0  \\
0 & 1 & 0 & -\alpha  \\
-\alpha & 0 & 1 & 0  \\
0 & -\alpha & 0 & 1 \\
 \end{array} \right)\]
Summing up these contributions gives
\begin{align}\label{ali:v4}
V_R&=\delta^3\sqrt{\frac{1}{8}}\;4\;\Big|\det\Big(\frac{\partial
X_S^a}{\partial(s,u^1,u^2)}\Big)\Big|
\;\;\Big|\Big(1\sqrt{|\det\big(E^a_j(u)\big)|}+\sqrt{|\det\big(E^a_j(u)\big)|}\Big(\big|1-\frac{3\alpha^2}{4}-
\frac{\alpha^3}{4}\,\big|\Big)^{\frac{1}{2}}\nonumber\\
&+\sqrt{|\det\big(E^a_j(u)\big)|}\Big(\big||1-\frac{3\alpha^2}{4}-
\frac{\alpha^3}{4}\,\big|\Big)^{\frac{1}{2}}+\sqrt{|\det\big(E^a_j(u)\big)|}\Big(\big|1-\alpha^2\,\big|\Big)^{\frac{1}{2}}
\end{align}
To first-order approximation we obtain
\begin{align}
V_R&=\delta^3
2\sqrt{2}\sqrt{|\det\big(E^a_j(u)\big)|}\;\Big|\det\Big(\frac{\partial
X_S^a}{\partial(s,u^1,u^2)}\Big)\;\;\Big|1 -\frac{1}{2}\Big(
\frac{3\alpha^2}{16} -\frac{\alpha^2}{4}\Big)+\mathcal{O}(3)
\end{align}

It should be noted that, although the term $\det(E^a_j(u))$ is
vertex dependent, we can safely assume that, to first-order in
$l/\delta$, the values will be the same for each vertex within
each periodicity cell that involves only an order of four
vertices. Thus this term can be factored out from the equation.
This first-order approximation will be used throughout the rest of
this paper. As mentioned previously this is justified since we
choose $\frac{l}{\delta}<<1$. It follows that the terms
$\alpha^2\propto \frac{l-x}{\delta}$, which are much smaller than
\emph{one} (see Section \ref{s4.2}).

The term proportional to $\alpha^2$ in the equation above
represents the $l/\delta$-correction for the expectation value of
the volume operator for a given region $R$. As in \cite{30b}, for
a general 4-valent graph, even in the zeroth-order approximation
the expectation value for the volume of a given region $R$ does
not coincide with the classical value for the volume of that
region. Notably, there is no linear correction in $l/\delta$!

\subsubsection{Expectation value of the volume operator for a rotated
4-valent graph} \label{s7.2.2} We now analyse how the results of
the calculations above depend on how the graph is embedded in
$\Rl^3$. Here we will consider rotational invariance;
translational invariance is discussed in the following subsection.

To analyse the rotational dependence of the expectation value of
the volume operator, we will perform an Euler rotation of the
graph with respect to some arbitrary Euler angles $\beta$, $\psi$,
$\alpha$ and then repeat the calculation. The transformation
matrix is
\[R= \left( \begin{array}{ccc}\label{a1}
\cos\p \cos\s-\sin\p \cos\te \sin\psi  & \cos\psi \sin\p+\cos\te \cos\p \sin\psi  & \sin\psi \sin\te   \\
-\sin\psi \cos\p-\cos\te \sin\p \cos\psi & -\sin\psi \sin\p+ \cos\te \cos\p \sin\psi & \cos\psi \sin\te \\
\sin\te \sin\p & -\sin\te \cos\p & \cos\te\end{array} \right)\]
The coordinates of the rotated vertices are then given by
$V^{\prime}_i=R\cdot\vec{V_i}$
\[\left(\begin{array}{c}
x_{V^{\prime}_i}\\
y_{V^{\prime}_i}\\
z_{V^{\prime}_i}
\end{array} \right)=
\left( \begin{array}{c}
R_{11}x_{v_i} + R_{12}y_{v_i} + R_{13}z_{v_i}  \\
R_{21}x_{v_i} + R_{22}y_{v_i} + R_{23}z_{v_i} \\
R_{31}x_{v_i} + R_{32}y_{v_i} + R_{33}z_{v_i}\end{array} \right)\]

Applying this transformation matrix to the 4-valent graph we
obtain the following new coordinates for the vertices:
\begin{eqnarray}
V^{\prime}_2&=&\Big((-R_{11}+R_{12}+R_{13})\frac{\delta}{\sqrt{3}},
(-R_{21}+R_{22}+R_{23})\frac{\delta}{\sqrt{3}},
(-R_{31}+R_{32}+R_{33})\frac{\delta}{\sqrt{3}}\Big)\\
V^{\prime}_8&=&\Big((R_{11}+3R_{12}+R_{13})\frac{\delta}{\sqrt{3}},
(R_{21}+3R_{22}+R_{23})\frac{\delta}{\sqrt{3}},(R_{31}+3R_{32}+R_{33})
\frac{\delta}{\sqrt{3}}\Big)\\
V^{\prime}_{13}&=&\Big(2(R_{12}-R_{13})\frac{\delta}{\sqrt{3}},
2(R_{22}-R_{23})\frac{\delta}{\sqrt{3}},2(R_{32}-R_{33})\frac{\delta}{\sqrt{3}}\Big)
\end{eqnarray}
The new angles between the rotated edges and the $x$,$y$,$z$-axes
can now easily be computed using elementary trigonometry.

As an explanatory example let us consider the edge $e_{0,2}$. To
find the angles this edge has with respect to the axes we need
first to compute the coordinates of the vector $\vec{e_{0,2}}$
starting at vertex $V_0$ and ending at vertex $V_2$. In this case,
the coordinates of $\vec{e_{0,2}}$ coincide with the coordinates
of the vertex $V_2$:  \be\vec{e_{2,0}} = \Big((−R_{11} + R_{12}
+ R_{13})\frac{\delta}{\sqrt{3}}, (−R_{21} + R_{22} + R_{23}
)\frac{\delta}{\sqrt{3}} , (−R_{31} + R_{32} + R_{33} )
\frac{\delta}{\sqrt{3}}\Big)\ee If instead we considered the edge
$e_{25}$ we would get \be \vec{e_{25}}=e_{02}-e_{05}=
\Big(-(R_{11}+R_{12}+R_{13})\frac{\delta}{\sqrt{3}},-(R_{21}+R_{22}+
R_{23})\frac{\delta}{\sqrt{3}},-(R_{31}+R_{32}+R_{33})
\frac{\delta}{\sqrt{3}}\Big)\ee

Once we have the coordinates for $\vec{e_{2,0}}$, the angle,
$\phi_{e_{2,0}}$, it forms with respect to the $z$-coordinate is
\be \label{equ:p}
\phi_{e_{2,0}}=\tan^{-1}\Big(\frac{\sqrt{x^2+y^2}}{z}\Big)=
\tan^{-1}\Big(\frac{\sqrt{(-R_{11}+R_{12}+R_{13})^2+(-R_{21}+
R_{22}+R_{23})^2}}{(-R_{31}+R_{32}+R_{33})}\Big)\ee The angle,
$\theta_{e_{2,0}}$, between the projection of the vector on the
$x$--$y$ plane and the $x$-axis is given by \be\label{equ:t}
\theta_{e_{2,0}}=\tan^{-1}\Big(\frac{y}{x}\Big)=
\tan^{-1}\left(\frac{-R_{21}+R_{22}+R_{23}}{-R_{11}+R_{12}+R_{13}}\right)\ee

In the same way we can obtain the angles for all the edges in our
graph in terms of the  elements of the transformation matrix. Thus
the orientation of each of the edges of the graph will depend on
the matrix elements of the transformation matrix, i.e., on the
Euler angles that parameterise the rotation.

In order to determine the rotational dependence of the expectation
value of the volume operator, we have performed a case study in
which the expectation values were computed for all possible
orientations of the graphs that have non-zero measure in $SO(3)$.
Such possible orientations were described in Section \ref{s4.4}.
To aid calculational simplicity, these sub-cases are  defined in
terms of possible ranges of values for the angles $\theta$ and
$\phi$ for each edge in the graph, rather than on possible values
for the Euler angles.

In order to keep our results as general as possible, we performed
our subdivisions so as to cover all possible situations. This is
less tedious than it might seem since we are dealing with regular
lattices and, therefore, once  the angles for the edges of one
vertex are fixed, we immediately know the orientation of the edges
of all other vertices.

Let us choose $V^{\prime}_0$ as our reference vertex with respect
to which the possible orientations of the edges are defined. The
edges incident at $V^{\prime}_0$ are $e_{0,1}$ , $e_{0,2}$ ,
$e_{0,3}$ , $e_{0,4}$. In defining the orientation we  use the
convention that both $0<\phi<2\pi$ and $0<\theta<2\pi$ increases
anti-clockwise.

Once  the rotational matrix has been applied, whatever the values
of the Euler angles might be, we will end up in a situation in
which two edges $e_i$, $e_j$ point upwards, i.e.,
$-\frac{\pi}{2}<\phi_{e_i}, \phi_{e_j}<\frac{\pi}{2}$, and the
remaining two edges point downwards, i.e.
$\frac{\pi}{2}<\phi_{e_k}, \phi_{e_l}<\frac{3\pi}{2}$. This is a
consequence of the geometry of the 4-valent graph. We will call
two edges pointing in the same up, or down, direction an `up', or
`down', couple respectively. Since we are considering only those
edge orientation with measure \emph{non-zero} in $SO(3)$, the angles
of the edges $e_i$, $e_j$ of each up/down couple will satisfy the
following conditions: $|\theta_{e_i}|=|90+\theta_{e_j}|$ and
$|\phi_{e_i}|=|\phi_{e_j}-54.75|$

Given a particular choice of up and down couple we have to
specify in which octant (see Figure \ref{figdivision})
 each edge lies. This is required since different octants
induce different values for the geometric factor $G_{\gamma,V}$.
The angles $\theta_{e_i}$ and $\phi_{e_i}$ required for an edge
$e_i$ to lie in each of the octants are listed in Table
\ref{tab:4g} where, again, we use the convention that
$0<\phi_{e_i}<2\pi$ and $0<\theta_{e_i}<2\pi$, with both angles increasing in
an anticlockwise direction.
\begin{table}[h] \footnotesize
\begin{center}
\begin{tabular}{l|l|l|l|l}
& A & B & C & D \\ \hline $\phi_{e_i}$&
$\frac{3\pi}{2}<\phi_{e_i}<2\pi$ & $0<\phi_{e_i}<\frac{\pi}{2}$
&$0<\phi_{e_i}<\frac{\pi}{2}$  &
$\frac{3\pi}{2}<\phi_{e_i}<2\pi$\\ \hline
$\theta_{e_i}$&$0<\theta_{e_i}<\frac{\pi}{2}$&
$\frac{\pi}{2}<\theta_{e_i}<\pi$
&$\pi<\theta_{e_i}<\frac{3\pi}{2}$ &
$\frac{3\pi}{2}<\theta_{e_i}<2\pi$\\ \hline
\end{tabular}
 \end{center}
 \begin{center}
\begin{tabular}{l|l|l|l|l}
&  E & F & G & H\\ \hline $\phi_{e_i}$&
$\pi<\phi_{e_i}<\frac{3\pi}{2}$ &$\frac{\pi}{2}<\phi_{e_i}<\pi$&
$\frac{\pi}{2}<\phi_{e_i}<\pi$ &$\pi<\phi_{e_i}<\frac{3\pi}{2}$\\
\hline $\theta_{e_i}$& $0<\theta_{e_i}<\frac{\pi}{2}$&
$\frac{\pi}{2}<\theta_{e_i}<\pi$
&$\pi<\theta_{e_i}<\frac{3\pi}{2}$ &
$\frac{3\pi}{2}<\theta_{e_i}<2\pi$\\ \hline
\end{tabular}
\caption{Angle-ranges for each octant} \label{tab:4g}
\end{center}
\end{table}
However, because of the geometry of a 4-valent graph, the allowed
angle-ranges have to be restricted to those listed in Table
\ref{tab:4}.
\begin{table}[h] \footnotesize
\begin{center}
\begin{tabular}{l|l|l|l|l}
& A & B & C & D \\ \hline $\phi_{e_i}$&
$\frac{3\pi}{2}<\phi_{e_i}<2\pi-\sin^{-1}(\frac{1}{3}$ &
$\sin^{-1}(\frac{1}{3}<\phi_{e_i}<\frac{\pi}{2}$
&$\sin^{-1}(\frac{1}{3}<\phi_{e_i}<\frac{\pi}{2}$  &
$\frac{3\pi}{2}<\phi_{e_i}<2\pi-\sin^{-1}(\frac{1}{3}$\\ \hline
$\theta_{e_i}$&$0<\theta_{e_i}<\frac{\pi}{2}$&
$\frac{\pi}{2}<\theta_{e_i}<\pi$
&$\pi<\theta_{e_i}<\frac{3\pi}{2}$ &
$\frac{3\pi}{2}<\theta_{e_i}<2\pi$\\ \hline
\end{tabular}
 \end{center}
 \begin{center}
\begin{tabular}{l|l|l|l|l}
&  E & F & G & H\\ \hline $\phi_{e_i}$&
$\pi+\sin^{-1}(\frac{1}{3})<\phi_{e_i}<\frac{3\pi}{2}$
&$\frac{\pi}{2}<\phi_{e_i}<\pi-\sin^{-1}(\frac{1}{3})$&
$\frac{\pi}{2}<\phi_{e_i}<\pi-\sin^{-1}(\frac{1}{3})$
&$\pi+\sin^{-1}(\frac{1}{3}<\phi_{e_i}<\frac{3\pi}{2}$\\ \hline
$\theta_{e_i}$& $0<\theta_{e_i}<\frac{\pi}{2}$&
$\frac{\pi}{2}<\theta_{e_i}<\pi$
&$\pi<\theta_{e_i}<\frac{3\pi}{2}$ &
$\frac{3\pi}{2}<\theta_{e_i}<2\pi$\\ \hline
\end{tabular}
\caption{4-valent graph angle-ranges for each octant \label{tab:4}
}
\end{center}
\end{table}

Our calculations show that for all possible sub-cases of angle
arrangements defined in Table \ref{tab:4}, the expectation value
for the volume operator is rotational invariant \emph{only}
at the zeroth-order\footnote{This is a consequence of the fact
that the geometric factors $G_{\gamma, V_i}$ for each of the
sub-cases in Table \ref{tab:4} will be the same (see Section
\ref{s 2.4.1})}, while higher-order terms \emph{are}
rotationally dependent. Therefore, in what follows, we will not
compute the expectation value for the volume operator as computed
for each possible orientation of the graph. Instead we will choose
a particular sub-case of Table \ref{tab:4} and compute the
expectation value for such a sub-case. Specifically we will choose
the case in which the  arrangement of edges, incident at the
vertex $V_0$, after a rotation, is given by the following ranges:
\begin{align}
\label{ali:angle}
&0<\theta_{e_{0,2}}<\frac{\pi}{2}\hspace{.5in}\frac{3\pi}{2}<\phi_{e_{0,2}}<2\pi-\sin^{-1}\frac{1}{3}\nonumber\\ &\pi<\theta_{e_{0,4}}<\frac{3\pi}{2}\hspace{.5in}\sin^{-1}(\frac{1}{3})<\phi_{e_{0,4}}<\frac{\pi}{2}\nonumber\\
&\frac{\pi}{2}<\theta_{e_{0,1}}<\pi\hspace{.5in}\frac{3\pi}{2}<\phi_{e_{0,1}}<\pi-\sin^{-1}(\frac{1}{3})\nonumber\\
&\frac{3\pi}{2}<\theta_{e_{0,3}}<2\pi\hspace{.5in}\pi+\sin^{-1}(\frac{1}{3})<\phi_{e_{0,3}}<\frac{3\pi}{2}
\end{align}
This implies that the edges $e_{0,2}$, $e_{0,3}$, $e_{0,4}$ and
$e_{0,1}$ lie in the octants $A$, $H$, $C$ and $F$. From the
geometry of the 4-valent lattice, the angles of the edges incident
at all the other vertices follow.

There is a vast range of Euler angles for which the case above is
obtained but, for the sake of brevity, we will not list them
here. What is important, though, is that such case has a non-zero
measure in $SO(3)$.

It should be noted that different combinations of angles within
the angle ranges in (\ref{ali:angle}) lead to different outcomes for
the expectation value of the volume operator, since they lead to
different values of the terms
$\frac{t_{e_ie_j}}{\sqrt{t_{e_i}t_{e_j}}}$. However, in
zeroth-order, the expectation value of the volume operator will be
the same irrespectively of which angles satisfying (\ref{ali:angle})
we decide to utilise. In fact, the rotational dependence of the
expectation value of the volume operator, in the zeroth-order in
$\frac{l}{\delta}$, is determined \emph{solely} by the geometric
factors $G_{\gamma, V_i}$. For the case which we are analysing
(\ref{ali:angle}), the values of $G_{\gamma, V_i}$ will be the same
irrespectively of which sub-case of (\ref{ali:angle}) we analyse. On
the other hand, the dependence of the expectation value of the
volume operator on higher orders of $\frac{l}{\delta}$ \emph{is}
determined by the terms $\frac{t_{e_ie_j}}{\sqrt{t_{e_i}t_{e_j}}}$
and, therefore, will depend on the sub-cases we analyse.

This discussion shows that for higher orders in $\frac{l}{\delta}$ the expectation value of the volume operator \emph{is} rotational dependent since, as stated above, for differing angle-ranges that lead to the same geometric factors, the values of the terms $\frac{t_{e_ie_j}}{\sqrt{t_{e_i}t_{e_j}}}$ will differ.\\

We will now compute the expectation value of the volume operator
for the periodicity cell in the 4-valent graph for the case in
which the angles of the edges incident at vertex $V_0$ satisfy
condition (\ref{ali:angle}).
The first step in order to compute the expectation value of the
volume operator is to define the matrix $\sqrt{A}^{-1}$, whose
off-diagonal entries are the terms
$\frac{t_{e_ie_j}}{\sqrt{t_{e_i}t_{e_j}}}$. This matrix
is given in Table \ref{tab:a4} in the Appendix. Although different combinations of angles satisfying condition (\ref{ali:angle}) will lead to different values of the terms $\frac{t_{e_ie_j}}{\sqrt{t_{e_i}t_{e_j}}}$, however, any such combination will lead to the same non-zero entries of the matrix $\sqrt{A}^{-1}$. This means that the pairs of edges commonly intersecting a plaquette in a given direction will coincide for \emph{any} combination of angles satisfying conditions (\ref{ali:angle}), even though the number $t_{e_i,e_j}$ of plaquettes they commonly intersect will differ in each case. Therefore in computing the matrix $\sqrt{A}^{-1}$ we will not determine the precise value of the individual entries, but we will leave them as general as possible. Their precise values can be computed once a specific combination of angles satisfying (\ref{ali:angle}) is chosen.

Given the matrix $\sqrt{A}^{-1}$ we are then able to apply formula
(\ref{4.29}) for computing the expectation value of the volume
operator. As in the aligned case, we first compute the expectation
value of the volume operator for each of the four vertices, and
then  sum their contributions. In what follows,  the term
$\frac{t_{e_i,e_j}}{\sqrt{t_{e_i}t_{e_j}}}$ is denoted by
$\alpha_{e_i,e_j}=\frac{t_{e_i,e_j}}{\sqrt{t_{e_i}t_{e_j}}}$. An
explicit form for these terms can be found in Section 11.1 of the
Appendix.

The expectation value for the volume operator for the entire
periodicity cell, up to first-order in $\frac{l}{\delta}$ is:
\begin{align}
V_R=&\delta^3\sqrt{\frac{1}{8}}\Big|\det\Big(\frac{\partial
X_S^a}{\partial(s,u^1,u^2)}\Big)\Big|\Big(
4\sqrt{|det\big(E^a_j(u)\big)|} \nonumber\\
&+ 4\sqrt{|1 +\frac{1}{2}(- \alpha_{e_{8,10},e_{8,11}}^2 -
\alpha_{e_{8,10},e_{8,12}}^2 - \alpha_{e_{8,10},e_{8,5}}^2 - \alpha_{e_{8,11},e_{8,12}}^2 - \alpha_{e_{8,11},e_{8,5}}^2 - \alpha_{e_{8,12},e_{8,5}}^2)|}\sqrt{|det\big(E^a_j(u)\big)|}\nonumber\\
&+ 4\sqrt{|1 +\frac{1}{2}(- \alpha_{e_{13,18},e_{13,15}}^2 - \alpha_{e_{13,19},e_{13,15}}^2 - \alpha_{e_{13,19},e_{13,18}}^2 -\alpha_{e_{13,3},e_{13,15}}^2 - \alpha_{e_{13,3},e_{13,18}}^2 - \alpha_{e_{13,3},e_{13,19}}^2)|} \sqrt{|det\big(E^a_j(u)\big)|}\nonumber\\
&+ 4\sqrt{|1 +\frac{1}{2}(- \alpha_{e_{2,0},e_{2,5}}^2 -
\alpha_{e_{2,0},e_{2,6}}^2 -  \alpha_{e_{2,0},e_{2,7}}^2 -
\alpha_{e_{2,5},e_{2,6}}^2 - \alpha_{e_{2,5},e_{2,7}}^2 -
\alpha_{e_{2,6},e_{2,7}}^2)|}
\sqrt{|det\big(E^a_j(u)\big)|}\;\Big)+\mathcal{O}(3)
 \end{align}
By performing a Taylor expansion for each of the roots present in
the above formula, we can factor out the term
$\sqrt{|\det\big(E^a_j(u)\big)|}$ since, in the first-order
approximation that we are considering,
 they turn out to be the same for each
 vertex. Such an approximation is justified by the analysis performed in Section (\ref{s4.2}).
We then obtain
 \begin{align}
V_R=&\delta^32\sqrt{2}\sqrt{|\det\big(E^a_j(u)\big)|}
\Big(1+\frac{1}{4}( - \alpha_{e_{8,10},e_{8,11}}^2 -
\alpha_{e_{8,10},e_{8,12}}^2 - \alpha_{e_{8,10},e_{8,5}}^2 - \alpha_{e_{8,11},e_{8,12}}^2 - \alpha_{e_{8,11},e_{8,5}}^2 - \alpha_{e_{8,12},e_{8,5}}^2-\nonumber\\
& - \alpha_{e_{13,18},e_{13,15}}^2 - \alpha_{e_{13,19},e_{13,15}}^2 - \alpha_{e_{13,19},e_{13,18}}^2 -\alpha_{e_{13,3},e_{13,15}}^2 - \alpha_{e_{13,3},e_{13,18}}^2 - \alpha_{e_{13,3},e_{13,19}}^2- \\
& \alpha_{e_{2,0},e_{2,5}}^2 - \alpha_{e_{2,0},e_{2,6}}^2 - \alpha_{e_{2,0},e_{2,7}}^2 - \alpha_{e_{2,5},e_{2,6}}^2 - \alpha_{e_{2,5},e_{2,7}}^2 - \alpha_{e_{2,6},e_{2,7}}^2)\Big)\\
&=\delta^3\sqrt{|\det\big(E^a_j(u)\big)|}2\sqrt{2}\bigg(1
+\frac{1}{4}\big(-\sum^{\frac{n}{2}(n-1)}_{i,j=1;j\neq
}\alpha_{ji}^2\big)\;\Big|\det\Big(\frac{\delta
X_S^a}{\delta(s,u^1,u^2)}\Big)\Big|\bigg)
 \end{align}
The term $\frac{1}{2}(-\sum^{\frac{n}{2}(n-1)}_{i,j=1;j\neq
}\alpha^2_{ji})$ represents $l/\delta$-corrections. Each term
$\alpha_{i,j}$ is proportional to $C\frac{l^{\prime}}{\delta}$ for
$l^{\prime}<l$; $C$ is a constant that depends on the Euler angles
we chose. On the other hand the geometric factors
$G_{\gamma,V_{i}}$ for cases (\ref{ali:angle}) coincide with the
geometric factors as computed for any of the sub-cases in Table
\ref{tab:4}, i.e., $G_{\gamma, V_i}=2\sqrt{2}$. This implies that
although for such cases the expectation value of the volume
operator \emph{is} rotational invariant in zeroth-order,
nonetheless, it does \emph{not} reproduce the correct
semiclassical limit.

For those embeddings whose measure is \emph{zero} in $SO(3)$, the
geometric factor turns out to be different and, in zeroth-order in
$l/\delta$, leads to a different value of the expectation value of
the volume operator as computed for 4-valent graphs.

\subsubsection{Expectation value of the volume operator for a translated
4-valent graph} \label{s7.2.3}

In this Section we will analyse whether the expectation value of
the volume operator for the 4-valent graph is translational
invariant with respect to the plaquette.

To perform this analysis we consider our original aligned graph
and  translate it by an arbitrary vector $\vec{p}=(\epsilon_x,
\epsilon_y, \epsilon_z)$. The new coordinates for the vertices
are:
\begin{eqnarray}
V^{''}_0&=&(\epsilon_x,\epsilon_y,\epsilon_z)\mbox{ with }
\epsilon_x>\epsilon_y>\epsilon_z\\
V^{''}_2&=&\Big(-\frac{\delta}{\sqrt{3}}+\epsilon_x,\epsilon_y+\frac{\delta}{\sqrt{3}},\epsilon_z+\frac{\delta}{\sqrt{3}}\Big)\\
V^{''}_8&=&\Big(\frac{\delta}{\sqrt{3}}+\epsilon_x,\epsilon_y+3\frac{\delta}{\sqrt{3}},\epsilon_z+\frac{\delta}{\sqrt{3}}\Big)\\
V^{''}_{13}&=&\Big(\epsilon_x,\epsilon_y+2\frac{\delta}{\sqrt{3}},\epsilon_z-2\frac{\delta}{\sqrt{3}}\Big)
\end{eqnarray}

Similarly to the analysis for rotational invariance,  the
computation of the expectation value of the volume operator can be
divided into different sub-cases, each of which would lead to
different outcomes.

The first division is given by the choice of the signs and the
relations between $\epsilon_x$, $\epsilon_y$ and $\epsilon_z$,
i.e., whether they are positive or negative and whether one
coordinate is bigger or equal to another. Each of these
cases can be ultimately subdivided into sub-cases depending on the
relation between the ratio $b=\frac{\delta}{\sqrt{3}}$ and the
coordinates of the translational vector.

To carry out our calculations we choose the following:
\begin{enumerate}
\item[1)]$b>\epsilon_x>\epsilon_y>\epsilon_z>0$
\item[2)]$|V^i_k|-|V_k^j|>n^i_kl-n^j_kl\mbox{ for all } |V_k^i|>|V_k^j|$
\end{enumerate}
Altogether, such conditions will allow to determine both the
sign of the coordinates for each of the vertices of the translated
graph and also the magnitude relation between the coordinates of
each vertex.

However, it will transpire that our result is independent of which
case we decide to use to perform the calculations. In fact, as we
will see, in zeroth-order, the expectation value of the volume
operator for a 4-valent graph is translation invariant up to
combinations of measure \emph{zero} in $SO(3)$. However, for
higher orders of approximation this will no longer be true.

As a first step in our calculations we need to specify the allowed
positions for each of the translated vertices. Due to the highly
symmetrical structure of the 4-valent graph, in order to determine
the allowed positions of each vertex, it suffices to find the
allowed positions of one reference vertex. We choose such a
reference vertex to be $V_0$, whose new coordinates are
$V_0=(\epsilon_x,\epsilon_y,\epsilon_z)$.

The number of stacks intersected by the vector that represents
vertex $V_0$ in the $x$, $y$ and $z$-directions are, respectively,
$n=[\frac{\epsilon_x}{l}]$, $m=[\frac{\epsilon_y}{l}]$ and
$p=[\frac{\epsilon_z}{l}]$ (where $[]$ indicates the Gauss
bracket). It follows that the allowed positions for vertex $V_0$
are given by the following ranges of each coordinate:
$nl<\epsilon_x<nl+l$, $ml<\epsilon_y<ml+l$ and
$pl<\epsilon_x<pl+l$.

It turns out that to carry out the calculations for the
expectation value of the volume operator we have to
restrict the value-range of the coordinates $\epsilon_x$,
$\epsilon_y$ and $\epsilon_z$. We choose $
nl<\epsilon_x<nl+\frac{l}{6}$, $ml<\epsilon_z<ml+\frac{l}{6}$ and
$pl<\epsilon_z<pl+\frac{l}{6}$

We will now compute the expectation value of the volume operator
of the periodicity lattice of the 4-valent graph. We will not give
the details of all the calculations involved since they are quite
lengthy. However, the method utilised is the same as for the non-translated case, namely, for each of the four vertices comprising the
periodicity cell we consider the sub-matrix of $\sqrt{A}^{-1}$
labelled by the four edges intersecting at the vertex. For each of
these sub-matrices, call them $M$, we compute the determinant of
the four $3\times 3$ sub-matrices of $M$ defined by the triplets
of linearly-independent edges. We then sum up the contributions
coming from each of the vertices. Similarly as for the aligned
4-valent graph we have
\begin{equation}
T^{\{x,y\}}_{e_i}=F^{\{x,y\}}_{e_i}=Z^{\{x,y\}}_{e_i}=1\hspace{.5in}
\forall  e_i\in \gamma
\end{equation}

The expression for the matrix $\sqrt{A}^{-1} $ is
\[ \left( \begin{array}{ccccccccccccccc}
1 & \frac{-C_0}{2} & \frac{-B_0}{2} & \frac{-C_0}{2} & 0 & 0 & 0 & 0 & 0 & 0 & 0 & 0 & 0 & 0 & 0  \\
\frac{-C_0}{2} & 1 & \frac{-C_0}{2} & \frac{-B_0}{2} & \frac{-A_2}{2} & \frac{-C_2}{2} & \frac{-A_2}{2} & 0 & 0 & 0 & 0 & 0 & 0 & 0 & 0 \\
\frac{-B_0}{2} & \frac{-C_0}{2} & 1 & \frac{-C_0}{2} & 0 & 0 & 0 & 0 & 0 & 0 & 0 & 0 & 0 & 0 & 0  \\
\frac{-C_0}{2} &\frac{-B_ 0}{2} & \frac{-C_0}{2} & 1 & 0 & 0 & 0 & 0 & 0 & 0 & 0 & 0 & 0 & 0 & 0  \\
0 & \frac{-A_2}{2}& 0 & 0 & 1 & \frac{-A_2}{2} & \frac{-C_2}{2} & 0 & 0 & 0 & 0 & 0 & 0 & 0 & 0  \\
0 &\frac{-C_2}{2}  & 0 & 0 & \frac{-A_2}{2} & 1 & \frac{-A_2}{2} & 0 & 0 & 0 & 0 & 0 & 0 & 0 & 0  \\
0 & \frac{-A_2}{2} & 0 & 0 & \frac{-C_2}{2} &\frac{-A_2}{2} & 1 &0 & 0 &0 & 0 & 0 & 0 & 0 & 0 \\
0 & 0 & 0 & 0 & 0 & 0 & 0 & 1 & \frac{-A_{13}}{2} & \frac{-C_{13}}{2} &\frac{-A_{13}}{2} & 0 & 0 & 0 & 0 \\
0 & 0 & 0 & 0 & 0 & 0 & 0 & \frac{-A_{13}}{2} & 1 & \frac{-A_{13}}{2} & \frac{-C_{13}}{2} & 0 & 0 & 0 & 0 \\
0 & 0 & 0 & 0 & 0 & 0 & 0 & \frac{-C_{13}}{2} & \frac{-A_{13}}{2} & 1 & \frac{-A_{13}}{2} & 0 & 0 & 0 & 0  \\
0 & 0 & 0 & 0 & 0 & 0 & 0 & \frac{-A_{13}}{2} & \frac{-C_{13}}{2} & \frac{-A_{13}}{2} & 1 &0&0& 0 & 0 \\
0 & 0 & 0 & 0 & 0 & 0 & 0 & 0 & 0 & 0 & 0 & 1 & \frac{-A_8}{2} & \frac{-C_8}{2} &\frac{-C_8}{2}  \\
0 & 0 & 0 & 0 & 0 & 0 & 0 & 0 & 0 & 0 & 0 &\frac{-A_8}{2} & 1 & \frac{-C_8}{2} & \frac{-C_8}{2}  \\
0 & 0 & 0 & 0 & 0 & 0 & 0 & 0 & 0 & 0 & 0 & \frac{-C_8}{2} & \frac{-C_8}{2} & 1 & \frac{-A_8}{2} \\
0 & 0 & 0 & 0 & 0 & 0 & 0 & 0 & 0 & 0 & 0 &\frac{-C_8}{2}& \frac{-C_8}{2} & \frac{-A_8}{2} & 1  \\
\end{array} \right)\]
where $A_i=(x_{V_i}-n_il)\frac{1}{\delta\sqrt{3}}$,
$B_i=(y_{V_i}-m_il)\frac{1}{\delta\sqrt{3}}$,
$C_i=(z_{V_i}-p_il)\frac{1}{\delta\sqrt{3}}$.

Applying the method described above we compute the expectation
value for the volume operator for one periodicity cell to be
\begin{align}
V_R=&\sqrt{\frac{1}{8}}\delta^3|\det\Big(\frac{\delta X_S^a}{\delta(s,u^1,u^2)}\Big)\Big|2\times\\
&\Bigg(\sqrt{|\det\Big(E^a_j(u)\Big)|}\sqrt{|4 - B_0^2 - 2 C_0^2 - B_0 C_0^2|}+\sqrt{|\det\Big(E^a_j(u)\Big)|}\sqrt{|4 - 2 A_2^2 - A_2^2 C_2 - C_2^2| }+\nonumber\\
&\sqrt{|\det\Big(E^a_j(u)\Big)|}\sqrt{|4 - 2 A_{13}^2 - A_{13}^2
C_{13} - C_{13}^2| }+\sqrt{|\det\Big(E^a_j(u)\Big)|}\sqrt{|4 -
A_8^2 - 2 C_8^2 - A_8 C_8^2|}\Bigg)
\end{align} In first-order approximation we then obtain
\begin{align}
V_R=&2\sqrt{\frac{1}{8}}\delta^3\sqrt{|\det\Big(E^a_j(u)\Big)|}
\Bigg(4 - \frac{1}{8}\Big(B_0^2 + 2 C_0^2 + B_0 C_0^2 + 2 A_2^2 + A_2^2 C_2 + C_2^2 +2 A_{13}^2+ A_{13}^2 C_{13} + C_{13}^2 +\nonumber\\
& A_8^2 + 2 C_8^2 + A_8 C_8^2\Big)\Bigg)+\mathcal{O}(4)\Bigg)\Big|\det\Big(\frac{\delta X_S^a}{\delta(s,u^1,u^2)}\Big)\Big|\nonumber\\
=&2\sqrt{2}\delta^3\sqrt{|\det\Big(E^a_j(u)\Big)|} \Bigg(1 -
\frac{1}{32}\Big(B_0^2+ 2 C_0^2 +  2 A_2^2 + C_2^2 +2A_{13}^2 +
  C_{13}^2  + A_8^2 + 2 C_8^2
\Big)+\mathcal{O}(3)\Bigg)\Big|\det\Big(\frac{\delta
X_S^a}{\delta(s,u^1,u^2)}\Big)\Big|
\end{align}
where in the last equation we have only considered first-order
contributions obtained by the usual Taylor series of the square
root (see Section \ref{s4.2}). Thus, we were able to factor out
the term $\sqrt{\det(E^a_j(u))}$. As it is evident, the
corrections of second order in $l/\delta$  are not translationally
invariant.

\subsection{Analysis of the expectation value of the volume operator
for 6-valent graphs} \label{s7.3}

 In this Section we will calculate the expectation value of
 the volume operator for a 6-valent graph. First we  consider the
 non-rotated graph, then we will analyse the rotational and
 translational dependence of the expectation value by performing a
 rotation of the graph, followed by  a translation of the graph. We will
 then recalculate the expectation value.

\subsubsection{Expectation value of the volume operator for a general
6-valent graph} \label{s7.3.1}

Similarly as  for the 4-valent graph, we will analyse
the case for which $\frac{\delta}{l}>0$; the motivation for such a
choice was given in Section \ref{s4.2}. For computational
simplicity we will position the graph so that the $(0,0,0)$
coordinates of the graph coincide with the $(0,0,0)$ coordinates
of the plaquette. We also need to align the graph in such a way
that each vertex is symmetrical with
respect to the axis. Therefore we will choose, for each vertex $V_k$, the value $\phi_{e_{k,i}}=45$ for all edges $e_{k,i}$ incident at $V_k$ and $\theta_{e_{k,i}}=45$ for four edges, while the remaining two will have $\theta_{e_{k,i}}=0$. This edge orientation corresponds to the limiting case (2) described in Section \ref{s 2.4.1}.

As for the diamond lattice, we  choose the vertex $V_2$ as our
reference vertex with respect to which we determine the allowed
positions of all the remaining vertices of the graph. We also
choose the allowed values of the $x$-coordinate of $V_2$ to be
$nl<x_{V_2}<nl+\frac{l}{6}$, where, in this case,
$n=[\frac{\delta\sqrt{2}}{2l}]$. Using the same method used
in Section 3.1 we can compute all the terms
$\frac{t_{e_{i}e_{j}}} {\sqrt{t_{e_{i}}t_{e_{j}}}}$ for the
periodicity cell of the 6-valent graph that contains nine
vertices. Given the geometry of the 6-valent graph we have the
following values for the term in equations (\ref{equ:te})--(\ref{equ:te3})
\begin{align}
Z^{\{x,y\}}_{e_i}&=\begin{cases}\frac{2}{\sqrt{2}}& \text{ iff } \theta_{e_i}\neq 0\\
1&\text{ iff } \theta_{e_i}= 0
\end{cases}
\hspace{.5in}
T^{\{x,y\}}_{e_i}=\begin{cases}\frac{\sqrt{2}}{2}& \text{ iff } \theta_{e_i}\neq 0\\
1&\text{ iff } \theta_{e_i}= 0
\end{cases}\nonumber\\
F^x_{e_i}&=\begin{cases}1 & \text{ iff } \theta_{e_i}\neq 0\\
\infty& \text{ iff } \theta_{e_i}= 0
\end{cases}\hspace{.5in}
F^x_{e_i}=1\hspace{.5in} \forall e_i\in \gamma
\end{align}

The coordinates of the vertices of the periodicity cell are
\begin{eqnarray*}
V_0&=&(0,0,0)\\
V_2&=&\big(\delta\frac{\sqrt{2}}{2},0,\delta\frac{\sqrt{2}}{2}\big)\\
V_3&=&\big((\delta\frac{\sqrt{2}}{2}-(\delta\frac{1}{2},-\delta\frac{1}{2},2\delta\frac{\sqrt{2}}{2}\big)\\
V_4&=&\big(2\delta\frac{\sqrt{2}}{2},-2\delta\frac{1}{2},2\delta\frac{\sqrt{2}}{2}\big)\\
V_5&=&\big(0,-2\delta\frac{1}{2},0\big)\\
V_{13}&=&\big(\delta\frac{\sqrt{2}}{2},-2\delta\frac{1}{2},\delta\frac{\sqrt{2}}{2}\big)\\
V_{18}&=&\big(2\delta\frac{\sqrt{2}}{2}+\delta\frac{1}{2},-1\delta\frac{1}{2},\delta\frac{\sqrt{2}}{2}\big)\\
V_{15}&=&\big(\delta\frac{\sqrt{2}}{2}-\delta,-2\delta\frac{1}{2},3\delta\frac{\sqrt{2}}{2}\big)\\
V_{14}&=&\big(\delta\frac{\sqrt{2}}{2}+\delta\frac{1}{2},-3\delta\frac{1}{2},0\big)
\end{eqnarray*}

The values for the terms $\frac{t_{e_{i}e_{j}}}
{\sqrt{t_{e_{i}}t_{e_{j}}}}$ are given in the section \ref{sa.2.1}
of the appendix

The value for the expectation value of the volume operator for one
periodicity cell consisting of \emph{nine} vertices is computed to
be
\begin{align}\label{ali:vr}
&V_R=\frac{1}{8}\delta^3\,\Big|\det\Big(\frac{\delta
X_S^a}{\delta(s,u^1,u^2)}\Big)
\Big|2\Bigg(\sqrt{8}\sqrt{|\det\Big(E^a_j(u)\Big)|}+\sqrt{|8 -
\frac{\alpha^2}{2} - \frac{23 (\alpha^{'})^2}{4} - \frac{\alpha
(\alpha^{'})^2}{4} -\frac{
 3 \alpha (\alpha^{'})^2}{4 \sqrt{2}}2}\sqrt{|\det\Big(E^a_j(u)\Big)|}\nonumber\\
&+\sqrt{8 - \frac{5 (A_{3}^{'})^2}{2} - \frac{(A_{3}^{'})^2
A_{3}}{4} - \frac{5 A_{3}^2}{2} - \sqrt{2} A_{3}^2 - \frac{
 A_{3}^{'} A_{3} \alpha^{'}}{2} - (\alpha^{'})^2 - \frac{(A_{3}^{'})^2 \beta}{4} - \frac{
( A_{3}^{'})^2 \beta}{2 \sqrt{2}} - \frac{3 \beta^2}{2} - \sqrt{2}
\beta^2 - \frac{
 3 A_{3}^{'} A_{3} \beta^{'}}{2}}\nonumber\\
&\overline{ -
 \sqrt{2} A_{3}^{'} A_{3} \beta^{'} - \frac{\alpha^{'} \beta \beta^{'}}{2} -
 \frac{\alpha^{'} \beta \beta^{'}}{2 \sqrt{2}} -  \frac{7 (\beta^{'})^2}{2} -  \frac{
 A_{3} (\beta^{'})^2}{2 \sqrt{2}}|}\sqrt{|\det\Big(E^a_j(u)\Big)|}+\sqrt{|8 - 12 \beta^2 - 8 \sqrt{2} \beta^2|}\sqrt{|\det\Big(E^a_j(u)\Big)|}\nonumber\\
&+\sqrt{|8 - 2 \alpha^2 - 9( \alpha^{'})^2 - 2 \sqrt{2} \alpha
(\alpha^{'})^2 -
 12 \beta^2 - 8 \sqrt{2} \beta^2 - 12 \alpha^{'} \beta \beta^{'} -
 9 \sqrt{2} \alpha^{'} \beta \beta^{'} - 12 (\beta^{'})^2 -
 4 \alpha (\beta^{'})^2|}\sqrt{|\det\Big(E^a_j(u)\Big)|}\nonumber\\
&+\sqrt{|8 - \frac{7 \alpha^2}{4} - \frac{7( \alpha^{'})^2}{2} -
\frac{13 \alpha (\alpha^{'})^2}{
 8 \sqrt{2}} - 2 \alpha \beta -
 2 \sqrt{2} \alpha \beta - \frac{(\alpha^{'})^2 \beta}{2} - \frac{
 3 (\alpha^{'})^2 \beta}{2 \sqrt{2}} - 6 \beta^2 - \frac{
 5 \alpha \alpha^{'} \beta^{'}}{2 \sqrt{2}} -
 2 \alpha^{'} \beta \beta^{'}- 4 (\beta^{'})^2|}\nonumber\\
&\times \sqrt{|\det\Big(E^a_j(u)\Big)|}+\sqrt{|8 - \frac{3
(A^{'}_{18})^2}{2} - \alpha^2/2 - \frac{A^{'}_{18} \alpha
\alpha^{'}}{
 2 \sqrt{2}} - \frac{3 (\alpha^{'})^2}{4} - 3 \beta^2 - 2 \sqrt{2} \beta^2 -
 A^{'}_{18} \beta \beta^{'} - \frac{3 A^{'}_{18} \beta \beta^{'}}{
 2 \sqrt{2}} - \frac{\alpha^{'} \beta \beta^{'}}{2}-}\nonumber\\
&\overline{ - \frac{
 3 \alpha^{'} \beta \beta^{'}}{4 \sqrt{2}} -
 3 (\beta^{'})^2 - \frac{\alpha (\beta^{'})^2}{2}|}\sqrt{|\det\Big(E^a_j(u)\Big)|}
+\sqrt{|8 - 3 (A^{'}_{15})^2 - \frac{(A^{'}_{15})^2
A_{15}}{\sqrt{2}} - 2 A_{15}^2 - \frac{
 3 A^{'}_{15} A_{15} \alpha^{'}}{\sqrt{2}} - 9 (\alpha^{'})^2 }\nonumber\\
&\overline{- \frac{9 A_{15} (\alpha^{'})^2}{
 4 \sqrt{2}} - \frac{3( A^{'}_{15})^2 \beta}{2} - \frac{(A^{'}_{15})^2 \beta}{\sqrt{2}} -
 3 \sqrt{2} A_{15} \beta - \frac{9 (\alpha^{'})^2 \beta}{2} - 10 \beta^2 -
 4 \sqrt{2} \beta^2 - \frac{A^{'}_{15} A_{15} \beta^{'}}{\sqrt{2}} -
 3 \alpha^{'} \beta \beta^{'} -
 3 \sqrt{2} \alpha^{'} \beta \beta^{'}}\nonumber\\
&\overline{ - 2
(\beta^{'})^2|}\sqrt{|\det\Big(E^a_j(u)\Big)|}+\sqrt{|8 - 2(
A^{'}_{14})^2 - A_{14}^2 - \frac{3 (A^{'}_{14})^2 \beta}{4} -
\frac{3 (A^{'}_{14})^2 \beta}{
 2 \sqrt{2}} - 3 \sqrt{2} A_{14} \beta} \nonumber\\
&\overline{- 18 \beta^2 -
 9 \sqrt{2} \beta^2 - \frac{3 A^{'}_{14} A_{14} \beta^{'}}{\sqrt{2}} - \frac{
 9 A^{'}_{14} \beta \beta^{'}}{2} - 9 (\beta^{'})^2|
}\sqrt{|\det\Big(E^a_j(u)\Big)|}\;\Bigg)
\end{align}
Expanding the square root (see the analysis in Section \ref{s4.2})
and considering first-order terms we obtain
\begin{align}
&V_R=2\delta^3\sqrt{|\det\big(E^a_j(u)\big)|}\,\Big|\det\Big(\frac{\delta
X_S^a}{\delta(s,u^1,u^2)}\Big)
\Big|\Bigg(9+\frac{1}{2}\times\frac{1}{8}\big(- 2 (A^{'}_{14})^2 - 3 (A^{'}_{15})^2 - \frac{3 (A^{'}_{18})^2}{2} - \frac{5 (A_{3}^{'})^2}{2} - A_{14}^2\nonumber\\
&  - 2 A_{15}^2  - \frac{5 A_{3}^2}{2} -
 \sqrt{2} A_{3}^2 - \frac{19 \alpha^2}{4}   -
 29 (\alpha^{'})^2  -
 3 \sqrt{2} A_{14} \beta - 3 \sqrt{2} A_{15} \beta - 2 \alpha \beta -
 2 \sqrt{2} \alpha \beta  - \frac{125 \beta^2}{2} -
 32 \sqrt{2} \beta^2  - \frac{67 (\beta^{'})^2}{2})\;\Bigg)
\end{align}
Contrary to \cite{30b} we find that, to zeroth-order in
$l/\delta$, the expectation value of the volume operator for a
6-valent graph does \textit{not} have the correct semiclassical
limit. On the other hand, if the graph is aligned to the
orientation of the plaquette we \emph{do} obtain the correct
semiclassical value. However, this embedding has measure
\emph{zero} in $SO(3)$.

\subsubsection{Expectation value of the volume operator for a rotated
6-valent graph} \label{s7.3.2}

We will now analyse the expectation value of the volume operator
for  a rotated 6-valent graph. As for the 4-valent graph,
different choices of Euler angles in the rotation give different
values of $t_{e_ie_j}$. Therefore, we will  once again have to
define sub-cases which are defined according to the possible
ranges of values for the angles $\phi_{e_i}$ and $\theta_{e_i}$
for each edge $e_i$. Because of the geometry of the 6-valent
lattice, we know that for any edge, $e$, of a given vertex there
exists a co-linear edge, $e^{\prime}$, which intersects the same
vertex. This implies that, given two co-linear edges $e$ and
$e^{\prime}$,  we can define the angles of $e$ (respectively
$e^{\prime}$) in terms of $e^{\prime}$ (respectively $e$) as
follows: $\phi_e=180^{\circ}-\phi_{e^{\prime}}$ and
$\theta_e=180^{\circ}-\theta_{e^{\prime}}$. Such relations reduce
the number of cases that need to be analysed.

We choose the vertex $V^{\prime}_0$ as our reference vertex. The
relations for the angles of the edges incident at $V^{\prime}_0$
are:
\begin{eqnarray}\label{eqn:cond}
\phi_{e_{0,7}}&=&180^{\circ}-\phi_{e_{0,2}}\nonumber\\
\phi_{e_{0,1}}&=&180^{\circ}-\phi_{e_{0,17}}\nonumber\\
\phi_{e_{0,6}}&=&180^{\circ}-\phi_{e_{0,8}} \nonumber\\
\theta_{e_{0,8}}&=&180^{\circ}-\theta_{e_{0,6}}\nonumber\\
\theta_{e_{0,17}}&=&180^{\circ}-\theta_{e_{0,1}}\nonumber\\
\theta_{e_{0,2}}&=&180^{\circ}-\theta_{e_{0,7}}
\end{eqnarray}
These relations imply that the allowed values of the angles of the
edges at a given vertex  fall into one of the following
groups:\begin{enumerate}
\item A given triplet of edges points upwards, i.e., their angle $\phi$ lies between $-\frac{\pi}{2}$ and $\frac{\pi}{2}$, and the triplet formed by their co-linear edges points downwards, i.e., their angle $\phi$ lies between $\frac{\pi}{2}$ and $\frac{3\pi}{2}$. This situation arises when none of the edges is aligned with one of the $x$, $y$, $z$-coordinates. However we have two distinct sub-cases which satisfy this arrangement of edges
\begin{itemize}
\item[a.] No edge lies in any plaquette.
\item[b.] Only one edge and its co-planar lie in a given plaquette (Figure \ref{fig6}).
\end{itemize}
\item A given couple of edges points upwards i.e., their angle $\phi$ lies between $-\frac{\pi}{2}$ and $\frac{\pi}{2}$, and their co-linear edges point downwards i.e., their angle $\phi$ lies between $\frac{\pi}{2}$ and $\frac{3\pi}{2}$. This situation arises when one edge (and subsequently its collinear edge) is aligned with one of the coordinates axis and, subsequently, the remaining two edges and their co-linear lie in two different plaquettes in the same direction (Figure \ref{fig:9})
\item Only one edge points upwards, i.e., its angle $\phi$ lies between $-\frac{\pi}{2}$ and $\frac{\pi}{2}$, and the co-linear edge points downwards i.e., its angle $\phi$ lies between $\frac{\pi}{2}$ and $\frac{3\pi}{2}$. This situation arises when all of the edges are aligned with the coordinate axis (Figure \ref{fig:orig}).
\end{enumerate}
A discussion of each of these cases and the respective value for
the geometric factor was carried out in Section \ref{s 2.4.1}. There it
was shown that only case 1a above has \emph{non-zero} measure in
SO(3), therefore we will restrict our analysis to such a case.

It is straight forward to see that case 1a can be divided into
further sub-cases according to the values of the $\theta$-angles
and the relations between the $\phi$-angles of each of the up/down
couples. In what follows, we will not give the results for all
possible choices. Instead, we will  choose a particular sub-case
and perform the calculations for the expectation value of the
volume operator with respect to this sub-case. As we will see,
these calculations show that, up to embeddings of measure zero in
$SO(3)$, the semiclassical behavior of the volume operator, in
zeroth-order, does not depend on how the graph is rotated: \emph{a
fortiori}, it is independent of the particular case we analysed.

\begin{figure}[htbp]
\begin{center}
\includegraphics[scale=0.6]{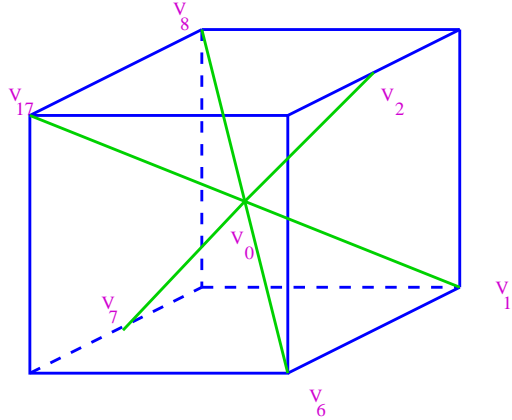}
 \caption{6-valent vertex\label{fig:64}}
\end{center}
\end{figure}

\begin{figure}[htbp]
\begin{center}
 \includegraphics[scale=0.6]{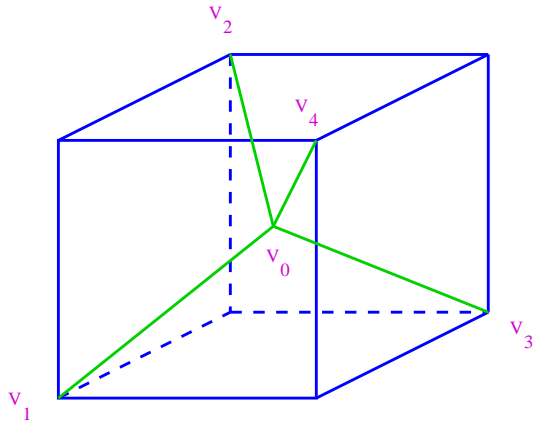}
\caption{4-valent vertex\label{fig:46}}
\end{center}
\end{figure}

In order to carry out a proper comparison between the
semiclassical behavior of the volume operator as applied to graphs
of different valence, we will apply the same Euler transformations
(i.e., with the same Euler angles) to each of the graphs we
consider. Since we have not specified the values of the Euler
angles, the only way to do this is to assume that after a
rotation, those edges which had the same angles used in both the
aligned 4-valent and 6-valent case will end up in the same octant.
For example, consider Figures \ref{fig:64} and \ref{fig:46}  which
depict both 6-valent and 4-valent vertices respectively. From such
pictures it is easy to see that the edges $e_{0,2}$ and $e_{0,3}$
of the 4-valent graph have the same $\theta$- and $\phi$-angles as
the edges, $e_{0,8}$, and, $e_{0,1}$, of the 6-valent graph.
Therefore, in the rotated case we will assume that  $e_{0,8}$,
and, $e_{0,1}$ lie in the same octants as $e_{0,2}$ and $e_{0,3}$
respectively. It follows that the angle-ranges for the edges
incident at vertex $V_0$ for a 6-valent graph are:
\begin{align}\label{ali:ranges6}
&\frac{3\pi}{2}<\theta_{e_{0,2}}<\frac{7\pi}{4}\hspace{.5in}\frac{3\pi}{2}<\phi_{e_{0,2}}<2\pi\nonumber\\
&\frac{\pi}{4}<\theta_{e_{0,8}}<\frac{\pi}{2}\hspace{.5in}0<\phi_{e_{0,8}}<\frac{\pi}{2}\nonumber\\
&\frac{3\pi}{4}<\theta_{e_{0,17}}<\pi\hspace{.5in}0< \phi_{e_{0,17}}<\frac{\pi}{2}\nonumber\\
&\frac{\pi}{2}<\theta_{e_{0,7}}<\frac{3\pi}{4}\hspace{.5in}\frac{\pi}{2}<\phi_{e_{0,7}}<\pi\nonumber\\
&\frac{\pi}{2}<\theta_{e_{0,6}}<\frac{3\pi}{4}\hspace{.5in}\pi<\phi_{e_{0,6}}<\frac{3\pi}{2}\nonumber\\
&\frac{7\pi}{4}<\theta_{e_{0,1}}<2\pi\hspace{.5in}\pi<\phi_{e_{0,1}}<\frac{3\pi}{2}
\end{align}
Such conditions of the angles implies that the edges $e_{0,1}$,
$e_{0,2}$, $e_{0,6}$, $e_{0,7}$, $e_{0,7}$ and $e_{0,17}$ lie in
the octants $H$, $D$, $G$, $F$, $A$ and $B$ respectively.

Since we are considering a regular 6-valent lattice, the above
ranges of angles induce a relation on all the other angle-ranges
of the edges  for each vertex in the graph.
The coordinates of the rotated vertices are:
\begin{eqnarray}
V^{\prime}_0&=&\Big(0,0,0\Big)\nonumber\\
V^{\prime}_2&=&\Big((R_{11}+R_{13})\frac{\delta
\sqrt{2}}{2},(R_{21}+R_{23})\frac{\delta
\sqrt{2}}{2},(R_{31}+2R_{33})
        \frac{\delta \sqrt{2}}{2}\Big)\nonumber\\
V^{\prime}_3&=&\Big((R_{11}(\frac{\sqrt{2}-1}{2})-\frac{R_{12}}{2}+R_{13} \sqrt{2})\delta,(R_{21}(\frac{\sqrt{2}-1}{2})-\frac{R_{22}}{2}+R_{23} \sqrt{2})\delta,(R_{31}(\frac{\sqrt{2}-1}{2})-\frac{R_{32}}{2}+R_{33} \sqrt{2})\delta\Big)\nonumber\\
V_{4}^{\prime}&=&\Big((\sqrt{2}R_{11}-R_{12}+\sqrt{2}R_{13})\delta
, (\sqrt{2}R_{21}-R_{22}+\sqrt{2}R_{23})\delta ,
(\sqrt{2}R_{31}-R_{32}+\sqrt{2}R_{33})\delta \Big)\nonumber\\
V_{5}^{\prime}&=&\Big(-R_{12}\delta ,
R_{22}\delta, R_{32}\delta\Big)\nonumber\\
V_{13}^{\prime}&=&\Big((-\frac{\sqrt{2}}{2}R_{11}-R_{12}+\frac{\sqrt{2}}{2}R_{13})\delta
,
(-\frac{\sqrt{2}}{2}R_{21}-R_{22}+\frac{\sqrt{2}}{2}R_{23})\delta,
(-\frac{\sqrt{2}}{2}R_{31}-R_{32}+\frac{\sqrt{2}}{2}R_{33})\delta\Big)\nonumber\\
V_{18}^{\prime}&=&\Big(((\sqrt{2}+\frac{1}{2})R_{11}-\frac{1}{2}R_{12}+\frac{\sqrt{2}}{2}R_{13})\delta
,((\sqrt{2}+\frac{1}{2})R_{21}-\frac{1}{2}R_{22}+\frac{\sqrt{2}}{2}R_{23})\delta,
((\sqrt{2}+\frac{1}{2})R_{31}-\frac{1}{2}R_{32}+\frac{\sqrt{2}}{2}R_{33})\delta\Big)\nonumber\\
V^{\prime}_{15}&=&\Big(((\frac{\sqrt{2}}{2}-1)R_{11}-R_{12}+\frac{3\sqrt{2}}{2}R_{13})\delta,
((\frac{\sqrt{2}}{2}-1)R_{21}-R_{22}+\frac{3\sqrt{2}}{2}R_{23})\delta,((\frac{\sqrt{2}}{2}-1)R_{31}-R_{32}+\frac{3\sqrt{2}}{2}R_{33})\delta\Big)\nonumber\\
V_{14}^{\prime}&=&\Big((\frac{\sqrt{2}+1}{2}R_{11}-\frac{3}{2}R_{12})\delta
, (\frac{\sqrt{2}+1}{2}R_{21}-\frac{3}{2}R_{22})\delta,
(\frac{\sqrt{2}+1}{2}R_{31}-\frac{3}{2}R_{32})\delta\Big)
\end{eqnarray}
Similarly, as was done for the 4-valent case, different choices of combination of angles satisfying conditions (\ref{ali:ranges6}) above will lead to different values for the terms $\frac{t_{e_i,e_j}}{\sqrt{t_{e_i}t_{e_j}}}$. However, the couples of edges commonly intersecting a given stack will coincide for any such combination. This implies that the matrix $\sqrt{A}^{-1}$ will have the same entries for any sub-case of (\ref{ali:ranges6}) but their specific values will be different.

Moreover, the geometric factor $G_{\gamma,V}$ of any sub-case of
(\ref{ali:ranges6}) will coincide. It follows that any combination
of angles that satisfies conditions (\ref{ali:ranges6}) will lead to the
same value in zeroth-order in $\frac{l}{\delta}$ of the
expectation value for the volume operator. Therefore, as was done for
the 4-valent case, in order to compute the expectation value for
the volume operator, we will not specify a particular sub-case of
(\ref{ali:ranges6}), but leave the result as general as possible.

Given conditions  \ref{ali:ranges6} the expectation value of the
volume for the periodicity cell is
\begin{align}
&V_R=\delta^3\sqrt{\frac{1}{8}} \Big|\det\Bigg(\frac{\delta
X_S^a}{\delta(s,u^1,u^2)}\bigg)\Big|2\Bigg\{
2\sqrt{2}\times\sqrt{|det\Big(E^a_j(u)\Big)|}\nonumber\\
&+\sqrt{|8  +\frac{1}{2}\Big(- \rho_{6_3,8_3}^2 - \rho_{6_3,2_3}^2 - \rho_{11_3,6_3}^2 - \rho_{11_3,8_3}^2 - \rho_{11_3,13_3}^2 - \rho_{11_3,2_3}^2 - \rho_{17_3,6_3}^2 - \rho_{17_3,8_3}^2 - \rho_{17_3,13_3}^2 - \rho_{17_3,2_3}^2}\nonumber\\
&\qquad \overline{  - \rho_{13_3,8_3}^2- \rho_{13_3,2_3}^2\Big)|}\times \sqrt{|det\Big(E^a_j(u)\Big)|}\nonumber\\
 &+ \sqrt{|8 +\frac{1}{2}\Big(- \rho_{14_13,21_13}^2 - \rho_{9_13,14_13}^2 - \rho_{3_13,21_13}^2 - \rho_{3_13,9_13}^2 - \rho_{4_13,9_13}^2 - \rho_{4_13,21_13}^2 - \rho_{4_13,3_13}^2 - \rho_{4_13,14_13}^2 - \rho_{5_13,9_13}^2 }\nonumber\\
&\qquad \overline{- \rho_{5_13,21_13}^2  - \rho_{5_13,3_13}^2- \rho_{5_13,14_13}^2\Big)|}\times \sqrt{|det\Big(E^a_j(u)\Big)|} \nonumber\\
 &+ \sqrt{|8+\frac{1}{2}\Big( - \rho_{0_2,10_2}^2 - \rho_{0_2,11_2}^2 - \rho_{0_2,3_2}^2 - \rho_{0_2,9_2}^2 - \rho_{11_2,3_2}^2 - \rho_{12_2,10_2}^2 - \rho_{12_2,11_2}^2 - \rho_{12_2,3_2}^2 - \rho_{12_2,9_2}^2 - \rho_{9_2,10_2}^2 }\nonumber\\
&\overline{-\rho_{11_2,10_2}^2- \rho_{9_2,3_2}^2\Big)|}\times \sqrt{|det\Big(E^a_j(u)\Big)|} \nonumber\\
 &+ \sqrt{|8+\frac{1}{2}\Big( - \rho_{12_12,4_12}^2 - \rho_{12_12,f_12}^2 - \rho_{9_12,12_12}^2 - \rho_{9_12,4_12}^2 - \rho_{9_12,e_12}^2 - \rho_{9_12,f_12}^2 - \rho_{d_12,12_12}^2 - \rho_{d_12,4_12}^2 - \rho_{d_12,e_12}^2 }\nonumber\\
&\qquad \overline{- \rho_{d_12,f_12}^2 - \rho_{e_12,4_12}^2 - \rho_{e_12,f_12}^2\Big)|}\times \sqrt{|det\Big(E^a_j(u)\Big)|} \nonumber\\
 &+ \sqrt{|8+\frac{1}{2}\Big( - 2\rho_{c^{'}_15,3_15}^2
- \rho_{d^{'}_15,3_15}^2 - \rho_{d^{'}_15,c_15}^2 -
\rho_{d^{'}_15,e^{'}_15}^2 - \rho_{d^{'}_15,f^{'}_15}^2
-\rho_{e^{'}_15,c^{'}_15}^2 - \rho_{f^{'}_15,3_15}^2 -
\rho_{f^{'}_15,e^{'}_15}^2 -
\rho_{a^{'}_15,3_15}^2 - \rho_{a^{'}_15,c^{'}_15}^2} \nonumber\\
&\overline{ - \rho_{a^{'}_15,e^{'}_15}^2 - \rho_{a^{'}_15,f^{'}_15}^2\Big)|}\times \sqrt{|det\Big(E^a_j(u)\Big)|}\nonumber\\
&+ \sqrt{|8+\frac{1}{2}\Big( - \rho_{13_4,18_4}^2 - \rho_{13_4,c_4}^2 - \rho_{16_4,13_4}^2 - \rho_{16_4,18_4}^2 - \rho_{16_4,b_4}^2 - \rho_{b_4,18_4}^2 - \rho_{b_4,c_4}^2 - \rho_{a_4,13_4}^2 - \rho_{a_4,18_4}^2 - \rho_{a_4,b_4}^2 - \rho_{a_4,c_4}^2} \nonumber\\
  &\overline{- \rho_{c_4,16_4}^2\Big)|}\times \sqrt{|det\Big(E^a_j(u)\Big)|}\nonumber\\
&+ \sqrt{|8+\frac{1}{2}\Big( - \rho_{13_14,a_14}^2 - \rho_{13_14,h_14}^2 - \rho_{13_14,l_14}^2 - \rho_{h_14,a_14}^2 - \rho_{j_14,a_14}^2 - \rho_{j_14,h_14}^2 - \rho_{j_14,l_14}^2 - \rho_{k_14,13_14}^2 - \rho_{k_14,h_14}^2 - \rho_{k_14,j_14}^2 }\nonumber\\
&\qquad \overline{ - \rho_{k_14,l_14}^2- \rho_{l_14,a_14}^2 \Big)|}\times \sqrt{|det\Big(E^a_j(u)\Big)|}\nonumber\\
 & + \sqrt{|8+\frac{1}{2}\Big(- \rho_{17_5,6_5}^2 -
\rho_{17_5,13_5}^2 - \rho_{17_5,r_5}^2 - \rho_{13_5,6_5}^2 -
\rho_{p_5,17_5}^2 - \rho_{p_5,6_5}^2 - \rho_{p_5,q_5}^2 -
\rho_{p_5,r_5}^2 - \rho_{q_5,6_5}^2 -
\rho_{q_5,13_5}^2}\nonumber\\
&\overline{-\rho_{q_5,r_5}^2 -
 \rho_{r_5,13_5}^2\Big)|}\times \sqrt{|det\Big(E^a_j(u)\Big)|}\;\Bigg\}
 \end{align}
 Expanding the square roots (see Section \ref{s4.2}) and considering just first-order terms we obtain
  \begin{align}\label{equ:right}
&V_R=\delta^3\sqrt{|\det\big(E^a_j(u)\big)|}\; 2\Bigg(9 +
\frac{1}{2}\times\frac{1}{16}\Big(
- \rho_{6_3,8_3}^2 - \rho_{6_3,2_3}^2 - \rho_{11_3,6_3}^2 - \rho_{11_3,8_3}^2 - \rho_{11_3,13_3}^2 - \rho_{11_3,2_3}^2 - \rho_{17_3,6_3}^2 - \rho_{17_3,8_3}^2 -\nonumber\\
& - \rho_{17_3,13_3}^2 - \rho_{17_3,2_3}^2 - \rho_{13_3,8_3}^2- \rho_{13_3,2_3}^2- \rho_{14_{13},21_{13}}^2 - \rho_{9_{13},14_{13}}^2 - \rho_{3_{13},21_{13}}^2 - \rho_{3_{13},9_{13}}^2 - \rho_{4_{13},9_{13}}^2 - \rho_{4_{13},21_{13}}^2 - \rho_{4_{13},3_{13}}^2\nonumber\\
 & - \rho_{4_{13},14_{13}}^2 - \rho_{5_{13},9_{13}}^2 - \rho_{5_{13},21_{13}}^2 - \rho_{5_{13},3_{13}}^2- \rho_{5_{13},14_{13}}^2- \rho_{0_2,10_2}^2 - \rho_{0_2,11_2}^2- \rho_{0_2,3_2}^2 - \rho_{0_2,9_2}^2- \rho_{11_2,3_2}^2 - \rho_{12_2,10_2}^2 \nonumber\\
 &    - \rho_{12_2,11_2}^2 - \rho_{12_2,3_2}^2 - \rho_{12_2,9_2}^2 - \rho_{9_2,10_2}^2 -\rho_{11_2,10_2}^2- \rho_{9_2,3_2}^2 - \rho_{12_{12},4_{12}}^2 - \rho_{12_{12},f_{12}}^2 - \rho_{9_{12},12_{12}}^2 - \rho_{9_{12},4_{12}}^2 - \rho_{9_{12},e_{12}}^2\nonumber\\
 &  - \rho_{9_{12},f_{12}}^2 - \rho_{d_{12},12_{12}}^2 - \rho_{d_{12},4_{12}}^2 - \rho_{d_{12},e_{12}}^2 - \rho_{d_{12},f_{12}}^2 - \rho_{e_{12},4_{12}}^2 - \rho_{e_{12},f_{12}}^2- 2\rho_{c^{'}_15,3_15}^2- \rho_{d^{'}_15,3_15}^2 - \rho_{d^{'}_15,c_15}^2 - \rho_{d^{'}_15,e^{'}_15}^2\nonumber\\
 & -\rho_{d^{'}_15,f^{'}_15}^2 -\rho_{e^{'}_15,c^{'}_15}^2 - \rho_{f^{'}_15,3_15}^2 - \rho_{f^{'}_15,e^{'}_15}^2 -
\rho_{a^{'}_15,3_15}^2 - \rho_{a^{'}_15,c^{'}_15}^2 - \rho_{a^{'}_15,e^{'}_15}^2 - \rho_{a^{'}_15,f^{'}_15}^2- \rho_{13_4,18_4}^2 - \rho_{13_4,c_4}^2 - \rho_{16_4,13_4}^2\nonumber\\
& - \rho_{16_4,18_4}^2 - \rho_{16_4,b_4}^2 - \rho_{b_4,18_4}^2 - \rho_{b_4,c_4}^2 - \rho_{a_4,13_4}^2 - \rho_{a_4,18_4}^2 - \rho_{a_4,b_4}^2 - \rho_{a_4,c_4}^2- \rho_{c_4,16_4}^2- \rho_{13_14,a_14}^2 - \rho_{13_14,h_14}^2 - \rho_{13_14,l_14}^2\nonumber\\
  &  - \rho_{h_14,a_14}^2 - \rho_{j_14,a_14}^2 - \rho_{j_14,h_14}^2 - \rho_{j_14,l_14}^2 - \rho_{k_14,13_14}^2 - \rho_{k_14,h_14}^2 - \rho_{k_14,j_14}^2 - \rho_{k_14,l_14}^2- \rho_{l_14,a_14}^2 - \rho_{17_5,6_5}^2 -
\rho_{17_5,13_5}^2 \nonumber\\
 & - \rho_{17_5,r_5}^2 - \rho_{13_5,6_5}^2 - \rho_{p_5,17_5}^2 -
\rho_{p_5,6_5}^2 - \rho_{p_5,q_5}^2 - \rho_{p_5,r_5}^2 -
\rho_{q_5,6_5}^2 - \rho_{q_5,13_5}^2-\rho_{q_5,r_5}^2 -
 \rho_{r_5,13_5}^2\Big)\Bigg)\;
\Big|\det\Big(\frac{\delta X_S^a}{\delta(s,u^1,u^2)}\Big)\Big|\nonumber\\
&=2\delta^3\sqrt{|\det\big(E^a_j(u)\big)|}\bigg(9
+\frac{1}{2}\Big(-\frac{1}{16}\sum^{\frac{n}{2}(n-1)}_{i,j=1;j\neq
i }\rho_{j_k,i_k}^2 \Big)\bigg)\;\Big|\det\Big(\frac{\delta
X_S^a}{\delta(s,u^1,u^2)}\Big)\Big|
 \end{align}
Here $\rho_{l_i,k_i}$ represent the off-diagonal entries of the
matrix $\sqrt{A}^{-1}$ (see Section \ref{sa.2.2}) which denote the
value of the term $\frac{\sqrt{t_{e_{i,l},
e_{i,k}}}}{t_{e_{i,l}}t_{e_{i,k}}}$ for the edges $e_{i,l}$ and
$e_{i,k}$ incident at the vertex $V_i$.

The term proportional to $\rho_{j_k,i_k}^2$ represents the
higher-order corrections to the expectation value of the volume
operator. Each term $\rho_{i_k,j_k}$ is proportional to $C\times
\frac{l^{\prime}}{\delta}$ for $l^{\prime}<l$ where $C$ is a
constant that  depends on the Euler angles we choose. It follows
that only the zeroth-order of the expectation value of the volume
operator for the 6-valent graph is rotationally invariant, up to
embeddings with measure zero in $SO(3)$. However, only embeddings
which have measure \emph{zero} in $SO(3)$ (when the edges are
aligned to the plaquettes) give the correct semiclassical limit.

\subsubsection{Expectation value of the volume operator for a
translated 6-valent graph} \label{s7.3.3}

In this Section we will calculate the expectation value of the
volume operator for a translated 6-valent graph. To make the
comparison as accurate as possible, we translate the 6-valent
graph by a vector with more or less the same properties as the
vector with respect to which we translated the 4-valent graph,
namely:
\begin{enumerate}
\item[1)] $\frac{\delta }{2}\ge\epsilon_x\ge\epsilon_y\ge\epsilon_z\ge 0$
\item[2)] $\frac{\delta\sqrt{2}}{2}<\epsilon_i+\epsilon_j$
\item[3)]$\frac{\delta\sqrt{2}}{2}-\frac{\delta 1}{2}<\epsilon_z$
\item[4)]$|V^i_k|-|V_k^j|>n^i_kl-n^j_kl\mbox{ for all } |V_k^i|>|V_k^j|$
\end{enumerate}

The coordinates of the translated vertices are
\begin{eqnarray}
V^{''}_0&=&(\epsilon_x,\epsilon_y,\epsilon_z)\\
V^{''}_2&=&\Big(\frac{\delta
\sqrt{2}}{2}+\epsilon_x,\epsilon_y,\frac{\delta
\sqrt{2}}{2}+\epsilon_z\Big)\\
V^{''}_3&=&\Big(\epsilon_x+\frac{\delta \sqrt{2}}{2}-\frac{\delta
}{2},\epsilon_y-\frac{\delta
}{2},\delta \sqrt{2}+\epsilon_z\Big)\\
V^{''}_4&=&\Big(\delta \sqrt{2}+\epsilon_x,\epsilon_y-\delta
,\delta \sqrt{2}+\epsilon_z\Big)\\
V^{''}_5&=&\Big(\epsilon_x,\epsilon_y-\delta ,
\epsilon_z\Big)\\
V^{''}_{13}&=&\Big(\frac{\delta
\sqrt{2}}{2}+\epsilon_x,\epsilon_y-\delta
,\frac{\delta \sqrt{2}}{2}+\epsilon_z\Big)\\
V^{''}_{18}&=&\Big(\delta
\sqrt{2}+\frac{\delta}{2}+\epsilon_x,\epsilon_y-\frac{\delta}{2},\epsilon_z+\frac{\delta
\sqrt{2}}{2}\Big)\\
V^{''}_{15}&=&\Big(\frac{\delta \sqrt{2}}{2}-\frac{\delta
2}{2}+\epsilon_x,\epsilon_y-\delta
,3\frac{\delta \sqrt{2}}{2}+\epsilon_z\Big)\\
V^{''}_{14}&=&\Big(\frac{\delta
\sqrt{2}}{2}+\frac{\delta}{2}+\epsilon_x,\epsilon_y-3\frac{\delta}{2},\epsilon_z\Big)
\end{eqnarray}
Similarly as for the aligned case we have
\begin{align}
Z^{\{x,y\}}_{e_i}&=\begin{cases}\frac{2}{\sqrt{2}}& \text{ iff } \theta_{e_i}\neq 0\\
1&\text{ iff } \theta_{e_i}= 0
\end{cases}
\hspace{.5in}
T^{\{x,y\}}_{e_i}=\begin{cases}\frac{\sqrt{2}}{2}& \text{ iff } \theta_{e_i}\neq 0\\
1&\text{ iff } \theta_{e_i}= 0
\end{cases}\nonumber\\
F^x_{e_i}&=\begin{cases}1 & \text{ iff } \theta_{e_i}\neq 0\\
\infty& \text{ iff } \theta_{e_i}= 0
\end{cases}\hspace{.5in}
F^x_{e_i}=1\hspace{.5in} \forall e_i\in\gamma
\end{align}

The expression for the matrix $\sqrt{A}^{-1}$ is given in Table
\ref{tab:a6t} in the Appendix.

Given the conditions above we obtain the following value for the
expectation value of the volume operator as applied to the
periodicity cell:
\begin{align}\label{ali:6t}
&V_R=
\sqrt{\frac{4}{8}}\delta^3
 \Bigg(\sqrt{|8 - \frac{3 (A^{'}_0)^2}{2} - \frac{5 B_0^2}{2} - \sqrt{2} B_0^2 - A^{'}_0 B_0 B^{'}_0 - \frac{A^{'}_0 B_0 B^{'}_0}{\sqrt{2}} - 2 (B^{'}_0)^2 - B_0 C_0 - \frac{A^{'}_0 B^{'}_0 C_0}{4}- \frac{ 3 C_0^2}{4}- \frac{A^{'}_0 C_0 C^{'}_0}{2 \sqrt{2}}}\nonumber\\
&\overline{   - \frac{B^{'}_0 C_0 C^{'}_0}{4}- \frac{
 B^{'}_0 C_0 C^{'}_0}{4 \sqrt{2}} - \frac{3 (C^{'}_0)^2}{2} -\frac{B_0 (C^{'}_0)^2}{4} - \frac{3 B_0 (C^{'}_0)^2}{
 4 \sqrt{2}} - \frac{C_0 (C^{'}_0)^2}{8 \sqrt{2}})|}\times\sqrt{|det\Big(E^a_j(u)\Big)|}\nonumber\\
 & + \sqrt{|8 - \frac{3 (A^{'}_2)^2}{2} - \frac{7 (B^{'}_2)^2}{2} - A^{'}_2 B^{'}_2 B_2 - \frac{3 A^{'}_2 B^{'}_2 B_2}{
 2 \sqrt{2}} - 3 B_2^2 - 2 \sqrt{2} B_2^2 - \frac{3 B^{'}_2 B_2 C^{'}_2}{4}- \frac{
 5 B^{'}_2 B_2 C^{'}_2}{4 \sqrt{2}} - (C^{'}_2)^2 - \frac{(B^{'}_2)^2 C_2}{4} }\nonumber\\
&\overline{ - \frac{(B^{'}_2)^2 C_2}{
 2 \sqrt{2}} - \frac{A^{'}_2 C^{'}_2 C_2}{2 \sqrt{2}} - C_2^2/2|} \times\sqrt{|det\Big(E^a_j(u)\Big)|}+ \sqrt{|8 - \frac{3 (A^{'}_3)^2}{2} - A_3^2 - \frac{5 (B^{'}_3)^2}{2}- \frac{A_3 (B^{'}_3)^2}{\sqrt{2} }- \frac{
 B^{'}_3 B_3}{2}- \frac{3 A^{'}_3 B^{'}_3 B_3}{4}} \nonumber\\
&\overline{  - \frac{A^{'}_3 B^{'}_3 B_3}{\sqrt{2}} - 3 B_3^2 -
 2 \sqrt{2} B_3^2 - \frac{A^{'}_3 B_3^2}{4} - \frac{A^{'}_3 B_3^2}{2 \sqrt{2}} -\frac{
 A^{'}_3 A_3 C^{'}_3}{2} - \frac{B^{'}_3 B_3 C^{'}_3}{2} - \frac{3 B^{'}_3 B_3 C^{'}_3}{4 \sqrt{2}} - \frac{
 3 (C^{'}_3)^2}{4}|}\times\sqrt{|det\Big(E^a_j(u)\Big)|}\nonumber\\
 &+ \sqrt{|8 - \frac{3 (A^{'}_4)^2}{2} -\frac{ A_4^2}{2} - 3 (B^{'}_4)^2 - \frac{A_4 (B^{'}_4)^2}{\sqrt{2}} -
 A^{'}_4 B^{'}_4 B_4 - \frac{3 A^{'}_4 B^{'}_4 B_4}{2 \sqrt{2}} - 3 B_4^2- 2 \sqrt{2} B_4^2 - \frac{
 B^{'}_4 B_4 C^{'}_4}{2} - \frac{3 B^{'}_4 B_4 C^{'}_4}{4 \sqrt{2}}}\nonumber\\
&\overline{  - \frac{3 (C^{'}_4)^2}{4} - \frac{
 A^{'}_4 C^{'}_4 C_4}{2 \sqrt{2}} - \frac{C_4^2}{4}|}\times\sqrt{|det\Big(E^a_j(u)\Big)|} +\sqrt{|8 - 3( A^{'}_5)^2 - \frac{(A^{'}_5)^2 A_5}{\sqrt{2}} - A_5^2 - \frac{(A^{'}_5)^2 B_5}{2} - \frac{(A^{'}_5)^2 B_5}{
 2 \sqrt{2}} - \sqrt{2} A_5 B_5 }\nonumber\\
&\overline{ - \frac{3 B_5^2}{2}- \frac{13 (C^{'}_5)^2}{8} - \frac{
 A^{'}_5 (C^{'}_5)^2}{4 \sqrt{2}} - \frac{A_5 (C^{'}_5)^2}{2 \sqrt{2}} - \frac{B_5 (C^{'}_5)^2}{4} - \frac{
 B_5 (C^{'}_5)^2}{4 \sqrt{2}} - \frac{(A^{'}_5)^2 C_5}{4 \sqrt{2}} - B_5 C_5 - \frac{
 3 A^{'}_5 C^{'}_5 C_5}{4 \sqrt{2}} - \frac{(C^{'}_5)^2 C_5}{8 \sqrt{2}} - \frac{5 C_5^2}{8}|}\nonumber\\
 & \times\sqrt{|det\Big(E^a_j(u)\Big)|}+ \sqrt{|8 - \frac{3 (A^{'}_{13})^2}{2} - \frac{11 (B^{'}_{13})^2}{4} - A^{'}_{13} B^{'}_{13} B_{13} - \frac{
 A^{'}_{13} B^{'}_{13} B_{13}}{\sqrt{2}} - 3 B_{13}^2 - 2 \sqrt{2} B_{13}^2 - \frac{
 B^{'}_{13} B_{13} C^{'}_{13}}{2}}\nonumber\\
&\overline{ - \frac{3 B^{'}_{13} B_{13} C^{'}_{13}}{4 \sqrt{2}} -
\frac{3 (C^{'}_{13})^2}{4} - \frac{
 (B^{'}_{13})^2 C_{13}}{2} - \frac{A^{'}_{13} C^{'}_{13} C_{13}}{2 \sqrt{2}} - \frac{C_{13}^2}{2}|} \times\sqrt{|det\Big(E^a_j(u)\Big)|}+ \sqrt{|8 - \frac{5 (A^{'}_{18})^2}{4} - \frac{A_{18}^2}{4} - 3 (B^{'}_{18})^2}\nonumber\\
 & \overline{- A^{'}_{18} B^{'}_{18} B_{18} - \frac{
 3 A^{'}_{18} B^{'}_{18} B_{18}}{2 \sqrt{2}} - 3 B_{18}^2 - 2 \sqrt{2} B_{18}^2- \frac{
 B^{'}_{18} B_{18} C^{'}_{18}}{2} - \frac{3 B^{'}_{18} B_{18} C^{'}_{18}}{4 \sqrt{2}} - \frac{3 (C^{'}_{18})^2}{4} - \frac{
 (B^{'}_{18})^2 C_{18}}{2} - \frac{A^{'}_{18} C^{'}_{18} C_{18}}{4 \sqrt{2}}}\nonumber\\
&\overline{  - \frac{A_{18} C^{'}_{18} C_{18}}{
 4 \sqrt{2}}- \frac{C_{18}^2}{2}|}\times\sqrt{|det\Big(E^a_j(u)\Big)|} + \sqrt{|8 - 3 (A^{'}_{15})^2 - \frac{(A^{'}_{15})^2 A_{15}}{\sqrt{2}} - 2 A_{15}^2 - \frac{A^{'}_{15} A_{15} B^{'}_{15}}{
 2 \sqrt{2}} - \frac{B^{'}_{15})^2}{2} - \frac{3 (A^{'}_{15})^2 B_{15}}{4}}\nonumber\\
&\overline{- \frac{(A^{'}_{15})^2 B_{15}}{2 \sqrt{2}} -
  \sqrt{2} A_{15} B_{15} - 2 B_{15}^2 - \sqrt{2} B_{15}^2 - \frac{
 A^{'}_{15} A_{15} C_{15}}{\sqrt{2}} - \frac{B^{'}_{15} B_{15} C_{15}}{4} - \frac{B^{'}_{15} B_{15} C_{15}}{
 2 \sqrt{2}} - C_{15}^2 - \frac{A_{15} C_{15}^2}{4 \sqrt{2}} - \frac{B_{15} C_{15}^2}{8}|} \nonumber\\
 &\times\sqrt{|det\Big(E^a_j(u)\Big)|}+ \sqrt{|8 - \frac{3 (A^{'}_{14})^2}{2} - 2 (B^{'}_{14})^2 - A^{'}_{14} B^{'}_{14} B_{14} - \frac{
 A^{'}_{14} B^{'}_{14} B_{14}}{\sqrt{2} }- \frac{5 B_{14}^2}{2} - \sqrt{2} B_{14}^2 - \frac{
 3 (C^{'}_{14})^2}{2} }\nonumber\\
&\overline{- \frac{B_{14} (C^{'}_{14})^2}{4} - \frac{3 B_{14}
(C^{'}_{14})^2}{4 \sqrt{2}}- \frac{
 A^{'}_{14} B^{'}_{14} C_{14}}{4} - B_{14} C_{14} - \frac{A^{'}_{14} C^{'}_{14} C_{14}}{2 \sqrt{2}}- \frac{
 B^{'}_{14} C^{'}_{14} C_{14}}{4} - \frac{B^{'}_{14} C^{'}_{14} C_{14}}{4 \sqrt{2}} - \frac{(C^{'}_{14})^2 C_{14}}{
 8 \sqrt{2}} - \frac{3 C_{14}^2}{4}|}\nonumber\\
&\times\sqrt{|det\Big(E^a_j(u)\Big)|}\;\Bigg)\Big|det\Big(\frac{\delta
X_S^a}{\delta(s,u^1,u^2)}\Big)\Big|
  \end{align}
Expanding the square root and considering only first-order
contributions (see Section \ref{s4.2}) we obtain
\begin{align}
V_R=&\delta^3\sqrt{|det\Big(E^a_j(u)\Big)|}\Big|det\Big(\frac{\delta
X_S^a}{\delta(s,u^1,u^2)}\Big)\Big|2\Bigg(9+\frac{1}{16}\Big( -
\frac{3 (A^{'}_0)^2}{2} - \frac{5 B_0^2}{2}- \sqrt{2} B_0^2   - 2
(B^{'}_0)^2 - B_0 C_0  - \frac{
 3 C_0^2}{4}\nonumber\\
 & - \frac{3 (C^{'}_0)^2}{2} - \frac{3 (A^{'}_2)^2}{2} - \frac{7 (B^{'}_2)^2}{2}  - 3 B_2^2 - 2 \sqrt{2} B_2^2  - (C^{'}_2)^2 - \frac{C_2^2}{2} - \frac{3 (A^{'}_3)^2}{2}- A_3^2 - \frac{5 (B^{'}_3)^2}{2}  - \frac{
 B^{'}_3 B_3}{2}  \nonumber\\
&- 3 B_3^2 -
 2 \sqrt{2} B_3^2  - \frac{
 3 (C^{'}_3)^2}{4} - \frac{3 (A^{'}_4)^2}{2} - \frac{A_4^2}{2} - 3 (B^{'}_4)^2  - 3 B_4^2 - 2 \sqrt{2} B_4^2- \frac{3 (C^{'}_4)^2}{4}  - \frac{C_4^2}{4}- 3 (A^{'}_5)^2\nonumber\\
 &   - A_5^2  - \sqrt{2} A_5 B_5 - \frac{3 B_5^2}{2} - \frac{13 (C^{'}_5)^2}{8} - B_5 C_5 - \frac{5 C_5^2}{8}  - \frac{3 (A^{'}_{13})^2}{2} - \frac{11 (B^{'}_{13})^2}{4} - 3 B_{13}^2 - 2 \sqrt{2} B_{13}^2  - \frac{3 (C^{'}_{13})^2}{4}\nonumber\\
 & - \frac{C_{13}^2}{2} - \frac{5 (A^{'}_{18})^2}{4} - \frac{A_{18}^2}{4} - 3 (B^{'}_{18})^2  - 3 B_{18}^2 - 2 \sqrt{2} B_{18}^2- \frac{3 (C^{'}_{18})^2}{4}  - \frac{C_{18}^2}{2} - 3 (A^{'}_{15})^2  - 2 A_{15}^2 - \frac{B^{'}_{15})^2}{2}\nonumber\\
 &-
  \sqrt{2} A_{15} B_{15} - 2 B_{15}^2 - \sqrt{2} B_{15}^2  - C_{15}^2 - \frac{B_{15} C_{15}^2}{8}- \frac{3 (A^{'}_{14})^2}{2} - 2 (B^{'}_{14})^2  - \frac{5 B_{14}^2}{2}- \sqrt{2} B_{14}^2 - \frac{
 3 (C^{'}_{14})^2}{2} \nonumber\\
 &- B_{14} C_{14} - \frac{3 C_{14}^2}{4}\Big)\;\Bigg)
\end{align}
 where the terms
 \begin{align}\label{ali:terms}
&A^{\prime}_i=\frac{|x_{V_i}|-n_i}{\delta(\sqrt{\sqrt{2}+1)}}\hspace{.5in}A_i=\frac{(|x_{V_i}|-n_il)}{\delta_e(1+\frac{\sqrt{2}}{2}}
\nonumber\\
&B^{\prime}_i=\frac{(|y_{V_i}|-m_il)}{\delta(\sqrt{\sqrt{2}+1)}}\hspace{.5in}B_i=\frac{(|y_{V_i}|-m_il)}{\delta_e(1+\frac{\sqrt{2}}{2}}
\nonumber\\
&C^{\prime}_i=\frac{(|z_{V_i}|-p_il)}{\delta(\sqrt{\sqrt{2}+1)}}\hspace{.5in}C_i=\frac{(|z_{V_i}|-p_il)}{\delta_e(1+\frac{\sqrt{2}}{2}}
\end{align}
are the off-diagonal matrix elements of $\sqrt{A}^{-1}$ (see
Section \ref{sa.2.3}). The quantities $x_{V_i}$, $y_{V_i}$,
$z_{V_i}$ represent are the $x$, $y$, $z$-coordinates of the vertex
$V_i$ respectively.

In the last line of the equation above we have performed a Taylor
expansion and factored out the term
$\sqrt{\det\big(E^a_j(u)\big)}$, since we can assume that,
although it is vertex dependent, the values of this term to
first-order will be the same for each vertex. Due to the
appearance of the terms (\ref{ali:terms}), which are proportional to
the Euler angles, equation (\ref{ali:6t}) is translational
invariant (up to embeddings of measure zero in $SO(3)$), only at
zeroth-order.

\subsection{Analysis of the expectation value of the volume operator
for 8-valent graphs} \label{s7.4}

In this Section we will calculate the expectation value of the
volume operator for an 8-valent graph. As in the case of 4- and
6-valent graphs, we  first consider the non-rotated graph. We
 then analyse the rotational and translational dependence of
the expectation value by performing a rotation and then a
translation of the graph; we then repeat the calculation. It
transpires that, even for the 8-valent graph, the off-diagonal
elements of the matrix have non-trivial contributions that cause
the expectation value of the volume operator to be translationally
and rotationally dependent, for higher orders than the zeroth-one.

\subsubsection{Expectation value of the volume operator for a general
8-valent graph} \label{s7.4.1}

As in the previous cases, we take the $(0,0,0)$ point of the
lattice to coincide with the $(0,0,0)$ point of the plaquette, and
each vertex to be symmetric with respect to the axis. The
coordinates of the vertices comprising the periodicity cell are
the following:
\begin{eqnarray}
V_1&=&\Big(\frac{-\delta}{\sqrt{3}},\frac{-\delta}{\sqrt{3}},\frac{\delta}{\sqrt{3}}\Big)\\
V_{9}&=&\Big(0,\frac{-2\delta}{\sqrt{3}},0\Big)\\
V_{12}&=&\Big(0,\frac{-2\delta}{\sqrt{3}},\frac{2\delta}{\sqrt{3}}\Big)\\
V_{4}&=&\Big(\frac{-\delta}{\sqrt{3}},\frac{-\delta}{\sqrt{3}},\frac{3\delta}{\sqrt{3}}\Big)
\end{eqnarray}
For the 8-valent lattice we choose $V_1$ as our reference vertex.
The allowed value for its $x$-coordinate is
$nl<|x_{V_2}|<nl+\frac{l}{4}$, where
$n=[\frac{\delta}{\sqrt{3}l}]$. Similarly to the cases of  4- and
6-valent graphs, the allowed positions of the remaining vertices
in the periodicity cell can be computed from the allowed positions
of $V_1$. Because of the geometry of the 8-valent graph we obtain
\begin{equation}
T^{\{x,y\}}_{e_i}=F^{\{x,y\}}_{e_i}=Z^{\{x,y\}}_{e_i}=1\hspace{.5in}
\forall  e_i\in \gamma
\end{equation}
This results in the following values for the terms
$\frac{t_{e_{i}e_{j}}} {\sqrt{t_{e_{i}}t_{e_{j}}}}$ as computed
for the above five vertices.
\begin{center}
\begin{tabular}{l|l}
$V_0$& $\text{all} \frac{t_{e_{i}e_{j}}}
{\sqrt{t_{e_{i}}t_{e_{j}}}}=0$\\ \hline $V_1$& $\text{14 terms
equal to }\beta; \text{ 12 terms equal to } 2\beta$\\ \hline
$V_9$& $\text{8 terms equal to }2\beta$\\ \hline $V_{12}$&
$\text{8 terms equal to }4\beta; \text{ 4 terms equal to }
2\beta$\\ \hline $V_{e}$& $\text{1 term equal to }6\beta; \text{ 9
terms equal to } 2\beta; \text{ 12 terms equal to } \beta; \text{
2 terms equal to } 4\beta$\\ \hline
\end{tabular}
\end{center}
Here $\beta:=(\frac{\delta}{\sqrt{3}}-nl)\frac{1}{\sqrt{3}\delta}$
and it is proportional to the off-diagonal entries of the matrix
$\sqrt{A}^{-1}$ whose values are given in Section \ref{sa.3.1} in
the Appendix.

The expectation value of the volume operator is:
\begin{align}
V_R=&\sqrt{\frac{1}{8}}\delta^3|det(\frac{\delta
X_S^a}{\delta(s,u^1,u^2)})|2\;\Bigg(
\sqrt{32}\sqrt{|det\Big(E^a_j(u)\Big)|}+ \sqrt{|det\Big(E^a_j(u)\Big)|}\times\sqrt{|32 - 60 \beta^2 - 28 \beta^3|}\nonumber\\&+\sqrt{|det\Big(E^a_j(u)\Big)|}\times\sqrt{|32- 16 \beta^2|}+\sqrt{|det\Big(E^a_j(u)\Big)|}\times\sqrt{|32- 144 b^2 - 64 \beta^3|}\nonumber\\
&+\sqrt{|det\Big(E^a_j(u)\Big)|}\sqrt{|32 - 104 \beta^2 - 32
\beta^3|}\;\Bigg)
\end{align}
By approximating to first-order (see Section \ref{s4.2}) we obtain
\begin{align}
V_R=&4\delta^3
\sqrt{|det\Big(E^a_j(u)\Big)|}\;\Big|det(\frac{\delta
X_S^a}{\delta(s,u^1,u^2)})\Big|\;\Bigg(5+\frac{1}{2}\times\frac{1}{32}\Big(
324 \beta^2\Big)\Bigg)
\end{align}
In this case, the deviation from the classical value of the volume
of a region, $R$, is of the order four, even to zeroth-order in
$l/\delta$.

 \subsubsection{Expectation value of the volume operator for a rotated
 8-valent graph}
\label{s7.4.2}

We will now analyse the expectation value of the volume operator
for a rotated 8-valent graph. In order to make the comparison with
the 4- and 6-valent graphs as accurate as possible, we will rotate
the 8-valent graph by the same amount the other valence
graphs were rotated. It follows that the angles of the edges incident at $V_0$
will satisfy the following conditions:
\begin{enumerate}
\item[1)] $0<\theta_{0,2},\theta_{e_{0,8}}<\frac{\pi}{2}$, $\frac{3\pi}{2}<\theta_{0,3},\theta_{0,7}<2\pi$ $\frac{\pi}{2}<\theta_{0,1},\theta_{0,6}<\pi$ and $\pi<\theta_{0,4},\theta_{0,5}<\frac{5\pi}{4}$.
\item[2)]$\frac{3\pi}{2}<\phi_{e_{0,2}},\phi_{e_{0,3}}<2\pi-\sin^{-1}\frac{1}{3}$, $\pi+\sin^{-1}(\frac{1}{3})<\phi_{e_{0,7}},\phi_{e_{0,8}}<\frac{3\pi}{2}$,$\frac{3\pi}{2}<\phi_{e_{0,6}},\phi_{e_{0,5}}<\pi-\sin^{-1}(\frac{1}{3})$ and $\sin^{-1}(\frac{1}{3})<\phi_{e_{0,4}}, \phi_{e_{0,1}}<\frac{\pi}{2}$
\end{enumerate}

The angles for the co-linear edges are defined through the formula
$\theta_{e}=180^{\circ}-\theta_{e_{\text{collinear}}}$ and
$\phi_{e}=180^{\circ}-\phi_{e_{\text{collinear}}}$ respectively.
It follows that the edges $e_{0,1}$, $e_{0,2}$, $e_{0,3}$,
$e_{0,4}$, $e_{0,5}$, $e_{0,6}$, $e_{0,7}$ and $e_{0,8}$ lie in
the octants $B$, $A$, $D$, $C$, $G$, $F$, $H$ and $E$
respectively. The coordinates of the rotated vertices are
\begin{eqnarray}
V^{\prime}_1&=&\Big((-R_{11}-R_{12}+R_{13})\frac{\delta}
{\sqrt{3}} , (-R_{21}-R_{22}+R_{23})\frac{\delta} {\sqrt{3}},
(-R_{31}-R_{32}+R_{33})\frac{\delta} {\sqrt{3}}\Big)\\
V^{\prime}_9&=&\Big(-2R_{12}\frac{\delta} {\sqrt{3}},
-2R_{22}\frac{\delta} {\sqrt{3}}, -2R_{32}\frac{\delta} {\sqrt{3}}\Big)\\
V^{\prime}_{12}&=&\Big((-R_{12}+2R_{13})\frac{\delta} {\sqrt{3}},
(-R_{22}+2R_{23})\frac{\delta} {\sqrt{3}},
(-2R_{32}+2R_{33})\frac{\delta} {\sqrt{3}}\Big)\\
V^{\prime}_{e}&=&\Big((-R_{11}-R_{12}+3R_{13})\frac{\delta}
{\sqrt{3}}, (-R_{21}-R_{22}+3R_{23})\frac{\delta} {\sqrt{3}},
(-R_{31}-R_{32}+3R_{33})\frac{\delta} {\sqrt{3}}\Big)
\end{eqnarray}

As was done for the 4- and 6-valent graphs, in order to carry out
the calculations for the expectation value of the volume operator
we would have to specify a particular combinations of angles
satisfying conditions 1) and 2) above. However, all combinations
satisfying 1) and 2) above lead to the same value, in zeroth-order
in $\frac{l}{\delta}$,  of the expectation value of the volume
operator. Rotational dependence will only appear for higher
orders in $\frac{l}{\delta}$. Moreover any sub-case of 1) and 2)
will lead to the same couples of edges commonly intersecting
surfaces $s_{\alpha,t}^I$ in a given stack. Therefore, to leave
the result as general as possible, we will not specify a
particular sub-case of 1) and 2) but simply derive a general
expression for the expectation value of the volume operator given
conditions 1)and 2).

The expectation value of the volume operator for a periodicity
region, $R$, is then computed as
\begin{equation}
\delta^3\sqrt{|\det\big(E^a_j(u)\big)|}\;4\bigg(5+\frac{1}{2}\Big(-\frac{1}{32}\sum^{\frac{n}{2}(n-1)}_{i,j=1;
j\neq i }\alpha_{ji}^2\Big)\,\Big|\det\Big(\frac{\delta
X_S^a}{\delta(s,u^1,u^2)}\Big)\Big|\bigg)
\end{equation}
where the terms $\alpha_{ij}$ are the off-diagonal entries of the
matrix $\sqrt{A}^{-1}$ (See Section \ref{sa.3.2} in the Appendix)
Evidently, the higher-order corrections are angle dependent, while
the zeroth-ones are not. Therefore, as for the 4- and 6-valent
case, the expectation value of the volume operator for the
8-valent graph is rotational invariant, in zeroth-order up to
measure \emph{zero} in $SO(3)$. However, it does \emph{not} reproduce the correct semiclassical limit.

\subsubsection{Expectation value of the volume operator for a
translated 8-valent graph} \label{s7.4.3}

We now consider the translated 8-valent graph. As for the  4- and
6-valent graphs we  choose the following conditions on the
components of the translation vector:
\begin{enumerate}
\item[1)]$b>\epsilon_x>\epsilon_y>\epsilon_z>0$
\item[2)]$|V^i_k|-|V_k^j|>n^i_kl-n^j_kl\mbox{ for all } |V_k^i|>|V_k^j|$
\end{enumerate}
Similarly as for the aligned 8-valent graph we have
\begin{equation}
T^{\{x,y\}}_{e_i}=F^{\{x,y\}}_{e_i}=Z^{\{x,y\}}_{e_i}=1\hspace{.5in}
\forall  e_i\in\gamma
\end{equation}

The coordinates of the translated vertices are
\begin{eqnarray}
V^{'}_0&=&(\epsilon_x,\epsilon_y,\epsilon_z)\\
V^{'}_1&=&\Big(-\frac{\delta} {\sqrt{3}}+\epsilon_x,-\frac{\delta} {\sqrt{3}}+\epsilon_y,\frac{\delta} {\sqrt{3}}+\epsilon_z\Big)\\
V^{'}_9&=&\Big(\epsilon_x,-2\frac{\delta} {\sqrt{3}}+\epsilon_y,\frac{\delta} {\sqrt{3}}+\epsilon_z\Big)\\
V^{'}_{12}&=&\Big(\epsilon_x,-\frac{\delta} {\sqrt{3}}+\epsilon_y,2\frac{\delta} {\sqrt{3}}+\epsilon_z\Big)\\
V^{'}_e&=&\Big(-\frac{\delta} {\sqrt{3}}+\epsilon_x,-\frac{\delta}
{\sqrt{3}}+\epsilon_y,3\frac{\delta} {\sqrt{3}}+\epsilon_z\Big)
\end{eqnarray}
\\
The expression for the matrix $\sqrt{A}^{-1}$ is given in Section
\ref{sa.3.3} in the Appendix.

The value obtained for the volume of a region $R$ is:
\begin{align}
&V_R=\sqrt{\frac{1}{8}}\delta^3
\Big|det\Big(\frac{\delta X_S^a}{\delta(s,u^1,u^2)}\Big)\Big|\times \nonumber\\
&\Bigg\{\sqrt{|4\Big(32 - 4 A_0^2 - 12 B_0^2 - 2 A_0 B_0^2 -
\frac{45 C_0^2}{2} - 2 A_0 C_0^2 - 6 B_0 C_0^2 - \frac{
 5 C_0^2 D_0}{2} - 3 D_0^2 - \frac{B_0 D_0^2}{2} - C_0^2 E}\nonumber\\
&\qquad \overline{ - \frac{B_0 D_0 E_0}{2} - E_0^2 - B_0^2 F_0 -
 C_0^2 F_0 - F_0^2\Big)|}\times\sqrt{|det\Big(E^a_j(u)\Big)|}  \nonumber\\
&+ \sqrt{|4\Big(32 - 16 A_1^2 - 2 A_1^3 - 24 B_1^2 - 8 A_1 B_1^2 -
B_1^2 D_1 - D_1^2 -
 A_1^2 E_1 - B_1^2 E_1 - E_1^2 - 3 B_1^2 F_1 -}\nonumber\\
&\qquad \overline{- \frac{A_1 D_1 F_1}{2} - 3 F_1^2 - \frac{
 A_1 F_1^2}{2}\Big)|}\times\sqrt{|det\Big(E^a_j(u)\Big)|} \nonumber\\
&+ \sqrt{|4\Big(32 - 24 B_9^2 - 4 B_9^2 C_9 - 8 C_9^2 - 6 B_9^2
D_9 - \frac{9 D_9^2}{2} - \frac{
 C_9 D_9^2}{2} - C_9^2 E_9 - \frac{3 E_9^2}{2} - \frac{C_9 E_9^2}{2} - B_9^2 F_9-} \nonumber\\
&\qquad \overline{ - \frac{ C_9 D_9 F_9}{2} - F_9^2
\Big)|}\times\sqrt{|det\Big(E^a_j(u)\Big)|} \nonumber\\
&+ \sqrt{|4\Big(32 - 24 A_{12}^2 - 6 A_{12}^2 B_{12} - 12 B_{12}^2
- 2 A_{12}^2 C_{12} - 2 B_{12}^2 C_{12} - 4 C_{12}^2 -
 A_{12}^2 D_{12} - B_{12}^2 D_{12} - D_{12}^2 - A_{12}^2 E_{12} - E_{12}^2 }\nonumber\\
&\qquad \overline{ - 3 A_{12}^2 F_{12} -\frac{
 B_{12} E_{12} F_{12}}{2}- 3 F_{12}^2 - \frac{B_{12} F_{12}^2}{2}\Big)|}\times\sqrt{|det\Big(E^a_j(u)\Big)|} \nonumber\\
&+ \sqrt{|4\Big(32 - \frac{29 A_e^2}{2} - \frac{43 B_e^2}{2} - 6
A_e B_e^2 - 2 A_e^2 C_e - 2 C_e^2 -
 B_e^2 D_e - D_e^2 - 2 A_e^2 E_e - B_e^2 E_e - \frac{3 E_e^2}{2} - 3 B_e^2 F_e -}\nonumber\\
&\qquad \overline{ \frac{
 A_e D_e F_e}{2} - 3 F_e^2 - \frac{A_e F_e^2}{2}\Big)|}\times\sqrt{|det\Big(E^a_j(u)\Big)|} \Bigg\}
\end{align}
Considering only contributions of first-order in
$\frac{l}{\delta}$ (see Section \ref{s4.2}) we obtain
\begin{align}
&V_R=4\delta^3\sqrt{|det\Big(E^a_j(u)\Big)|} \Big|det\Big(\frac{\delta X_S^a}{\delta(s,u^1,u^2)}\Big)\Big|\times \Bigg\{ \sqrt{1+\frac{1}{32}\Big( - 4 A_0^2 - 12 B_0^2  - \frac{45 C_0^2}{2}  - 3 D_0^2 - E_0^2 -  F_0^2\Big)}\nonumber\\
&+ \sqrt{1+\frac{1}{32}\Big(-16 A_1^2 -  24 B_1^2 - D_1^2 -   -
E_1^2   - 3 F_1^2 \Big)}+\sqrt{1+\frac{1}{32}\Big( - 24 B_9^2
- 8 C_9^2  - \frac{9 D_9^2}{2}  - \frac{3 E_9^2}{2} - F_9^2 \Big)} \nonumber\\
&+ \sqrt{1\frac{1}{32}\Big(-24 A_{12}^2  - 12 B_{12}^2  - 4
C_{12}^2   - D_{12}^2  - E_{12}^2
 - 3 F_{12}^2 \Big)}+ \sqrt{1+\frac{1}{32}\Big( - \frac{29 A_e^2}{2} - \frac{43 B_e^2}{2} - 2 C_e^2  - D_e^2  - \frac{3 E_e^2}{2}
- 3 F_e^2\Big)}\Bigg\}\nonumber\\
&=4\delta^3
\sqrt{|det\Big(E^a_j(u)\Big)|} \Big|det\Big(\frac{\delta X_S^a}{\delta(s,u^1,u^2)}\Big)\Big|\times\Bigg\{ 5+\frac{1}{2}\times\frac{1}{32}\Big( - 4 A_0^2 - 12 B_0^2  - \frac{45 C_0^2}{2}  - 3 D_0^2 - E_0^2 -  F_0^2\nonumber\\
&-16 A_1^2 -  24 B_1^2 - D_1^2 -   - E_1^2   - 3 F_1^2  - 24 B_9^2
- 8 C_9^2  - \frac{9 D_9^2}{2}  - \frac{3 E_9^2}{2} - F_9^2 \Big) \nonumber\\
&-24 A_{12}^2  - 12 B_{12}^2  - 4 C_{12}^2   - D_{12}^2  -
E_{12}^2
 - 3 F_{12}^2- \frac{29 A_e^2}{2} - \frac{43 B_e^2}{2} - 2 C_e^2  - D_e^2  - \frac{3 E_e^2}{2}
- 3 F_e^2 \Big)\Bigg\}
\end{align}

where $A_i$, $B_i$, $C_i$ $D_i$, $E_i$ and $F_i$ are the matrix
elements of $\sqrt{A}^{-1}$ as defined in  Section \ref{sa.3.3} of
the Appendix. Again, translational invariance holds only at
zeroth-order up to measure \emph{zero} in $SO(3)$. However, it does \emph{not} reproduce the correct semiclassical limit.

\section{Summary and Conclusions}
\label{s8}

We have shown that if we use semiclassical states derived from the
area complexifier, then we do \emph{not} obtain the correct semiclassical
value of the volume, operator unless we perform an artificial
re-scaling of the coherent state label and we restrict our
calculation to the following special cases:
\begin{enumerate}
\item[1)] The edges of the graph are aligned with the orientation of the plaquettes (6-valent graph).
\item[2)] Two or more edges lie in a given plaquette (4-valent graph).
\item[3)] One edge is aligned with a given plaquette while a second edge lies in a given plaque (4-valent graph).
\end{enumerate}
However, such combination of edges have measure zero in $SO(3$).
For embeddings whose measure in $SO(3)$ is non-trivial we do not
obtain the correct semiclassical behavior for the volume operator
for any valence of the graph.

This result suggests strongly that the area complexifier coherent
states are not the correct states with which to analyse
semiclassical properties in LQG. Moreover, as previously
mentioned, if embedding independence (staircase problem) is to be
eliminated, area complexifier coherent states should be ruled out
as semiclassical states altogether.
%
%
%
\\
\\
{\large\bf Acknowledgments}\\
\\I would like to thank very much Thomas Thiemann for illuminating discussions and useful comments on the manuscript. I am also grateful to the Perimeter Institute for Theoretical Physics for hospitality and financial support where parts of the present work were carried out.
Research performed at Perimeter Institute for Theoretical Physics
is supported in part by the Government of Canada through NSERC and
by the Province of Ontario through MRI.

\begin{appendix}

\section{Edge metric calculations}
\label{sa}

\subsection{Edge metric components for the 4-valent graph}
\label{sa.1} In this Section we list the values for each of the
terms $\frac{t_{e_{i}e_{j}}} {\sqrt{t_{e_{i}}t_{e_{j}}}}$ which
comprise the matrix $\sqrt{A}^{-1}$ that appears in the
computation of the expectation value of the volume operator. The
method for computing such terms was described in Section
\ref{s7.1} with the aid of a two-dimensional example. It is worth
recalling that the term
$t_{e_{i}e_{j}}=t_{e_{i}e_{j}}^x+t_{e_{i}e_{j}}^y+t_{e_{i}e_{j}}^z$
represents the total number of plaquettes in each direction that
are  intersected by both the edges $e_i$ and $e_j$. On the other hand,  the
terms $t_{e_{i}}=t_{e_{i}}^x+t_{e_{i}}^y+t_{e_{i}}^z$ and
$t_{e_{j}}=t_{e_{j}}^x+t_{e_{j}}^y+t_{e_{j}}^z$ represent are the total number of plaquettes in each direction
intersected by the edges $e_i$ and $e_j$ respectively.
\subsubsection{Rotated 4-valent graph}
\label{sa.1.2}

In this Section we will list the values of all the terms,
$\frac{t_{e_{i}e_{j}}} {\sqrt{t_{e_{i}}t_{e_{j}}}}$, that
 appear in the matrix $\sqrt{A}^{-1}$ for the rotated 4-valent
graph. This is a $15 \times 15$ matrix labelled by the edges
belonging to \emph{four} vertices, as for the aligned case. At this point, we recall the meaning of the following symbols
\begin{align}
&Z^x_{e_i}=|\frac{1}{\cos\theta_{e_i}}\cot\phi_{e_i}|\hspace{.5in}Z^y_{e_i}=|\frac{1}{\sin\theta_{e_i}}\cot\phi_{e_i}|\nonumber\\
&F^x_{e_i}=|\cot\theta_{e_i}|\hspace{.7in}F^y_{e_i}=|\tan\theta_{e_i}|\nonumber\\
&T^x_{e_i}=|\tan\phi_{e_i}\cos\theta_{e_i}|\hspace{.5in}T^y_{e_i}=|\tan\phi_{e_i}\sin\theta_{e_i}|\nonumber\\
&t_{e_i}=t_{e_i}^x+t_{e_i}^y+t_{e_i}^z=\delta
\sin\phi_{e_i}\cos\theta_{e_i}+\delta
\sin\phi_{e_i}\sin\theta_{e_i}+\delta
\cos\phi_{e_i}=\alpha_{e_i}\delta
\end{align}
which represent the angular dependence of the terms $^xt^z$, $^yt^z$, $^yt^x$, $^xt^y$, $^zt^x$ and $^zt^y$ respectively (see Section \ref{s7.1}).

The arrangement of angles for which we choose to perform our calculations
will be such that the edges incident at vertex $V_0$---$e_{0,1}$,
$e_{0,2}$, $e_{0,3}$ and $e_{0,4}$---lie in the octants $F$, $A$,
$H$ and $C$ respectively. However, as mentioned in previous
Sections, in order to compute the specific value for the terms
$t_{e_i,e_j}$, we would need to choose a particular value for each
of the edge's angles. Nonetheless, we will refrain from doing so
here, since we would like to leave our result as general as
possible. What we will do, instead, is to write the terms
$t_{e_i,e_j}$ as functions depending on the angles of the edges
and on the position of the vertex at which they are commonly incident.
Specifically, we recall from Section \ref{s7.1} that the number of
plaquettes a given edge $e_i$ intersects in direction $z$ is given
by \begin{equation}\label{equ:int} t^z_{e_i}:=^xt^z_{e_i}\cap
^yt^z_{e_i}:=\Gamma^z_{e_i}\end{equation} where
\begin{equation*}
^xt^z_{e_i}=\begin{cases}
(x_{V_j}-n_{x_{V_j}}l)Z^x_{e_i} & \text{iff the edge points in the negative $x$-direction} \\
(n_{x_{V_j}}l+l-x_{V_j})Z^x_{e_i} & \text{iff the edge points in
the positive $x$-direction}
\end{cases}
\end{equation*}
and
 \begin{equation*}
^yt^z_{e_i}=\begin{cases}
(x_{V_j}-n_{x_{V_j}}l)Z^y_{e_i} & \text{iff the edge points in the negative $z$-direction} \\
(n_{x_{V_j}}l+l-x_{V_j})Z^y_{e_i} & \text{iff the edge points in
the positive $z$-direction}
\end{cases}
\end{equation*}
Similarly, for intersections in the $x$- and $y$-direction we have the
following:
\begin{equation}t^x_{e_i}:=^zt^x_{e_i}\cap ^yt^x_{e_i}:=\Gamma^x_{e_i}\end{equation}
where
\begin{equation*}
^yt^x_{e_i}=\begin{cases}
(y_{V_j}-n_{y_{V_j}}l)F^x_{e_i} & \text{iff the edge points in the negative $y$-direction} \\
(n_{y_{V_j}}l+l-y_{V_j})F^x_{e_i} & \text{iff the edge points in
the positive $y$-direction}
\end{cases}
\end{equation*}

 \begin{equation*}
^zt^x_{e_i}=\begin{cases}
(z_{V_j}-n_{z_{V_j}}l)T^x_{e_i} & \text{iff the edge points in the negative $z$-direction} \\
(n_{z_{V_j}}l+l-z_{V_j})T^x_{e_i} & \text{iff the edge points in
the positive $z$-direction}
\end{cases}
\end{equation*}
and
\begin{equation}t^y_{e_i}:=^zt^y_{e_i}\cap ^xt^y_{e_i}:=\Gamma^y_{e_i}\end{equation}
where
\begin{equation*}
^xt^y_{e_i}=\begin{cases}
(x_{V_j}-n_{x_{V_j}}l)F^y_{e_i} & \text{iff the edge points in the negative $y$-direction} \\
(n_{x_{V_j}}l+l-x_{V_j})F^y_{e_i} & \text{iff the edge points in
the positive $y$-direction}
\end{cases}
\end{equation*}

 \begin{equation*}
^zt^y_{e_i}=\begin{cases}
(z_{V_j}-n_{z_{V_j}}l)T^y_{e_i} & \text{iff the edge points in the negative $z$-direction} \\
(n_{z_{V_j}}l+l-z_{V_j})T^y_{e_i} & \text{iff the edge points in
the positive $z$- direction}
\end{cases}
\end{equation*}
From the above formulae it is straightforward to deduce that the
precise value for the terms $^it^k_{e_j}$ will differ according to
which angles we choose for each edge. However, which couple of
edges commonly intersect surfaces in a given direction only
depends on the octant in which the edges lie. Since we do not want to
restrict our calculations to any specific angle, we will simply
define the terms $t_{e_i,e_j}$ as follows
\begin{equation}
t_{e_i,e_j}:=\Gamma^z_{e_i}\cap\Gamma^z_{e_j}+\Gamma^x_{e_i}\cap\Gamma^x_{e_j}+\Gamma^y_{e_i}\cap\Gamma^y_{e_j}:=\Xi_{e_i}\cap\Xi_{e_i}
\end{equation}
and
\begin{equation}
\alpha_{e_i,e_j}:=\frac{t_{e_{i}e_{j}}}
{\sqrt{t_{e_{i}}t_{e_{j}}}}:=\frac{\Xi_{e_i}\cap\Xi_{e_i}}{\sqrt{t_{e_{i}}t_{e_{j}}}}
\end{equation}

To simplify the notation, we denote the edge, $e_i$, going from
vertex $V_k$ to vertex $V_j$ as $e_{k,j}$. We can then write the
terms above in the following simplified way:
$\alpha_{e_{k,i},e_{k,j}}=\alpha_{k_i,l_i}$.

Given the above, the non-zero entries of the upper-half entries of
$\sqrt{A}^{-1}$ for the rotated 4-valent graph are listed below.
\begin{center}
\begin{tabular}{lll}\label{tab:a4}
$(\sqrt{A})^{-1}_{5,2}=-\frac{1}{2}\alpha_{7_2,0_2}\qquad $&$\qquad (\sqrt{A})^{-1}_{6,2}=-\frac{1}{2}\alpha_{6_2,0_2}\qquad $&$\qquad (\sqrt{A})^{-1}_{7,2}=-\frac{1}{2}\alpha_{5_2,0_2}$\\
$(\sqrt{A})^{-1}_{6,5}=-\frac{1}{2}\alpha_{6_2,7_2}\qquad $&$\qquad (\sqrt{A})^{-1}_{7,5}=-\frac{1}{2}\alpha_{5_2,7_2}\qquad $&$\qquad (\sqrt{A})^{-1}_{7,6}=-\frac{1}{2}\alpha_{5_2,6_2}$\\
$(\sqrt{A})^{-1}_{9,8}=-\frac{1}{2}\alpha_{5_8,12_8}\qquad $&$\qquad (\sqrt{A})^{-1}_{10,8}=-\frac{1}{2}\alpha_{11_{8},12_{8}}\qquad $&$\qquad (\sqrt{A})^{-1}_{11,8}=-\frac{1}{2}\alpha_{10_8,12_8}$\\
$(\sqrt{A})^{-1}_{10,9}=-\frac{1}{2}\alpha_{11_8,5_8}\qquad $&$\qquad (\sqrt{A})^{-1}_{11,9}=-\frac{1}{2}\alpha_{10_{8},5_{8}}\qquad $&$\qquad (\sqrt{A})^{-1}_{11,10}=-\frac{1}{2}\alpha_{10_8,11_8}$\\
$(\sqrt{A})^{-1}_{13,12}=-\frac{1}{2}\alpha_{15_{13},3_{13}}\qquad $&$\qquad (\sqrt{A})^{-1}_{14,12}=-\frac{1}{2}\alpha_{18_{13},3_{13}}\qquad $&$\qquad (\sqrt{A})^{-1}_{15,12}=-\frac{1}{2}\alpha_{19_{13},3_{13}}$\\
$(\sqrt{A})^{-1}_{14,13}=-\frac{1}{2}\alpha_{18_{13},15_{13}}\qquad $&$\qquad (\sqrt{A})^{-1}_{15,13}=-\frac{1}{2}\alpha_{19_{13},15_{13}}\qquad $&$\qquad (\sqrt{A})^{-1}_{15,14}=-\frac{1}{2}\alpha_{19_{13},18_{13}}$\\
\end{tabular}
\end{center}
Here $(\sqrt{A})^{-1}_{i,j}$ indicates the $i$-th column and the
$j$-th row entry of the matrix $(\sqrt{A})^{-1}$.

\subsubsection{Translated 4-valent graph}
\label{sa.1.3} In this Section we  give the matrix entries of
$\sqrt{A}^{-1}$ obtained after translating the 4-valent graph. As
for the aligned and rotated graphs discussed above, there are
four
vertices in the periodicity cell. The values for the terms $t_{e_ie_j}=t_{e_ie_j}^x+t_{e_ie_j}^y+t_{e_ie_j}^z$ as obtained for each vertex are given below. The (Gauss brackets) terms $n_i=[\frac{V_i^x}{l}]$, $m_i=[\frac{V_i^y}{l}]$ and $p_i=[\frac{V_i^z}{l}]$ represent the number of stacks which each edge $e_i$ intersects in the $x$-, $y$- and $z$-direction respectively (see Section \ref{s7.1}). \\[15pt]
1) $V^{''}_0=(\epsilon_x,\epsilon_y,\epsilon_z)$ with
$\epsilon_x>\epsilon_y>\epsilon_z$
\begin{center}
\begin{tabular}{l|l|l}
$(V^{''}_0)^z$& $(V^{''}_0)^x$&$(V^{''}_0)^y$\\\hline
$t^z_{e_{0,1}e_{0,3}}=t^z_{e_{0,2}e_{0,4}}=(\epsilon_y-m_0l)$&
$t^x_{e_{0,3}e_{0,4}}=t^x_{e_{0,1}e_{0,2}}=(\epsilon_z-p_0l)$
&$=t^y_{e_{0,3}e_{0,2}}=t^y_{e_{0,1}e_{0,4}}=(\epsilon_z-p_0l)$
\end{tabular}
\end{center}
2) $V^{''}_2=(-\frac{\delta_e}{
\sqrt{3}}+\epsilon_x,\epsilon_y+\frac{\delta_e}{
\sqrt{3}},\epsilon_z+\frac{\delta_e}{ \sqrt{3}})$
\begin{center}
\begin{tabular}{l|l|l}
$(V^{''}_2)^z$& $(V^{''}_2)^x$&$(V^{''}_2)^y$\\\hline
$t^z_{e_{2,6}e_{2,5}}=t^z_{e_{2,0}e_{2,7}}=(|\epsilon_x-\frac{\delta_e}{
\sqrt{3}}|-n_2l)$&$t^x_{e_{2,0}e_{2,5}}=t^x_{e_{2,7}e_{2,6}}=(\frac{\delta_e}{
\sqrt{3}}+\epsilon_z-p_2l)$&$t^y_{e_{2,0}e_{2,6}}=t^y_{e_{2,5}e_{2,7}}=(|\epsilon_x-\frac{\delta_e}{
\sqrt{3}}|-n_2l)$
\end{tabular}
\end{center}

3) $V^{''}_8=(\frac{\delta_e}{
\sqrt{3}}+\epsilon_x,\epsilon_y+3\frac{\delta_e}{
\sqrt{3}},\epsilon_z+\frac{\delta_e}{ \sqrt{3}})$
\begin{center}
\begin{tabular}{l|l|l}
$(V^{''}_8)^z$& $(V^{''}_8)^x$&$(V^{''}_8)^y$\\\hline
$t^z_{e_{8,10}e_{8,11}}=t^z_{e_{8,5}e_{8,12}}=(\epsilon_x+\frac{\delta_e}{
\sqrt{3}}-m_8l)$&
$t^x_{e_{8,5}e_{8,10}}=t^x_{e_{8,12}e_{8,11}}=(\epsilon_z+\frac{\delta_e}{
\sqrt{3}}-p_8l)$
&$t^y_{e_{8,11}e_{8,5}}=t^y_{e_{8,10}e_{8,12}}=(\frac{\delta_e}{
\sqrt{3}}+\epsilon_z-p_8l)$
\end{tabular}
\end{center}

5) $V^{''}_{13}=(\epsilon_x,\epsilon_y+2\frac{\delta_e}{
\sqrt{3}},\epsilon_z-2\frac{\delta_e}{ \sqrt{3}})$
\begin{center}
\begin{tabular}{l|l|l}
$(V^{''}_{13})^z$& $(V^{''}_{13})^x$&$(V^{''}_{13})^y$\\\hline
$t^z_{e_{13,3}e_{13,15}}=t^z_{e_{13,19}e_{13,18}}=(\epsilon_x-n_{13}l)$&
$t^x_{e_{13,19}e_{13,15}}=t^x_{e_{13,18}e_{13,3}}=(|\epsilon_z-2\frac{\delta_e}{ \sqrt{3}}|-p_{13}l)$ &$t^y_{e_{13,15}e_{13,18}}=t^y_{e_{13,19}e_{13,3}}=$\\
& & $=(\epsilon_x-n_{13}l)$
\end{tabular}
\end{center}

The upper-half entries of the
matrix $\sqrt{A}^{-1}$ are listed in the following table. It should be noted that the lower indices refer to the matrix entries, while the upper ones indicate the edges that we are considering. Therefore $(\sqrt{A}^{-1})^{1_0,2_0}_{2,1}$ denotes the matrix entry at row 1 column 2, which is the term $\frac{t_{e_{0,1},e_{0,2}}}{\sqrt{t_{e_{0,1}}t_{e_{0,2}}}}$ corresponding to the edges $e_{0,1}$ and $e_{0,2}$, which join vertex $V_0$ to $V_1$ and $V_0$ to $V_2$ respectively.\\[5pt]
\begin{center}
\begin{tabular}{lll}\label{tab:a4t}
$(\sqrt{A}^{-1})^{1_0,2_0}_{2,1}=-\frac{1}{2}C_0\qquad $&$\qquad (\sqrt{A}^{-1})^{1_0,3_0}_{3,1}=-\frac{1}{2}B_0\qquad $&$\qquad (\sqrt{A}^{-1})^{1_0,4_0}_{4,1}=-\frac{1}{2}C_0$\\
$(\sqrt{A}^{-1})^{2_0,3_0}_{3,2}=-\frac{1}{2}C_0\qquad $&$\qquad (\sqrt{A}^{-1})^{2_0,4_0}_{4,2}=-\frac{1}{2}B_0\qquad $&$\qquad (\sqrt{A}^{-1})^{0_2,7_2}_{5,2}=-\frac{1}{2}A_2$\\
$(\sqrt{A}^{-1})^{0_2,5_2}_{6,2}=-\frac{1}{2}C_2\qquad $&$\qquad (\sqrt{A}^{-1})^{0_2,6_2}_{7,2}=-\frac{1}{2}A_2\qquad $&$\qquad (\sqrt{A}^{-1})^{3_0,4_0}_{4,3}=-\frac{1}{2}C_0$\\
$(\sqrt{A}^{-1})^{7_2,5_2}_{6,5}=-\frac{1}{2}A_2\qquad $&$\qquad (\sqrt{A}^{-1})^{7_2,6_2}_{7,5}=-\frac{1}{2}C_2\qquad $&$\qquad (\sqrt{A}^{-1})^{5_2,6_2}_{7,6}=-\frac{1}{2}B_2$\\
$(\sqrt{A}^{-1})^{19_{13},18_{13}}_{9,8}=-\frac{1}{2}A_{13}\qquad $&$\qquad (\sqrt{A}^{-1})^{19_{13},15_{13}}_{10,8}=-\frac{1}{2}C_{13}\qquad $&$\qquad (\sqrt{A}^{-1})^{19_{13},3_{13}}_{11,8}=-\frac{1}{2}A_{13}$\\
$(\sqrt{A}^{-1})^{18_{13},15_{13}}_{10,9}=-\frac{1}{2}A_{13}\qquad $&$\qquad (\sqrt{A}^{-1})^{18_{13},3_{13}}_{11,9}=-\frac{1}{2}C_{13}\qquad $&$\qquad (\sqrt{A}^{-1})^{15_{13},3_{13}}_{11,10}=-\frac{1}{2}A_{13}$\\
$(\sqrt{A}^{-1})^{10_{8},11_{8}}_{13,12}=-\frac{1}{2}A_{8}\qquad $&$\qquad (\sqrt{A}^{-1})^{10_{8},12_{8}}_{14,12}=-\frac{1}{2}C_{8}\qquad $&$\qquad (\sqrt{A}^{-1})^{10_{8},5_{8}}_{15,12}=-\frac{1}{2}C_{8}$\\
$(\sqrt{A}^{-1})^{11_{8},12_{8}}_{14,13}=-\frac{1}{2}C_{8}\qquad $&$\qquad (\sqrt{A}^{-1})^{11_{8},5_{8}}_{15,13}=-\frac{1}{2}C_{8}\qquad $&$\qquad (\sqrt{A}^{-1})^{12_{8},5_{8}}_{15,14}=-\frac{1}{2}A_{8}$\\
\end{tabular}
\end{center}
where
$A_i=(|x_{V_i}|-n_il)\frac{1}{\delta\sqrt{3}}$,
$B_i=(|y_{V_i}|-m_il)\frac{1}{\delta\sqrt{3}}$,
$C_i=(|z_{V_i}|-p_il)\frac{1}{\delta\sqrt{3}}$.

In all the above formulae, the term $\frac{1}{\delta_e\sqrt{3}}$
comes from the fact that
$t_{e_i}=t^x_{e_i}+t^y_{e_i}+t^z_{e_i}=\delta_e\sqrt{3}$ for all
edges $e_i$
\subsection{Edge metric components for the 6-valent graph}
\label{sa.2} In this Section we give the values of the terms
$\frac{t_{e_{i,j}e_{i,j}}} {\sqrt{t_{e_{i,j}}t_{e_{i,j}}}}$ that
represent the entries of the matrix $\sqrt{A}^{-1}$ for the
6-valent graph. We do this for all non-rotated, rotated and
translated cases respectively.
\subsubsection{Aligned 6-valent graph}
\label{sa.2.1}
For the 6-valent graph, the periodicity cell contains \emph{nine} vertices. Thus the matrix $\sqrt{A}^{-1}$ is a $46\times 46$ matrix.

We will now give the values for the terms
$\frac{t_{e_{k,i},e_{k,j}}}{\sqrt{t_{e_{k,i}}t_{e_{k,j}}}}$ which
we denote as $\Gamma_{i_k,j_k}$. We will also use the following
short-hand notations
\begin{align}
&\alpha:=\frac{\frac{\sqrt{2}}{2}-nl}{\delta (1+\frac{\sqrt{2}}{2})}\hspace{.5in}\alpha^{'}:=\frac{\frac{\sqrt{2}}{2}-nl}{\delta \sqrt{\sqrt{2}+1}}\nonumber\\
&\beta:=\frac{\frac{1}{2}-n^{''}l}{\delta (1+\frac{\sqrt{2}}{2}})\hspace{.5in}\beta^{'}:=\frac{\frac{1}{2}-n^{''}l}{\delta \sqrt{\sqrt{2}+1}}\nonumber\\
&A_k:=\frac{|V^x_k|-n^{'}l}{\delta
(1+\frac{\sqrt{2}}{2})}\hspace{.5in}A_k^{'}:=\frac{|V^x_k|-n^{'}l}{\delta
\sqrt{\sqrt{2}+1}}
\end{align}
1) $V_0$ all terms $\Gamma_{i_k,j_k}=0$\\[5pt]
2) $V_2$
\begin{center}
\begin{tabular}{l|l|l}
$\Gamma_{0_2,11_2}=-\frac{\alpha^{'}}{2}$&$\Gamma_{0_2,3_2}=-\frac{\alpha^{'}}{2}$&$\Gamma_{0_2,10_2}=-\frac{\alpha^{'}}{2}$\\
$\Gamma_{0_2,9_2}=-\frac{\alpha^{'}}{\sqrt{2}}$&$\Gamma_{11_2,12_2}=-\frac{\alpha^{'}}{\sqrt{2}}$&$\Gamma_{11_2,10_2}=-\frac{\alpha\sqrt{2}}{4}$\\
$\Gamma_{12_2,3_2}=-\frac{\alpha^{'}}{\sqrt{2}}$&$\Gamma_{12_2,10_2}=-\frac{\alpha^{'}\sqrt{2}}{4}$&$\Gamma_{12_2,9_2}=-\frac{\alpha^{'}\sqrt{2}}{4}$\\
$\Gamma_{3_2,9_2}=-\frac{\alpha\sqrt{2}}{4}$& &
\end{tabular}
\end{center}
3) $V_3$
\begin{center}
\begin{tabular}{l|l|l}
$\Gamma_{2_3,17_3}=-\frac{A_3^{'}}{2}$&$\Gamma_{2_3,13_3}=-\beta(\frac{1}{\sqrt{2}}+\frac{1}{2})$&$\Gamma_{2_3,19_3}=-\frac{A_3}{2}$\\
$\Gamma_{2_3,16_3}=-\frac{\beta}{2}$&$\Gamma_{17_3,13_3}=-\frac{A_3^{'}}{2}$&$\Gamma_{17_3,19_3}=-\frac{A_3^{'}}{2}$\\
$\Gamma_{17_3,15_3}=-\beta^{'}$&$\Gamma_{13_3,15_3}=-\frac{A_3}{2}$&$\Gamma_{16_3,13_3}=-\frac{\alpha^{'}\sqrt{2}}{2}$\\
$\Gamma_{19_3,15_3}=-A_3(\frac{1}{\sqrt{2}}-\frac{1}{2})$&$\Gamma_{19_3,16_3}=-\frac{\beta^{'}}{\sqrt{2}}$&$\Gamma_{16_3,15_3}=-\frac{A_3^{'}}{\sqrt{2}}$
\end{tabular}
\end{center}
4) $V_4$
\begin{center}
\begin{tabular}{l|l|l}
$\Gamma_{a_4,18_4}=-\beta(\frac{2}{\sqrt{2}}+1)$&$\Gamma_{a_4,13_4}=-\alpha^{'}$&$\Gamma_{a_4,c_4}=-\frac{\alpha\sqrt{2}}{2}$\\
$\Gamma_{a_4,b_4}=-\frac{\alpha^{'}\sqrt{2}}{2}$&$\Gamma_{18_4,13_4}=-\frac{\beta^{'}2}{\sqrt{2}}$&$\Gamma_{18_4,b_4}=-\beta^{'}$\\
$\Gamma_{18_4,16_4}=-\frac{\alpha\sqrt{2}}{2}$&$\Gamma_{13_4,c_4}=-\alpha^{'}$&$\Gamma_{13_4,16_4}=-\beta^{'}$\\
$\Gamma_{b_4,c_4}=-\frac{\alpha^{'}2}{\sqrt{2}}$&$\Gamma_{16_4,c_4}=-\beta(\frac{2}{\sqrt{2}}+1)$&$\Gamma_{b_4,16_4}=-\frac{\beta^{'}2}{\sqrt{2}}$
\end{tabular}
\end{center}
5) $V_5$
\begin{center}
\begin{tabular}{l|l}
$\Gamma_{r_5,17_5}=-\beta(\frac{2}{\sqrt{2}}+1)$&$\Gamma_{6_5,q_5}=-\beta(\frac{2}{\sqrt{2}}+1)$
\end{tabular}
\end{center}
6) $V_{13}$
\begin{center}
\begin{tabular}{l|l|l}
$\Gamma_{5_{13},9_{13}}=-\frac{\alpha^{'}}{2}$&$\Gamma_{5_{13},3_{13}}=-\frac{\alpha^{'}}{2}$&$\Gamma_{5_{13},21_{13}}=-\frac{\alpha^{'}}{2}$\\
$\Gamma_{5_{13},14_{13}}=-\frac{\alpha^{'}}{2}$&$\Gamma_{4_{13},9_{13}}=-\frac{\sqrt{2}\alpha^{'}}{4}$&$\Gamma_{4_{13},3_{13}}=-\frac{2\beta^{'}}{\sqrt{2}}$\\
$\Gamma_{4_{13},21_{13}}=-\frac{\alpha^{'}}{\sqrt{2}}$&$\Gamma_{4_{13},14_{13}}=-\frac{\alpha^{'}\sqrt{2}}{4}$&$\Gamma_{3_{13},9_{13}}=-\frac{\alpha\sqrt{2}}{4}$\\
$\Gamma_{9_{13},14_{13}}=-\frac{2\beta}{\sqrt{2}}-\frac{\alpha\sqrt{2}}{4}$&$\Gamma_{3_{13},21_{13}}=-\beta-\frac{\alpha}{\sqrt{2}}$&$\Gamma_{21_{13},14_{13}}=-\frac{\alpha\sqrt{2}}{4}$
\end{tabular}
\end{center}
7) $V_{18}$
\begin{center}
\begin{tabular}{l|l|l}
$\Gamma_{4_{18},9_{18}}=-\frac{\alpha^{'}}{2}$&$\Gamma_{4_{18},12_{18}}=-\beta(\frac{1}{\sqrt{2}}+\frac{1}{2})$&$\Gamma_{4_{18},d_{18}}=-\frac{A_{18}^{'}}{\sqrt{2}}$\\
$\Gamma_{4_{18},e_{18}}=-\frac{\sqrt{2}\alpha}{4}$&$\Gamma_{12_{18},9_{18}}=-\frac{\beta^{'}}{2}$&$\Gamma_{f_{18},9_{18}}=-\frac{\beta^{'}}{\sqrt{2}}$\\
$\Gamma_{e_{18},9_{18}}=-\frac{A_{18}^{'}}{2}$&$\Gamma_{12_{18},f_{18}}=-\frac{\alpha\sqrt{2}}{4}$&$\Gamma_{d_{18},12_{18}}=-\frac{\beta^{'}}{\sqrt{2}}$\\
$\Gamma_{d_{18},f_{18}}=-\frac{\beta^{'}}{2}$&$\Gamma_{e_{18},f_{18}}=-\beta(\frac{1}{\sqrt{2}}-\frac{1}{2})$&$\Gamma_{e_{18},d_{18}}=-\frac{\alpha^{'}\sqrt{2}}{4}$
\end{tabular}
\end{center}
8) $V_{15}$
\begin{center}
\begin{tabular}{l|l|l}
$\Gamma_{3_{15},d^{'}_{15}}=-\frac{A_{15}}{2}$&$\Gamma_{3_{15},c^{'}_{15}}=-\frac{3\alpha^{'}\sqrt{2}}{4}$&$\Gamma_{3_{15},f^{'}_{15}}=-\frac{A^{'}_{15}}{\sqrt{2}}$\\
$\Gamma_{3_{15},a^{'}_{15}}=-\frac{A_{15}}{\sqrt{2}}-\beta$&$\Gamma_{e^{'}_{15},d^{'}_{15}}=-\beta(\frac{2}{\sqrt{2}}+1)$&$\Gamma_{e^{'}_{15},c^{'}_{15}}=-\frac{A^{'}_{15}}{2}$\\
$\Gamma_{e^{'}_{15},f^{'}_{15}}=-\frac{3\alpha^{'}}{2}$&$\Gamma_{e^{'}_{15},a^{'}_{15}}=-\frac{A_{15}}{2}$&$\Gamma_{c^{'}_{15},d^{'}_{15}}=-\frac{A^{'}_{15}}{2}$\\
$\Gamma_{f^{'}_{15},d^{'}_{15}}=-\beta^{'}$&$\Gamma_{c^{'}_{15},a^{'}_{15}}=-\frac{3\alpha^{'}\sqrt{2}}{4}$&$\Gamma_{a^{'}_{15},f^{'}_{15}}=-\frac{A^{'}_{15}}{\sqrt{2}}$
\end{tabular}
\end{center}
9) $V_{14}$
\begin{center}
\begin{tabular}{l|l|l}
$\Gamma_{13_{14},a_{14}}=-\frac{3\beta^{'}}{\sqrt{2}}$&$\Gamma_{h_{14},a_{14}}=-\frac{A_{14}^{'}}{\sqrt{2}}$&$\Gamma_{13_{14},h_{14}}=-\frac{A_{14}}{\sqrt{2}}-\frac{3\beta}{\sqrt{2}}$\\
$\Gamma_{j_{14},l_{14}}=-\beta(\frac{3}{\sqrt{2}}+\frac{3}{2})$&$\Gamma_{j_{14},k_{14}}=-\frac{A^{'}_{14}}{2}$&$\Gamma_{l_{14},k_{14}}=-\frac{A^{'}_{14}}{2}$
\end{tabular}
\end{center}
The upper-diagonal entries of the matrix $\sqrt{A}^{-1}$ are
listed below. Similarly, as for the 4-valent case, the lower
indices represent the matrix entries while the top indices
indicate the edges we are considering.
\begin{center}
\begin{tabular}{lll}\label{tab:a6a}
$(\sqrt{A}^{-1})^{0_2, 11_2}_{7,2}=-\frac{\alpha^{'}}{2}\qquad $&$\qquad (\sqrt{A}^{-1})^{0_2, 3_2}_{9,2}=-\frac{\alpha^{'}}{2} \qquad $&$\qquad (\sqrt{A}^{-1})^{0_2, 10_2}_{10,2}=-\frac{\alpha^{'}}{2}$\\
$(\sqrt{A}^{-1})^{0_2, 9_2}_{11,2}=-\frac{\alpha^{'}}{\sqrt{2}}\qquad $&$\qquad (\sqrt{A}^{-1})^{12_2, 12_2}_{8,7}=-\frac{\alpha^{'}}{\sqrt{2}}\qquad $&$\qquad (\sqrt{A}^{-1})^{11_2, 10_2}_{10,7}=-\frac{\alpha\sqrt{2}}{4}$\\
$(\sqrt{A}^{-1})^{12_2, 3_2}_{9,8}=-\frac{\alpha^{'}}{\sqrt{2}}\qquad $&$\qquad (\sqrt{A}^{-1})^{12_2, 10_2}_{10,8}=-\frac{\alpha^{'}\sqrt{2}}{4}\qquad $&$\qquad (\sqrt{A}^{-1})^{12_2, 9_2}_{11,8}=-\frac{\alpha^{'}\sqrt{2}}{4}$\\
$(\sqrt{A}^{-1})^{3_2, 9_2}_{11,9}=-\frac{\alpha\sqrt{2}}{4}\qquad $&$\qquad (\sqrt{A}^{-1})^{2_3, 17_3}_{12,9}=-\frac{A^{'}_3}{2}\qquad $&$\qquad (\sqrt{A}^{-1})^{2_3, 13_3}_{13,9}=-\beta(\frac{1}{\sqrt{2}}+\frac{1}{2})$\\
$(\sqrt{A}^{-1})^{2_3, 19_3}_{14,9}=-\frac{A_3}{2}\qquad $&$\qquad (\sqrt{A}^{-1})^{2_3, 16_3}_{16,9}=-\frac{\beta^{'}}{2}\qquad $&$\qquad (\sqrt{A}^{-1})^{17_3, 13_3}_{13,12}=-\frac{A^{'}_3}{2}$\\
$(\sqrt{A}^{-1})^{17_3, 19_3}_{14,12}=-\frac{A^{'}_3}{2}\qquad $&$\qquad (\sqrt{A}^{-1})^{17_3, 15_3}_{15,12}=-\beta^{'}\qquad $&$\qquad (\sqrt{A}^{13_3, 15_3})^{-1}_{15,13}=-\frac{A_3^{'}}{2}$\\
$(\sqrt{A}^{-1})^{13_3, 16_3}_{16,13}=-\frac{\alpha^{'}\sqrt{2}}{2}\qquad $&$\qquad (\sqrt{A}^{-1})^{3_{13}, 4_{13}}_{17,13}=-\frac{2\beta^{'}}{\sqrt{2}}\rho\qquad $&$\qquad (\sqrt{A}^{-1})^{3_{13}, 5_{13}}_{18,13}=-\frac{\alpha^{'}}{2}$\\
$(\sqrt{A}^{-1})^{3_{13}, 9_{13}}_{20,13}=-\frac{\alpha\sqrt{2}}{2}\qquad $&$\qquad (\sqrt{A}^{-1})^{3_{13}, 21_{13}}_{21,13}=-\frac{\alpha\sqrt{2}}{4}\qquad $&$\qquad (\sqrt{A}^{-1})^{19_3, 15_3}_{15,14}=-A_3(\frac{1}{2}+\frac{1}{\sqrt{2}})$\\
$(\sqrt{A}^{-1})^{19_3, 16_3}_{16,14}=-\frac{\beta^{'}}{\sqrt{2}}\qquad $&$\qquad (\sqrt{A}^{-1})^{15_3, 16_3}_{17,15}=-\frac{A^{'}_3}{\sqrt{2}}\qquad $&$\qquad (\sqrt{A}^{-1})^{3_{15}, d^{'}_{15}}_{38,15}=-\frac{A_{15}}{2}$\\
$(\sqrt{A}^{-1})^{3_{15}, c^{'}_{15}}_{39,15}=-\frac{3\alpha^{'}\sqrt{2}}{4}\qquad $&$\qquad (\sqrt{A}^{-1})^{3_{15}, f^{'}_{15}}_{40,15}=-\frac{A_{15}}{\sqrt{2}}-\beta\rho\qquad $&$\qquad (\sqrt{A}^{-1})^{a_{4}, 18_{4}}_{18,17}=-\frac{2\beta}{\sqrt{2}}-\beta$\\
$(\sqrt{A}^{-1})^{a_{4}, 13_{4}}_{19,17}=-\alpha^{'}\qquad $&$\qquad (\sqrt{A}^{-1})^{a_{4}, c_{4}}_{20,17}=-\frac{\alpha\sqrt{2}}{4}\qquad $&$\qquad (\sqrt{A}^{-1})^{a_{4}, b_{4}}_{21,17}=-\frac{\alpha^{'}\sqrt{2}}{2}$\\
$(\sqrt{A}^{-1})^{18_{4}, 13_{4}}_{19,18}=-\frac{\beta^{'}2}{\sqrt{2}} \qquad $&$\qquad (\sqrt{A}^{-1})^{18_{4}, b_{4}}_{21,18}=-\beta^{'}\qquad $&$\qquad (\sqrt{A}^{-1})^{18_{4}, 16_{4}}_{22,18}=-\frac{\alpha\sqrt{2}}{2}$\\
$(\sqrt{A}^{-1})^{4_{18},
9_{18}}_{32,18}=-\frac{\alpha^{'}}{2}\qquad $&$\qquad
(\sqrt{A}^{-1})^{4_{18},
12_{18}}_{33,18}=-\beta(\frac{1}{\sqrt{2}}+\frac{1}{2})\qquad
$&$\qquad
(\sqrt{A}^{-1})^{4_{18}, d_{18}}_{35,18}=-\frac{A^{'}_{18}}{\sqrt{2}}$\\
$(\sqrt{A}^{-1})^{4_{18},
e_{18}}_{36,18}=-\frac{\alpha\sqrt{2}}{4}\qquad $&$\qquad
(\sqrt{A}^{-1})^{13_{4}, c_{4}}_{20,19}=-\alpha^{'}\qquad $&$\qquad (\sqrt{A}^{-1})^{13_{4}, 16_{4}}_{22,19}=-\beta^{'} $\\
$ (\sqrt{A}^{-1})^{4_{13},
14_{13}}_{29,19}=-\frac{\alpha^{'}\sqrt{2}}{4}\rho\qquad $&$\qquad
(\sqrt{A}^{-1})^{4_{13}, 9_{13}}_{30,19}=-\frac{\alpha^{'}\sqrt{2}}{4}\qquad $&$\qquad (\sqrt{A}^{-1})^{4_{13}, 21_{13}}_{31,19}=-\frac{\alpha^{'}}{\sqrt{2}}$\\
$(\sqrt{A}^{-1})^{c_{4}, b_{4}}_{21,20}=-\frac{2\alpha^{'}}{\sqrt{2}}\qquad $&$\qquad(\sqrt{A}^{-1})^{c_{4}, 16_{4}}_{22,20}=-\beta(\frac{2}{\sqrt{2}}+1)\qquad $&$\qquad (\sqrt{A}^{-1})^{b_{4}, 16_{4}}_{22,21}=-\frac{2\beta^{'}}{\sqrt{2}} $\\
$(\sqrt{A}^{-1})^{r_{15}, 17_{15}}_{23,24}=-\beta(\sqrt{2}+1)\qquad $&$\qquad(\sqrt{A}^{-1})^{5_{13}, 14_{13}}_{29,25}=-\frac{\alpha^{'}}{2}\qquad $&$\qquad (\sqrt{A}^{-1})^{5_{13}, 9_{13}}_{30,25}=-\frac{\alpha^{'}}{2}$\\
$ (\sqrt{A}^{-1})^{5_{13}, 21_{13}}_{31,25}=-\frac{\alpha^{'}}{2}\qquad $&$\qquad(\sqrt{A}^{-1})^{q_{5}, 6_{5}}_{28,27}=-\beta(\sqrt{2}+1)\qquad $&$\qquad (\sqrt{A}^{-1})^{14_{13}, 9_{13}}_{30,29}=-\frac{2\beta}{\sqrt{2}}-\frac{\sqrt{2}\alpha}{4}$\\
$ (\sqrt{A}^{-1})^{14_{13},
21_{13}}_{31,29}=-\frac{\alpha\sqrt{2}}{4}\qquad
$&$\qquad(\sqrt{A}^{-1})^{13_{14},
a_{14}}_{43,29}=-\frac{2\beta^{'}}{\sqrt{2}}\qquad $&$\qquad
(\sqrt{A}^{-1})^{13_{14}, h_{14}}_{46,29}=-\frac{A_{14}}{\sqrt{2}}-\frac{3\beta}{2}$\\
$(\sqrt{A}^{-1})^{9_{18}, 12_{18}}_{33,32}=-\frac{\beta^{'}}{2}\qquad $&$\qquad (\sqrt{A}^{-1})^{9_{18}, f_{18}}_{34,32}=-\frac{\beta^{'}}{2}\qquad $&$\qquad (\sqrt{A}^{-1})^{9_{18}, e_{18}}_{36,33}=-\frac{A^{'}_{18}}{2}\rho$\\
$(\sqrt{A}^{-1})^{12_{18}, f_{18}}_{35,34}=-\frac{\alpha\sqrt{2}}{4}\qquad $&$\qquad (\sqrt{A}^{-1})^{12_{18}, d_{18}}_{36,34}=-\frac{\beta^{'}}{\sqrt{2}}\qquad $&$\qquad (\sqrt{A}^{-1})^{f_{18}, d_{18}}_{36,35}=-\frac{\beta^{'}}{2}$\\
$(\sqrt{A}^{-1})^{f_{18}, e_{18}}_{37,35}=-\beta(\frac{1}{\sqrt{2}}+\frac{1}{2})\qquad $&$\qquad (\sqrt{A}^{-1})^{d_{18}, e_{18}}_{37,36}=-\frac{\alpha^{'}\sqrt{2}}{4}\qquad $&$\qquad (\sqrt{A}^{-1})^{e^{'}_{15}, d^{'}_{15}}_{39,38}=-\beta(\sqrt{2}+1)$\\
$(\sqrt{A}^{-1})^{e^{'}_{15}, c^{'}_{15}}_{40,38}=-\frac{A^{'}_{15}}{2}\qquad $&$\qquad (\sqrt{A}^{-1})^{e^{'}_{15}, f^{'}_{15}}_{41,39}=-\frac{3\alpha^{'}}{2}\qquad $&$\qquad (\sqrt{A}^{-1})^{e^{'}_{15}, a^{'}_{15}}_{42,39}=-\frac{A_{15}}{2}$\\
$(\sqrt{A}^{-1})^{d^{'}_{15}, c^{'}_{15}}_{41,40}=-\frac{A^{'}_{15}}{2}\rho\qquad $&$\qquad (\sqrt{A}^{-1})^{d^{'}_{15}, f^{'}_{15}}_{42,40}=-\beta^{'}\qquad $&$\qquad (\sqrt{A}^{-1})^{c^{'}_{15}, a^{'}_{15}}_{43,41}=-\frac{3\alpha^{'}\sqrt{2}}{4}$\\
$(\sqrt{A}^{-1})^{f^{'}_{15}, a^{'}_{15}}_{43,42}=-\frac{A^{'}_{15}}{\sqrt{2}}\qquad $&$\qquad (\sqrt{A}^{-1})^{k_{14}, l_{14}}_{45,43}=-\frac{A^{'}_{14}}{2}\qquad $&$\qquad (\sqrt{A}^{-1})^{k_{14}, j_{14}}_{46,43}=-\frac{A^{'}_{14}}{2}$\\
$(\sqrt{A}^{-1})^{a_{14},
h_{14}}_{47,44}=-\frac{A^{'}_{14}}{\sqrt{2}}\qquad $&$\qquad
(\sqrt{A}^{-1})^{l_{14},
j_{14}}_{46,45}=-3\beta(\frac{1}{\sqrt{2}}+\frac{1}{2})$&
\end{tabular}
\end{center}

\subsubsection{Rotated 6-valent graph}
\label{sa.2.2}

In this Section we give the values of all the terms $\frac{t_{e_{i}e_{j}}} {\sqrt{t_{e_{i}}t_{e_{j}}}}$ that appear in the matrix $\sqrt{A}^{-1}$ for the rotated 6-valent graph. In order to carry out the calculations we will rotate the graph such that the edges incident at vertex $V_0$: $e_{0,1}$, $e_{0,2}$, $e_{0,6}$, $e_{0,7}$, $e_{0,8}$ and $e_{0,17}$ lie in the octants $H$, $D$, $G$, $F$, $A$ and $B$ respectively. As for the aligned graph, we consider the periodicity cell composed of nine vertices. Therefore $\sqrt{A}^{-1}$ will be a $46\times 46$ matrix.

In what follows the terms $\rho_{i_k,j_k}$ represent the values for $\frac{t_{e_{k,i}e_{k,j}}} {\sqrt{t_{e_{k,i}}t_{e_{k,j}}}}$ and are defined as for the 4-valent case (see \ref{sa.1.2}) \\
Given the above, the upper-half elements of the matrix
$\sqrt{A}^{-1}$ are:
\begin{center}
\begin{tabular}{lll}\label{tab:a6}
$(\sqrt{A})^{-1}_{20,2}=-\frac{1}{2}\rho_{0_2,9_2}\qquad $&$\qquad (\sqrt{A})^{-1}_{22,2}=-\frac{1}{2}\rho_{0_2,11_2}\qquad $&$\qquad (\sqrt{A})^{-1}_{24,2}=-\frac{1}{2}\rho_{0_2,3_2}$\\
$(\sqrt{A})^{-1}_{25,2}=-\frac{1}{2}\rho_{0_2,10_2}\qquad $&$\qquad (\sqrt{A})^{-1}_{8,7}=-\frac{1}{2}\rho_{a_4,18_4}\qquad $&$\qquad (\sqrt{A})^{-1}_{9,7}=-\frac{1}{2}\rho_{a_4,13_4}$\\
$(\sqrt{A})^{-1}_{10,7}=-\frac{1}{2}\rho_{a_4,c_4}\qquad $&$\qquad (\sqrt{A})^{-1}_{11,7}=-\frac{1}{2}\rho_{a_4,b_4}\qquad $&$\qquad (\sqrt{A})^{-1}_{9,8}=-\frac{1}{2}\rho_{13_4,18_4}$\\
$(\sqrt{A})^{-1}_{11,8}=-\frac{1}{2}\rho_{b_4,18_4}\qquad $&$\qquad (\sqrt{A})^{-1}_{12,8}=-\frac{1}{2}\rho_{16_4,18_4}\qquad $&$\qquad (\sqrt{A})^{-1}_{21,8}=-\frac{1}{2}\rho_{9_{18},4_{18}}$\\
$(\sqrt{A})^{-1}_{35,8}=-\frac{1}{2}\rho_{d_{18},4_{18}}\qquad $&$\qquad (\sqrt{A})^{-1}_{36,8}=-\frac{1}{2}\rho_{4_{18},e_{18}}\qquad $&$\qquad (\sqrt{A})^{-1}_{10,9}=-\frac{1}{2}\rho_{c_4,13_4}$\\
$(\sqrt{A})^{-1}_{12,9}=-\frac{1}{2}\rho_{16_{4},13_{4}}\qquad $&$\qquad (\sqrt{A})^{-1}_{19,9}=-\frac{1}{2}\rho_{4_{13},9_{13}}\qquad $&$\qquad (\sqrt{A})^{-1}_{27,9}=-\frac{1}{2}\rho_{3_{13},4_{13}}$\\
$(\sqrt{A})^{-1}_{31,9}=-\frac{1}{2}\rho_{4_{13},21_{13}}\qquad $&$\qquad (\sqrt{A})^{-1}_{32,9}=-\frac{1}{2}\rho_{4_{13},14_{13}}\qquad $&$\qquad (\sqrt{A})^{-1}_{33,9}=-\frac{1}{2}\rho_{4_{18},12_{18}}$\\
$(\sqrt{A})^{-1}_{11,10}=-\frac{1}{2}\rho_{b_{4},c_{4}}\qquad $&$\qquad (\sqrt{A})^{-1}_{12,10}=-\frac{1}{2}\rho_{c_{4},16_{4}}\qquad $&$\qquad (\sqrt{A})^{-1}_{12,11}=-\frac{1}{2}\rho_{b_{4},18_{4}}$\\
$(\sqrt{A})^{-1}_{14,13}=-\frac{1}{2}\rho_{r_{5},17_{5}}\qquad $&$\qquad (\sqrt{A})^{-1}_{15,13}=-\frac{1}{2}\rho_{r_{5},13_{5}}\qquad $&$\qquad (\sqrt{A})^{-1}_{16,13}=-\frac{1}{2}\rho_{r_{5},p_{5}}$\\
$(\sqrt{A})^{-1}_{17,13}=-\frac{1}{2}\rho_{r_{5},q_{5}}\qquad $&$\qquad (\sqrt{A})^{-1}_{15,14}=-\frac{1}{2}\rho_{13_{5},17_{5}}\qquad $&$\qquad (\sqrt{A})^{-1}_{16,14}=-\frac{1}{2}\rho_{p_{5},17_{5}}$\\
$(\sqrt{A})^{-1}_{18,14}=-\frac{1}{2}\rho_{6_{5},17_{5}}\qquad $&$\qquad (\sqrt{A})^{-1}_{17,15}=-\frac{1}{2}\rho_{q_{5},13_{5}}\qquad $&$\qquad (\sqrt{A})^{-1}_{18,15}=-\frac{1}{2}\rho_{6_{5},13_{5}}$\\
$(\sqrt{A})^{-1}_{19,15}=-\frac{1}{2}\rho_{9_{13},5_{5}}\qquad $&$\qquad (\sqrt{A})^{-1}_{27,15}=-\frac{1}{2}\rho_{3_{13},5_{13}}\qquad $&$\qquad (\sqrt{A})^{-1}_{31,15}=-\frac{1}{2}\rho_{21_{13},5_{13}}$\\
$(\sqrt{A})^{-1}_{32,15}=-\frac{1}{2}\rho_{14_{13},5_{13}}\qquad $&$\qquad (\sqrt{A})^{-1}_{17,16}=-\frac{1}{2}\rho_{q_{5},p_{5}}\qquad $&$\qquad (\sqrt{A})^{-1}_{18,16}=-\frac{1}{2}\rho_{6_{5},p_{5}}$\\
$(\sqrt{A})^{-1}_{18,17}=-\frac{1}{2}\rho_{6_{5},9_{5}}\qquad $&$\qquad (\sqrt{A})^{-1}_{27,19}=-\frac{1}{2}\rho_{3_{13},9_{13}}\qquad $&$\qquad (\sqrt{A})^{-1}_{32,19}=-\frac{1}{2}\rho_{14_{13},9_{13}}$\\
$(\sqrt{A})^{-1}_{26,20}=-\frac{1}{2}\rho_{12_{2},9_{2}}\qquad $&$\qquad (\sqrt{A})^{-1}_{27,20}=-\frac{1}{2}\rho_{3_{2},9_{2}}\qquad $&$\qquad (\sqrt{A})^{-1}_{28,20}=-\frac{1}{2}\rho_{10_{2},9_{2}}$\\
$(\sqrt{A})^{-1}_{33,21}=-\frac{1}{2}\rho_{12_{18},9_{18}}\qquad $&$\qquad (\sqrt{A})^{-1}_{34,21}=-\frac{1}{2}\rho_{f_{18},9_{18}}\qquad $&$\qquad (\sqrt{A})^{-1}_{36,21}=-\frac{1}{2}\rho_{e_{18},9_{18}}$\\
$(\sqrt{A})^{-1}_{23,22}=-\frac{1}{2}\rho_{12_{2},11_{2}}\qquad $&$\qquad (\sqrt{A})^{-1}_{24,22}=-\frac{1}{2}\rho_{3_{2},11_{2}}\qquad $&$\qquad (\sqrt{A})^{-1}_{25,22}=-\frac{1}{2}\rho_{10_{2},11_{2}}$\\
$(\sqrt{A})^{-1}_{24,23}=-\frac{1}{2}\rho_{3_{2},12_{2}}\qquad $&$\qquad (\sqrt{A})^{-1}_{25,23}=-\frac{1}{2}\rho_{10_{2},12_{2}}\qquad $&$\qquad (\sqrt{A})^{-1}_{26,24}=-\frac{1}{2}\rho_{17_{3},2_{3}}$\\
$(\sqrt{A})^{-1}_{27,24}=-\frac{1}{2}\rho_{13_{3},2_{3}}\qquad $&$\qquad (\sqrt{A})^{-1}_{28,24}=-\frac{1}{2}\rho_{19_{3},2_{3}}\qquad $&$\qquad (\sqrt{A})^{-1}_{30,24}=-\frac{1}{2}\rho_{16_{3},2_{3}}$\\
$(\sqrt{A})^{-1}_{27,26}=-\frac{1}{2}\rho_{13_{3},17_{3}}\qquad $&$\qquad (\sqrt{A})^{-1}_{28,26}=-\frac{1}{2}\rho_{19_{3},17_{3}}\qquad $&$\qquad (\sqrt{A})^{-1}_{29,26}=-\frac{1}{2}\rho_{15_{3},17_{3}}$\\
$(\sqrt{A})^{-1}_{29,27}=-\frac{1}{2}\rho_{15_{3},13_{3}}\qquad $&$\qquad (\sqrt{A})^{-1}_{30,27}=-\frac{1}{2}\rho_{16_{3},13_{3}}\qquad $&$\qquad (\sqrt{A})^{-1}_{31,27}=-\frac{1}{2}\rho_{21_{13},3_{13}}$\\
$(\sqrt{A})^{-1}_{29,28}=-\frac{1}{2}\rho_{15_{3},19_{3}}\qquad $&$\qquad (\sqrt{A})^{-1}_{30,28}=-\frac{1}{2}\rho_{16_{3},19_{3}}\qquad $&$\qquad (\sqrt{A})^{-1}_{30,29}=-\frac{1}{2}\rho_{16_{3},15_{3}}$\\
$(\sqrt{A})^{-1}_{38,29}=-\frac{1}{2}\rho_{d^{'}_{15},3_{15}}\qquad $&$\qquad (\sqrt{A})^{-1}_{39,29}=-\frac{1}{2}\rho_{c^{'}_{15},3_{15}}\qquad $&$\qquad (\sqrt{A})^{-1}_{40,29}=-\frac{1}{2}\rho_{f^{'}_{15},3_{15}}$\\
$(\sqrt{A})^{-1}_{41,29}=-\frac{1}{2}\rho_{a^{'}_{15},3_{15}}\qquad $&$\qquad (\sqrt{A})^{-1}_{32,31}=-\frac{1}{2}\rho_{14_{13},21_{13}}\qquad $&$\qquad (\sqrt{A})^{-1}_{42,32}=-\frac{1}{2}\rho_{a_{14},13_{14}}$\\
$(\sqrt{A})^{-1}_{43,32}=-\frac{1}{2}\rho_{h_{14},13_{14}}\qquad $&$\qquad (\sqrt{A})^{-1}_{45,32}=-\frac{1}{2}\rho_{k_{14},13_{14}}\qquad $&$\qquad (\sqrt{A})^{-1}_{46,32}=-\frac{1}{2}\rho_{l_{14},13_{14}}$\\
$(\sqrt{A})^{-1}_{34,33}=-\frac{1}{2}\rho_{f_{18},12_{18}}\qquad $&$\qquad (\sqrt{A})^{-1}_{35,33}=-\frac{1}{2}\rho_{d_{18},12_{18}}\qquad $&$\qquad (\sqrt{A})^{-1}_{35,34}=-\frac{1}{2}\rho_{d_{18},f_{18}}$\\
$(\sqrt{A})^{-1}_{36,34}=-\frac{1}{2}\rho_{e_{18},f_{18}}\qquad $&$\qquad (\sqrt{A})^{-1}_{36,35}=-\frac{1}{2}\rho_{e_{18},d_{18}}\qquad $&$\qquad (\sqrt{A})^{-1}_{38,37}=-\frac{1}{2}\rho_{d^{'}_{15},e^{'}_{15}}$\\
$(\sqrt{A})^{-1}_{39,37}=-\frac{1}{2}\rho_{c^{'}_{15},e^{'}_{15}}\qquad $&$\qquad (\sqrt{A})^{-1}_{40,37}=-\frac{1}{2}\rho_{f^{'}_{15},e^{'}_{15}}\qquad $&$\qquad (\sqrt{A})^{-1}_{41,37}=-\frac{1}{2}\rho_{a^{'}_{15},e^{'}_{15}}$\\
$(\sqrt{A})^{-1}_{39,38}=-\frac{1}{2}\rho_{c^{'}_{15},d^{'}_{15}}\qquad $&$\qquad (\sqrt{A})^{-1}_{40,38}=-\frac{1}{2}\rho_{f^{'}_{15},d^{'}_{15}}\qquad $&$\qquad (\sqrt{A})^{-1}_{41,39}=-\frac{1}{2}\rho_{a^{'}_{15},c^{'}_{15}}$\\
$(\sqrt{A})^{-1}_{41,40}=-\frac{1}{2}\rho_{a^{'}_{15},f^{'}_{15}}\qquad $&$\qquad (\sqrt{A})^{-1}_{45,42}=-\frac{1}{2}\rho_{h_{14},a_{14}}\qquad $&$\qquad (\sqrt{A})^{-1}_{46,42}=-\frac{1}{2}\rho_{j_{15},a_{15}}$\\
$(\sqrt{A})^{-1}_{47,42}=-\frac{1}{2}\rho_{l_{14},a_{14}}\qquad $&$\qquad (\sqrt{A})^{-1}_{44,43}=-\frac{1}{2}\rho_{j_{14},h_{14}}\qquad $&$\qquad (\sqrt{A})^{-1}_{45,43}=-\frac{1}{2}\rho_{j_{15},k_{15}}$\\
$(\sqrt{A})^{-1}_{45,44}=-\frac{1}{2}\rho_{k_{14},j_{14}}\qquad $&$\qquad (\sqrt{A})^{-1}_{46,44}=-\frac{1}{2}\rho_{l_{14},j_{14}}\qquad $&$\qquad (\sqrt{A})^{-1}_{46,45}=-\frac{1}{2}\rho_{l_{15},k_{15}}$\\
\end{tabular}
\end{center}

\subsubsection{Translated 6-valent graph}
\label{sa.2.3}
We will now give the values for the terms $t_{e_ie_j}=t_{e_ie_j}^x+t_{e_ie_j}^y+t_{e_ie_j}^z$ as computed for each of the nine vertices comprising the periodicity cell for the 6-valent graph. The (Gauss brackets) terms $n_i=[\frac{V_i^x}{l}]$, $m_i[\frac{V_i^y}{l}]$ and $p_i=[\frac{V_i^z}{l}]$ represent the number of stacks which each edge $e_i$ intersects in the $x$-, $y$- and $z$-direction respectively (see Section \ref{s7.1}). \\[10pt]
1) $V^{''}_0=(\epsilon_x,\epsilon_y,\epsilon_z)$ with
$\epsilon_x>\epsilon_y>\epsilon_z$
\begin{center}
\begin{tabular}{l|l|l}
$(V^{''})_0^z$&$(V^{''}_0)^x$&$(V^{''}_0)^y$\\\hline
$t^z_{e_{0,1}e_{0,6}}=(\epsilon_y-m_0l)Z^y_6$&$t^x_{e_{0,1}e_{0,6}}=(\epsilon_z-p_0l)T^x_1$
&$t^y_{e_{0,1}e_{0,8}}=(\epsilon_z-p_0l)T^y_1$\\
$t^z_{e_{0,6}e_{0,7}}=(\epsilon_y-m_0l)Z^y_6$
&$t^x_{e_{0,2}e_{0,6}}=(\epsilon_z-p_0l)T^x_6$&$
t^y_{e_{0,6}e_{0,17}}=(\epsilon_z-p_0l)T^y_6$\\
$t^z_{e_{0,1}e_{0,7}}=(\epsilon_x-n_0l)Z^x_7$&$t^x_{e_{0,2}e_{0,1}}=(\epsilon_z-p_0l)T^x_1$ &\\
$t^z_{e_{0,2}e_{0,8}}=(\epsilon_x-n_0l)Z^x_8$&
$t^x_{e_{0,8}e_{0,7}}=(\epsilon_z-p_0l)T^x_7$&
\\
$t^z_{e_{0,2}e_{0,17}}=(\epsilon_y-m_0l)Z^y_{17}$&
$t^x_{e_{0,8}e_{0,17}}=(\epsilon_y-m_0l)F^x_{17}$ &\\
$t^z_{e_{0,8}e_{0,17}}=(\epsilon_y-m_0l)Z^y_{17}$&$t^x_{e_{0,7}e_{0,17}}=(\epsilon_z-p_0l)T^x_7$&

\end{tabular}
\end{center}
2) $V^{''}_2=(\frac{\delta_e
\sqrt{2}}{2}+\epsilon_x,\epsilon_y,\frac{\delta_e
\sqrt{2}}{2}+\epsilon_z)$
\begin{center}
\begin{tabular}{l|l|l}
$(V^{''}_2)^z$&$(V^{''}_2)^x$&$(V^{''}_2)^y$\\\hline
$t^z_{e_{2,0}e_{2,10}}=(\epsilon_x+\frac{\delta_e
\sqrt{2}}{2}-n_2l)Z^x_0$&$ t^x_{e_{2,12}e_{2,10}}=(\frac{\delta_e
\sqrt{2}}{2}+\epsilon_z-p_2l)T^x_{10}$&
$t^y_{e_{2,10}e_{2,11}}=(\frac{\delta_e \sqrt{2}}{2}+\epsilon_z-p_2l)T^y_{10}$\\
$t^z_{e_{2,9}e_{2,0}}=(\epsilon_y-m_2l)Z^y_9$
&$t^x_{e_{2,9}e_{2,12}}=(\epsilon_y-m_2p)F^x_9$&
$t^y_{e_{2,9}e_{2,3}}=(\frac{\delta_e \sqrt{2}}{2}+\epsilon_z-p_2l)T^y_9$\\
$t^z_{e_{2,9}e_{2,10}}=(\epsilon_y-m_2l)Z^y_9$&$
t^x_{e_{2,9}e_{2,10}}=(\epsilon_y-m_2p)F^x_9$&
\\
$t^z_{e_{2,11}e_{2,12}}=(\epsilon_x+\frac{\delta_e
\sqrt{2}}{2}-n_2l)Z^x_{11}$&
$t^x_{e_{2,11}e_{2,0}}=(\frac{\delta_e \sqrt{2}}{2}+\epsilon_z-p_2l)T^x_0$&\\
$t^z_{e_{2,3}e_{2,12}}=(\epsilon_y-m_2l)Z^y_3$&
$t^x_{e_{2,11}e_{2,3}}=(\epsilon_y-m_2l)F^x_3$&\\
$t^z_{e_{2,11}e_{2,3}}=(\epsilon_x+\frac{\delta_e
\sqrt{2}}{2}-n_2l)Z^x_3$&
$t^x_{e_{2,3}e_{2,0}}=(\epsilon_y-m_2p)F^x_3$&
\end{tabular}
\end{center}

3) $V^{''}_3=(\epsilon_x+\delta_e(\frac{ \sqrt{2}}{2}-\frac{1
}{2}),\epsilon_y-\frac{\delta_e }{2},2\frac{\delta_e
\sqrt{2}}{2}+\epsilon_z)$
\begin{center}
\begin{tabular}{l|l|l}
$(V^{''}_3)^z$&$(V^{''}_3)^x$&$(V^{''}_3)^y$\\\hline
$t^z_{e_{3,2}e_{3,13}}=(|\epsilon_y-\frac{\delta_e
}{2}|-m_3l)Z^y_{2}$&$
t^x_{e_{3,16}e_{3,2}}=(|\epsilon_y+\frac{\delta_e
}{2}-m_3l|)F^x_2$&
$t^y_{e_{3,19}e_{3,2}}=(\epsilon_x+\delta_e(\frac{\sqrt{2}}{2}-\frac{1}{2})-n_3l)F^x_{19}$\\
$t^z_{e_{3,17}e_{3,2}}=(|\epsilon_y-\frac{\delta_e
}{2}-m_3l|)Z^y_{2}$&
$t^x_{e_{3,16}e_{3,13}}=(\epsilon_z+2\frac{\delta_e
\sqrt{2}}{2}-p_3l)T^x_{13}$&
$t^y_{e_{3,15}e_{3,13}}=(\epsilon_x+\delta_e(\frac{\sqrt{2}}{2}-\frac{1 }{2})-n_3l)F^x_{15}$\\
$t^z_{e_{3,17}e_{3,13}}=(\epsilon_x+\delta_e(\frac{\sqrt{2}}{2}-\frac{1
}{2})-n_3l)Z^x_{17}$&
$t^x_{e_{3,13}e_{3,2}}=(|\epsilon_y+\frac{\delta_e
}{2}-m_3l|)F^x_2$&
\\
$t^z_{e_{3,16}e_{3,15}}=(\epsilon_x+\delta_e(\frac{\sqrt{2}}{2}-\frac{1
}{2})-n_3l)Z^x_{15}$&
$t^x_{e_{3,15}e_{3,17}}=(\epsilon_z+2\frac{\delta_e
\sqrt{2}}{2}-p_3l)T^x_{17}$&
\\
$t^z_{e_{3,16}e_{3,19}}=(|\epsilon_y-\frac{\delta_e
}{2}-m_3l|)Z^y_{19}$&
$t^x_{e_{3,15}e_{3,19}}=(|\epsilon_y-\frac{\delta_e
}{2}-m_3l|)F^x_{19}$&
\\
$t^z_{e_{3,19}e_{3,15}}=(|\epsilon_y-\frac{\delta_e
}{2}-m_3l|)Z^y_{19}$&
$t^x_{e_{3,19}e_{3,17}}=(|\epsilon_y-\frac{\delta_e
}{2}-m_3l|)F^x_{19}$&

\end{tabular}
\end{center}
4) $V^{''}_4=(2\frac{\delta_e
\sqrt{2}}{2}+\epsilon_x,\epsilon_y-2\frac{\delta_e
}{2},2\frac{\delta_e \sqrt{2}}{2}+\epsilon_z)$
\begin{center}
\begin{tabular}{l|l|l}
$(V^{''}_4)^z$&$(V^{''}_4)^x$&$(V^{''}_4)^y$\\\hline
$t^z_{e_{4,a}e_{4,18}}=(|\epsilon_y-\delta_e |-m_4l)Z^y_{18}$&
$t^x_{e_{4,a}e_{4,18}}=(|\epsilon_y-\delta_e |-m_4l)F^x_{18}$&
$t^y_{e_{4,16}e_{4,18}}=(\epsilon_z+2\frac{\delta_e \sqrt{2}}{2}-p_4l)T^y_{18}$\\
$t^z_{e_{4,13}e_{4,a}}=(\epsilon_x+2\frac{\delta_e
\sqrt{2}}{2}-m_4l)Z^x_{13}$&
$t^x_{e_{4,b}e_{4,a}}=(\epsilon_z+2\frac{\delta_e
\sqrt{2}}{2}-p_4l)T^x_a$&
$t^y_{e_{4,a}e_{4,c}}=(\epsilon_z+2\frac{\delta_e \sqrt{2}}{2}-p_4l)T^y_a$\\
$t^z_{e_{4,13}e_{4,18}}=(|\epsilon_y-\delta_e |-m_4l)Z^y_{18}$&
$t^x_{e_{4,18}e_{4,b}}=(|\epsilon_y-\delta_e |-m_4l)F^x_{18}$&
\\
$t^z_{e_{4,b}e_{4,16}}=(|\epsilon_y-\delta_e |-m_4l)Z^y_{16}$&
$t^x_{e_{4,16}e_{4,13}}=(|\epsilon_y-\delta_e |-m_4l)F^x_{16}$&
\\
$t^z_{e_{4,c}e_{4,16}}=(|\epsilon_y-\delta_e |-m_4l)Z^y_{16}$&
$t^x_{e_{4,16}e_{4,c}}=(|\epsilon_y-\delta_e |-m_4l)F^x_{16}$&
\\
$t^z_{e_{4,c}e_{4,b}}=(\epsilon_x+2\frac{\delta_e
\sqrt{2}}{2}-n_4l)Z^x_{c}$&
$t^x_{e_{4,13}e_{4,c}}=(\epsilon_z+2\frac{\delta_e
\sqrt{2}}{2}-p_4l)T^x_{13}$&

\end{tabular}
\end{center}
5) $V^{''}_5=(\epsilon_x,\epsilon_y-\delta_e , \epsilon_z)$
\begin{center}
\begin{tabular}{l|l|l}
$(V^{''}_5)^z$&$(V^{''}_5)^x$&$(V^{''}_5)^y$\\\hline
$t^z_{e_{5,q}e_{5,6}}=(|\epsilon_y-\delta_e |-m_5l)Z^y_6$&
$t^x_{e_{5,q}e_{5,6}}=( \epsilon_z-p_5l)T^x_6$&
$t^y_{e_{5,17}e_{5,6}}=( \epsilon_z-p_5l)T^y_6$\\
$t^z_{e_{5,q}e_{5,p}}=(\epsilon_x-n_5l)Z^x_p$&
$t^x_{e_{5,q}e_{5,13}}=( \epsilon_z-p_5l)T^x_q$&
$t^y_{e_{5,r}e_{5,q}}=(\epsilon_z-p_5l)T^y_q$\\
$t^z_{e_{5,6}e_{5,p}}=(\epsilon_x-n_5l)Z^x_p$&
$t^x_{e_{5,6}e_{5,13}}=( \epsilon_z-p_5l)T^x_6$&
\\
$t^z_{e_{5,13}e_{5,17}}=(\epsilon_x-n_5l)Z^x_{17}$&
$t^x_{e_{5,p}e_{5,17}}=( \epsilon_z-p_5l)T^x_p$&
\\
$t^z_{e_{5,r}e_{5,17}}=(\epsilon_x-n_5l)Z^x_{r}$&
$t^x_{e_{5,r}e_{5,17}}=(|\epsilon_y-\delta_e |-m_5l)F^x_r$&
\\
$t^z_{e_{5,r}e_{5,13}}=(\epsilon_x-n_5l)Z^x_{r}$&
$t^x_{e_{5,r}e_{5,p}}=(\epsilon_z-p_5l)T^x_p$&
\end{tabular}
\end{center}
6) $V^{''}_{13}=(\epsilon_x+\frac{\delta_e
\sqrt{2}}{2},\epsilon_y-\delta_e ,\epsilon_z+\frac{\delta_e
\sqrt{2}}{2})$
\begin{center}
\begin{tabular}{l|l|l}
$(V^{''}_{13})^z$&$(V^{''}_{13})^x$&$(V^{''}_{13})^y$\\\hline
$t^z_{e_{13,14}e_{13,9}}=(|\epsilon_y-\delta_e|-m_{13}l)Z^y_{9}$&
$t^x_{e_{13,9}e_{13,14}}=(|\epsilon_y-\delta_e|-m_{13}l)F^x_9$&
$t^y_{e_{13,3}e_{13,9}}=(\epsilon_z+\frac{\delta_e \sqrt{2}}{2}-p_{13}l)T^y_9$\\
$t^z_{e_{13,5}e_{13,14}}=(\epsilon_-n_{13}l)Z^x_{5}$&
$t^x_{e_{13,4}e_{13,14}}=(\epsilon_z+\frac{\delta_e
\sqrt{2}}{2}-p_{13}l)T^x_{14}$&
$t^y_{e_{13,14}e_{13,21}}=(\epsilon_z+\frac{\delta_e \sqrt{2}}{2}p_{13}l)T^y_{14}$\\
$t^z_{e_{13,5}e_{13,9}}=(|\epsilon_y-\delta_e|-m_{13}l)Z^y_9$&
$t^x_{e_{13,4}e_{13,9}}=(|\epsilon_y-\delta_e|-m_{13}l)F^x_9$&
\\
$t^z_{e_{13,3}e_{13,4}}=(|\epsilon_y-\delta_e|-m_{13}l)Z^y_3$ &
$t^x_{e_{13,3}e_{13,5}}=(|\epsilon_y-\delta_e|-m_{13}l)F^x_3$&
\\
$t^z_{e_{13,3}e_{13,21}}=(|\epsilon_y-\delta_e|-m_{13}l)Z^y_{3}$ &
$t^x_{e_{13,3}e_{13,21}}=(|\epsilon_y-\delta_e|-m_{13}l)F^x_{3}$&
\\
$t^z_{e_{13,4}e_{13,21}}=(\epsilon_x-n_{13}l)Z^x_{21}$ &
$t^x_{e_{13,5}e_{13,21}}=(\epsilon_z+\frac{\delta_e
\sqrt{2}}{2}-p_{13}l)T^x_{5}$&
\end{tabular}
\end{center}
7) $V^{''}_{18}=(\epsilon_x+ \delta_e
(2\frac{\sqrt{2}}{2}+\frac{1}{2}),\epsilon_y-\frac{\delta_e
}{2},\epsilon_z+\frac{\delta_e \sqrt{2}}{2})$
\begin{center}
\begin{tabular}{l|l|l}
$(V^{''}_{18})^z$&$(V^{''}_{18})^x$&$(V^{''}_{18})^y$\\\hline
$t^z_{e_{18,f}e_{18,e}}=(|\epsilon_y-\frac{\delta_e }{2}|
-m_{18}l)Z^y_f$&
$t^x_{e_{18,f}e_{18,e}}=(|\epsilon_y-\frac{\delta_e }{2}|
-m_{18}l)F^x_f$&
$t^y_{e_{18,12}e_{18,f}}=(\epsilon_z+\frac{\delta_e \sqrt{2}}{2}-p_{18}l)T^y_f$\\
$t^z_{e_{18,e}e_{18,9}}=(\epsilon_x+ \delta_e
(2\frac{\sqrt{2}}{2}+\frac{1}{2}-n_{18}l)Z^x_9$&
$t^x_{e_{18,d}e_{18,e}}=(\epsilon_z+\frac{\delta_e
\sqrt{2}}{2}-p_{18}l)T^x_e$&
$t^y_{e_{18,e}e_{18,4}}=(\epsilon_z+\frac{\delta_e \sqrt{2}}{2}-p_{18}l)T^y_e$\\
$t^z_{e_{18,f}e_{18,9}}=(|\epsilon_y-\frac{\delta_e }{2}|
-m_{18}l)Z^y_f$&
$t^x_{e_{18,f}e_{18,d}}=(|\epsilon_y-\frac{\delta_e }{2}|
-m_{18}l)F^x_f$ &
\\
$t^z_{e_{18,12}e_{18,d}}=(|\epsilon_y-\frac{\delta_e }{2}|
-m_{18}l)Z^y_{12}$ &
$t^x_{e_{18,12}e_{18,9}}=(|\epsilon_y-\frac{\delta_e }{2}|
-m_{18}l)F^x_{12}$&
\\
$t^z_{e_{18,12}e_{18,4}}=(|\epsilon_y-\frac{\delta_e }{2}|
-m_{18}l)Z^y_{12}$ &
$t^x_{e_{18,12}e_{18,4}}=(|\epsilon_y-\frac{\delta_e }{2}|
-m_{18}l)F^x_{12}$&
\\
$t^z_{e_{18,d}e_{18,4}}=(\epsilon_x+ \delta_e
(2\frac{\sqrt{2}}{2}+\frac{1}{2}-n_{18}l)Z^x_{4}$ &
$t^x_{e_{18,9}e_{18,4}}=(\epsilon_z+\frac{\delta_e
\sqrt{2}}{2}-p_{18}l)T^y_9$&
\end{tabular}
\end{center}
8) $V^{''}_{15}=(\epsilon_x+\delta_e
(\frac{\sqrt{2}}{2}-1),\epsilon_y-\delta_e
,\epsilon_z+3\frac{\delta_e \sqrt{2}}{2})$
\begin{center}
\begin{tabular}{l|l|l}
$(V^{''}_{15})^z$&$(V^{''}_{15})^x$&$(V^{''}_{15})^y$\\\hline
$t^z_{e_{15,3}e_{15,a^{'}}}=(\epsilon_x+\delta_e
(\frac{\sqrt{2}}{2}-1)-n_{15}l)Z^x_{3}$&
$t^x_{e_{15,a^{'}}e_{15,3}}=(\epsilon_y-\delta_e-m_{15}l)F^x_3$&
$t^y_{e_{15,d^{'}}e_{15,3}}=(\epsilon_x+\delta_e (\frac{\sqrt{2}}{2}-1)-n_{15}l)F^y_3$\\
$t^z_{e_{15,f^{'}}e_{15,a^{'}}}=(\epsilon_x+\delta_e
(\frac{\sqrt{2}}{2}-1)-n_{15}l)Z^x_{a^{'}}$&
$t^x_{e_{15,a^{'}}e_{15,c^{'}}}=(\epsilon_z+3\frac{\delta_e
\sqrt{2}}{2}-p_{15}l)T^x_{a^{'}}$&
$t^y_{e_{15,a^{'}}e_{15,e^{'}}}=(\epsilon_x+\delta_e (\frac{\sqrt{2}}{2}-1)-n_{15}l)F^y_{a^{'}}$\\
$t^z_{e_{15,f^{'}}e_{15,3}}=(\epsilon_x+\delta_e
(\frac{\sqrt{2}}{2}-1)-n_{15}l)Z^x_{3}$&
$t^x_{e_{15,3}e_{15,c^{'}}}=(\epsilon_y-\delta_e-m_{15}l)F^x_{3}$&
\\
$t^z_{e_{15,d^{'}}e_{15,c^{'}}}=(\epsilon_x+\delta_e
(\frac{\sqrt{2}}{2}-1)-n_{15}l)Z^x_{c^{'}}$ &
$t^x_{e_{15,d^{'}}e_{15,f^{'}}}=(\epsilon_y-\delta_e-m_{15}l)F^x_{d^{'}}$&
\\
$t^z_{e_{15,d^{'}}e_{15,e^{'}}}=(\epsilon_y-\delta_e-m_{15}l)Z^y_{d^{'}}$
&
$t^x_{e_{15,d^{'}}e_{15,e^{'}}}=(\epsilon_y-\delta_e-m_{15}l)F^x_{d^{'}}$&
\\
$t^z_{e_{15,c^{'}}e_{15,e^{'}}}=(\epsilon_x+\delta_e
(\frac{\sqrt{2}}{2}-1)-n_{15}l)Z^x_{c^{'}}$ &
$t^x_{e_{15,f^{'}}e_{15,e^{'}}}=(\epsilon_z+3\frac{\delta_e
\sqrt{2}}{2}-p_{15}l)T^x_{f^{'}}$&
\end{tabular}
\end{center}
9) $V^{''}_{14}=(\epsilon_x+\delta_e(\frac{\delta_e
\sqrt{2}}{2}+\frac{1}{2}),\epsilon_y-3\frac{\delta_e
}{2},\epsilon_z)$
\begin{center}
\begin{tabular}{l|l|l}
$(V^{''}_{14})^z$&$(V^{''}_{14})^x$&$(V^{''}_{14})^y$\\\hline
$t^z_{e_{14,j}e_{9,l}}=(|\epsilon_y-3\frac{\sqrt{2}}{2}|-m_{14}l)Z^y_l$&
$t^x_{e_{14,j}e_{14,l}}=(\epsilon_z-p_{14}l)T^x_l$&
$t^y_{e_{14,13}e_{14,l}}=(\epsilon_z-p_{14}l)T^y_l$\\
$t^z_{e_{14,k}e_{14,j}}=(\epsilon_x+\delta_e(\frac{\delta_e
\sqrt{2}}{2}+\frac{1}{2})-n_{14}l)Z^x_k$&
$t^x_{e_{14,a}e_{14,j}}=(\epsilon_z-p_{14}l)T^x_j$&
$t^y_{e_{14,h}e_{14,j}}=(\epsilon_z-p_{14}l)T^y_j$\\
$t^z_{e_{14,k}e_{14,l}}=(|\epsilon_y-3\frac{\sqrt{2}}{2}|-m_{14}l)Z^y_l$&
$t^x_{e_{14,l}e_{14,a}}=(\epsilon_z-p_{14}l)T^x_l$&
\\
$t^z_{e_{14,a}e_{14,13}}=(|\epsilon_y-3\frac{\sqrt{2}}{2}|-m_{14}l)Z^y_{13}$&
$t^x_{e_{14,13}e_{14,k}}=(\epsilon_z-p_{14}l)T^x_k$&
\\
$t^z_{e_{14,h}e_{14,13}}=(|\epsilon_y-3\frac{\sqrt{2}}{2}|-m_{14}l)Z^y_{13}$&
$t^x_{e_{14,13}e_{14,h}}=(|\epsilon_y-3\frac{\delta_e
}{2}|-m_{14}l)F^x_{13}$&
\\
$t^z_{e_{14,h}e_{14,a}}=(\epsilon_x+\delta_e(\frac{\delta_e
\sqrt{2}}{2}+\frac{1}{2})-n_{14}l)Z^x_{h}$&
$t^x_{e_{14,k}e_{14,h}}=(\epsilon_z-p_{14}l)T^x_k$&
\end{tabular}
\end{center}
We now need to define the off-diagonal elements of the matrix
$\sqrt{A}^{-1}$ given by
$\frac{t_{e_i,e_j}}{\sqrt{t_{e_i}t_{e_j}}}$. The terms
$t_{e_i}=t^z_{e_i}+t^x_{e_i}+t^y_{e_i}$ represent the total number
of stacks intersected by the edge $e_i$ in the $z$-, $x$- and $y$-direction respectively. Therefore we obtain
\begin{equation}
t_{e_i}=\delta(\sin(\phi_{e_i})\cos(\theta_{e_i})+\sin(\phi_{e_i})\sin(\theta_{e_i})+\sin(\phi_{e_i}))=
\begin{cases}
\delta_e(1+\frac{\sqrt{2}}{2}) \text{ for all edges $e_i$ such
that $\theta_{e_i}\neq 0$}\\[3pt] \delta(\sqrt{2}+1)\text{ for those
edges such that $\theta_{e_i}= 0$}
\end{cases}
\end{equation} \\
Throughout we will use the following short-hand notation
\begin{align}
&A^{\prime}_i=\frac{|x_{V_i}|-n_i}{\delta(\sqrt{\sqrt{2}+1)}}\hspace{.5in}A_i=\frac{(|x_{V_i}|-n_il)}{\delta_e(1+\frac{\sqrt{2}}{2}}
\nonumber\\
&B^{\prime}_i=\frac{(|y_{V_i}|-m_il)}{\delta(\sqrt{\sqrt{2}+1)}}\hspace{.5in}B_i=\frac{(|y_{V_i}|-m_il)}{\delta_e(1+\frac{\sqrt{2}}{2}}
\nonumber\\
&C^{\prime}_i=\frac{(|z_{V_i}|-p_il)}{\delta(\sqrt{\sqrt{2}+1)}}\hspace{.5in}C_i=\frac{(|z_{V_i}|-p_il)}{\delta_e(1+\frac{\sqrt{2}}{2}}
\nonumber
\end{align}
\text{Here, $e_{k,i}$ denotes  the edge starting at vertex $V_i$ and
ending at vertex $V_j$}

We will now give the values of the upper-diagonal entries of the matrix
$\sqrt{A}^{-1}$.

\begin{center}
\begin{tabular}{lll}\label{tab:a6t}
$(\sqrt{A}^{-1})^{1_0,2_0}_{2,1}=\frac{-C^{'}_{0}\sqrt{2}}{4}\qquad $&$\qquad (\sqrt{A}^{-1})^{1_0,8_0}_{3,1}=\frac{-C_{0}\sqrt{2}}{4}\qquad $&$\qquad (\sqrt{A}^{-1})^{1_0,6_0}_{5,1}=-\frac{C_{0}\sqrt{2}}{4}-\frac{B_0}{\sqrt{2}}$\\
$(\sqrt{A}^{-1})^{1_0,7_0}_{6,1}=\frac{-A^{'}_{0}}{2}\qquad $&$\qquad (\sqrt{A}^{-1})^{2_0,3_0}_{3,2}=\frac{-A^{'}_{0}}{\sqrt{2}}\qquad $&$\qquad (\sqrt{A}^{-1})^{2_0,4_0}_{4,2}=\frac{-B^{'}_{0}}{\sqrt{2}}$\\
$(\sqrt{A}^{-1})^{2_0,5_0}_{5,2}=\frac{-C^{'}_{0}\sqrt{2}}{4}\qquad $&$\qquad (\sqrt{A}^{-1})^{0_2,10_2}_{7,2}=\frac{A^{'}_{2}}{2}\qquad $&$\qquad (\sqrt{A}^{-1})^{0_2,9_2}_{8,2}=\frac{-B^{'}_{2}}{\sqrt{2}}$\\
$(\sqrt{A}^{-1})^{0_2,3_2}_{9,2}=\frac{-B^{'}_{2}}{\sqrt{2}}\qquad $&$\qquad (\sqrt{A}^{-1})^{0_2,11_2}_{11,2}=\frac{-C^{'}_{2}}{2}\qquad $&$\qquad (\sqrt{A}^{-1})^{8_0,17_0}_{4,3}=\frac{-B_{0}}{2}(1+\frac{2}{\sqrt{2}})$\\
$(\sqrt{A}^{-1})^{8_0,7_0}_{6,3}=\frac{-C^{'}_{0}}{2}\qquad $&$\qquad (\sqrt{A}^{-1})^{17_0,6_0}_{5,4}=\frac{-C_{0}\sqrt{2}}{4}\qquad $&$\qquad (\sqrt{A}^{-1})^{17_0,7_0}_{6,4}=\frac{-C^{'}_{0}}{2}$\\
$(\sqrt{A}^{-1})^{6_0,7_0}_{7,5}=\frac{-B^{'}_{0}}{\sqrt{2}}\qquad $&$\qquad (\sqrt{A}^{-1})^{10_2,9_2}_{8,7}=\frac{-B_{2}}{2}(1+\frac{2}{\sqrt{2}})\qquad $&$\qquad (\sqrt{A}^{-1})^{10_2,12_2}_{10,7}=\frac{-C^{'}_{2}\sqrt{2}}{4}$\\
$(\sqrt{A}^{-1})^{10_2,11_2}_{11,7}=\frac{-C_{2}\sqrt{2}}{4}\qquad $&$\qquad (\sqrt{A}^{-1})^{9_2,3_2}_{9,8}=\frac{-C_{2}\sqrt{2}}{4}\qquad $&$\qquad (\sqrt{A}^{-1})^{9_2,12_2}_{10,8}=\frac{-B^{'}_{2}}{2}$\\
$(\sqrt{A}^{-1})^{3_2,12_2}_{10,9}=\frac{-B^{'}_{2}}{\sqrt{2}}\qquad $&$\qquad (\sqrt{A}^{-1})^{3_2,11_2}_{11,9}=\frac{-B_2}{2}(1+\frac{2}{\sqrt{2}})\qquad $&$\qquad (\sqrt{A}^{-1})^{2_3,16_3}_{13,9}=\frac{-B_{3}}{2}$\\
$(\sqrt{A}^{-1})^{2_3,19_3}_{14,9}=\frac{-A_{3}}{2}\qquad $&$\qquad (\sqrt{A}^{-1})^{2_3,17_3}_{15,9}=\frac{-B_{3}}{\sqrt{2}}\qquad $&$\qquad (\sqrt{A}^{-1})^{2_3,13_3}_{16,9}=\frac{-B_{3}}{2}(\frac{2}{\sqrt{2}}+1)$\\
$(\sqrt{A}^{-1})^{12_2,11_2}_{11,10}=\frac{-A^{'}_{2}}{\sqrt{2}}\qquad $&$\qquad (\sqrt{A}^{-1})^{15_3,16_3}_{13,12}=\frac{-A_{3}}{\sqrt{2}}\qquad $&$\qquad (\sqrt{A}^{-1})^{15_3,19_3}_{14,12}=\frac{-B_{3}}{2}(\frac{2}{\sqrt{2}}+1)$\\
$(\sqrt{A}^{-1})^{15_3,17_3}_{15,12}=\frac{-C_{3}}{2}\qquad $&$\qquad (\sqrt{A}^{-1})^{15_3,13_3}_{16,12}=\frac{-A_{3}}{2}\qquad $&$\qquad (\sqrt{A}^{-1})^{3_{15},d_{15}}_{38,12}=\frac{-A_{15}}{2}$\\
$(\sqrt{A}^{-1})^{3_{15},c_{15}}_{39,12}=\frac{-C^{'}_{15}\sqrt{2}}{4}\qquad $&$\qquad (\sqrt{A}^{-1})^{3_{15},f_{15}}_{40,12}=\frac{-A^{'}_{15}}{\sqrt{2}}\qquad $&$\qquad (\sqrt{A}^{-1})^{3_{15},a_{15}}_{41,12}=-\frac{B_{15}}{2}-\frac{A_{15}}{\sqrt{2}}$\\
$(\sqrt{A}^{-1})^{16_3,19_3}_{14,13}=\frac{-B_{3}}{\sqrt{2}}\qquad $&$\qquad (\sqrt{A}^{-1})^{16_3,13_3}_{16,13}=\frac{-C_{3}\sqrt{2}}{4}\qquad $&$\qquad (\sqrt{A}^{-1})^{19_3,17_3}_{15,14}=\frac{-B_{3}}{2}$\\
$(\sqrt{A}^{-1})^{17_3,13_3}_{16,15}=\frac{-A_{3}}{2}\qquad $
&$\qquad (\sqrt{A}^{-1})^{3_{13},4_{13}}_{19,16}=\frac{-B^{'}_{13}}{\sqrt{2}}\qquad $&$\qquad (\sqrt{A}^{-1})^{3_{13},5_{13}}_{27,16}=\frac{-B^{'}_{13}}{2}$\\
$(\sqrt{A}^{-1})^{3_{13},9_{13}}_{29,16}=\frac{-C_{13}\sqrt{2}}{4}\qquad $&$\qquad (\sqrt{A}^{-1})^{3_{13},21_{13}}_{31,16}=\frac{-B_{13}}{2}(\frac{2}{\sqrt{2}}+1)\qquad $&$\qquad (\sqrt{A}^{-1})^{a_{4},18_{4}}_{18,17}=\frac{-B_{4}}{2}(\frac{2}{\sqrt{2}}+1)$\\
$(\sqrt{A}^{-1})^{a_{4},13_{4}}_{19,17}=\frac{-A^{'}_{4}}{2}\qquad $&$\qquad (\sqrt{A}^{-1})^{a_{4},c_{4}}_{20,17}=\frac{-C_{4}\sqrt{2}}{4}\qquad $&$\qquad (\sqrt{A}^{-1})^{a_{4},b_{4}}_{21,17}=\frac{-C^{'}_{4}\sqrt{2}}{4}$\\
$(\sqrt{A}^{-1})^{18_{4},13_{4}}_{19,18}=\frac{-B^{'}_{4}}{\sqrt{2}}\qquad $&$\qquad (\sqrt{A}^{-1})^{18_{4},b_{4}}_{21,18}=\frac{-B^{'}_{4}}{2}\qquad $&$\qquad (\sqrt{A}^{-1})^{18_{4},16_{4}}_{22,18}=\frac{-A_{4}}{2}$\\
$(\sqrt{A}^{-1})^{4_{18},e_{18}}_{32,18}=\frac{-C_{18}\sqrt{2}}{4}\qquad $&$\qquad (\sqrt{A}^{-1})^{4_{18},12_{18}}_{33,18}=\frac{-B_{18}}{2}(\frac{2}{\sqrt{2}}+1)\qquad $&$\qquad (\sqrt{A}^{-1})^{4_{18},9_{18}}_{35,18}=\frac{-C^{'}_{18}}{2}$\\
$(\sqrt{A}^{-1})^{4_{18},d_{18}}_{36,18}=\frac{-A^{'}_{4}}{\sqrt{2}}\qquad
$&
$\qquad(\sqrt{A}^{-1})^{13_{4},c_{4}}_{20,19}=\frac{-C^{'}_{4}}{2}\qquad $&$\qquad (\sqrt{A}^{-1})^{13_{4},16_{4}}_{22,19}=\frac{-B^{'}_{4}}{4}$\\
$(\sqrt{A}^{-1})^{4_{13},9_{13}}_{29,19}=\frac{-B^{'}_{13}}{2}\qquad
$&$\qquad
(\sqrt{A}^{-1})^{4_{13},14_{13}}_{30,19}=\frac{-C^{'}_{13}\sqrt{2}}{2}\qquad $&$\qquad (\sqrt{A}^{-1})^{4_{13},21_{13}}_{31,19}=\frac{-B_{13}}{2}(\frac{2}{\sqrt{2}}+1) $\\
$(\sqrt{A}^{-1})^{c_{4},b_{4}}_{21,20}=\frac{-A^{'}_{4}}{\sqrt{2}}\qquad
$&$\qquad
(\sqrt{A}^{-1})^{c_{4},16_{4}}_{22,20}=\frac{-B_{4}}{2}(\frac{2}{\sqrt{2}}+1)\qquad
$&$\qquad
(\sqrt{A}^{-1})^{b_{4},16_{4}}_{22,21}=\frac{-B^{'}_{4}}{\sqrt{2}}$\\
$(\sqrt{A}^{-1})^{6_{5},p_{5}}_{24,23}=\frac{-A^{'}_{5}}{2}\qquad
$&$\qquad
(\sqrt{A}^{-1})^{6_{5},q_{5}}_{25,23}=-\frac{B_{5}}{2}-\frac{C_5\sqrt{2}}{4}\qquad $&$\qquad (\sqrt{A}^{-1})^{6_{5},17_{5}}_{26,23}=\frac{-C_{5}\sqrt{2}}{4} $\\
$(\sqrt{A}^{-1})^{6_{5},13_{5}}_{27,23}=\frac{-C^{'}_{5}\sqrt{2}}{4}\qquad
$&$\qquad
(\sqrt{A}^{-1})^{p_{5},q_{5}}_{25,24}=\frac{-A^{'}_{5}}{2}\qquad $&$\qquad (\sqrt{A}^{-1})^{p_{5},17_{5}}_{26,24}=\frac{-C^{'}_{5}}{2}$\\
$(\sqrt{A}^{-1})^{p_{5},r_{5}}_{28,24}=\frac{-C^{'}_{5}}{2}\qquad
$&$\qquad
(\sqrt{A}^{-1})^{q_{5},13_{5}}_{27,25}=\frac{-C^{'}_{5}\sqrt{2}}{4}\qquad $&$\qquad (\sqrt{A}^{-1})^{q_{5},r_{5}}_{28,25}=\frac{-C_{5}\sqrt{2}}{4}$\\
$(\sqrt{A}^{-1})^{17_{5},13_{5}}_{27,26}=\frac{-A^{'}_{5}}{\sqrt{2}}\qquad
$&$\qquad
(\sqrt{A}^{-1})^{17_{5},r_{5}}_{28,26}=-\frac{A_{5}}{\sqrt{2}}-\frac{B_5}{2}\qquad $&$\qquad (\sqrt{A}^{-1})^{13_{5},r_{5}}_{28,27}=\frac{-A^{'}_{5}}{\sqrt{2}}$\\
$(\sqrt{A}^{-1})^{5_{13},9_{13}}_{29,27}=\frac{-B^{'}_{13}}{\sqrt{2}}\qquad
$&$\qquad
(\sqrt{A}^{-1})^{5_{13},14_{13}}_{30,27}=\frac{-C^{'}_{13}\sqrt{2}}{4}\qquad
$&$\qquad
(\sqrt{A}^{-1})^{5_{13},21_{13}}_{31,27}=\frac{-C^{'}_{13}}{2}$\\
$(\sqrt{A}^{-1})^{9_{13},14_{13}}_{30,29}=\frac{-B_{13}}{2}(\frac{2}{\sqrt{2}}+1)\qquad
$&$\qquad
(\sqrt{A}^{-1})^{14_{13},21_{13}}_{31,30}=\frac{-C_{13}\sqrt{2}}{4}\qquad $&$\qquad (\sqrt{A}^{-1})^{13_{14},k_{14}}_{42,30}=\frac{-C^{'}_{14}}{2}$\\
$(\sqrt{A}^{-1})^{13_{14},a_{14}}_{43,30}=\frac{-B^{'}_{14}}{\sqrt{2}}\qquad
$&$\qquad
(\sqrt{A}^{-1})^{13_{14},l_{14}}_{44,30}=\frac{-C_{14}\sqrt{2}}{4}\qquad $&$\qquad(\sqrt{A}^{-1})^{13_{14},l_{14}}_{46,30}=\frac{-B_{14}}{2}(\frac{2}{\sqrt{2}}+1)$\\
$(\sqrt{A}^{-1})^{e_{18},f_{18}}_{34,32}=\frac{-B_{18}}{2}(\frac{2}{\sqrt{2}}+1)\qquad$&
$\qquad(\sqrt{A}^{-1})^{e_{18},9_{18}}_{35,32}=\frac{-A^{'}_{18}}{2}\qquad
$&$\qquad(\sqrt{A}^{-1})^{e_{18},d_{18}}_{36,32}=\frac{-C^{'}_{18}\sqrt{2}}{4}$
\\$ (\sqrt{A}^{-1})^{12_{18},f_{18}}_{34,33}=\frac{-C_{18}\sqrt{2}}{4}\qquad$&
$\qquad(\sqrt{A}^{-1})^{12_{18},9_{18}}_{35,33}=\frac{-B^{'}_{18}}{2}\qquad $&$\qquad(\sqrt{A}^{-1})^{12_{18},d_{18}}_{36,33}=\frac{-B^{'}_{18}}{\sqrt{2}}$\\
$(\sqrt{A}^{-1})^{f_{18},9_{18}}_{35,34}=\frac{-B^{'}_{18}}{\sqrt{2}}\qquad$&
$\qquad(\sqrt{A}^{-1})^{f_{18},d_{18}}_{36,34}=\frac{-B^{'}_{18}}{2} \qquad $&$\qquad(\sqrt{A}^{-1})^{e^{'}_{15},d^{'}_{15}}_{38,37}=\frac{-B_{15}}{2}(\frac{2}{\sqrt{2}}+1) $\\
$
(\sqrt{A}^{-1})^{e^{'}_{15},c^{'}_{15}}_{39,37}=\frac{-A^{'}_{15}}{2}\qquad$&
$\qquad(\sqrt{A}^{-1})^{e^{'}_{15},f^{'}_{15}}_{40,37}=\frac{-C^{'}_{15}}{2}\qquad $&$\qquad(\sqrt{A}^{-1})^{e^{'}_{15},a^{'}_{15}}_{41,37}=\frac{-A_{15}}{2} $\\
$
(\sqrt{A}^{-1})^{d^{'}_{15},c^{'}_{15}}_{39,38}=\frac{-A^{'}_{15}}{2}\qquad$&
$\qquad(\sqrt{A}^{-1})^{d^{'}_{15},f^{'}_{15}}_{40,38}=\frac{-B^{'}_{15}}{2}\qquad $&$\qquad(\sqrt{A}^{-1})^{c^{'}_{15},a^{'}_{15}}_{41,39}=\frac{-C^{'}_{15}\sqrt{2}}{2}$\\
$
(\sqrt{A}^{-1})^{f^{'}_{15},a^{'}_{15}}_{41,40}=\frac{-A^{'}_{15}}{\sqrt{2}}\qquad$&
$\qquad(\sqrt{A}^{-1})^{k_{14},l_{14}}_{44,42}=\frac{-B^{'}_{14}}{\sqrt{2}}\qquad $&$\qquad (\sqrt{A}^{-1})^{k_{14},j_{14}}_{45,42}=\frac{-A^{'}_{14}}{2} $\\
$
(\sqrt{A}^{-1})^{k_{14},h_{14}}_{46,42}=\frac{-C^{'}_{14}}{2}\qquad$&
$\qquad(\sqrt{A}^{-1})^{a_{14},l_{14}}_{44,43}=\frac{-C^{'}_{14}\sqrt{2}}{4}\qquad $&$\qquad (\sqrt{A}^{-1})^{a_{14},j_{14}}_{45,43}=\frac{-C^{'}_{14}\sqrt{2}}{4}$\\
$
(\sqrt{A}^{-1})^{a_{14},h_{14}}_{46,43}=\frac{-A^{'}_{14}}{\sqrt{2}}\qquad$&
$\qquad(\sqrt{A}^{-1})^{l_{14},j_{14}}_{45,44}=-\frac{C_{14}\sqrt{2}}{4}-\frac{B_{14}}{\sqrt{2}}\qquad
$&$\qquad
(\sqrt{A}^{-1})^{j_{14},h_{14}}_{46,45}=\frac{-C_{14}\sqrt{2}}{4}$
\end{tabular}
\end{center}

\subsection{Edge-metric components for the 8-valent graph}
\label{sa.3} In this Section we give the values for the terms
$\frac{t_{e_{i,j}e_{i,j}}} {\sqrt{t_{e_{i,j}}t_{e_{i,j}}}}$ for
the 8-valent graph as computed for aligned, rotated, and
translated graphs respectively.
\subsubsection{Aligned 8-valent graph}
\label{sa.3.1} For the case of an 8-valent graph, the periodicity
cell contains five vertices, therefore  $\sqrt{A}^{-1}$ is
a $36\times36$ matrix. Below we give the non-zero upper-diagonal
elements of the matrix $\sqrt{A}^{-1}$. The lower indices refer to
the matrix elements while, the upper indices, indicate which edges
we are taking into consideration.
\begin{center}
\begin{tabular}{lll}\label{tab:a8a}
$(\sqrt{A}^{-1})^{0_1,13_1}_{10,1}=-\beta\qquad $& $\qquad
(\sqrt{A}^{-1})^{0_1,12_1}_{11,1}=-\frac{1}{2}\beta\qquad$&
$\qquad (\sqrt{A}^{-1})^{0_1,11_1}_{12,1}=-\beta$\\
$ (\sqrt{A}^{-1})^{0_1,9_1}_{13,1}=-\frac{1}{2}\beta\qquad$&
$\qquad(\sqrt{A}^{-1})^{0_1,15_1}_{14,1}=-\beta\qquad $&$\qquad (\sqrt{A}^{-1})^{0_1,10_1}_{15,1}=-\frac{1}{2}\beta $\\
$ (\sqrt{A}^{-1})^{14_1,13_1}_{10,9}=-\frac{1}{2}\beta\qquad$&
$\qquad(\sqrt{A}^{-1})^{14_1,12_1}_{11,9}=-\beta\qquad $&$\qquad (\sqrt{A}^{-1})^{14_1,11_1}_{12,9}=-\frac{1}{2}\beta $\\
$ (\sqrt{A}^{-1})^{14_1,9_1}_{13,9}=-\beta\qquad$&
$\qquad(\sqrt{A}^{-1})^{14_1,15_1}_{14,9}=-\beta\qquad $&$\qquad (\sqrt{A}^{-1})^{14_1,10_1}_{15,9}=-\beta $\\
$ (\sqrt{A}^{-1})^{13_1,12_1}_{11,10}=-\frac{1}{2}\beta\qquad$&
$\qquad(\sqrt{A}^{-1})^{13_1,11_1}_{12,10}=-\beta\qquad $&$\qquad (\sqrt{A}^{-1})^{13_1,15_1}_{14,10}=-\beta $\\
$ (\sqrt{A}^{-1})^{13_1,10_1}_{15,10}=-\frac{1}{2}\beta\qquad$&
$\qquad(\sqrt{A}^{-1})^{12_1,11_1}_{12,11}=-\frac{1}{2}\beta\qquad $&$\qquad (\sqrt{A})^{-1}_{13,11}=-\beta $\\
$(\sqrt{A}^{-1})^{12_1,9_1}_{14,11}=-\beta\qquad
$&$\qquad(\sqrt{A}^{-1})^{10_1,9_1}_{16,11}=-\beta\qquad$&
$\qquad(\sqrt{A}^{-1})^{1_{12},m_{12}}_{23,11}=-\beta$\\
$ (\sqrt{A}^{-1})^{1_{12},a_{12}}_{24,11}=-2\beta\qquad$&
$\qquad(\sqrt{A}^{-1})^{1_{12},e_{12}}_{28,11}=-2\beta\qquad $&$\qquad (\sqrt{A}^{-1})^{11_{1},9_{1}}_{13,12}=-\frac{1}{2}\beta $\\
$ (\sqrt{A}^{-1})^{11_{1},15_{1}}_{14,12}=-\beta\qquad$&
$\qquad(\sqrt{A}^{-1})^{9_{1},15_{1}}_{14,13}=-\frac{1}{2}\beta\qquad $&$\qquad (\sqrt{A}^{-1})^{9_{1},10_{1}}_{15,13}=-\beta $\\
$ (\sqrt{A}^{-1})^{1_{9},a_{9}}_{16,13}=-\beta\qquad$&
$\qquad(\sqrt{A}^{-1})^{15_{1},10_{1}}_{15,14}=-\frac{1}{2}\beta\qquad $&$\qquad (\sqrt{A}^{-1})^{4_{9},b_{9}}_{18,17}=-\beta $\\
$ (\sqrt{A}^{-1})^{5_{9},c_{9}}_{20,19}=-\beta\qquad$&
$\qquad(\sqrt{A}^{-1})^{6_{9},d_{9}}_{22,21}=-\beta\qquad $&$\qquad (\sqrt{A}^{-1})^{m_{12},a_{12}}_{24,23}=-2\beta $\\
$ (\sqrt{A}^{-1})^{m_{12},e_{12}}_{28,23}=-2\beta\qquad$&
$\qquad(\sqrt{A}^{-1})^{a_{12},e_{12}}_{28,24}=-\beta\qquad $&$\qquad (\sqrt{A}^{-1})^{b_{12},4_{12}}_{26,25}=-2\beta $\\
$ (\sqrt{A}^{-1})^{b_{12},h_{12}}_{27,25}=-\beta\qquad$&
$\qquad(\sqrt{A}^{-1})^{b_{12},k_{12}}_{29,25}=-2\beta\qquad $&$\qquad (\sqrt{A}^{-1})^{4_{12},h_{12}}_{27,26}=-2\beta $\\
$ (\sqrt{A}^{-1})^{4_{12},k_{12}}_{29,26}=-\beta\qquad$&
$\qquad(\sqrt{A}^{-1})^{h_{12},k_{12}}_{29,27}=-2\beta\qquad$&$\qquad(\sqrt{A}^{-1})^{11_{e},12_{e}}_{31,30}=-\beta $\\
$
(\sqrt{A}^{-1})^{11_{e},14_{e}}_{32,30}=-\frac{1}{2}\beta\qquad$&
$\qquad(\sqrt{A}^{-1})^{11_{e},p_{e}}_{33,30}=-2\beta\qquad$&$\qquad(\sqrt{A}^{-1})^{11_{e},q_{e}}_{34,30}=-\frac{1}{2}\beta $\\
$ (\sqrt{A}^{-1})^{11_{e},g_{e}}_{36,30}=-\frac{1}{2}\beta\qquad$&
$\qquad(\sqrt{A}^{-1})^{13_{e},14_{e}}_{32,31}=-\beta\qquad$&
$\qquad(\sqrt{A}^{-1})^{13_{e},p_{e}}_{33,31}=-\frac{1}{2}\beta$\\
$(\sqrt{A}^{-1})^{13_{e},q_{e}}_{34,31}=-\beta\qquad$&
$\qquad(\sqrt{A}^{-1})^{13_{e},r_{e}}_{35,31}=-\frac{1}{2}\beta\qquad$&
$\qquad(\sqrt{A}^{-1})^{14_{e},q_{e}}_{34,32}=-\frac{1}{2}\beta$\\
$(\sqrt{A}^{-1})^{14_{e},r_{e}}_{35,32}=-2\beta\qquad$&
$\qquad(\sqrt{A}^{-1})^{14_{e},g_{e}}_{36,32}=-\frac{1}{2}\beta\qquad$&
$\qquad(\sqrt{A}^{-1})^{p_{e},q_{e}}_{34,33}=-\beta$\\
$(\sqrt{A}^{-1})^{p_{e},r_{e}}_{35,33}=-\frac{1}{2}\beta\qquad$&
$\qquad(\sqrt{A}^{-1})^{p_{e},g_{e}}_{36,33}=-\beta\qquad$&
$\qquad(\sqrt{A}^{-1})^{q_{e},r_{e}}_{35,34}=-\beta$\\
$\qquad(\sqrt{A}^{-1})^{q_{e},g_{e}}_{36,34}=-\frac{1}{2}\beta\qquad$&
$\qquad(\sqrt{A}^{-1})^{r_{e},g_{e}}_{36,35}=-\beta$&
\end{tabular}
\end{center}
where $\beta=\frac{(\frac{\delta}{\sqrt{3}}-nl)}{\delta\sqrt{3}}$.

\subsubsection{Rotated 8-Valent graph}
\label{sa.3.2}

In this Section we  give the values of all the terms $\frac{t_{e_{k,i}e_{k,j}}} {\sqrt{t_{e_{k,i}}t_{e_{k,j}}}}$ that  appear in the matrix $\sqrt{A}^{-1}$ for the rotated 8-valent graph. As in the aligned case, the matrix $\sqrt{A}^{-1}$ is a $36\times36$ matrix labelled by the edges of the five vertices that comprise the periodicity cell. The following short-hand notation is used: $\mu_{k_i,k_j}=\frac{t_{e_{k,i}e_{k,j}}} {\sqrt{t_{e_{k,i}}t_{e_{k,j}}}}$ which is defined in a similar manner as that done for the 4, and 6-valent graphs.

The elements of the upper-half of the matrix $\sqrt{A}^{-1}$ are:
\begin{center}
\begin{tabular}{lll}\label{tab:a8r}
$(\sqrt{A})^{-1}_{11,1}=-\frac{1}{2}\mu_{12_1,0_1}\qquad $&$\qquad (\sqrt{A})^{-1}_{12,1}=-\frac{1}{2}\mu_{11_1,0_1}\qquad $&$\qquad (\sqrt{A})^{-1}_{10,1}=-\frac{1}{2}\mu_{13_1,0_1}$\\
$(\sqrt{A})^{-1}_{13,1}=-\frac{1}{2}\mu_{9_1,0_1}\qquad $&$\qquad (\sqrt{A})^{-1}_{15,1}=-\frac{1}{2}\mu_{10_1,0_1}\qquad $&$\qquad (\sqrt{A})^{-1}_{15,1}=-\frac{1}{2}\mu_{15_1,0_1}$\\
$(\sqrt{A})^{-1}_{12,11}=-\frac{1}{2}\mu_{11_1,12_1}\qquad $&$\qquad (\sqrt{A})^{-1}_{10,9}=-\frac{1}{2}\mu_{13_1,14_1}\qquad $&$\qquad (\sqrt{A})^{-1}_{11,9}=-\frac{1}{2}\mu_{14_1,12_1}$\\
$(\sqrt{A})^{-1}_{13,11}=-\frac{1}{2}\mu_{9_1,12_1}\qquad $&$\qquad (\sqrt{A})^{-1}_{15,11}=-\frac{1}{2}\mu_{10_1,12_1}\qquad $&$\qquad (\sqrt{A})^{-1}_{27,11}=-\frac{1}{2}\mu_{4_{12},1_{12}}$\\
$(\sqrt{A})^{-1}_{24,11}=-\frac{1}{2}\mu_{a_{12},1_{12}}\qquad $&$\qquad (\sqrt{A})^{-1}_{25,11}=-\frac{1}{2}\mu_{b_{12},1_{12}}\qquad $&$\qquad (\sqrt{A})^{-1}_{28,11}=-\frac{1}{2}\mu_{e_{12},1_{12}}$\\
$(\sqrt{A})^{-1}_{27,11}=-\frac{1}{2}\mu_{h_{12},1_{12}}\qquad $&$\qquad (\sqrt{A})^{-1}_{23,11}=-\frac{1}{2}\mu_{m_{12},1_{12}}\qquad $&$\qquad (\sqrt{A})^{-1}_{12,10}=-\frac{1}{2}\mu_{13_1,11_1}$\\
$(\sqrt{A})^{-1}_{9,12}=-\frac{1}{2}\mu_{14_1,11_1}\qquad $&$\qquad (\sqrt{A})^{-1}_{13,12}=-\frac{1}{2}\mu_{9_1,11_1}\qquad $&$\qquad (\sqrt{A})^{-1}_{14,12}=-\frac{1}{2}\mu_{15_1,11_1}$\\
$(\sqrt{A})^{-1}_{36,32}=-\frac{1}{2}\mu_{14_{e},g_{e}}\qquad$&$\qquad (\sqrt{A})^{-1}_{10,9}=-\frac{1}{2}\mu_{14_{1},13_{1}}\qquad $&$\qquad (\sqrt{A})^{-1}_{15,10}=-\frac{1}{2}\mu_{10_{1},13_{1}}$\\
$(\sqrt{A})^{-1}_{14,10}=-\frac{1}{2}\mu_{15_{1},13_{1}}\qquad $&$\qquad (\sqrt{A})^{-1}_{13,9}=-\frac{1}{2}\mu_{9_{1},14_{1}}\qquad $&$\qquad (\sqrt{A})^{-1}_{15,9}=-\frac{1}{2}\mu_{10_{1},14_{1}}$\\
$(\sqrt{A})^{-1}_{14,9}=-\frac{1}{2}\mu_{15_{1},14_{1}}\qquad $&$\qquad (\sqrt{A})^{-1}_{15,13}=-\frac{1}{2}\mu_{10_{1},9_{1}}\qquad $&$\qquad (\sqrt{A})^{-1}_{14,15}=-\frac{1}{2}\mu_{15_{1},9_{1}}$\\
$(\sqrt{A})^{-1}_{17,13}=-\frac{1}{2}\mu_{4_{9},1_{9}}\qquad $&$\qquad (\sqrt{A})^{-1}_{16,13}=-\frac{1}{2}\mu_{a_{9},1_{9}}\qquad $&$\qquad (\sqrt{A})^{-1}_{18,13}=-\frac{1}{2}\mu_{b_{9},1_{9}}$\\
$(\sqrt{A})^{-1}_{21,13}=-\frac{1}{2}\mu_{6_{9},1_{9}}\qquad $&$\qquad (\sqrt{A})^{-1}_{19,13}=-\frac{1}{2}\mu_{5_{9},1_{9}}\qquad $&$\qquad (\sqrt{A})^{-1}_{22,13}=-\frac{1}{2}\mu_{d_{9},1_{9}}$\\
$(\sqrt{A})^{-1}_{15,14}=-\frac{1}{2}\mu_{15_{1},10_{1}}\qquad $&$\qquad (\sqrt{A})^{-1}_{17,16}=-\frac{1}{2}\mu_{a_{9},4_{9}}\qquad $&$\qquad (\sqrt{A})^{-1}_{18,17}=-\frac{1}{2}\mu_{b_{9},4_{9}}$\\
$(\sqrt{A})^{-1}_{21,17}=-\frac{1}{2}\mu_{6_{9},4_{9}}\qquad $&$\qquad (\sqrt{A})^{-1}_{19,17}=-\frac{1}{2}\mu_{5_{9},4_{9}}\qquad $&$\qquad (\sqrt{A})^{-1}_{20,17}=-\frac{1}{2}\mu_{c_{9},4_{9}}$\\
$(\sqrt{A})^{-1}_{18,16}=-\frac{1}{2}\mu_{b_{9},a_{9}}\qquad $&$\qquad (\sqrt{A})^{-1}_{21,16}=-\frac{1}{2}\mu_{6_{9},a_{9}}\qquad $&$\qquad (\sqrt{A})^{-1}_{22,16}=-\frac{1}{2}\mu_{d_{9},a_{9}}$\\
$(\sqrt{A})^{-1}_{20,16}=-\frac{1}{2}\mu_{c_{9},a_{9}}\qquad $&$\qquad (\sqrt{A})^{-1}_{19,18}=-\frac{1}{2}\mu_{5_{9},b_{9}}\qquad $&$\qquad (\sqrt{A})^{-1}_{22,18}=-\frac{1}{2}\mu_{d_{9},b_{9}}$\\
$(\sqrt{A})^{-1}_{20,18}=-\frac{1}{2}\mu_{c_{9},b_{9}}\qquad $&$\qquad (\sqrt{A})^{-1}_{21,19}=-\frac{1}{2}\mu_{5_{9},6_{9}}\qquad $&$\qquad (\sqrt{A})^{-1}_{22,21}=-\frac{1}{2}\mu_{d_{9},6_{9}}$\\
$(\sqrt{A})^{-1}_{21,22}=-\frac{1}{2}\mu_{c_{9},6_{9}}\qquad $&$\qquad (\sqrt{A})^{-1}_{22,19}=-\frac{1}{2}\mu_{d_{9},5_{9}}\qquad $&$\qquad (\sqrt{A})^{-1}_{20,19}=-\frac{1}{2}\mu_{c_{9},5_{9}}$\\
$(\sqrt{A})^{-1}_{22,20}=-\frac{1}{2}\mu_{c_{9},d_{9}}\qquad $&$\qquad(\sqrt{A})^{-1}_{32,31}=-\frac{1}{2}\mu_{13_{e},14_{e}}\qquad$&$\qquad (\sqrt{A})^{-1}_{30,28}=-\frac{1}{2}\mu_{12_{e},11_{e}}$\\
$(\sqrt{A})^{-1}_{33,30}=-\frac{1}{2}\mu_{p_{e},11_{e}}\qquad $&$\qquad (\sqrt{A})^{-1}_{34,30}=-\frac{1}{2}\mu_{q_{e},11_{e}}\qquad $&$\qquad (\sqrt{A})^{-1}_{36,30}=-\frac{1}{2}\mu_{g_{e},11_{e}}$\\
$(\sqrt{A})^{-1}_{32,30}=-\frac{1}{2}\mu_{14_{e},11_{e}}\qquad $&$\qquad (\sqrt{A})^{-1}_{31,30}=-\frac{1}{2}\mu_{13_{e},11_{e}}\qquad $&$\qquad (\sqrt{A})^{-1}_{26,24}=-\frac{1}{2}\mu_{a_{12},4_{12}}$\\
$(\sqrt{A})^{-1}_{26,25}=-\frac{1}{2}\mu_{b_{12},4_{12}}\qquad $&$\qquad (\sqrt{A})^{-1}_{28,26}=-\frac{1}{2}\mu_{e_{12},4_{12}}\qquad $&$\qquad (\sqrt{A})^{-1}_{27,26}=-\frac{1}{2}\mu_{h_{12},4_{12}}$\\
$(\sqrt{A})^{-1}_{29,26}=-\frac{1}{2}\mu_{k_{12},4_{12}}\qquad $&$\qquad (\sqrt{A})^{-1}_{25,24}=-\frac{1}{2}\mu_{b_{12},a_{12}}\qquad $&$\qquad (\sqrt{A})^{-1}_{28,24}=-\frac{1}{2}\mu_{e_{12},a_{12}}$\\
$(\sqrt{A})^{-1}_{29,33}=-\frac{1}{2}\mu_{m_{12},a_{12}}\qquad $&$\qquad (\sqrt{A})^{-1}_{29,24}=-\frac{1}{2}\mu_{k_{12},a_{12}}\qquad $&$\qquad (\sqrt{A})^{-1}_{27,25}=-\frac{1}{2}\mu_{h_{12},b_{12}}$\\
$(\sqrt{A})^{-1}_{25,23}=-\frac{1}{2}\mu_{m_{12},b_{12}}\qquad $&$\qquad (\sqrt{A})^{-1}_{29,25}=-\frac{1}{2}\mu_{b_{12},k_{12}}\qquad $&$\qquad (\sqrt{A})^{-1}_{28,27}=-\frac{1}{2}\mu_{h_{12},e_{12}}$\\
$(\sqrt{A})^{-1}_{28,23}=-\frac{1}{2}\mu_{m_{12},e_{12}}\qquad $&$\qquad (\sqrt{A})^{-1}_{29,28}=-\frac{1}{2}\mu_{k_{12},e_{12}}\qquad $&$\qquad (\sqrt{A})^{-1}_{35,28}=-\frac{1}{2}\mu_{r_{e},12_{e}}$\\
$(\sqrt{A})^{-1}_{33,28}=-\frac{1}{2}\mu_{p_{e},12_{e}}\qquad $&$\qquad (\sqrt{A})^{-1}_{36,23}=-\frac{1}{2}\mu_{g_{e},12_{e}}\qquad $&$\qquad (\sqrt{A})^{-1}_{32,28}=-\frac{1}{2}\mu_{14_{e},12_{e}}$\\
$(\sqrt{A})^{-1}_{31,28}=-\frac{1}{2}\mu_{13_{e},12_{e}}\qquad $&$\qquad (\sqrt{A})^{-1}_{27,23}=-\frac{1}{2}\mu_{m_{12},h_{12}}\qquad $&$\qquad (\sqrt{A})^{-1}_{29,27}=-\frac{1}{2}\mu_{k_{12},h_{12}}$\\
$(\sqrt{A})^{-1}_{29,23}=-\frac{1}{2}\mu_{k_{12},m_{12}}\qquad $&$\qquad (\sqrt{A})^{-1}_{35,33}=-\frac{1}{2}\mu_{p_{e},r_{e}}\qquad $&$\qquad (\sqrt{A})^{-1}_{35,34}=-\frac{1}{2}\mu_{q_{e},r_{e}}$\\
$(\sqrt{A})^{-1}_{36,35}=-\frac{1}{2}\mu_{g_{e},r_{e}}\qquad $&$\qquad (\sqrt{A})^{-1}_{35,32}=-\frac{1}{2}\mu_{14_{e},r_{e}}\qquad $&$\qquad (\sqrt{A})^{-1}_{35,31}=-\frac{1}{2}\mu_{13_{e},r_{e}}$\\
$(\sqrt{A})^{-1}_{34,33}=-\frac{1}{2}\mu_{q_{e},p_{e}}\qquad $&$\qquad (\sqrt{A})^{-1}_{36,33}=-\frac{1}{2}\mu_{g_{e},p_{e}}\qquad $&$\qquad (\sqrt{A})^{-1}_{33,31}=-\frac{1}{2}\mu_{13_{e},p_{e}}$\\
$(\sqrt{A})^{-1}_{36,33}=-\frac{1}{2}\mu_{g_{e},q_{e}}\qquad $&$\qquad (\sqrt{A})^{-1}_{34,32}=-\frac{1}{2}\mu_{14_{e},q_{e}}\qquad $&$\qquad (\sqrt{A})^{-1}_{34,31}=-\frac{1}{2}\mu_{13_{e},q_{e}}$\\
\end{tabular}
\end{center}

\subsubsection{Translated 8-valent graph}
\label{sa.3.3}

In this Section we will first give the values of all the terms
$t_{e_ie_j}=t_{e_ie_j}^x+t_{e_ie_j}^y+t_{e_ie_j}^z$ as computed
for all the five vertices comprising the periodicity cell
for the 8-valent graph. We then give the values for the
terms $\frac{t_{e_{i,k}e_{j,k}}} {\sqrt{t_{e_{i,k}}t_{e_{j,k}}}}$
that appear in the matrix $\sqrt{A}^{-1}$. As for the aligned
case, the matrix $\sqrt{A}^{-1}$ is a $36\times 36$ matrix
labelled by the edges of the five vertices
that comprise the periodicity cell.

The values for $t_{e_ie_j}$ are listed below. The (Gauss brackets) terms $n_i=[\frac{V_i^x}{l}]$, $m_i[\frac{V_i^y}{l}]$ and $p_i=[\frac{V_i^z}{l}]$ are the number of stacks that each edge $e_i$ intersects in the $x$-, $y$- and $z$-direction respectively (see Section \ref{s7.1})\\[10pt]
1) $V^{''}_0=(\epsilon_x,\epsilon_y,\epsilon_z)$\\[5pt]
\begin{center}
\begin{tabular}{l|l|l}
$(V^{''}_0)^z$&$(V^{''}_0)^x$&$(V^{''}_0)^y$\\\hline
$t^z_{e_{0,5}e_{0,6}}=t^z_{e_{0,5}e_{0,7}}=$&$t^x_{e_{0,5}e_{0,7}}=t^x_{e_{0,5}e_{0,4}}$& $t^y_{e_{0,5}e_{0,6}}=t^y_{e_{0,5}e_{0,4}}=$\\
$t^z_{e_{0,8}e_{0,6}}=t^z_{e_{0,5}e_{0,8}}=$&$t^x_{e_{0,5}e_{0,3}}=t^x_{e_{0,4}e_{0,7}}$ &$t^y_{e_{0,4}e_{0,6}}=t_{e_{0,5}e_{0,1}}=$\\
$t^z_{e_{0,3}e_{0,1}}=t^z_{e_{0,7}e_{0,6}}$& $t^x_{e_{0,7}e_{0,3}}=t^x_{e_{0,6}e_{0,8}}$&$t^y_{e_{0,6}e_{0,1}}=t^y_{e_{0,8}e_{0,7}}=$\\
$t^z_{e_{0,4}e_{0,3}}=t^z_{e_{0,4}e_{0,2}}=$&$t^x_{e_{0,6}e_{0,1}}=t^x_{e_{0,6}e_{0,2}}$&$t^y_{e_{0,8}e_{0,3}}=t^y_{e_{0,8}e_{0,2}}=$\\
$t^z_{e_{0,2}e_{0,1}}=t^z_{e_{0,1}e_{0,4}}=(\epsilon_y-m_0l)$ &$t^x_{e_{0,8}e_{0,1}}=t^x_{e_{0,8}e_{0,2}}=(\epsilon_z-p_0l)$ &$t^y_{e_{0,7}e_{0,3}}=t^y_{e_{0,7}e_{0,2}}=(\epsilon_z-p_0l)$\\
$t^z_{e_{0,3}e_{0,2}}=t^z_{e_{0,7}e_{0,8}}=(\epsilon_x-n_0l)$&$t^x_{e_{0,1}e_{0,2}}=t^x_{e_{0,4}e_{0,3}}=(\epsilon_y-m_0l)$ & $t^y_{e_{0,2}e_{0,3}}=t^y_{e_{0,4}e_{0,1}}=(\epsilon_x-n_0l)$\\
\end{tabular}
\end{center}
2) $V^{''}_1=(\epsilon_x-\frac{\delta_e
\sqrt{2}}{2},\epsilon_y-\frac{\delta_e
\sqrt{2}}{2},\epsilon_z+\frac{\delta_e \sqrt{2}}{2})$
\begin{center}
\begin{tabular}{l|l|l}
$(V^{''}_1)^z$&$(V^{''}_1)^x$&$(V^{''}_1)^y$\\\hline
$t^z_{e_{1,11}e_{1,13}}=t^z_{e_{1,9}e_{1,0}}=$&$t^x_{e_{1,9}e_{1,0}}=t^x_{e_{1,14}e_{1,13}}$& $t^y_{e_{1,9}e_{1,10}}=t^y_{e_{1,9}e_{1,12}}=$\\
$t^z_{e_{1,15}e_{1,10}}=t^z_{e_{1,9}e_{1,15}}=$&$t^x_{e_{1,9}e_{1,11}}=t^x_{e_{1,12}e_{1,0}}$ &$t^y_{e_{1,12}e_{1,10}}=t_{e_{1,9}e_{1,14}}=$\\
$t^z_{e_{1,11}e_{1,14}}=t^z_{e_{1,0}e_{1,10}}$& $t^x_{e_{1,0}e_{1,11}}=t^x_{e_{1,10}e_{1,15}}$&$t^y_{e_{1,12}e_{1,14}}=t^y_{e_{1,15}e_{1,0}}=$\\
$t^z_{e_{1,12}e_{1,11}}=t^z_{e_{1,12}e_{1,13}}=$&$t^x_{e_{1,12}e_{1,11}}=t^x_{e_{1,10}e_{1,13}}$&$t^y_{e_{1,15}e_{1,11}}=t^y_{e_{1,13}e_{1,11}}=$\\
$t^z_{e_{1,13}e_{1,14}}=t^z_{e_{1,0}e_{1,15}}=(|\epsilon_y-\frac{\delta}{\sqrt{3}}|-m_1l)$ &$t^x_{e_{1,15}e_{1,14}}=t^x_{e_{1,15}e_{1,13}}=(|\epsilon_y-\frac{\delta}{\sqrt{3}}|-m_1l)$ &$t^y_{e_{1,0}e_{1,11}}=t^y_{e_{1,0}e_{1,13}}=$\\
$t^z_{e_{1,9}e_{1,10}}=t^z_{e_{1,14}e_{1,12}}=(|\epsilon_x-\frac{\delta}{\sqrt{3}}|-n_1l)$&$t^x_{e_{1,9}e_{1,12}}=t^x_{e_{1,10}e_{1,14}}=(|\epsilon_x-\frac{\delta}{\sqrt{3}}|-n_1l)$ &$=(|\epsilon_x-\frac{\delta}{\sqrt{3}}|-n_1l)$\\
& & $t^y_{e_{1,15}e_{1,13}}=t^y_{e_{1,10}e_{1,14}}=$\\
& & $=(\epsilon_z+\frac{\delta}{\sqrt{3}}-m_1l)$\\
\end{tabular}
\end{center}

3) $V^{''}_9=(\epsilon_x,\epsilon_y-2\frac{\delta_e
\sqrt{2}}{2},\epsilon_z+\frac{\delta_e \sqrt{2}}{2})$
\begin{center}
\begin{tabular}{l|l|l}
$(V^{''}_9)^z$&$(V^{''}_9)^x$&$(V^{''}_9)^y$\\\hline
$t^z_{e_{9,4}e_{9,1}}=t^z_{e_{9,c}e_{9,5}}=$&$t^x_{e_{9,c}e_{9,5}}=t^x_{e_{9,a}e_{9,1}}$& $t^y_{e_{9,c}e_{9,d}}=t^y_{e_{9,c}e_{9,b}}=$\\
$t^z_{e_{9,6}e_{9,d}}=t^z_{e_{9,c}e_{9,6}}=$&$t^x_{e_{9,c}e_{9,4}}=t^x_{e_{9,b}e_{9,5}}$ &$t^y_{e_{9,b}e_{9,d}}=t_{e_{9,c}e_{9,a}}=$\\
$t^z_{e_{9,4}e_{9,a}}=t^z_{e_{9,5}e_{9,d}}$& $t^x_{e_{9,5}e_{9,4}}=t^x_{e_{9,d}e_{9,6}}$&$t^y_{e_{9,b}e_{9,a}}=t^y_{e_{9,6}e_{9,5}}=$\\
$t^z_{e_{9,b}e_{9,4}}=t^z_{e_{9,b}e_{9,1}}=$&$t^x_{e_{9,b}e_{9,4}}=t^x_{e_{9,d}e_{9,1}}$&$t^y_{e_{9,6}e_{9,4}}=t^y_{e_{9,6}e_{9,1}}=$\\
$t^z_{e_{9,1}e_{9,a}}=t^z_{e_{9,5}e_{9,6}}=(|\epsilon_y-2\frac{\delta}{\sqrt{3}}|-m_9l)$ &$t^x_{e_{9,6}e_{9,a}}=t^x_{e_{9,6}e_{9,1}}=(|\epsilon_y-2\frac{\delta}{\sqrt{3}}|-m_9l)$ &$t^y_{e_{9,5}e_{9,4}}=t^y_{e_{9,5}e_{9,1}}=(\epsilon_z-p_9l)$\\
$t^z_{e_{9,c}e_{9,d}}=t^z_{e_{9,a}e_{9,b}}=(\epsilon_x-n_9l)$&$t^x_{e_{9,c}e_{9,b}}=t^x_{e_{9,d}e_{9,a}}=(\epsilon_z-p_9l)$ & $t^y_{e_{9,d}e_{9,a}}=t^y_{e_{9,1}e_{9,4}}=(\epsilon_x-n_9l)$\\
\end{tabular}
\end{center}
4) $V^{''}_{12}=(\epsilon_x,\epsilon_y-\frac{\delta_e
\sqrt{2}}{2},\epsilon_z+2\frac{\delta_e \sqrt{2}}{2})$
\begin{center}
\begin{tabular}{l|l|l}
$(V^{''}_{12})^z$&$(V^{''}_{12})^x$&$(V^{''}_{12})^y$\\\hline
$t^z_{e_{12,b}e_{12,a}}=t^z_{e_{12,4}e_{12,1}}=$&$t^x_{e_{12,b}e_{12,4}}=t^x_{e_{12,m}e_{12,e}}$& $t^y_{e_{12,b}e_{12,a}}=t^y_{e_{12,e}e_{12,h}}=$\\
$t^z_{e_{12,1}e_{12,a}}=t^z_{e_{12,b}e_{12,1}}=$&$t^x_{e_{12,b}e_{12,h}}=t^x_{e_{12,k}e_{12,4}}$ &$t^y_{e_{12,k}e_{12,a}}=t^y_{e_{12,b}e_{12,m}}=$\\
$t^z_{e_{12,h}e_{12,m}}=t^z_{e_{12,4}e_{12,a}}$& $t^x_{e_{12,4}e_{12,h}}=t^x_{e_{12,a}e_{12,1}}$&$t^y_{e_{12,a}e_{12,m}}=t^y_{e_{12,1}e_{12,4}}=$\\
$t^z_{e_{12,h}e_{12,e}}=t^z_{e_{12,k}e_{12,e}}=$&$t^x_{e_{12,k}e_{12,h}}=t^x_{e_{12,a}e_{12,e}}$&$t^y_{e_{12,1}e_{12,h}}=t^y_{e_{12,1}e_{12,e}}=$\\
$t^z_{e_{12,e}e_{12,m}}=t^z_{e_{12,m}e_{12,k}}=(\epsilon_x-n_12l)$ &$t^x_{e_{12,1}e_{12,m}}=t^x_{e_{12,1}e_{12,e}}=(|\epsilon_y-2\frac{\delta}{\sqrt{3}}|-m_12l)$ &$t^y_{e_{12,k}e_{12,m}}=t^y_{e_{12,4}e_{12,e}}=$\\
$t^z_{e_{12,k}e_{12,h}}=t^z_{e_{12,b}e_{12,4}}=(|\epsilon_y-2\frac{\delta}{\sqrt{3}}|-m_12l)$&
$t^x_{e_{12,b}e_{12,k}}=t^x_{e_{12,a}e_{12,m}}=(\epsilon_z+2\frac{\delta}{\sqrt{3}}-p_12l)$
& $=(\epsilon_x-n_12l)$\\& &
$t^y_{e_{12,b}e_{12,k}}=t^y_{e_{12,4}e_{12,h}}=$\\
& & $(\epsilon_z+2\frac{\delta}{\sqrt{3}}-p_12l)$\\
\end{tabular}
\end{center}
5) $V^{''}_e=(\epsilon_x-\frac{\delta_e
\sqrt{2}}{2},\epsilon_y-\frac{\delta_e
\sqrt{2}}{2},\epsilon_z+3\frac{\delta_e \sqrt{2}}{2})$
\begin{center}
\begin{tabular}{l|l|l}
$(V^{''}_e)^z$&$(V^{''}_e)^x$&$(V^{''}_e)^y$\\\hline
$t^z_{e_{e,p}e_{e,q}}=t^z_{e_{e,12}e_{e,11}}=$&$t^x_{e_{e,12}e_{e,11}}=t^x_{e_{e,r}e_{e,q}}$& $t^y_{e_{e,12}e_{e,14}}=t^y_{e_{e,12}e_{e,g}}=$\\
$t^z_{e_{e,13}e_{e,14}}=t^z_{e_{e,12}e_{e,13}}=$&$t^x_{e_{e,12}e_{e,p}}=t^x_{e_{e,g}e_{e,11}}$ &$t^y_{e_{e,g}e_{e,14}}=t_{e_{e,12}e_{e,r}}=$\\
$t^z_{e_{e,p}e_{e,r}}=t^z_{e_{e,11}e_{e,14}}$& $t^x_{e_{e,11}e_{e,p}}=t^x_{e_{e,14}e_{e,13}}$&$t^y_{e_{e,g}e_{e,r}}=t^y_{e_{e,13}e_{e,11}}=$\\
$t^z_{e_{e,g}e_{e,p}}=t^z_{e_{e,g}e_{e,q}}=$&$t^x_{e_{e,g}e_{e,p}}=t^x_{e_{e,14}e_{e,q}}$&$t^y_{e_{e,13}e_{e,p}}=t^y_{e_{e,q}e_{e,p}}=$\\
$t^z_{e_{e,q}e_{e,r}}=t^z_{e_{e,11}e_{e,13}}=(|\epsilon_y-\frac{\delta}{\sqrt{3}}|-m_el)$ &$t^x_{e_{e,13}e_{e,r}}=t^x_{e_{e,13}e_{e,q}}=(|\epsilon_y-\frac{\delta}{\sqrt{3}}|-m_el)$ &$t^y_{e_{e,11}e_{e,p}}=t^y_{e_{e,11}e_{e,q}}=$\\
$t^z_{e_{e,12}e_{e,14}}=t^z_{e_{e,r}e_{e,g}}=(|\epsilon_x-\frac{\delta}{\sqrt{3}}|-n_el)$&$t^x_{e_{e,12}e_{e,g}}=t^x_{e_{e,14}e_{e,r}}=(\epsilon_z+3\frac{\delta}{\sqrt{3}}-p_el)$
& $=(|\epsilon_x-\frac{\delta}{\sqrt{3}}|-n_el)$\\& &
$t^y_{e_{e,13}e_{e,q}}=t^y_{e_{e,14}e_{e,r}}=$\\& &
$=(\epsilon_z+3\frac{\delta}{\sqrt{3}}-p_el)$
\end{tabular}
\end{center}
The values for the upper-half of the matrix $\sqrt{A}^{-1}$ are
given in the following table. As for the 4- and
6-valent graphs, the lower indices represent the matrix entries
while the upper indices indicate the edges under
consideration.
\begin{center}
\begin{tabular}{lll}\label{tab:a8t}
$(\sqrt{A}^{-1})^{5_0,6_0}_{2,1}=-\frac{1}{2}D_0\qquad $&$\qquad (\sqrt{A}^{-1})^{5_0,7_0}_{3,1}=-\frac{1}{2}D_0\qquad $&$\qquad (\sqrt{A}^{-1})^{5_0,8_0}_{4,1}=-\frac{1}{2}B_0$\\
$(\sqrt{A}^{-1})^{5_0,4_0}_{5,1}=-C_0\qquad $&$\qquad (\sqrt{A}^{-1})^{5_0,3_0}_{6,1}=-\frac{1}{2}C_0\qquad $&$\qquad (\sqrt{A}^{-1})^{5_0,1_0}_{8,1}=-\frac{1}{2}C_0$\\
$(\sqrt{A}^{-1})^{6_0,7_0}_{3,2}=-\frac{1}{2}B_0\qquad $&$\qquad (\sqrt{A}^{-1})^{6_0,8_0}_{4,2}=-\frac{1}{2}D_0\qquad $&$\qquad (\sqrt{A}^{-1})^{6_0,4_0}_{5,2}=-\frac{1}{2}C_0$\\
$(\sqrt{A}^{-1})^{6_0,2_0}_{7,2}=-\frac{1}{2}C_0\qquad $&$\qquad (\sqrt{A}^{-1})^{6_0,1_0}_{8,2}=-2C_0\qquad $&$\qquad (\sqrt{A}^{-1})^{7_0,8_0}_{4,3}=-\frac{1}{2}E_0$\\
$(\sqrt{A}^{-1})^{7_0,4_0}_{5,3}=-\frac{1}{2}C_0\qquad $&$\qquad (\sqrt{A}^{-1})^{7_0,3_0}_{6,3}=-C_0\qquad $&$\qquad (\sqrt{A}^{-1})^{7_0,2_0}_{7,3}=-C_0$\\
$(\sqrt{A}^{-1})^{8_0,3_0}_{6,4}=-\frac{1}{2}C_0\qquad $&$\qquad (\sqrt{A}^{-1})^{8_0,2_0}_{7,4}=-C_0\qquad $&$\qquad (\sqrt{A}^{-1})^{8_0,1_0}_{8,4}=-\frac{1}{2}C_0$\\
$(\sqrt{A}^{-1})^{4_0,3_0}_{6,5}=-B_0\qquad $&$\qquad (\sqrt{A}^{-1})^{4_0,2_0}_{7,5}=-\frac{1}{2}B_0\qquad $&$\qquad (\sqrt{A}^{-1})^{4_0,1_0}_{8,5}=-\frac{1}{2}F_0$\\
$(\sqrt{A}^{-1})^{3_0,2_0}_{7,6}=-A_0\qquad $&$\qquad (\sqrt{A}^{-1})^{3_0,1_0}_{8,6}=-\frac{1}{2}B_0\qquad $&$\qquad (\sqrt{A}^{-1})^{2_0,1_0}_{8,7}=-B_0$\\
$(\sqrt{A}^{-1})^{0_1,9_1}_{9,8}=-B_1\qquad $&$\qquad (\sqrt{A}^{-1})^{0_1,10_1}_{10,8}=-\frac{1}{2}B_1\qquad $&$\qquad (\sqrt{A}^{-1})^{0_1,15_1}_{11,8}=-F_1$\\
$(\sqrt{A}^{-1})^{0_1,12_1}_{13,8}=-\frac{1}{2}B_1\qquad $&$\qquad (\sqrt{A}^{-1})^{0_1,11_1}_{14,8}=-\frac{1}{2}F_1\qquad $&$\qquad (\sqrt{A}^{-1})^{0_1,13_1}_{15,8}=-\frac{1}{2}A_1$\\
$(\sqrt{A}^{-1})^{9_1,10_1}_{10,9}=-A_1\qquad $&$\qquad (\sqrt{A}^{-1})^{9_1,15_1}_{11,9}=-\frac{1}{2}B_1\qquad $&$\qquad (\sqrt{A}^{-1})^{9_1,14_1}_{12,9}=-\frac{1}{2}A_1$\\
$(\sqrt{A}^{-1})^{9_1,12_1}_{13,9}=-A_1\qquad $&$\qquad (\sqrt{A}^{-1})^{9_1,11_1}_{14,9}=-\frac{1}{2}B_1\qquad $&$\qquad (\sqrt{A}^{-1})^{1_9,d_9}_{16,9}=-\frac{1}{2}B_9$\\
$(\sqrt{A}^{-1})^{1_9,5_9}_{18,9}=-\frac{1}{2}C_9\qquad $&$\qquad (\sqrt{A}^{-1})^{1_9,6_9}_{19,9}=-\frac{1}{2}D_9\qquad $&$\qquad (\sqrt{A}^{-1})^{1_9,4_9}_{19,9}=-\frac{1}{2}F_9$\\
$(\sqrt{A}^{-1})^{1_9,b_9}_{20,9}=-\frac{1}{2}B_9\qquad $&$\qquad (\sqrt{A}^{-1})^{1_9,a_9}_{21,9}=-B_9\qquad $&$\qquad (\sqrt{A}^{-1})^{10_1,15_1}_{11,10}=-B_1$\\
$(\sqrt{A}^{-1})^{10_1,14_1}_{12,10}=-\frac{1}{2}E_1\qquad $&$\qquad (\sqrt{A}^{-1})^{10_1,12_1}_{13,10}=-\frac{1}{2}A_1\qquad $&$\qquad (\sqrt{A}^{-1})^{10_1,13_1}_{15,10}=-\frac{1}{2}B_1$\\
$(\sqrt{A}^{-1})^{15_1,14_1}_{12,11}=-\frac{1}{2}B_1\qquad $&$\qquad (\sqrt{A}^{-1})^{15_1,11_1}_{14,11}=-\frac{1}{2}A_1\qquad $&$\qquad (\sqrt{A}^{-1})^{15_1,13_1}_{15,11}=-\frac{1}{2}D_1$\\
$(\sqrt{A}^{-1})^{14_1,12_1}_{13,12}=-A_1\qquad $&$\qquad (\sqrt{A}^{-1})^{14_1,11_1}_{14,12}=-\frac{1}{2}B_1\qquad $&$\qquad (\sqrt{A}^{-1})^{14_1,13_1}_{15,12}=-B_1$\\
$(\sqrt{A}^{-1})^{12_1,11_1}_{14,13}=-B_1\qquad $&$\qquad (\sqrt{A}^{-1})^{12_1,13_1}_{15,13}=-\frac{1}{2}B_1\qquad $&$\qquad (\sqrt{A}^{-1})^{1_{12},a_{12}}_{23,13}=-\frac{1}{2}F_{12}$\\
$(\sqrt{A}^{-1})^{1_{12},b_{12}}_{24,13}=-\frac{1}{2}A_{12}\qquad $&$\qquad (\sqrt{A}^{-1})^{1_{12},4_{12}}_{25,13}=-A_{12}\qquad $&$\qquad (\sqrt{A}^{-1})^{1_{12},m_{12}}_{26,13}=-\frac{1}{2}B_{12}$\\
$(\sqrt{A}^{-1})^{1_{12},h_{12}}_{28,13}=-\frac{1}{2}A_{12}\qquad
$&$\qquad
(\sqrt{A}^{-1})^{1_{12},e_{12}}_{29,13}=-\frac{1}{2}F_{12}\qquad
$&$\qquad
(\sqrt{A}^{-1})^{11_{1},13_{1}}_{15,14}=-\frac{1}{2}F_{1}$\\
$(\sqrt{A}^{-1})^{d_{9},c_{9}}_{15,16}=-\frac{1}{2}E_{9}\qquad
$&$\qquad
(\sqrt{A}^{-1})^{d_{9},5_{9}}_{18,16}=-\frac{1}{2}B_{9}\qquad
$&$\qquad
(\sqrt{A}^{-1})^{d_{9},6_{9}}_{19,16}=-B_{9}$\\
$ (\sqrt{A}^{-1})^{d_{9},b_{9}}_{21,16}=-\frac{1}{2}C_{9}\qquad $&$\qquad (\sqrt{A}^{-1})^{d_{9},a_{9}}_{22,16}=-\frac{1}{2}E_{9}\qquad $&$\qquad(\sqrt{A}^{-1})^{c_{9},5_{9}}_{18,17}=-B_{9}$\\
$ (\sqrt{A}^{-1})^{c_{9},6_{9}}_{19,17}=-\frac{1}{2}B_{9}\qquad $&$\qquad (\sqrt{A}^{-1})^{c_{9},4_{9}}_{20,17}=-\frac{1}{2}B_{9}\qquad $&$\qquad(\sqrt{A}^{-1})^{c_{9},b_{9}}_{21,17}=-C_{9} $\\
$ (\sqrt{A}^{-1})^{c_{9},a_{9}}_{22,17}=-\frac{1}{2}C_{9}\qquad $&$\qquad(\sqrt{A}^{-1})^{5_{9},6_{9}}_{19,18}=-\frac{1}{2}D_{9}\qquad $&$\qquad (\sqrt{A}^{-1})^{5_{9},4_{9}}_{20,18}=-\frac{1}{2}D_{9}$\\
$ (\sqrt{A}^{-1})^{5_{9},b_{9}}_{21,18}=-\frac{1}{2}B_{9}\qquad $&$\qquad(\sqrt{A}^{-1})^{6_{9},4_{9}}_{20,19}=-\frac{1}{2}C_{9}\qquad $&$\qquad (\sqrt{A}^{-1})^{6_{9},a_{9}}_{22,19}=-\frac{1}{2}B_{9}$\\
$ (\sqrt{A}^{-1})^{4_{9},b_{9}}_{21,20}=-B_{9}\qquad $&$\qquad(\sqrt{A}^{-1})^{4_{9},a_{9}}_{22,20}=-\frac{1}{2}B_{9}\qquad $&$\qquad (\sqrt{A}^{-1})^{b_{9},a_{9}}_{22,21}=-\frac{1}{2}E_{9}$\\
$(\sqrt{A}^{-1})^{a_{12},b_{12}}_{24,23}=-A_{12}\qquad $&$\qquad(\sqrt{A}^{-1})^{a_{12},4_{12}}_{25,23}=-\frac{1}{2}A_{12}\qquad $&$\qquad (\sqrt{A}^{-1})^{a_{12},m_{12}}_{26,23}=-\frac{1}{2}E_{12}$\\
$(\sqrt{A}^{-1})^{a_{12},k_{12}}_{27,23}=-\frac{1}{2}A_{12}\qquad $&$\qquad(\sqrt{A}^{-1})^{a_{12},e_{12}}_{29,23}=-\frac{1}{2}B_{12}\qquad $&$\qquad (\sqrt{A}^{-1})^{b_{12},14_{12}}_{25,24}=-B_{12} $\\
$(\sqrt{A}^{-1})^{b_{12},m_{12}}_{26,24}=-\frac{1}{2}A_{12}\qquad $&$\qquad(\sqrt{A}^{-1})^{b_{12},k_{12}}_{27,24}=-C_{12}\qquad $&$\qquad (\sqrt{A}^{-1})^{b_{12},h_{12}}_{28,24}=-\frac{1}{2}B_{12}$\\
$ (\sqrt{A}^{-1})^{14_{12},k_{12}}_{27,25}=-\frac{1}{2}B_{12}\qquad $&$\qquad(\sqrt{A}^{-1})^{14_{12},h_{12}}_{28,25}=-\frac{1}{2}D_{12}\qquad $&$\qquad (\sqrt{A}^{-1})^{14_{12},e_{12}}_{29,25}=-\frac{1}{2}A_{12} $\\
$ (\sqrt{A}^{-1})^{m_{12},k_{12}}_{27,26}=-A_{12}\qquad $&$\qquad(\sqrt{A}^{-1})^{m_{12},h_{12}}_{28,26}=-\frac{1}{2}A_{12}\qquad $&$\qquad (\sqrt{A}^{-1})^{m_{12},e_{12}}_{29,26}=-\frac{1}{2}F_{12}$\\
$ (\sqrt{A}^{-1})^{k_{12},h_{12}}_{28,27}=-B_{12}\qquad $&$\qquad(\sqrt{A}^{-1})^{k_{12},e_{12}}_{29,27}=-\frac{1}{2}A_{12}\qquad $&$\qquad (\sqrt{A}^{-1})^{h_{12},e_{12}}_{29,28}=-A_{12} $\\
$ (\sqrt{A}^{-1})^{12_{e},11_{e}}_{30,29}=-B_{e}\qquad $&$\qquad(\sqrt{A}^{-1})^{12_{e},14_{e}}_{31,29}=-A_{e}\qquad $&$\qquad (\sqrt{A}^{-1})^{12_{e},13_{e}}_{32,29}=-\frac{1}{2}B_{e} $\\
$ (\sqrt{A}^{-1})^{12_{e},g_{e}}_{33,29}=-\frac{1}{2}E_{e}\qquad $&$\qquad(\sqrt{A}^{-1})^{12_{e},p_{e}}_{34,29}=-\frac{1}{2}B_{e}\qquad $&$\qquad (\sqrt{A}^{-1})^{12_{e},r_{e}}_{36,29}=-\frac{1}{2}A_{e} $\\
$(\sqrt{A}^{-1})^{11_{e},14_{e}}_{31,30}=-\frac{1}{2}B_{e}\qquad $&$\qquad(\sqrt{A}^{-1})^{11_{e},13_{e}}_{32,30}=-\frac{1}{2}F_{e}\qquad $&$\qquad (\sqrt{A}^{-1})^{11_{e},g_{e}}_{33,30}=-\frac{1}{2}B_{e} $\\
$ (\sqrt{A}^{-1})^{11_{e},p_{e}}_{34,30}=-\frac{1}{2}F_{e}\qquad
$&$\qquad(\sqrt{A}^{-1})^{11_{e},q_{e}}_{35,30}=-\frac{1}{2}A_{e}\qquad
$
&$\qquad (\sqrt{A}^{-1})^{14_{e},13_{e}}_{32,31}=-B_{e}$\\
$(\sqrt{A}^{-1})^{14_{e},g_{e}}_{33,31}=-\frac{1}{2}A_{e}\qquad $&$\qquad (\sqrt{A}^{-1})^{14_{e},q_{e}}_{35,31}=-\frac{1}{2}B_{e}\qquad $&$\qquad (\sqrt{A}^{-1})^{14_{e},r_{e}}_{36,31}=-C_{e}$\\
$(\sqrt{A}^{-1})^{13_{e},p_{e}}_{34,32}=-\frac{1}{2}A_{e}\qquad $&$\qquad (\sqrt{A}^{-1})^{13_{e},q_{e}}_{35,32}=-\frac{1}{2}D_{e}\qquad $&$\qquad (\sqrt{A}^{-1})^{13_{e},r_{e}}_{36,32}=-\frac{1}{2}B_{e}$\\
$(\sqrt{A}^{-1})^{g_{e},p_{e}}_{34,33}=-B_{e}\qquad $&$\qquad (\sqrt{A}^{-1})^{g_{e},q_{e}}_{35,33}=-\frac{1}{2}B_{e}\qquad $&$\qquad (\sqrt{A}^{-1})^{g_{e},r_{e}}_{36,33}=-A_{e}$\\
$(\sqrt{A}^{-1})^{p_{e},q_{e}}_{35,34}=-\frac{1}{2}F_{e}\qquad $&$\qquad (\sqrt{A}^{-1})^{p_{e},r_{e}}_{36,34}=-\frac{1}{2}B_{e}\qquad $&$\qquad (\sqrt{A}^{-1})^{q_{e},r_{e}}_{36,35}=-B_{e}$\\
\end{tabular}
\end{center}
where $A_i=(x_{V_i}-n_il)\frac{1}{\delta\sqrt{3}}$,
$B_i=(y_{V_i}-m_il)\frac{1}{\delta\sqrt{3}}$,
$C_i=(z_{V_i}-p_il)\frac{1}{\delta\sqrt{3}}$, $D_i=B_i+C_i$,
$E_i=A_i+C_i$, $F_i=B_i+A_i$, and the quantities $x_{V_i}$,
$y_{V_i}$, $z_{V_i}$ are the $x$-, $y$-, $z$-coordinates of the vertex
$V_i$. In all the above formulae the term
$\frac{1}{\delta_e\sqrt{3}}$ comes from the fact that
$t_{e_i}=t^x_{e_i}+t^y_{e_i}+t^z_{e_i}=\delta_e\sqrt{3}$ for all
edges $e_i$.

\end{appendix}

\end{document}